\documentclass[rmp,aps,nofootinbib,twocolumn]{revtex4}

\usepackage{amsmath}
\usepackage{amssymb}
\usepackage{graphics}
\usepackage{epsfig}

\def\inbar{\,\vrule height1.5ex width.4pt depth0pt}
\def\IR{\relax{\rm I\kern-.18em R}}
\def\IC{\relax\hbox{$\inbar\kern-.3em{\rm C}$}}

\def\q{{\bf q}}
\def\k{{\bf k}}

\def\rr{{\bf r}}
\def\R{{\bf R}}
\def\Q{{\bf Q}}
\def\S{{\bf S}}
\newcommand{\nn}{\nonumber}
\newcommand{\beqa}{\begin{eqnarray}}
\newcommand{\eeqa}{\end{eqnarray}}
\newcommand{\bk}{{\bf k}}
\newcommand{\br}{{\bf r}}
\newcommand{\BSCCO}{{Bi$_2$Sr$_2$CaCu$_2$O$_8$ }}

\newcommand{\YBCO}{{YBa$_2$Cu$_3$O$_{7+\delta}$}}

\newcommand{\YBCOtwofoureight}{{YBa$_2$Cu$_4$O$_{8}$}}

\newcommand{\bea}{\begin{eqnarray}}
 \newcommand{\eea}{\end{eqnarray}}
 \newcommand{\be}{\begin{equation}} \newcommand{\ee}{\end{equation}}

 \renewcommand{\a}{\mbox{$\alpha$}}

 \newcommand{\bfR}{\mbox{${{\bfR}}$}}

\newcommand{\s}{\sigma}

\begin{document}
\title{Defects in correlated metals and superconductors}

\author{H. Alloul}
\email{alloul@lps.u-psud.fr} \affiliation{Laboratoire de Physique
des Solides,  Univ. Paris-Sud, UMR 8502, CNRS, F-91405 Orsay, Cedex,
France}

\author{J. Bobroff}
\email{bobroff@lps.u-psud.fr} \affiliation{Laboratoire de Physique
des Solides,  Univ. Paris-Sud, UMR 8502, CNRS, F-91405 Orsay, Cedex,
France}

\author{M. Gabay}
\email{gabay@lps.u-psud.fr} \affiliation{Laboratoire de Physique
des Solides,  Univ. Paris-Sud, UMR 8502, CNRS, F-91405 Orsay, Cedex,
France}

\author{P.J. Hirschfeld}
\email{pjh@phys.ufl.edu}\affiliation{Department of Physics,
University of Florida, Gainesville, Florida 32611-8440, USA {\rm
and} Laboratoire de Physique des Solides, 
Univ. Paris-Sud, UMR 8502, CNRS, F-91405 Orsay, Cedex, France
\\~\\~\\\today \
%, draft
}

\begin{abstract}
In materials with strong local Coulomb interactions, simple
defects such as atomic substitutions strongly affect both
macroscopic and local properties of the system. A nonmagnetic
impurity, for instance, is seen to induce  magnetism nearby.
 Even without disorder,
models of such correlated systems are generally not soluble in 2
or 3 dimensions, and so few exact results are known for the
properties of such impurities.   Nevertheless, some simple
physical ideas have emerged from experiments and approximate
theories. Here, we first review what we can learn about this
problem from 1D antiferromagnetically correlated systems.  We then
discuss experiments on the high $T_c$ cuprate normal state which
probe the effect of impurities on local charge and spin degrees of
freedom, and compare with theories of single impurities in
correlated hosts, as well as phenomenological effective Kondo
descriptions.  Subsequently, we review theories of impurities in
$d$-wave superconductors including residual quasiparticle
interactions, and compare with experiments in the superconducting
state. We argue that existing data exhibit a remarkable similarity
to impurity-induced magnetism in the 1D case, implying the
importance of electronic correlations for the understanding of
these phenomena, and suggesting that impurities may provide
excellent probes of the still poorly understood ground state of
the cuprates.
\end{abstract}

%\date{May 2001}
\maketitle
\tableofcontents

\section{Introduction}

\label{sec:intro}

\subsection{Aim and scope of this article}

The study of ``strongly correlated electron systems'', materials
where the typical electronic interaction energies are of order or
larger than the kinetic energies, underwent a renaissance in the
1980's and 1990's after the discovery of superconductivity in
organic conductors, in heavy fermions and then in copper oxides.
In particular the cuprate family has had an enduring fascination
for the condensed matter community, and not simply because of the
allure of room temperature superconductivity. Most of the
important unsolved problems of quantum condensed matter physics --
metal insulator transition in low dimensions, breakdown of Fermi
liquid theory, origin and behavior of unconventional
superconductivity, quantum critical points, electronic
inhomogeneities and localization in interacting systems, and
quantum antiferromagnetism in low dimensions--are represented in
the set of questions currently under active consideration in this
lively field. Two factors are currently believed to underly the
anomalous properties of these materials: low dimensions and strong
correlations,
%({\bf comment by Marc: bare interactions need
%not be strong, but fluctuations drive these to the strongly
%correlated regime} ,
%
% I agree, but do you have a simple way of formulating this concept?
% I could not think of one.
both of which appear to be necessary to stabilize the new types of
electronic states present.

\begin{figure}[tbph]
\begin{center}
\includegraphics[width=1.\columnwidth]{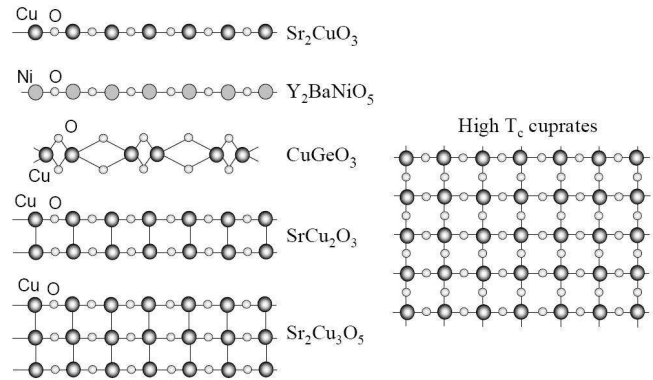}
\end{center}
\caption{A selection of actively studied inorganic oxides.  The
figure only shows the chain (Sr$_{2}$CuO$_{3}$,
Y$_{2}$BaNiO$_{5}$, CuGeO$_{3}$), ladder (SrCu$_{2}$O$_{3}$,
Sr$_{2}$Cu$_{3}$O$_{5}$) or planar (cuprates) unit in the
molecular structure.}

\label{fig:structures}
\end{figure}

 The cuprates contain
as common elements CuO$_{2}$ planes (Fig.\ref{fig:structures}),
which are sometimes considered to contain all the important
physics. Their structure is a stacking of such planes separated by
other oxide layers, which maintain charge neutrality and the
cohesion of the structure essentially through ionic interactions.
High temperature superconductivity (HTSC) has been of course the
physical property which has driven the interest for these systems
initially both for its fundamental aspects and its potential
applications.\ Another aspect responsible for their appeal has
certainly been the ease with which the carrier concentration can
be changed by chemical heterovalent substitutions or oxygen
insertion in the layers separating the CuO$_{2}$\ planes, which
play the important role of charge reservoirs.\ Electron or hole
doping can then be fairly continuously controlled from zero to
about 0.3\ charges per unit cell, which allows one to study the
evolution of the physical properties of these materials with
doping. Despite the fact that the phase diagrams of electron- or hole- doped
materials look qualitatively similar, their electronic properties in the normal and in the superconducting
states show significant differences. To date, the
hole doped family has been more extensively studied, particularly as far as the issue of impurities
is concerned and in this review we chose to focus on this type of systems.
\ In most other classes of correlated electron materials such as quasi -1D or -2D
organic compounds, 1D\ chains, or fullerides, the number of
carriers is difficult to modify, except close to some specific
compositions, where pressure must be applied to modify electronic
properties.

Impurities were used initially to test their pair-breaking
influence on superconductivity. The observation that nonmagnetic
impurities such as Zn reduced $T_{c}$ drastically triggered a
further interest in the electronic properties of impurity states
in the \ normal metallic phases as well, and led to the discovery
that even nonmagnetic impurities induce local magnetism, the
central theme of this review. In the following introductory
paragraphs, we highlight general features which characterize the
physics of correlated electron systems,  illustrating them using
the phase diagram of the cuprates. We then recall why point defect
impurities can be used as specific probes of the physical
properties of these complex materials, where understanding the
role of disorder has proven so crucial.\

\subsubsection{ Correlated electrons systems at a glance}

\begin{figure}[tbph]
\begin{center}
\includegraphics[width=0.9\columnwidth]{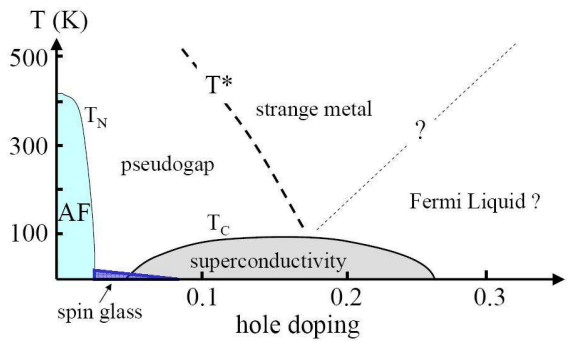}
\end{center}
\caption{Schematic phase diagram of cuprate HTSC materials:
temperature $T$ versus hole doping $n_h$.  AF refers to a state
with long-range antiferromagnetic order, which disappears at the
Neel temperature $T_N$.  Superconductivity with $d$-wave symmetry
sets in below a critical temperature $T_c$. }
\label{fig:cuprate_phase_diag}
\end{figure}

%Strongly correlated materials display very rich phase diagrams when the
%carrier concentration is changed or when the bandwidths are tuned with
%chemical pressure.
Generically at low $T$ an insulating, magnetic phase is found in
some region of the phase diagram of correlated systems, while a
metallic or superconducting state is present in a neighboring (or
in some cases overlapping) region.\ This situation can be
illustrated with the cuprate phase diagram of  Fig.
\ref{fig:cuprate_phase_diag}, which displays the different phases
versus the hole content in the CuO$_{2}$\ plane. Notice that here
zero doping corresponds to half filling, i.e.  a situation where
one hole is localized on each Cu atom, yielding
 an AF\ insulating ground state. Such a state is
usually nonmetallic when the Coulomb interaction is large enough
with respect to the bandwith.\ In some easily
compressible materials, for instance organic systems
% \textbf{%
%(ETCu2....???) }
such as the $\kappa$-(ET)$_2$CuX salts, applying pressure increases the bandwidth and
yields a Mott insulator to metal transition.\ In the case of
cuprates and some heavy fermion compounds, such as
CeRh$_{1-x}$Ir$_{x}$In$_{5}$, the AF order is suppressed by doping
and is replaced by a metallic or a superconducting state.\ These
types of metal insulator transitions \ (MIT)\ or superconductor
insulator transitions (SI) are archetypes of the problems
encountered in correlated electron physics, which are not yet
fully understood.

The phase diagram of correlated electron systems then presents
many novel states of matter, which in the case of cuprates
includes a spin glass, a pseudogap regime, a ``strange metal", a
$d$-wave superconductor and  a putative Fermi liquid, with only
the last   thought to be reasonably well understood (see Fig.
\ref{fig:cuprate_phase_diag}) theoretically. The nature of these
phases has been reviewed elsewhere and will be briefly discussed
in Section \ref{sec4}
%\ref{sec:expts_normal}
 below.   At $T=0$ the boundaries
between some of these phases may correspond to Quantum Critical
Points (QCP).
%A large interest has
%been triggered for those states which occur when quantum fluctuations govern
%the behavior of the electronic system and prohibit the establishment of a
%canonical ordered state describable by usual phase transition approaches.\
%Those Quantum Critical Points (QCP)
%might occur
In the cuprate phase diagram of Fig. \ref{fig:cuprate_phase_diag}
QCP might be present at the end points of the superconducting
dome,  or at the extensions to $T=0$ of the dashed lines shown
in Fig. \ref{fig:cuprate_phase_diag}, which have however not yet
been established as true phase transitions.
%\ For pure spin
%systems quantum fluctuations might allow to stabilize so called
%spin liquid states, a situation which can be highly favored by
%lattice frustration effects, such as in kagome, pyrochlores or
%even triangular lattices.

Much can be learned from  studies of systems with dimensions less
than two, since many theoretical results are available for 1D and
quasi-1D systems.  Furthermore, it is natural to imagine
approaching the 2D system, where few exact results are known, by
 adjoining chains to form ladders and to successively increase the
number of legs. Many physical situations encountered in one
dimension, such as the separation of spin and charge degrees of
freedom yielding the concepts of spinons and holons, which prevail
in 1D \ can be conceptually followed if one progressively
increases the dimensionality of the system from 1 to 2.\ Although
this was initially  considered  as a purely theoretical approach,
the skills of researchers in materials science have allowed the
synthesis of  systems implementing the theoretical strategy.\

Indeed, inorganic realizations of quasi-1D spin chains are
represented, e.g. by Sr$_{2}$CuO$_{3}$ (spin-1/2),
Y$_{2}$BaNiO$_{5}$ (spin-1), and CuGeO$_{3}$ (spin-1/2,
spin-Peierls). The Sr$_{2}$CuO$_{3}$ material contains CuO chains
which are similar to those found in the 2D cuprate YBCO. The
ladder compounds consisting of coupled spin-1/2 CuO chains, e.g.
 SrCu$_{2}$O$_{3}$ (2-leg)
and Sr$_{2}$Cu$_{3}$O$_{5}$ (3-leg) are natural low-D analogs of
the cuprates (see Fig. \ref{fig:structures}).  We discuss these
systems  in Sec. \ref{sec:1D}.

\subsubsection{Impurities as probes of correlated electron systems}

To gain insight into the complex problems which arise in
correlated electron systems, all  possible tools and techniques
have been extensively used and improved. We shall show below that
a specific approach, given a privileged place in this review, has
been to study the modifications of the physical properties induced
by the controlled introduction of point defects in the materials.
We shall report studies of impurity substitutions, of defects
created by electron irradiation, or of defects which occur
naturally even in the purest available materials synthesized so
far. This approach combines material science investigations with
the use of the most advanced local probe techniques.\

Impurities are known to modify  superconductivity in crucial
ways.\ In conventional $s$-wave BCS materials, disorder causes
changes in the penetration depth $\lambda \,$and in the coherence
\ length $\xi $. %, leading usually to an evolution from type I to
%type II, but it does not modify the pairing energy.\
For superconductors with non $s$-wave symmetry, very different
effects are expected since impurities break Cooper pairs.\ It is
therefore quite natural to probe the influence of defects on most
matters linked with superconductivity.\

\begin{figure}[tbph]
\begin{center}
\includegraphics[width=0.9\columnwidth]{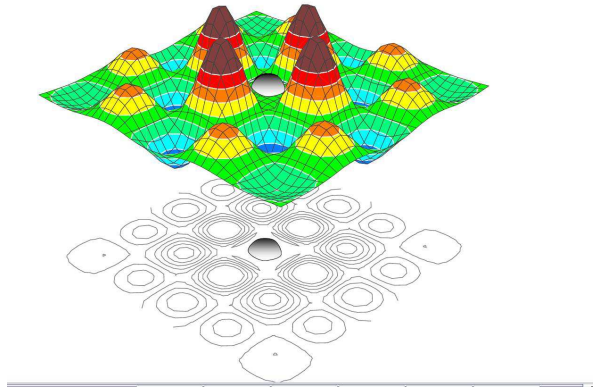}
\end{center}
\caption{
%Illustration
%of the paramagnetic RKKY oscillations
%induced by a magnetic impurity in a conventional metal (top), and
Cartoon of the staggered paramagnetic polarization induced by a
nonmagnetic defect (gray dot at center) in a correlated 2D
material. \ Such oscillations of the local susceptibilities are
directly detectable by spin probe experiments such as NMR, since
the NMR shifts of the various shells of neighbors of the defects
can be  resolved. } \label{fig:oscillations_schematic}
\end{figure}

In conventional metals, impurity effects constitute a relatively
well understood subject. The essence of the observed phenomena is
that an impurity is a local screened Coulomb potential, which
ideally is a uniform perturbation in \textbf{q} space,  inducing a
response which is inhomogeneous in real space but which reflects
the response to all \textbf{q} values.\ So if some singularity
occurs at a specific \textbf{q} value for a system, this singular
response will dominate the modifications introduced by the
impurity.\ In a classical\ metallic
system, since the response is homogeneous up to $\left| \mathbf{q}\right| =k_{F}$%
, the main detected feature comes from this truncation of the response at $%
k_{F}$, which yields the well known Friedel oscillations in the local
density of states for a charge defect, and the RKKY oscillation for a spin
defect.
%(see figure\ref{fig:oscillations_schematic}).
This idea can be
naturally generalized to any type of material.\ For instance, in magnetic
materials for which AF correlations can be important at a wave vector $%
\mathbf{q}=\mathbf{q}_{AF}$, a staggered magnetic response at this
wave vector is expected (Fig. \ref {fig:oscillations_schematic}).
\ This is exemplified in the case of an AF spin chain, in which a
staggered paramagnetic response appears near chain ends.
%(Fig. \ref {fig:oscillations_schematic}).
\ So quite
generally an impurity potential is a fundamental tool to probe the
specific response of a system.

%It is clear that this approach is all the more interesting when some
%physical properties might be hardly measurable directly in the pure system,
%or when some of the hidden physical properties can be revealed by the
%impurity potential.\ This interesting possibility has been progressively
The potential importance of this property was captured by experimentalists
%perceived from the experimental viewpoint using
who developed local probe techniques, such as NMR and STM, to map
out the spatial changes occurring around  extrinsic defects. For
instance, NMR experiments performed on the cuprates over the past
fifteen years or so, using a variety of nuclei, have developed a
fairly clear picture of how a nominally ``nonmagnetic'' impurity
creates a cloud of local staggered polarization on nearby Cu
sites, and have characterized the response as a function of doping
and temperature,  for different materials.

These %spectacular
striking results triggered a growing interest in the theory
community as well.
%Impurities in interacting Fermi systems manifest a wide range of
%startling phenomena which are quite different from their counterparts in
%weakly interacting cases. In particular, nonmagnetic ions are known to
%induce local magnetic states in strongly correlated systems, in contrast
%with the weakly interacting case.
The enormous theoretical effort devoted to solving problems of
homogeneous 2D correlated Fermi systems by approximate methods has
been occasionally applied as well to impurity problems or even to
finite disorder. While at this writing there is no generally
accepted theoretical description of an impurity in a strongly
correlated 2D system (as there is no generally accepted theory for
the homogeneous problem), a qualitative picture captured by many
different approaches has emerged and will be summarized in this
review. By contrast, the impurity problem has been quantitatively
studied  in the case of 1D systems, since exact analytical
techniques are available, and the defect's influence on transport,
 magnetism and thermal properties is known.
% It is related to boundary effects of finite size systems
%in numerical simulations treating the homogeneous problem.
The best studied and understood example of this phenomenon is the
nonmagnetic impurity in spin chain systems, which is equivalent to
a chain end in a non infinite spin chain. Here it can be shown in
a controlled fashion that a single (even weak) potential defect
induces a paramagnetic net moment and staggered magnetism in its
vicinity,
% (Fig. \ref{fig:oscillations_schematic}),
 and that the decay of this
magnetism directly measures the spin correlation functions of the
\textit{pure } state. Observations done on 2D cuprates have
%suggested
been successfully extended to quasi-1D systems where
% use of
impurities
have been used to probe magnetic
correlations in a wide variety of materials. The behavior
of these states depend on whether the pure host system manifests a
spin gap, which in turn depends on the number of coupled chains
and whether the spins are integer or half-integer.

%Let us finally point out that these
Studies of impurities in correlated electron systems are also
important to help us understand the influence of (intrinsic)
disorder that is present in nominally pure systems. As pointed out
before, in cuprate materials and in heavy fermion systems, one can
easily change the carrier concentration and thus explore a wide
range of the phase diagram using the same system. However the very
doping process introduces disorder.
%this great facility to change doping in such systems, has to be paid as a
%counterpart by all the inconveniences linked with the disorder intimately
%associated with the doping process. Although, some researchers have clearly
%pointed out early on
The importance of this phenomenon  has only been recently
appreciated  by much of the community, as local techniques such as
STM have imaged impurity states directly  and  brought to the fore
the study of disorder  as a key issue in correlated electron
systems. The efforts which have been made so far to understand the
influence of controlled disorder\ and impurities are very good
reference points to try to understand those due to the intrinsic
disorder.\ So although this is not the main aim of the present
review, we shall try hereafter to recall as well some of the
established facts which should help to separate out in the
materials problems those which might be attributed to the
intrinsic disorder.\

\subsection{Structure of the manuscript}

%I\ DID\ NOT\ MOFDIFY\ THIS SECTION\ WHCH\ SHOULD\ BE\ ADAPTED\ TO\ THE\
%ACTUAL\ CONTENTS\ OF\ THE\ VARIOUS\ SECTIONS

%Here we describe the structure of the paper.
% and, in addition to
%pedagogical aspects, the important questions in this field which
%we attempt to pose and partially answer here.
We will first give some preliminary background in Section
\ref{sec:models}, discussing the basic ideas of strongly
correlated electron physics, introducing the simplest and most
popular models like the Hubbard and $t-J$ Hamiltonians, along with
some brief words about current methods employed to solve them. We
then begin our main theme in the strange world of one dimension:
in Section \ref{sec:1D} we summarize progress which has been made
in recent years on  the conceptually simplest problem in the
field, that of an impurity in a 1D spin chain. We first attempt to
explain qualitatively the physical phenomena expected, discuss the
theoretical background for this expectation, and justify the use
of NMR to study such effects. We describe the most studied
materials and models first in their idealized homogeneous form in
terms of their physical properties (Haldane gap, quasi-long range
order). In the disorder-free limit, one finds correlated quantum ground
states, called spin liquids, and we define the concepts of
Resonating Valence Bond (RVB) phase, and of spinons or holons excitations.
 These notions will
be quite useful when we turn to the 2D case.
 We then introduce models of
defects and the concept of end-chain excitations, and show why NMR
is particularly useful for studying such effects. The discussion
is then broadened to include ladder compounds and models, which
are in certain well-defined senses steps on the road to two
dimensions. For the most part we restrict ourselves to pure spin
models, as there are few relevant experiments on doped ladder or
chain compounds.

In passing to two dimensions in Section \ref{sec4},
%\ref{sec:expts_normal},
 we
focus on the cuprate family, and first discuss the various types
of point defects which occur or can be induced in these materials.
We then attempt to make contact with what we have learned from
spin chains, by focussing on the undoped antiferromagnetic cuprate
insulators. The analogy is not completely straightforward,
however, since although the cuprate planes are sometimes regarded
as idealized 2D Heisenberg spin systems, the interplane couplings
and small spin anisotropies must be accounted for as well. In
addition, few experiments have been performed on local properties
of defects in the insulating systems, and so we pass quickly to
the underdoped pseudogap state, in the metallic regime.
 Here the suppression of the
single-particle DOS and magnetic susceptibility suggest that
impurity-induced magnetism should have a large relative effect,
and indeed the first observation of moment formation around
impurities was made in underdoped YBCO. We review these
experiments, the measurement and interpretation of the temperature
dependence of the local susceptibilities. We compare the
similarities and differences between impurities, in the cases of
Zn, nominally nonmagnetic, and Ni, nominally spin 1, and Li,
nominally nonmagnetic but charged. We review the changes in local
susceptibility found to occur when the system is doped further to
optimal doping, and discuss the behavior of the impurity
Curie-Weiss temperature, and a simple interpretation in terms of
Kondo screening. Finally, we discuss other measurements which do
not give local information but which are relevant to understanding
the physics of disorder in these materials, particularly transport
measurements.

In Section \ref{sec:thy2D} we review theoretical efforts to understand the
remarkable phenomenology just described in higher dimensions. Some of the
ideas and relevant calculations about moment formation in correlated systems
are not new, and we therefore begin by reminding the reader of early work on
nearly magnetic alloys and phosphorous doped silicon. More recent work
related to the properties of Zn and Ni in cuprates which describes these
ions as localized potentials embedded in a 2D Hubbard treated within the
random phase approximation are the intellectual heirs of these early works.
We describe the basic ideas, which involve the buildup of antiferromagnetic
correlations on the lattice due to the Hubbard repulsion, and the mode-mode
coupling induced by the impurity between this staggered response and the
homogeneous response actually measured in NMR in the normal state. These
theories provide a remarkably good description at optimal doping, but it is
not clear how to extend them to describe the effects in underdoped samples.
We then discuss several theoretical approaches which begin at strong
coupling with a $t-J$ model, and then discuss moment formation in the spin
gap state. For the most part, such theories have described spontaneous
moment formation, enhanced by the presence of a pseudogap, and do not
discuss the paramagnetic state in detail.

Section \ref{sec:SC}  tries to understand the impact of
superconductivity
%pair correlations
on the effects described above. To place the discussion in
context, we begin by reviewing briefly the basic ideas concerning the
properties of a single impurity and many-impurity interference in a $d$-wave
superconductor. We then describe how the predictions of theories described
in Section \ref{sec:thy2D} are modified in the presence of a
%superconducting
gap. Finally, we discuss experiments on impurities in the superconducting
state which directly probe local density of states and
magnetic correlations and we attempt to
compare the results to these ideas and predictions.
%
%\ DO\ WE\ KEEP\ QUESTIONS\ OR\ DO\ WE \ GIVE\ SOME\ OF\ THE\ ANSWERS\ IN\
%THE\ STRUCTURE\ OF\ THE\ MANUSCRIPT\ SECTION?
\section{Models of correlated Fermi systems }

\label{sec:models}

 To set the stage for our discussion of defects, we first give a lightning
account of the physics of homogeneous correlated systems and the
models used to describe them. We refer the reader to more
extensive reviews on the subject
\cite{PFulde:2006,PFazekas:1999}.
%\cite{PFulde:2006},\cite{PFazekas:1999}.
% Peter Fulde and Peter Thalmeier and Gertrud Zwicknagl, Solid State Physics
% Volume 60, Elsevier 2006
% Patrick Fazekas, Lecture Noteson Electron Correlation and Magnetism
% Series in Modern Condensed Matter Physics Vol 5, World Scientific 1999
 From a theoretical perspective, transport, thermal and magnetic observables can be obtained in strongly correlated systems once one has constructed a
Hamiltonian containing all the essential ingredients
characterizing these systems, yet simple enough to be amenable to
analytical or numerical computations. By and large, the model of
choice is the Hubbard model, which was introduced in the early
sixties to describe magnetism and metal-insulator transitions in
narrow band transition metals (\cite{MGutzwiller:1963,JHubbard:1963,JKanamori:1963}).
%(\cite{MGutzwiller:1963}, \cite{JHubbard:1963}).
 It consists of a kinetic term describing
the hopping dynamics of electrons (or holes) with site creation
operators ${a^{\dag}}_{i,\s}$ and a potential term accounting for
on-site Coulomb interactions. \bea\label{Hubbard1} {\cal
H}_{Hubb}=\sum_{<i,j>,\s}t_{i,j}\Bigl({a^{\dag}}_{i,\s}\;a_{j,\s}
+ h.c\Bigr) + {1\over 2}U\sum_{i,\s}n_{i,\s}n_{i,\overline{\s}}
\eea In the simplest version, the hopping amplitude $t_{i,j}$ is
limited to nearest-neighbors and $\s$ is one of the spin-$1\over
2$ eigenstates, while $\overline{\s}$ stands for the other one.
$n_i=\sum_{\s} n_{i,\s}$ fermions occupy site $i$.
The case $U>0$ corresponds to repulsive interactions, while $U<0$
mimics effective attractive interactions (mediated for
instance by electron-phonon or electron-paramagnon processes). For
the special case $U=0$ the state is either metallic or (band)
insulating, depending on the filling factor. The Hubbard model was
introduced originally to study the Mott transition which occurs
when the bandwidth of a system narrows sufficiently in the
presence of repulsive $U$.  The basic physics can be understood by
considering a lattice of atomic orbitals which can overlap to
allow conduction.  If there is one orbital per site, and one
electron per orbital, the system is a metal if $U=0$ according to
conventional electronic structure theory. Thus electrons may hop
from orbital to orbital, provided all states satisfy the Pauli
principle allowing only one $\uparrow $ and one $\downarrow$ spin
per orbital, i.e. the Hilbert space for each site consists of
$|0\rangle$, $|\uparrow\rangle$, $|\downarrow\rangle$, and
$|\uparrow\downarrow\rangle$. As $U>0$ is increased, however, the
Coulomb cost of doubly occupied sites $|\uparrow\downarrow\rangle$
increases, such that in the limit $U\rightarrow +\infty$ such
configurations are no longer allowed, or are ``projected out" of
the Hilbert space. If no doubly occupied sites are present,
 electrons are effectively prevented from hopping in the half-filled case,
and if a small number of holes are introduced one can easily see
that, although conduction can now take place, it is drastically
reduced due to the Coulomb repulsion. This is the primary effect
of electronic ``correlations"
 which will be considered here.

 More generally, if $U$ is non zero,
depending on its sign, on the size of the ratio ${U/ W}$ ($W$ is
the bandwidth) and on the filling factor one expects to promote
various
 instabilities of magnetic, metal-insulator or superconducting types, or even
phase separation. These transitions may occur at finite
temperature or at zero temperature depending on the spatial
dimensionality.
% Only a few exact results
%are available for this model.
%Most of them pertain to one-dimensional systems \cite{AMontorsi:1993} when exact
%analytical solutions can be obtained (see next section).
Only a few exact results are available for this model, mostly in
the one-dimensional case \cite{AMontorsi:1993}. In that limit,
fluctuations prevent the occurrence of ordered phases at finite
temperature. At $T=0$ and for half filling, the ground state is
insulating and has quasi-long range antiferromagnetic (AF) order
when $U>0$, or on-site singlet ordering when $U<0$
%\cite{TGiamarchi:2004}, \cite{AAuerbach:1994}.
\cite{TGiamarchi:2004,AAuerbach:1994}.
 By quasi-long range
order we mean that the expectation value of the staggered
magnetization is zero in the thermodynamic limit, but that the
corresponding coherence length is infinite (the alterning part of
the spin-spin correlation function decays as a power law at large
distances). For any other band filling,
%{\bf with thermodynamically finite density of carriers},
the chain is a non Fermi liquid paramagnetic metal, when Coulomb
interactions are repulsive,
% Marc--I don't think Nagaoka is important since we are not trying to be
%complete here
%(-- The so-called
%Nagaoka limit, when a ferromagnetic instability occurs in the
%presence of a single hole is not generic \cite{YNagaoka:1966}
%--),
or a quasi ordered s-wave superconductor when interactions are
attractive. For higher dimensions, only a handful of methods allow
analytic treatments of the Hubbard model. The simplest is the
Hartree-Fock approximation, including RPA extensions and beyond to
include fluctuation effects
%\cite{JKanamori:1963}, \cite{JHertz:1976}, \cite{JHirsch:1980},
%\cite{NEBickers:1989}, \cite{CHPao:1995}, \cite{KAryanpour:2002}.
\cite{JKanamori:1963,JHertz:1976,JHirsch:1980,NEBickers:1989,CHPao:1995,KAryanpour:2002}.
Its validity corresponds essentially to the weakly correlated
regime (${U/W}$ much less than one). When $U$ is large and
positive, one may use the Gutzwiller variational wave function to
determine the $T=0$ properties of the model
\cite{MGutzwiller:1965}, the Hubbard X-operators
\cite{SGOvchinnikov:2004},
% S.G. Ovchinnikov and V.V. Val'kov, Hubbard operators in the theory of
%strongly correlated electrons, World Scientific, 2004
 or alternatively auxiliary field methods developed for instance
by Kotliar and Ruckenstein \cite{GKotliar:1986} and by Lee and
Nagaosa
 \cite{PALee:2006},
% P.A. Lee and N. Nagaosa and X.G. Wen, Rev. Mod. Phys. 78, 17 (2006)
based on slave bosons
%\cite{SEBarnes:1976}, \cite{SEBarnes:1977},
%\cite{PColeman:1984}, \cite{NRead:1983}, \cite{NRead:1985},
%\cite{NEBickers:1987}.
 \cite{SEBarnes:1976,SEBarnes:1977,PColeman:1984,NRead:1983,NRead:1985,NEBickers:1987}.
 Note however that the Gutzwiller wave function accounts for on-site correlations
but neglect inter-site correlations, such as the Heisenberg AF interactions, which play an important role in the large $U$ limit \cite{TAKaplan:1982}.
%Marc---I didn't think this was useful:
%which allows to compute%observables at any $T$ and is such that its i
%predictions agree with
%those of the Gutzwiller variational method at $T=0$.
In the $d=\infty$ limit, many results pertaining to the Hubbard
model can be readily obtained by a diagrammatic treatment, thanks
to the fact that many terms vanish \cite{WMetzner:1989}, and
%once%again
the predictions are in agreement with those obtained by Gutzwiller
and by Kotliar-Ruckenstein.  Closely related is
% Marc-changed this:%One should also mention
dynamical mean field theory, a combination of analytic
and numerical tools allowing quantitative and accurate
determinations of metal insulator transitions
\cite{AGeorges:1996,YKakehashi:2004,BKyung:2006,GKotliar:2006}.
%\cite{AGeorges:1996},
%\cite{BKyung:2006}, \cite{GKotliar:2006}.
%Marc--I found this too vague.
% at zero and finite%temperature for various dimensions and dopings.

In the $U=+\infty$ limit, double occupancy of the sites becomes
energetically unfavorable, and carriers perform only virtual hops
onto neighboring sites. A canonical transformation then maps
Hamiltonian Eq.(\ref{Hubbard1}) onto the so-called $t-J$ model
%\cite{ABHarris:1967}, \cite{KAChao:1977},
% \cite{JHirsch:1985}
\cite{ABHarris:1967,KAChao:1977,JHirsch:1985}
%(A.B Harris and %R.V Lange Phys. Rev 157, 295, 1967 - K.A. Chao, J. Spalek and, A.M. Oles, J. Phys C10, L271, 1977 - J.E. Hirsch Phys. Rev. Lett. 54, 1317,
%1985)

\bea\label{t-J1} {\cal
H}_{t-J}&=&\sum_{<i,j>,\s}t_{i,j}\Bigl({a^{\dag}}_{i,\s}\;a_{j,\s}\;
+ h.c\Bigr )\nonumber \\&&+
J\sum_{<i,j>}\Bigl(\overrightarrow{S_i}\cdot\overrightarrow{S_j}
-{1\over4}n_i\;n_j\Bigr ) \eea Here, creation and annihilation
operators are not the canonical ones, but are projected onto the
subspace of non doubly occupied states. The exchange $J=$
${4t^2}$/$U$ is the AF coupling between the spins. In the half
filled case, the kinetic term has to be zero, and one is left with
a Heisenberg Hamiltonian for localized spins. As already
mentioned,
 in $1D$ one gets quasi-long range AF order at $T=0$. After
much debate, it is now generally agreed that the ground state is
antiferromagnetic in $2D$ at $T=0$ (fluctuations destroy the order
at finite temperatures). Away from half filling, one may treat the
kinetic term as a perturbation \cite{WFBrinkman:1970,APKampf:1994}.
 The bandwidth is strongly narrowed due to the
interaction between the carrier motion and spin dynamics, and the
energy scale is not governed by $t$, as one might expect, but
rather by $J$ \cite{JHXu:1991}. This remarkable property is also seen
in exact diagonalization of finite size clusters \cite{EDagotto:1994,PWLeung:1995}
and experiments are consistent with the theoretical prediction \cite{BOWells:1995,ADamascelli:2003}.
We note that the Hamiltonian
(\ref{t-J1}), if derived from Eq.(\ref{Hubbard1}), should also include an additional kinetic
contribution of order $t^2 / U$ involving three sites $<ijk>$
\bea\label{t-J2} {\cal H}_{3sites}&=&-2\;{t^2\over U}
\sum_{<i,j,k>,\s}\Bigl({a^{\dag}}_{i,\s}\;a_{k,\s}n_{j,\overline{\s}}\nonumber
\\&&
-{a^{\dag}}_{i,\s}\;a_{k,\overline{\s}}{a^{\dag}}_{j,\overline{\s}}\;a_{j,\s}
\;\Bigr)
 \eea
 For non zero doping and when carriers are mobile, one can rewrite
Hamiltonian Eq.(\ref{t-J2}) in terms of effective Heisenberg
interactions involving second and third neighbor spin couplings
\cite{MInui:1988}. A wide variety of phases can be obtained when
the model -- in its Hubbard or $t-J$ form -- is generalized to
include more complicated kinetic or correlation terms, or else to
include several orbitals or hybridizations between different
bands. Restricting Coulomb interactions to on-site terms is a
rather drastic approximation, and several authors have examined
the influence of an additional coupling between nearest neighbors
in Hubbard chains and ladders \cite{MFabrizio:1993,EOrignac:1997};
%\cite{MFabrizio:1993},\cite{EOrignac:1997};
%, \cite{TGiamarchi:2004};
 in addition to
Hamiltonian (\ref{Hubbard1}),
 where (in the case of ladder compounds) the hopping amplitude $t_{i,j}$
can take on different values along the chains and on the rungs,
one includes an extra Coulomb contribution:
\bea\label{nnCoulomb}{\cal H}_{nnc}= V\sum_{i,p} {n_{i}} {n_{i+p}}
\eea Here $p$ is the neighbor of $i$ on the {\it same} chain. For
chains, this term stabilizes insulating commensurate phases for
less than half filling. For ladders, it
%it favors d-wave
%superconductivity and also
 promotes an interesting orbital
antiferromagnetic order.
%, reminiscent of the loop current phase
%advocated by Varma in the pseudogap phase of the cuprates.
% (see
%below).
%\cite{BFauque:2006},
%\cite{CMVarma:2006}.
%\cite{BFauque:2006,CMVarma:2006}.
 In organic
materials, the molecular bonding is responsible for a pronounced
quasi-1D character (for instance in the Bechgaard or Fabre salt
series \cite{CBourbonnais:1999}),
 or a quasi -2D band structure character
(for instance in the BEDT series \cite{PLimelette:2003}).
 In the former case, the hopping
amplitude $t_{ij}$ in Eq.(\ref{Hubbard1}) has a value $t_a\simeq
1000$K along the conducting chains, $t_b\simeq {{t_a}/ 10}$ in one
of the directions perpendicular to the chains and $t_c\simeq
{{t_a}/100}$ in the third direction. In the latter case $t_a\simeq
t_b$.
%For some of these compounds, the chemical bonding between
%the molecules causes distortions which change the lattice
%periodicity along the chains and produce dimerization gaps. In
%terms of the electronic properties the dimerization effectively
%modifies the filling of the band. To model this effect, one adds
%Umklapp terms to Hamiltonian Eq.(\ref{Hubbard1}).
%{\bf [I don't understand the use of  ``Besides", and I am afraid general
%reader will miss the point here.]}
Since doping has not been successfully achieved for these
materials, the main tuning parameter is pressure: it is either
applied externally or internally when (halogen) atoms are
chemically substituted into the structure and this changes the ratio $t/ U$.
%increases, but also the dimerization gap closes. This changes the
%band filling (thus modifying the Umklapp terms or making them
%irrelevant), and Fermi surface nesting properties.

Heavy fermions compounds are generally discussed in the framework
of a periodic Anderson Hamiltonian which reads
\cite{GStewart:2001,PSun:2005} \bea\label{Heavyferm} {\cal
H}_{HF}&=& \sum_{i,N}{E^f}_N\;{n^f}_{i,N} +{1\over 2}U\hskip
-.3cm\sum_{i,N,N',N'\neq
N}{n^f}_{i,N}\;{n^f}_{i,N'}\nonumber\\~~~~~~ +\sum_{{\vec
k},N}&&\hskip-.5cm\epsilon_{\vec k}\; {c^{\dag}}_{{\vec
k},N}\;c_{{\vec k},N} +V\Bigl
(\sum_{i,N}{f^{\dag}}_{i,N}\;c_{i,N}\; + h.c\Bigr ) \eea The first
two terms represent the limiting form of ${\cal H}_{Hubb}$
Eq.(\ref{Hubbard1}) for f-electrons states with creation operators
${f^{\dag}}_{i,N}$, since the band is so narrow that it can be
approximated by a simple atomic level. The last term models the
hybridization of the f state on site $i$ with a wide band of
electrons with creation operators ${c^{\dag}}_{{\vec k},N}$. Their
Coulomb interactions can be neglected and the kinetic part
corresponds to the third term on the right hand side of
Eq.(\ref{Heavyferm}). Note that the summation over $N$ runs over
an {\it a priori} degenerate spin-orbit multiplet (although
crystal field effects and Hund's rule might of course reduce the
number of terms).

\begin{figure}[htbp]
\begin{center}
\includegraphics[width=1.\columnwidth]{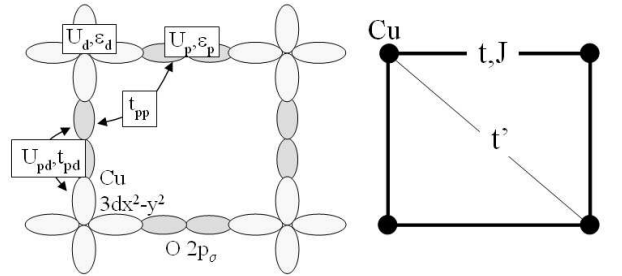}
\protect\caption{Cu and O orbitals in the copper oxide planes;
relevant site, hopping and correlation energies are included.
Integrating out the O variables yields a $t$,$t'$-$J$ model for the
effective $Cu$ lattice (right part of the picture). The notation is
defined in the text.} \label{CuO1}
\end{center}
\end{figure}

An important extension of the previous model concerns cuprate
materials. We will dwell upon this case in a little more detail,
since these compounds will serve as our preferred example for our
discussion of impurities in strongly correlated material. It is
generally accepted that modelling the electronic states in the
copper oxide planes allows one to capture the most remarkable and
relevant properties of these high temperature superconductors. In
this case, a narrow Cu d band hybridizes with a wide O p band. In
the undoped compound, the (electron) d$_{x^2-y^2}$ orbital is half
filled while the p orbitals are fully occupied and we denote their
energies  by $E_d$ and $E_p$, respectively. Since the difference
$E_d-E_p$ is small, the d$_{x^2-y^2}$, p$_x$ and $p_y$ orbitals
hybridize strongly (Fig. \ref{CuO1}) and form the basis of the
3-band model.
 The p$_z$ and d$_{3z^2 - r^2}$ wavefunctions couple to the buffer
(or reservoir) out of plane oxide layers and are usually not
included in a model of the CuO$_2$ planes.  However, they have been
shown to significantly renormalize the effective parameters in a
1-band model
 \cite{EPavarini:2001}. Conventionally, one introduces {\it hole} creation
operators ${d^{\dag}}_{i,\s}$ and ${p^{\dag}}_{j,\s}$ for the Cu
and O orbitals respectively.
 The most relevant site energies, hopping amplitudes and Coulomb $U$ are
shown in Fig. \ref{CuO1}; their values have been obtained through
computer studies
% are listed in Table (\ref{CuO2})
%\cite{MSHybertsen:1989}, \cite{AKMcMahan:1990}, \cite{HEskes:1990}, %\cite{APKampf:1994}.
\cite{MSHybertsen:1989,AKMcMahan:1990,HEskes:1990,APKampf:1994}.
Using \bea\label{paramCuO} \epsilon_d=-E_d-U_d-2U_{pd} \\
\nonumber \epsilon_p=-E_p-U_p-2U_{pd} \eea the model Hamiltonian
reads: \bea\label{H3band} {\cal
H}_{3band}&=&\epsilon_d\sum_{i,\s}{n_{i,\s}}^d
+\epsilon_p\sum_{j,\s}{n_{j,\s}}^p
\\ \nonumber && +\sum_{<i,j>,\s}{t_{i,j}}^{pd}\Bigl ({p^{\dag}}_{j,\s}\;d_{i,\s}\;
+ h.c\Bigr )   \nonumber
\\+\sum_{<j,j'>,\s}&& \hskip -.8cm{t_{j,j'}}^{pp}\Bigl
({p^{\dag}}_{j,\s}\;p_{j',\s} \; + h.c\Bigr )+
U_d\sum_i{n_{i,\uparrow}}^d{n_{i,\downarrow}}^d \nonumber
\\&&
+U_p\sum_j{n_{j,\uparrow}}^p{n_{j,\downarrow}}^p
+U_{pd}\sum_{<i,j>}{n_{i}}^d{n_{j}}^p\nonumber \eea

\begin{figure}[htbp]
\begin{center}
\includegraphics[width=1.\columnwidth]{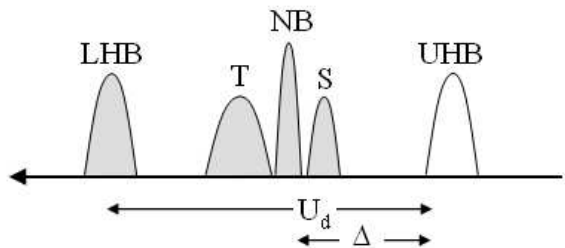}
\protect\caption{A schematic rendition of the main energy hole
states for the 3-band and t-J models. The UHB (upper Hubbard
band), NB(nonbonding band),  and LHB (lower Hubbard band) pertain
to the former while the $S$ (Zhang-Rice singlet band) and $T$
(Zhang-Rice triplet band) refer to the latter. The effective
charge transfer gap between the UHB and S bands is about 2eV. For
the undoped material pictured here, the UHB is completely filled
with holes.
%{\bf the arrow on the axis should
%be located at the opposite end, since the energy is lowest for the UHB in hole
%language)}
}
\label{CuO3}
\end{center}
\end{figure}

%\begin{table}
%  \centering
%  \begin{tabular}{|c|c|}
%    \hline
%    % after \\: \hline or \cline{col1-col2} \cline{col3-col4} ...
%    $U_d$ & $9.5\pm 1$ \\\hline
%    $U_p$ & $5\pm 1$ \\\hline
%    $U_{pd}$ & $1\pm 0.5$ \\\hline
%    $\Delta=\epsilon_p-\epsilon_d$ & $ 3.5 \pm 0.5$ \\\hline
%    $t_{pd}$ & $ 1.5\pm 0.2$ \\\hline
%    $t_{pp}$ & $0.65\pm 0.05$ \\
%    \hline
%  \end{tabular}
%\begin{figure}[htbp]
%\begin{center}
%\includegraphics[width=.9\columnwidth]{dataCuO.eps}
%\protect\caption{Table of the main parameter values (in eV)
%{\bf(MG:
% from a compilation of computer studies; the listed range
%for each parameter reflects differences in the values quoted by
%various authors
%\cite{MSHybertsen:1989,AKMcMahan:1990,HEskes:1990,APKampf:1994}.}
%\cite{MSHybertsen:1989}, \cite{AKMcMahan:1990},
%\cite{HEskes:1990},
% \cite{APKampf:1994}}
%{\bf
%It seems to me that all of these values are not universal enough
%to list without specifying the material and even, perhaps, the
%sources of the numbers.  Were they  taken from another table which
%we can cite?  Also, if we include it, perhaps we should add a
%column for $t$,$t'$, $J$,...? Or remove it completely (PH).}
%}
%\label{CuO2}
%\end{center}
%\end{figure}
%\end{table}
 The on-site and kinetic parts of ${\cal H}_{3band}$ yield three bands, as
shown in Fig. \ref{CuO1}. In the hole basis, the lowest energy
band (called antibonding in reference to electron states) is
mainly of d character with a small admixture of p. It is half
full, since one has one hole per CuO$_2$ unit. The highest
(bonding) band, mainly of p character, is empty and the charge
transfer gap between the two is $\Delta=\epsilon_p -\epsilon_d$.
An empty (nonbonding) p band is centered around $\epsilon_p$.
Since $U_d$ is very large, the splitting of the antibonding states
induced by the Coulomb energy $U_d$ pushes the 3d$^9\to$ 3d$^8$
band (again in reference to electron states) far above the bonding
band. Since $U_p$ and $ U_{pd}$ are  significantly smaller than
$U_d$, they do not affect the bonding and nonbonding bands too
strongly.
 Thus, the "upper" (3d$^9\to$ 3d$^{10}$) Hubbard band is filled with holes
 and constitutes the ground state; it is separated from the nonbonding excited
 states by a charge transfer gap of order $\Delta$ (Fig. \ref{CuO3}).
%Zhang and Rice
\cite{FCZhang:1988} proposed a further reduction of
the 3-band
 model to an effective single band model: a hole injected into the plane goes
 into an oxygen orbital since $U_d$ is more than two times larger than
$\Delta$. In fact, it lives on the four oxygens surrounding a
copper site, but its spin forms a so-called ``Zhang-Rice singlet"
with the copper spin.
 On that particular plaquette, one has a charge but no spin. Only one singlet
 can occupy a given plaquette at a time. The Zhang-Rice singlet then hops
from one plaquette to a neighboring one, carrying charge and
restoring a spin-$1\over 2$ on the first plaquette. The low energy
physics is then minimally described in terms of a $t-J$ model
Eq.(\ref{t-J1}) on a lattice (each site corresponds to the original
CuO$_2$ unit), where $t$ is of order $0.4$ eV and $J$ of order
$0.1$ eV, as shown in Fig. \ref{CuO1} \cite{APKampf:1994,EDagotto:1994}.
%\cite{APKampf:1994}, \cite{EDagotto:1994}.
The
Zhang-Rice states are shown in Fig. \ref{CuO3}.
Such a
reduction is not universally accepted, and %For instance, as we mentioned
%earlier, a model proposed by C. Varma
%%\cite{CVarma:2006}
%to explain the pseudogap phase of cuprate
%materials requires considering all three bands.
 there are many proposals which require  three or
more bands \cite{VJEmery:1987,CMVarma:2006}.  On the other hand, early NMR experiments suggested
that magnetic fluctuations on Cu and on O sites
 track each other, upon varying the temperature or the doping, lending
support to the Zhang-Rice scenario
\cite{HAlloul:1989,MTakigawa:1991}.
% \cite{HAlloul:1989}, \cite{MTakigawa:1991}.
 Still,
ARPES experiments which are used to assign values to the
$t_{ij}$ reveal that, in general,
 one needs to include at least second neighbor  Cu-Cu hoppings
%(note that the sign of the second neighbor hopping amplitude is
%negative)
\cite{MCSchabel:1998,ADamascelli:2003}.
% \cite{MCSchabel:1998}, \cite{ADamascelli:2003}.
 Neutron and Raman scattering allow one to determine $J$ and show that
nearest-neighbor couplings dominate. So a popular model used to
discuss the strongly correlated limit is the $t-t'-J$ Hamiltonian
\bea\label{t-t'-J} {\cal H}_{t-t'-J}&=&-t\sum_{<i,j>,\s}\Bigl
({c^{\dag}}_{i,\s}\;c_{j,\s}\; + h.c\Bigr ) \\
-t'\sum_{<i,i'>,\s}\Bigl({c^{\dag}}_{i,\s}\;c_{i',\s}\;\hskip
-.3cm &+ &\hskip -.2cm h.c\Bigr ) +J\sum_{<i,j>}\Bigl
(\overrightarrow{S_i}\cdot\overrightarrow{S_j} -{1\over
4}n_i\;n_j\Bigr ) \nonumber\eea where $i'$ is a second neighbor of
$i$ and where
 the ${c^{\dag}}_{i,\s}$ are hard core
operators indicating that any site can be occupied either by a
charge or by a spin, but not both.
% and where, {\bf in the hole doped case,} $t'<0$.
 The sign of $t'$ controls
the topology of the Fermi surface \cite{GBMartins:2001,ADamascelli:2003} and
when $t'<0$ one obtains a shape qualitatively similar to that seen in ARPES for the hole doped
cuprates. A negative value also lowers the probability for phase separation
(between hole rich and hole poor regions) to occur.
%Here, the ${c^{\dag}}_{i,\s}$ are hard core
%operators indicating that any site can be occupied either by a
%charge or by a spin, but not both.
 While exact
 solutions of the homogeneous strongly correlated models introduced above
 are not available in 2D, most practitioners in the correlated systems field
 agree on the qualitative aspects of their intermediate to high-energy
spectroscopic features.  In the crucial low-energy regions,
attention is currently focused on questions such as whether or
not they support superconductivity without the addition of
couplings to the third dimension, or other ingredients.  Enough
understanding has been developed in recent years to embolden some
authors to study inhomogeneous states within such models, using
the approximate techniques cited above. The great need for such
studies is driven by fascinating phenomena displayed by defects in
1D systems, as well as by a compelling series of experiments on 2D
cuprates, which we discuss below.

\section{Impurities in spin chains and ladders }
\label{sec:1D}

\subsection{Overview}
One of the biggest challenges that condensed matter physicists are
still facing today is the lack of a clear understanding of the low
energy properties of correlated two dimensional electronic
systems.
A route to the study of
correlations uses the spatial dimension as a parameter. In 1D,
generic interacting electronic systems belong to the strong
coupling limit (Luttinger or Luther-Emery)
\cite{JLuttinger:1963,ALuther:1974,ALuther:1976}.
% \cite{JLuttinger:1963},
%\cite{ALuther:1974}, \cite{ALuther:1976}.
 In 3D, weak coupling
physics appears to describe electronic systems well.
%Again
Two dimensional systems  stand at a dimensional crossroads between
the strongly and weakly correlated limits. Since the scope of this
review is focused on strongly correlated systems, we will work our
way from 1D to 2D: we recall key features of 1D physics and,
adding chains in parallel, forming ladders, we gradually construct
planar systems.
 This approach will help us gain insight into the
physics of the 2D case, providing properties evolve smoothly as a
function of the interchain couplings.

\subsection{The 1D world}\label{1Dworld}

One of the great advantages of considering the 1D case is that
many exact or quasi-exact results can be obtained in that limit,
using different techniques. Bethe-Ansatz or transfer matrix
%(Cardy Les Houches 88, den Nijs Domb Leibowitz)
%\cite{JCardy:1988},\cite{MDenNijs:1988}
allow  us to solve such
seminal
 models
as the XXZ chain or the Hubbard model,
%(Lieb-Wu68)
%\cite{ELieb:1968}
revealing the nature of the ground state and of the elementary
excitations. They allow  us to compute thermodynamic properties,
and in some instances correlations functions
%\cite{JCardy:1988}, \cite{MDenNijs:1988}, \cite{ELieb:1968}.
\cite{JCardy:1988,MDenNijs:1988,ELieb:1968}.
Bosonization or mapping to non-linear sigma models are methods
which give access to the low energy properties of 1D quantum
problems, when the energy spectrum is linearized in the vicinity
of the Fermi points \cite{TGiamarchi:2004,DCabra:2004};
% \cite{TGiamarchi:2004},\cite{DCabra:2004};
combined with conformal theory, these two methods allow  one  to
identify universality classes, to compute  critical  indices and
correlations functions accurately, and to give expressions for the
corrections to scaling due to boundary conditions or to thermal
effects. Functional renormalization group gives access to multiple energy scales
and is thus able to capture important features when impurities are present \cite{VMeden:2002,SAndergassen:2004}.
%In addition, conformal theory shows that bosons and
%fermions are two representations of the same algebra and thus
%explains why one can express a fermion in terms of bosonic
%operators in 1D.
In conjunction with the above analytical methods,
a number of powerful numerical techniques have been developed:
Density Matrix Renormalization Group (DMRG)
%(Schollw\"oek Rev. Mod. Phys2005)
%\cite{USchollwoeck:2005}
 or exact diagonalization
%\cite{EDavidson:1993}
provide information on the ground state
properties and on correlations of ordered or quasi-ordered phases.
Quantum Monte-Carlo (QMC)
% \cite{JGrotendorst:2002}
yields the
temperature dependence of various observables and correlation
functions, but, for fermionic variables, the infamous "sign
problem"
%does not allow to simulate
 prevents one from simulating temperatures much less than the
bare kinetic energy
 \cite{USchollwoeck:2005,EDavidson:1993,JGrotendorst:2002}.
%\cite{USchollwoeck:2005}, \cite{EDavidson:1993},
% \cite{JGrotendorst:2002}.
>From these studies, one may identify two generic types of systems:
those which fall into the Luttinger universality class and those
which belong to the Luther-Emery class. The former case
corresponds to a critical regime where correlation functions decay
as power laws of the distance; the system is compressible and
excitations are gapless. The latter case pertains to a gapped
regime; some or all of the excitations are gapped (incompressible
state) and the corresponding correlation functions decay
exponentially with distance. Many of the concepts that apply to 1D
systems and that one may hope to extend to higher dimensions can
be expressed fairly simply in the language of bosonization. We
restrict our presentation to the material that is necessary to
that end and we further refer the reader to some excellent reviews
on one dimensional problems
 \cite{IAffleck:1988,DMattis:1993,JVoit:1995,ATsvelik:1995,AGogolin:1999,TGiamarchi:2004,QM:2004,PLecheminant:2004}.
% \cite{IAffleck:1988}, \cite{DMattis:1993},
%\cite{JVoit:1995}, \cite{ATsvelik:1995}, \cite{AGogolin:1999},
%\cite{TGiamarchi:2004}, \cite{QM:2004}, \cite{PLecheminant:2004}.
%(ex Mattis, Voit, Affleck, Giamarchi, Tsvelik, Gogolin,
% Quantum Magnetism, U. Schollw\"ock, J. Richter, D.J.J Farnell,
%R.F Bishop Eds. Lect. Notes Phys. 645
%(Springer, Berlin, Heidelberg 2004) Chaps. 1-6-8).
 In the case of
a single band fermionic problem, low energy modes correspond to
excitations close to the Fermi surface which consists of two
points. One then defines (real space) anihilation operators with
spin $\sigma$ ($\sigma = \uparrow, \downarrow$),
$\psi_{r,\sigma}(x)$, for right moving fermions whose momentum is
close to $+k_F$ ($r=+1$) and for left moving fermions whose
momentum is close to $-k_F$ ($r=-1$).
%label
%{\bf momentum index?} $r$ ($r=+1$ for fermions near $+k_F$ and
%$r=-1$ for fermions near $-k_F$)
\bea\label{anihilop1_sigma}
\psi_{r,\sigma}(x)={{1}\over{\sqrt{2\pi\alpha}}}\eta_{r,\sigma}e^{irk_Fx}~
e^{-i[r\phi_{\sigma}(x)-\theta_{\sigma}(x)]} \eea
%{\bf (I changed exp to e.)}
 The bosonic phase operators $\phi_{\sigma}(x)$ and
$\Pi(x)={1\over{\pi}}\nabla\theta_{\sigma}(x)$ are canonically
conjugated.
%In
%the above expression, we have not written the usual Klein factor which allows
%for a change in the total number of fermions. This term is not needed if
%We include the zero modes in the definition of $\phi_{\sigma}(x)$
%and of $\theta_{\sigma}(x)$.
The Klein factors (Majorana fermions)
$\eta_{r,\sigma}$ ensure proper anticommutation of the $\psi$'s
with different $\sigma$ or $r$. In Eq.(\ref{anihilop1_sigma}) only
the lowest harmonics ($+k_F$ and $-k_F$) are included.  To study
commensurability effects or the influence of a non linear
energy-momentum dispersion one would include higher harmonics.
%(Pham, Giamarchi)
%\cite{KPham:2000}, \cite{TGiamarchi:2004}.
 At this stage let us
highlight some similarities and differences between 1D and 2D:

${\mathbf \bullet}$ Interaction terms in the Hamiltonian mix low
energy fluctuations associated with topologically distinct
portions of the Fermi surface ($+k_F$ and $-k_F$). These backward
scattering contributions are often relevant, i.e affect
significantly the thermodynamic and transport properties of the
system.
%change the nature of the fixed
%point flow.

${\mathbf \bullet}$ Introducing the  particle density operator
$\widehat{\rho}_{\sigma}(x)=({k_F}-\nabla\phi_{\sigma}(x))/\pi$,
one may rewrite Eq.(\ref{anihilop1_sigma}) as
\be\label{Chern-Simons}
\psi_{r,\sigma}(x)={{1}\over{\sqrt{2\pi\alpha}}}\eta_{r,\sigma}
e^{i[r\int{dy\;{\widehat{\rho}_{\sigma}(y)}}-\theta_{\sigma}(x)]}
\ee This form is the 1D analog of the expression for fermion
operators used in the context of the 2D fractional quantum Hall
effect \cite{EFradkin:1991,KPham:2000}.
%\cite{EFradkin:1991}, \cite{KPham:2000}.
%( Fradkin, Pham).
 The "Jordan- Wigner" term
\bea\label{ChernSimons}
e^{i[r\int{dy\;{\widehat{\rho}_{\sigma}(y)}}]}\eea is equivalent to the
topological Chern-Simons
 term. This analogy is made more precise in \cite{KPham:2000}
%(Pham)
 which shows that
the eigenfunctions of the Luttinger liquid are 1D Laughlin
wavefunctions \cite{ASeidel:2006}.

${\mathbf \bullet}$  If one defines charge (c) and spin (s) phase
operators for the $\phi$ fields \be
\phi_c(x)={1\over\sqrt{2}}[\phi_{\uparrow}(x)+\phi_{\downarrow}(x)]_;\;\;
\phi_s(x)={1\over\sqrt{2}}[\phi_{\uparrow}(x)-\phi_{\downarrow}(x)]
\ee and similarly for the $\theta$ fields,
%one may rewrite
%Eq.(\ref{anihilop1_sigma}) as \bea\label{anihilopchargespin}
%\psi_{r,\sigma}(x)&=&{{1}\over{\sqrt{2\pi\alpha}}}\eta_{r,\sigma}e^{ik_Fx}
%\\
%&&\times
%e^{{{-i}\over{\sqrt{2}}}[r\phi_c(x)-\theta_c(x)+\sigma(r\phi_{s}(x)-
%\theta_{s}(x))]} \nn\eea
%{\bf (should the next sentence follow
%after Eq. \ref{Hspin} since one needs to establish additivity of
%the Hamiltonian?)} Since the charge and spin phase fields commute
%we see that we have spin-charge separation.
one finds that the bosonized form of the low-energy Hamiltonian of
the interacting fermionic system is simply the sum of two terms,
one describing charge fluctuations

\bea\label{Hcharge}
H_c&=&{1\over{2\pi}}u_c\int_0^L{dx\;\Bigl( K_c[\pi\Pi_c]^2+{{[\nabla
\phi_c(x)]^2}\over{K_c}}}\Bigr)\nn\\
&&+{{2g_3}\over{(2\pi\alpha)^2}}\int_0^L{dx\;\cos{(\sqrt{8}\phi_c(x)-4k_F\;x)}}
\eea and one describing spin fluctuations

\bea\label{Hspin} H_s
&=&{1\over{2\pi}}u_s\int_0^L{dx\;\Bigl( K_s[\pi\Pi_s]^2+{{[\nabla
\phi_s(x)]^2}\over{K_s}}}\Bigr) \nn\\
&&+{{2g_1}\over{(2\pi\alpha)^2}}\int_0^L{dx\;\cos{\sqrt{8}\phi_s(x)}}.
\eea

%{\bf (should the next sentence follow
%after Eq. \ref{Hspin} since one needs to establish additivity of
%the Hamiltonian?)}
Since the charge and spin phase fields commute, we see that we
have spin-charge separation. Furthermore, one may rewrite
Eq.(\ref{anihilop1_sigma}) as \bea\label{anihilopchargespin}
\psi_{r,\sigma}(x)&=&{{1}\over{\sqrt{2\pi\alpha}}}\eta_{r,\sigma}e^{ik_Fx}
\\
&&\times
e^{{{-i}\over{\sqrt{2}}}[r\phi_c(x)-\theta_c(x)+\sigma(r\phi_{s}(x)-
\theta_{s}(x))]} .\nn\eea The fermion operator
Eq.(\ref{anihilopchargespin}) is the product of a charge field
times a spin field. In the Luttinger regime, if the spin part has
SU(2) symmetry, one may identify the spin field with the spinon
operator,  such that applying it to the ground state wavefunction
creates the elementary quantum of spin excitation.
%{\bf (define ``spinon" and
%``holon")}
By contrast, the charge part is not a simple holon operator
(applying the holon operator to the ground state wavefunction
would create the elementary quantum of charge excitation).
%\cite{KPham:2000}.

%(Pham).
% This feature is reminiscent of
%the slave-boson decomposition of the electron in 2D,
%$c_{\sigma}(x)=b^{\dagger}(x)f_{\sigma}(x)$, but we have to keep
%in mind
%notwithstanding
%the fact that the pseudo boson (b) and the pseudo fermion (f)
%operators are not independent (they satisfy local constraints).
 This feature is reminiscent of
the slave-boson ($c_{\sigma}(x)=b^{\dagger}(x)f_{\sigma}(x)$)
 or slave-fermion ($c_{\sigma}(x)=f^{\dagger}(x)b_{\sigma}(x)$) decomposition of the electron in 2D,
 but we have to keep
in mind
%notwithstanding
the fact that the pseudo boson (b) and the pseudo fermion (f)
operators are not independent (they satisfy local constraints) so that
these representations do not necessarily imply spin-charge separation.
Let us emphasize that spin-charge separation in 1D is mostly an
asymptotic property. For instance, the Bethe ansatz solution of
the Hubbard model shows that it only holds in the low energy
limit, except when $U$ goes to infinity (the t-J model) when it
applies at all energy scales \cite{MOgata:1990}.
%(Ogata, Shiba).
 In
the low energy sector, we can discuss magnetic and (charge)
transport properties separately and in the folowing we concentrate
on the former aspect.
\subsection{1D physics in the spin sector}\label{spinsector}

{\it 1. Spin-$1\over 2$ systems.} In the presence of a
commensurate potential that pins one charge per lattice site, the
spin Hamiltonian Eq.(\ref{Hspin}) can be mapped onto the low
energy representation of the XXZ spin-${1\over 2}$ chain,
\bea\label{HXXZ} H_{XXZ}=J\sum_i {S_i^xS_{i+1}^x+S_i^yS_{i+1}^y
+\Delta S_i^zS_{i+1}^z} \eea This is done using a gauge
transformation that changes $S^x$ and $S^y$ to $-S^x$ and $-S^y$
every other site. Since $2k_F={\pi}$,
%$2k_F={\pi\over a}$,
the $g_1$ term (see Eq. (\ref{Hspin}))
%becomes
corresponds to an Umklapp term $g_3=-J\Delta$ as in Eq.
(\ref{Hcharge}) (we set the lattice spacing to unity).
%$g_3=-J\Delta\;a$.
 Keeping the
$2k_F$ harmonics, the bosonized expression of the $S^z(x)$
operator at point $x$ reads (see for instance \cite{TGiamarchi:2004})
%\bea\label{Sz}
%S^z(x)=-{1\over{\sqrt{\pi}}}\nabla\phi +
%{{(-1)^x}\over{\pi\a}}\cos{\sqrt{4\pi}\phi} \eea
\bea\label{Sz}
S^z(x)=-{1\over{\pi}}\nabla\phi +
{{(-1)^x}\over{\pi\a}}\cos{2\phi} \eea

 The first piece
is the uniform ($q\sim 0$) part while the second piece is the
staggered ($q\sim {\pi}$)
%($q\sim {\pi\over a}$)
part. The isotropic limit
corresponds to $\Delta=1$ where one recovers the Heisenberg model.
For translationally and rotationally invariant spin-${1\over 2}$
systems with
%"short enough"
 sufficiently short range antiferromagnetic interactions, the
Marshall and Lieb-Schultz-Mattis (LSM)
% \cite{ELieb:1961}
 theorems
state that the ground state has total $S^z=0$ and is either non
degenerate with gapless excitations of odd parity, or has
degenerate ground states and breaks parity (reflection about a site),
with a gap to the first excited triplet state \cite{ELieb:1961}.
Clearly, the Heisenberg model falls into the former category and
represents the paradigm of spin liquids for spin-${1\over 2}$
spins (Haldane and Affleck generalized the previous theorems to
arbitrary spins, see below). Elementary excitations of this spin
liquid consist of pairs of (asymptotically free) spin-${1\over 2}$
spinons \cite{HBethe:1931,JdesCloizeaux:1962}.
%\cite{HBethe:1931}, \cite{JdesCloizeaux:1962}.
% (Bethe ansatz,
%Descloizeaux- Pearson, Pham).
Spin rotational symmetry implies
 $K_s=1$; the $g_1$
term then corresponds to a marginal operator which gives
logarithmic corrections to the staggered part of the correlation
functions
%Eq.(\ref{correlXXZ})
%{\bf (Is this correct eqn.?  Why cite them here when they are so
%far away in text?).}
 \cite{JKosterlitz:1974},
%\cite{TGiamarchi:2004}
%(Kosterlitz 74, Giamarchi p131)
\bea\label{correlXXX} \langle S^z(x,0)S^z(0,0)\rangle =C_1{1\over
x^2}+C_2(-1)^x {\Bigl( {1\over x} \Bigr)}\log{x}^{1/2}. \eea These
correlations can also be calculated at finite temperature, and
they allow  one to compute the uniform susceptibility per atomic
site, $\chi$ \cite{JBonner:1964,SEggert:1994}.
%\cite{SEggert:1994}, \cite{JBonner:1964}.
%(Eggert, Affleck,
%Takahashi 94, Bonner-Fisher 64).
At low T, setting $k_B=1$ and in units of ${{(g\mu_B)^2}/J}$ one
finds that
%CITER AUSSI BONNER FISHER ET DIRE QUE L'EXPRESSION SUIVANTE N'EST OK
%QU'A BASSE T
\bea\label{susceptXXX} \chi(T)\sim
{1\over\pi^2}(1+{1\over{2\log{{{7.7J}\over T}}}}). \eea
%{\bf ( Need to define
%$\chi$ so that reader can  get dimensions right.  No obvious point
%to include $\pi^2$ in formula).}
The correlation length is given by
\cite{KNomura:1991,MGreven:1996}
%\cite{KNomura:1991},\cite{MGreven:1996}
%(Nomura, Yamada 91, Greven 96)
 \bea\label{ksiXXX} \xi_{1D}\sim {1\over
T}\Bigl({1\over{2-[\log{({J\over{.3733 T}})}]^{-1}}}\Bigr). \eea

Very few experimental realizations of a
truly 1D Heisenberg $%
S=1/2$ chain have been achieved up to now. The best known example
is the Cu chain Sr$_{2}$CuO$_{3}$
% \cite{AKeren:1993} \cite{TAmi:1995} \cite{NMotoyama:1996}.
\cite{AKeren:1993,TAmi:1995,NMotoyama:1996}.
 Magnetic susceptibility measurements
by magnetization techniques
% \cite{NMotoyama:1996}
and by NMR
%\cite{KRThurber:2001}
 show good agreement with Eq.(\ref{susceptXXX}) at low temperatures
 and with a Bonner-Fisher behavior over the whole temperature range
%\cite{NMotoyama:1996} \cite{KRThurber:2001}.
\cite{NMotoyama:1996,KRThurber:2001}.
 The
$T$ dependence allows one to extract an antiferromagnetic coupling
between adjacent Cu sites of $J=2200\pm 200K$. Long range order
is observed only below $T_{N}=5K$,  confirming the strongly one
dimensional character of the chain. As will be shown in the
following section, impurity effects enable one to determine
correlation lengths compatible with Eq.(\ref{ksiXXX}).
The rotational symmetry of the XXZ spin-$1\over 2$ chain (Eq.(\ref{HXXZ})) is broken when $\Delta\ne 1$ in the expression
for $H_{XXZ}$. For $\Delta>1$ a long range antiferromagnetic state
is obtained in the $z$ direction.
% {\bf (for $J_2$, $\delta$=0?)}.
For, $|\Delta|<1$, the XXZ chain is in the Luttinger liquid
regime. Bethe ansatz methods allow one to obtain exact expressions
for the spin velocity \bea\label{velocity} v_s={J\over
2}{K_s\over{K_s-1}}\sin{\pi(1-{1\over K_s})} \eea
 and the
compressibility $K_s={\pi/{{\arccos{(-\Delta)}}}}$ for all values
of $\Delta$ \cite{ALuther:1975,DHaldane:1980}.
%\cite{ALuther:1975}, \cite{DHaldane:1980}.
%(Luther-Peschel 75, Haldane 80).
Equal time spin-spin correlation functions decay as power laws at
large distances $x$, implying a critical behavior and infinite
correlation length.
%\bea\label{correlXXZ}
%<S^z(x,0)S^z(0,0)>&=&C_1{1\over x^2}+C_2(-1)^x{\Bigl( {1\over x}
%\Bigr)}^{K_s}\\ \nonumber <S^+(x,0)S^-(0,0)>&=&C_3{\Bigl( {1\over
%x}\Bigr)}^{K_s+{1\over K_s}}+C_4(- 1)^x{\Bigl( {1\over x}
%\Bigr)}^{{1\over K_s}} \eea

%Another type of anisotropy that breaks rotational invariance is
%the Dzyaloshinskii-Moriya term. Adding $\sum_i \vec D\cdot(\vec
%S_i\times\vec S_{i+1})$ to the Heisenberg Hamiltonian,
%{\bf where $\vec D$ is an arbitrary vector},
% one can show that this model can
%be mapped onto an XXZ chain with $\Delta<1$ once an incommensurate
%spin rotation is performed in the ${xy}$ plane.
 \vskip .2cm

 {\it 2. Spin-$1$ systems.} The situation here
is qualitatively different from that of the previous section.
Based on the study of the low energy physics of the spin 1
Heisenberg chain, Haldane \cite{DHaldane:1983}
%(Haldane 83)
showed that the ground state of rotationally and translationally
invariant integer spin systems is a non degenerate singlet and
that excitations to the first odd-parity state (triplet) are
gapped. He established these properties using a non linear sigma
model which showed that the difference with the half-integer case
stems from the presence of a Berry phase in the latter case
%\cite{DHaldane:1983}, \cite{IAffleck:1989}.
\cite{DHaldane:1983,IAffleck:1989}.
% (Haldane 83, Affleck 89).
 This Berry phase is the generalization of the 1D Chern-Simons term (see Eq.(\ref{ChernSimons}) that we
wrote down in the introduction to Section(\ref{1Dworld}).
Correlation functions are given by \bea\label{correl_spin1}
<\vec{S}(x,0)\cdot\vec{S}(0,0)>\sim (-1)^x{{\exp{- |x|/\xi}}\over
{\sqrt{|x|}}} \eea Here $\xi= v_s/\Delta_{Hal}$ where
$\Delta_{Hal}=0.4105\;J$ and, as $T\to 0$, \be\label{xi_Hal} \xi\sim 6\ee
%\be\label{xi_Hal} \xi\sim 6\;a \ee
 \cite{IAffleck:1989,OGolinelli:1994}.
% \cite{IAffleck:1989}, \cite{OGolinelli:1994}.
%(Affleck 89, Golinelli 94).

 Accordingly, the susceptibility varies as
\bea\label{susc_S1} \chi(T)\sim
{1\over\sqrt{T}}exp{-{\Delta_{Hal}\over T}} \eea
\cite{ATsvelik:1987,KDamle:1998,YKim:1998}.
%\cite{ATsvelik:1987}, \cite{KDamle:1998}, \cite{YKim:1998}.
% (Tsvelik JETP 66 221 87, Damle-Sachdev 98, Kim 98)

Experimental realizations of the $S=1$ antiferromagnetic chain
could be obtained in the Ni chains
[Ni(C$_{2}$H$_{8}$N$_{2}$)$_{2}$(NO$_{2}$)]CIO$_{4}$ (NENP)
 and
YBa$_{2}$NiO$_{5}$ \cite{DJButtrey:1990}. In both cases, a gap
could be observed, confirming Haldane's conjecture
 \cite{JPRenard:1987,JPRenard:1988,JDarriet:1993,BBattlog:1994,TSakaguchi:1996,GXu:1996}.
% \cite{JPRenard:1987} \cite{JPRenard:1988}
%\cite{JDarriet:1993} \cite{BBattlog:1994} \cite{TSakaguchi:1996}
%\cite{GXu:1996}.
% confirming Haldane's conjecture.
In YBa$_{2}$NiO$_{5}$,
the large intrachain coupling $J=280K$%
, the small single ion anisotropy and the small ratio between
interchain and intrachain coupling ($J^{\prime }/J<5$ $10^{-4}$)
gives this compound a very good  1D Heisenberg chain character,
with a Haldane gap $\Delta =100K$
%\cite{JDarriet:1993}\cite{BBattlog:1994}\cite{JDas:2004}.
\cite{JDarriet:1993,BBattlog:1994,JDas:2004}. In both
componds, the temperature dependence of the susceptibility follows
that given by Eq.(\ref{susc_S1}).

\begin{figure}[htbp]
      \begin{center}
\includegraphics[width=\columnwidth]{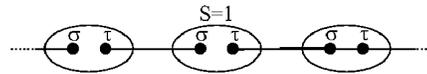}
      \protect\caption{Artist's conception of a valence bond state.
     Each spin-1 $\vec S$ is written as the sum of two spin-${1\over 2}$
      states, $\vec \sigma +\vec \tau$; valence bonds are formed between a $\sigma$
      and a $\tau$ belonging to neighboring sites.
%      {\bf Add more description--what is what?}
       }
        \label{VBS3}
         \end{center}
          \end{figure}

 Haldane's conjecture received convincing theoretical confirmation via the mapping of the
spin Hamiltonian \bea\label{Hs1}
H_{s1}=J\sum_i{S_i^xS_{i+1}^x+S_i^yS_{i+1}^y +\Delta
S_i^zS_{i+1}^z}+{D\over J}\Bigl(S_i^z\Bigr)^2 \eea onto a classical 2D
restricted solid on solid model \cite{MdenNijs:1989}.
%(den Nijs, Rommelse 89).
 These
authors showed that the massive phase can be understood in terms
of the ordering of a Jordan-Wigner like order parameter. The
nature of the ground state and of the excitations were revealed
thanks to the bosonization analysis of this model
\cite{HSchulz:1986}.
%(Schulz 86).
 Each spin is written as the sum
of two spin-1/2, $\vec{S}_i=\vec{\sigma}_i+\vec{\tau}_i$
 (Fig. \ref{VBS3}).
Insofar as singlet states on any given site are not relevant in
the low energy sector, $H_{s1}$ is now the sum of two coupled
spin-$1\over 2$ systems. The most relevant coupling is of the form
\bea\label{Josephson} g\int{dx\cos{(2\theta_o(x))}} \eea where
$\theta_0$ is an "optical mode"- like phase field, related to an
antisymmetric combination of $\theta$ operators (see
Eq.(\ref{anihilop1_sigma})). Since the renormalized $g$ flows to
minus infinity,
%\cite{HSchulz:1986}
$\theta_0$ has to be either $0$
or $\pi$ in order to minimize the energy in the ground state.
 In the context of bosonization, the $\theta$ field characterizes
a superfluid phase. When it acquires a finite expectation value, a
Josephson superfluid current flows across the system. Thus we dub
the term Eq.(\ref{Josephson}) a Josephson coupling. In spin
language this term comes from kinetic contributions of the form
$\sigma_i^+\tau_{i+1}^-$. The long range ordering of this quantity
is responsible for a spin gap proportional to $g$. It signals the
formation of an RVB phase, i.e a non degenerate, rotationally and
translationally invariant valence bond gapped phase.
%     \begin{figure}[htbp]
%      \begin{center}
%\includegraphics[width=1.\columnwidth]{Bobroff_1.eps}
%      \protect\caption{Decomposition of a spin-1 into two spin-$1/2$ forming
%an RVB (valence bond) state (Bobroff1) }
%        \label{VBS1}
%         \end{center}
%          \end{figure}

%\begin{figure}[htbp]
%      \begin{center}
%\includegraphics[width=\columnwidth]{Bobroff_2.eps}
%      \protect\caption{Decomposition of a spin-1 into two spin-$1/2$ forming
%an RVB (valence bond) state (Bobroff2)
%       }
%        \label{VBS2}
%         \end{center}
%          \end{figure}
\begin{figure}[htbp]
      \begin{center}
\includegraphics[width=0.9\columnwidth]{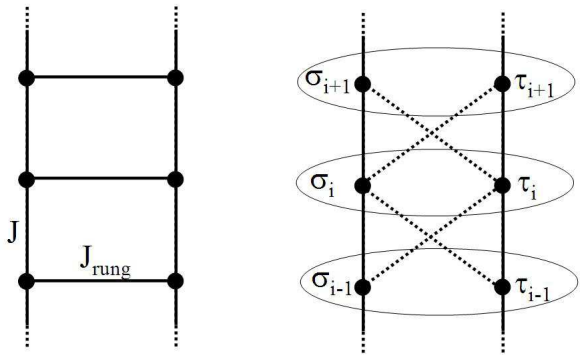}
      \protect\caption{
      Left: a two-leg spin-1/2 ladder. Dots represent the spins
with couplings $J$ along the legs and $J_{rung}$ along the rungs.
Right: a spin-$1$ chain can be viewed as a two-leg  spin-1/2
ladder with diagonal couplings (dotted lines)
       }
        \label{ladder-spin1}
         \end{center}
          \end{figure}

 This term is the counterpart of the spinon hopping
amplitude order parameter introduced in the framework of the 2D
slave-boson representation of the $t-J$ model
\cite{GKotliar:1988,PLee:1992}.
%\cite{GKotliar:1988}, \cite{PLee:1992}.
%(Nagaosa-Lee 92,Kotliar 88...).
Excitations break the valence bond singlets promoting those to
the triplet state. This picture emerges from the exact solution of
the AKLT model \cite{IAffleck:1987}.
%(Affleck,Kennedy,Lieb, Tasaki87).
It is the spin-1 Heisenberg model with an additional term
${1\over3}J\sum_i\Bigl(\vec{S}_i\cdot\vec{S}_{i+1}\Bigr)^2 $.
 Valence bonds are formed between nearest neighbor sites and spin-spin
correlation
 functions decay exponentially with distance, with a correlation length
$\xi_{AKLT}=1/\log{3}$ \cite{DArovas:1988,DArovas:1991}.
%\cite{DArovas:1988}, \cite{DArovas:1991}.
%(Arovas 88, 91).
 The inclusion of the
additional term does not affect the low energy physics of the
Heisenberg model.
 These particular valence bond states are sometimes considered to be characteristic of a bosonic RVB phase \cite{DArovas:1988,TNg:1992} and to be relevant to the physics of cuprate or other strongly correlated materials.
 It is noteworthy that in the $\sigma$-$\tau$ language, the Heisenberg
Hamiltonian can be viewed as a two-leg ladder model, with
intrachain and interchain couplings (Fig. \ref{ladder-spin1}).

%{\bf I have nothing against whimsical section headings, but I
%don't understand ``Up, up and away"} \vskip .2cm
%{\it Up, Up, and away}
{\it 3. Higher spin states.} Affleck presented a physicist's proof
extending the ground state and excitations properties of the
spin-$1\over 2$ case to general half-integer  spins
\cite{IAffleck:1989}.
%(Affleck 89).
Haldane also predicted that general integer spins fall into the
same class as the spin-1 case. These characteristics are difficult
to prove rigorously, for want of exact solutions for general spin
$S$, but one can use the decomposition of $S$ into ${2S}$
%spin-$1\over 2$
half-integer spins for integer $S$
%spins
and ${2(S-1/2)} +1$
%spin-$1 \over 2$
 half-integer spins for
%the
half integer $S$
%case
\cite{HSchulz:1986}.
%(Schulz 86).
Forming valence bonds between different "$\sigma$" and "$\tau$"
%{\bf ( note changes)}
 species on
neighboring sites, one is left with a "single" spin-$1\over 2$
chain for the half-integer case. This gives an overall gapless
regime, with massless spinons. For the integer spin case, on can
generalize the AKLT model \cite{MOshikawa:1992}
%(Oshikawa 92)
for any S, and this gives an exact RVB (valence bond) gapped
state, with a correlation length $\xi_{AKLT}=1/\log{(1+{2/ S})}$.
Again we can view this latter case as a model for an $S$-leg
ladder system in the low energy limit ($S$ chains with interchain
couplings along rungs).

{\it 4. Lord of the rungs.} On the road to 2D, we now turn to
spin-$1\over 2$ ladders. Two and three leg Cu ladder structures
were experimentally realized in SrCu$_{2}$O$_{3}$ and
Sr$_{2}$Cu$_{3}$O$_{5}$ oxides \cite{THiroi:1991,MAzuma:1994}.
%\cite{THiroi:1991}\cite{MAzuma:1994}.
 Couplings along the rung and
along the leg are similar, of the order of 1000 to 2000K. In the
two-leg compound, the ladders are almost decoupled from each
others since the system does not show any magnetic ordering down to
T=50 mK \cite{KKojima:1995a,KKojima:1995b}.
%\cite{KKojima:1995a}, \cite{KKojima:1995b}.
 This is not the
case in the three-leg compound, which orders at 52 K, probably due
to weak transverse couplings. A gap is observed only for the
two-leg system, as shown by macroscopic measurements
%\cite{MAzuma:1994}
 and NMR \cite{MAzuma:1994,KIshida:1994,KIshida:1996}.
%\cite{MAzuma:1994} \cite{KIshida:1994} \cite{KIshida:1996}.
In a similar ladder cuprate Sr$_{14-x}$Ca$_{x}$Cu$_{24}$O$_{41}$,
hole doping could be achieved through Ca-Sr substitution, leading
for high enough pressure to a superconducting behavior at x = 13.6
(Uehara et al., 1996).

In these spin ladders, exact solutions are not available,
but, for low energies, we can use some of the techniques that we discussed
above to determine ground states and excitation properties
\cite{SDellAringa:1997,EOrignac1:1998,ANersesyan:1998,AGogolin:1999,RCitro:2002}.
%\cite{SDellAringa:1997}, \cite{EOrignac1:1998},
%\cite{ANersesyan:1998}, \cite{AGogolin:1999}, \cite{RCitro:2002}.
%\cite{TGiamarchi:2004}.
% (Giam04, Gogolin 99, Orignac-Giam 98, Nersesyan 98, Citro-Orignac 02,
%Dell' Aringa 97).
In addition, DMRG, exact diagonalization and Quantum Monte-Carlo
(no sign problem for spins!) also add useful
information \cite{SWhite:1994,DPoilblanc:1994,MGreven:1996,EDagotto:1999,DJohnston:2000}.
%\cite{SWhite:1994}, \cite{DPoilblanc:1994},
%\cite{MGreven:1996}, \cite{EDagotto:1999}, \cite{DJohnston:2000}.
%(White 94, Poilblanc
%94, Greven 96, Dagotto Rept. Prog. Phys 99, Johnston 00).
In the case of the 2-leg ladder, bosonization is easy to perform
% \cite{SStrong:1994}
%(Strong-Millis 94)
 after introducing acoustic (symmetric under the exchange of
the two legs) and optical (antisymmetric) modes
\cite{SStrong:1994}. Remarkably, the form is precisely that
obtained by Schulz
%\cite{HSchulz:1986}
%(Schulz 86)
in his
analysis of the Haldane spin-$1$ chain in terms of two coupled
spin-$1\over 2$ legs (see subsection {\it 2.}).
 Accordingly, the spin sector is fully gapped as soon as any interchain
coupling is turned on.
 Once again (see Eq.(\ref{Josephson}) and subsequent comments), a Josephson coupling between
the two chains of the form $\sigma_i^+\tau_{i+1}^-$
% (FIGURE allowing to understand that i-i+1 config. is $J_{rung}\sigma_i\tau_i$)
is relevant and provides the gap between the valence bond RVB
ground state and the first triplet excited state, see
Fig. \ref{ladder-spin1}.
%{\bf One should be careful however,
When comparing the spin-$1$ chain with a two-leg spin-$1\over 2$ ladder, one should bear in mind the particular topology (see right panel of Fig. \ref{ladder-spin1}); it plays an important role, for instance in the presence of defects or of open boundary conditions \cite{TNg:1994,SQin:1995}.

Similarly, the analysis of the 3-leg
ladder within bosonization leads to the result
 that one sector is gapless. As previously mentioned, a gap was experimentally observed
  in the two-leg compound SrCu$_{2}$O$_{3}$ while no gap was detected in the three-leg
 material Sr$_{2}$Cu$_{3}$O$_{5}$.
  It is possible to analyse the ladder
problem with arbitrary number of legs ($n_l$), using the non
linear sigma model,
%\cite{SDellAringa:1997},
% (Dell' Aringa 97),
 and one finds indeed that an odd number of legs
leads to a gapless state with power law decay of the spin-spin
correlation functions, while for an even number of legs correlations
fall off exponentially, such that $\xi_{lad}\sim \exp{({\pi\over
2}n_l)}$ \cite{SDellAringa:1997}.
 In the weak coupling limit ($J_{rung}<<J$), the gap decreases exponentially to
zero with increasing
 number of legs: $\Delta_{lad}\sim \xi^{-1}$. These results are confirmed, and
extended to the isotropic limit ($J_{rung}=J)$ using numerical
work \cite{MGreven:1996}:
% (Greven 96):
 $\Delta_{lad}=0.50, 0.16, 0.05$ for $n_l=2, 4, 6$. At large distances,
 spin-spin correlation functions behave exactly as in the spin-1 case
(Eq.(\ref{correl_spin1})) and the susceptibility is given by
Eq.(\ref{susc_S1})
%{\bf (check eqn ref I put in)}
 where  one substitutes $\Delta_{lad}$  for
$\Delta_{Hal}$. Here $\xi=3.2, 10$ for $n_l=2,4$.

{\it 5. Summary of 1D physics in the spin sector.} A picture emerges
from the above description: for (spin) rotationally and
translationally invariant systems, the ground state of 1D spins
chains and ladders is either a degenerate dimer phase where
reflection symmetry about a site is broken, or  a non degenerate
valence bond phase, a.k.a
 a Resonating
Valence Bond phase (RVB), where parity symmetry exists for {\it
any} site \cite{PAnderson:1997}.
%(Anderson 97).
 It consists of
singlets, which resonate coherently on a length scale $\xi$. Since
the spectrum is linearized near the Fermi points, the binding
energy in this RVB state is $\Delta\sim v_s\xi^{-1}$. Within the
RVB phase, quasiparticles consist of spinons. These are thus
confined within a distance $\xi$ of each other, forming bound
state singlets. An excitation breaks the singlet, leading to a
triplet state. This triplet can propagate through the lattice as a
magnon excitation. For strictly 1D chains at $T=0$, $\xi$ is
infinite in the spin-$1\over 2$ case, and we have a gapless spin
liquid with asymptoptically free spinons while $\xi$
%{\bf (forgot backslash)}
is finite in the spin-$1$ case. For ladders, we have a gapped RVB
phase and $\xi_{lad}\sim \exp{({\pi\over 2}n_l)}$ for even numbers
($n_l$) of legs. The limit $n_l\to\infty$ is singular, since the
ground state of a 2D spin-$1\over 2$ system is believed to sustain
long range antiferromagnetic order \cite{SChakravarty:1989}.
% (Chakravarty 89,...).
 The RVB phase that constitutes the ground state of spin-$1$ and
even-leg ladder systems is a strongly
correlated phase in that it has long range phase coherence. This
property is quite independent of the strength of the interactions.
Can we actually see spinons? Simulations are of course one way to
do so: a 1D ring with an odd number of spins shows a spinon
extending over the entire circumference. Open chains  are another
way of revealing these quasiparticles. From an experimental
perspective, there is a way to
%implement
realize these situations: adding nonmagnetic impurities breaks the
1D chain. We will discuss this point in the next section. Another
way to probe spinons -- akin to the odd number of spins case -- is
to consider ladders with odd number of legs. In that case, the
"unpaired" spinons form a 1D spin liquid phase, which is revealed
in experiments.
% The case of dimerized systems is a little bit
%different . In the spin-Peierls problem, translational invariance
%is broken, selecting one of two possible ground states.
%Excitations turn the dimer singlets into triplets. In the
%$J_1-J_2$ model, frustration causes degeneracy of the ground state
%and excitations are pairs of massive spinons. However, in both
%instances, it is the $\phi$ field that becomes ordered. Since the
%$\theta$ field is canonically conjugated to $\phi$, it is totally
%disordered (correlation functions for the $\theta$ field decay
%exponentially with distance). By contrast,
If we associate a
Josephson phase ($\theta$) with each singlet in the RVB
phase, this $\theta$ field is either quasi-ordered in the cases of
1D spin liquid and odd-leg ladders or ordered for even-leg
ladders: in the former situation, correlation functions of the
$\theta$ field decay algebraically with distance, while in the
latter they tend to a constant value for $|x|\to\infty$.

%  {\bf can this be expanded or illustrated?  not clear as it is now.}
%\subsubsection{1D physics in the charge sector}
%This is a monster, so I don't want to say much except, possibly two things:
%d-wave correlations and orbital antiferromagnetism.
\subsection{Defects in 1D spin chains}
>From an experimental perspective, intrinsic defects will modify
the magnetic properties of materials. Examples of impurities are
weak- (magnetic coupling) links, broken chains, magnetic or
nonmagnetic atomic substitutions
\cite{SEggert:1994,SEggert:1995,JIgarashi:1995,WZhang:1997,WZhang:1998,SRommer:1999,VBrunel:1999,SRommer:2000}.
%\cite{SEggert:1994}, \cite{SEggert:1995},
%\cite{SRommer:1999}, \cite{VBrunel:1999}, \cite{SRommer:2000}.
%(Rommer-Eggert 99, 2000, Eggert-Affleck 95, Eggert PhD thesis 94, Brunel-Bocquet-Jolicoeur 99,..).
They produce spectacular effects which are direct consequences of
the correlated 1D states so that we may actually use defects in a
controlled way in order to probe the ground state and the
excitations in the spin chains that we introduced in the previous
section. Nonmagnetic atoms supress the couplings $J$ between
neighboring spins. Thus, in 1D, a finite concentration of such
atoms will break the chain into disconnected segments. At low
temperature, each  segment is in its ground state. In the case of
spin-$S$ Heisenberg chains, Marshall's theorem tells us that the
total spin is zero for an even number of spins in a chain or $S$
for an odd number of spins in the chain. Thus odd chains will give
a Curie contribution to the uniform magnetic susceptibility
% \cite{SWessel:2000}
% (Wessel-Haas 2000),
 proportional to the
concentration of impurities (assuming small concentrations and
equal probability of getting even or odd chains)
\cite{SWessel:2000, JSirker:2006}. This property does not tell
much about the nature of the correlated state {\it per se}. In
contrast, we show below that boundaries cause additional
% spectacular
phenomena which reveal the nature of the low
temperature phases; the reason is because the spin density in the
z direction is the sum of two terms (see Eq.(\ref{Sz})): a uniform
part ($\propto \nabla \phi$)
%{\bf (should be made clearer earlier that this is true)}
which is essentially unaffected by the breaking of translational
invariance, except in the vicinity of the edges plus a staggered
(Umklapp) part giving two types of contributions, one stemming
from finite size effects
% \cite{SEggert:2002}
%(Eggert-Affleck-Horton 2002)
 and one from the boundaries \cite{SEggert:2002}. We first
consider spin-$1\over 2$ compounds, then discuss spin-$1$ chains
and ladders and conclude with quasi-1D systems \cite{SEggert:2004}.
% (Eggert-Affleck 2004).

\subsubsection{Spin-$1\over 2$ systems}
\label{subsub:spinonehalf} Let us first discuss the influence of
boundaries on the magnetic properties of a spin-$1\over 2$
isotropic Heisenberg chain. Bethe ansatz,
% \cite{FEssler:1996},
%\cite{SFujimoto:2000},
% (Essler 96, Fujimoto 2000),
 conformal theory
%\cite{JCardy:1991}, \cite{SLukyanov:1998}
%(Cardy-Lewellen 91, Lukyanov 98)
 or bosonization techniques
% \cite{SEggert:1995},
%\cite{SFujimoto:2004}
%(Eggert-Affleck 95, Fujimoto-Eggert 2004)
allow  one to compute all relevant observables
\cite{FEssler:1996,SFujimoto:2000,JCardy:1991,SLukyanov:1998,SEggert:1995,SFujimoto:2004}.
%\cite{FEssler:1996}, \cite{SFujimoto:2000},
%\cite{JCardy:1991}, \cite{SLukyanov:1998},
% \cite{SEggert:1995}, \cite{SFujimoto:2004}.
In
bosonization language, a boundary forces a condition on the
$r=-1$ and $r=+1$ modes (Eq.(\ref{anihilop1_sigma})) (see for
instance \cite{TGiamarchi:2004}, p 306). Breaking translational
invariance affects spin-spin correlation functions.
% Setting
%$u_1=x_1+x_2+v_st$, $u_2=x_1+x_2-v_st$, $u_3=x_1-x_2+v_st$,
%$u_4=x_1- x_2-v_st$, one finds that the uniform part is given by
%\bea\label{correl-unif} <S^z(x_1,0)S^z(x_2,t)>\sim
%-{K\over{4\pi^2}}\Bigl(\sum_i{T^2\over{\sinh{(\pi Tu_i)}^2}}\Bigr)
%\eea while the alternating part is given by
%\bea\label{correl-altern} &&<S^z(x_1,0)S^z(x_2,t)>~~\sim~~ \nn\\
%&&~~~~~~ (-1)^{x_1-x_2}{\pi T\over v_s}\Bigl(\sinh{(2\pi
%Tx_1)}\sinh{(2\pi Tx_2)}\\ \nonumber &&\times\Bigl[\sinh{(\pi
%Tu_1)}\sinh{(\pi Tu_2)}\sinh{(\pi Tu_3)}\sinh{(\pi
%Tu_4)}\Bigr]^{-1}\Bigr)^{1\over 2} \eea
Far from the boundary, one recovers the usual behavior; for
instance, if one applies a {\it uniform} magnetic field $H$, one
gets a
% constant
magnetization equal to $\chi (T)H$, where the uniform
susceptibility is given by Eq.(\ref{susceptXXX}).
Close to the boundary, the susceptibility becomes site dependent; it
is the sum of two contributions. One of those has a
magnitude close to $\chi (T)H$, such that in
Fourier space its components are centered around $q=0$ (see \cite{SEggert:1995}).
In addition, because
of the boundary, the alternating part of the spin-spin correlation
function also contributes to the uniform susceptibility. This is an
impurity induced effect. For finite temperature,
% in addition to
%a "uniform" susceptibility (
%displayed in Fig.(\ref{spin0.5imp2}),
one gets a site dependent, staggered magnetization of the form
\cite{SEggert:1995}
%(Eggert-Affleck 95)
% \bea\label{altsuscimp}
%M(x)\sim {(-1)^x{{{{0.58\;x H}\over v_s}\sqrt{{{\pi T}\over
%v_s}}}\over{\sqrt{\sinh{({{2\pi x\;T}\over v_s})}}}}} \eea where
%$v_s$ is obtained from Eq.(\ref{velocity}) in the limit $K_s=1$.
%PAR RAPPORT A LA PARTIE THEORIQUE, et les formules de
%susceptibilite correspondantes, (susceptXXX) ,JE CROIS QU'IL
%SERAIT BON DE GARDER MA FORMULATION AVEC J ET PAS vs (ci apres).
\begin{equation}\label{altsuscimp}
M (x)=H(-1)^{x}\frac{1.16}{\pi }\frac{x}{\sqrt{\frac{J}{2T}%
\sinh (\frac{4Tx}{J})}}
\end{equation}
\vskip .2cm
%{\bf do we need to add homogeneous term $m_0$, or
%define $\delta M$?  Also, we should stick to either small $m$ or
%big $M$ for magnetization. }
 %\vskip .2cm

\begin{figure}[htbp]
      \begin{center}
\includegraphics[width=0.9\columnwidth]{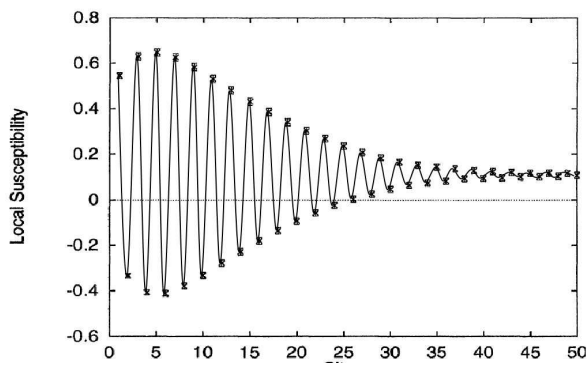}
      \protect\caption{Alternating magnetization caused by a boundary in a spin-$1\over 2$
chain
%. The local susceptibility vs. distance from the open end
according to Monte Carlo simulations at
$T=J/15$ \cite{SEggert:1995}. }
        \label{spin0.5imp1}
         \end{center}
          \end{figure}

At low enough temperature $T\ll J$, the envelope of the
magnetization increases initially with the distance to the
impurity site, peaks at $x=0.48J/T$ in cell units, then decreases
with an exponential asymptote of typical extension $\xi _{imp}$
%{\bf which} follows a $1/T$ behavior
as shown in Fig. \ref{spin0.5imp1}.
%{\bf (Fig. \ref{spin0.5imp1} doesn't show this)}.
From Eq.(\ref{altsuscimp}) one finds that $\xi _{imp}\sim T^{-1}$.

%\begin{figure}[htbp]
%      \begin{center}
%\includegraphics[width=\columnwidth]{Eggert-Affleck95.eps}
%      \protect\caption{The uniform and alternating parts of the local
%susceptibility according to Monte Carlo simulations at $T=J/15$
%compared to field theory. Note that the theoretical curve does not
%lie on top of the MC curve; see \cite{SEggert:1995} }
%        \label{spin0.5imp2}
%         \end{center}
%          \end{figure}

One needs to be a bit careful when taking the $T=0$ limit. The
expression for the magnetization given by Eq.(\ref{altsuscimp})
with $T$ set to zero is restricted to "small enough" values of
$x$. The proper form reads \bea\label{altsuscimpT0} M(x)\sim
(-1)^x\sqrt{{2\over x}}\sin{({H x\over {2 v_S}})} \eea and shows
that a finite field $H$ produces an incommensurate oscillating
magnetization (on top of the (quasi-) uniform part).
% given by
%\bea\label{altsuscimpT0} M(x)\sim (-1)^x\sqrt{{2\over x}}\sin{({H
%x\over {2 v_S}})} \eea {\bf ( should we say what physics of
%discommensuration is?)}
 On sites
close to the boundary, the effect of the "impurity" is seen
through extra contributions to the magnetization
%and to the
%susceptibility
%{\bf (I removed ``and to the susceptibility")}
 which are
%{\it finite}
 nonzero even at $T=0$ \cite{SRommer:1999}.
%(Rommer-Eggert99).
 Recently, \cite{SFujimoto:2004}
% Fujimoto and Eggert
%\cite{SFujimoto:2004}
%(Fujimoto-Eggert 2004)
showed that one needs to include a term generated by  marginally
irrelevant bulk operators.
% \cite{SFujimoto:2004}.
 It gives a
quasi-Curie contribution to the susceptibility of the form
$\chi_B\sim \Bigl({ {T\log{(T)}}}\Bigr)^{-1}$. Note that this
additional piece exists both for even and odd chains, since this
is a "surface" effect.

In practice, creating a boundary does not require one to cut the
chain. Indeed, a weak link (a smaller exchange coupling between
two neighboring spins) is a strong perturbation and ultimately, as
the temperature is reduced, renormalization and numerical
techniques show that it produces similar effects on the
susceptibility to the case of the open chain \cite{SRommer:1999}.
%(Rommer-Eggert 99)
%(see Fig.(\ref{spin0.5imp3})).
%\begin{figure}[htbp]
%      \begin{center}
%\includegraphics[width=\columnwidth]{Eggertthesis3.eps}
%      \protect\caption{Alternating magnetization caused by a weak link in a spin-$1/2$ %chain.
%      The local susceptibility as a function of distance from a weakened link
%$J' = 0.75J$ at $T = J/15$ \cite{SEggert:1994} (S. Eggert,
%thesis). }
%        \label{spin0.5imp3}
%         \end{center}
%          \end{figure}

\begin{figure}[htbp]
      \begin{center}
\includegraphics[width=0.9\columnwidth]{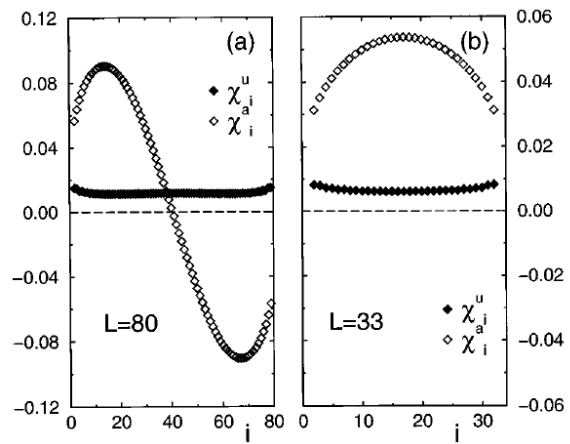}
      \protect\caption{
%Alternating magnetization in the case of spin-$1/2$ systems
%with open boundaries, for even and odd chains.
 Uniform ($\chi^u_i$) and
staggered ($\chi^a_i$) components of the
envelope of the local susceptibility $\chi_i$ obtained
with the DMRG method applied to the $S=1/2$ Heisenberg model on a
chain with $L$ sites and open boundaries, using $m$ states in the iterations and
studying the subspace with a total spin in the $z$ direction equal
to 1. (a) corresponds to $L=80$ and $m=16$, and (b) to $L=33$ and
$m=32$  \cite{MLaukamp:1998}.}
        \label{boundarymag1}
         \end{center}
          \end{figure}

As we stated in the introduction to this section, defects break a
chain into disconnected segments. The boundaries of each segment
give rise to extra "surface" contributions to the magnetization.
In addition, there are "volume" contributions associated with
finite size effects: since the staggered coherence length diverges
at $T=0$ in the thermodynamic limit, one gets a low $T$ finite
alternating moment per site $\propto {1\over\sqrt{L}}$ for a chain
of length $L$ \cite{SEggert:2002}.
%(Eggert-Affleck-Horton 2002).
This size dependence is a direct consequence of the $r^{-1}$
fall-off of the staggered correlation function. For odd-$L$ chains
this magnetization is a manifestation of the presence of a
spin-$1/2$ in the ground state, and for even-$L$ chains it
accompanies a spin-$1$ excitation above the ground state (see
Fig. \ref{boundarymag1}).
The local magnetic susceptibility
$\chi_i$ on site $i$ is the sum of a uniform contribution $\chi^u_i$ and
an alternating contribution. Fig. \ref{boundarymag1} shows $\chi^u_i$
and the staggered part of the alternating term, namely
$\chi^a_i=(-1)^i (\chi_i - \chi^u_i)$.

\begin{figure}[htbp]
      \begin{center}
\includegraphics[width=\columnwidth]{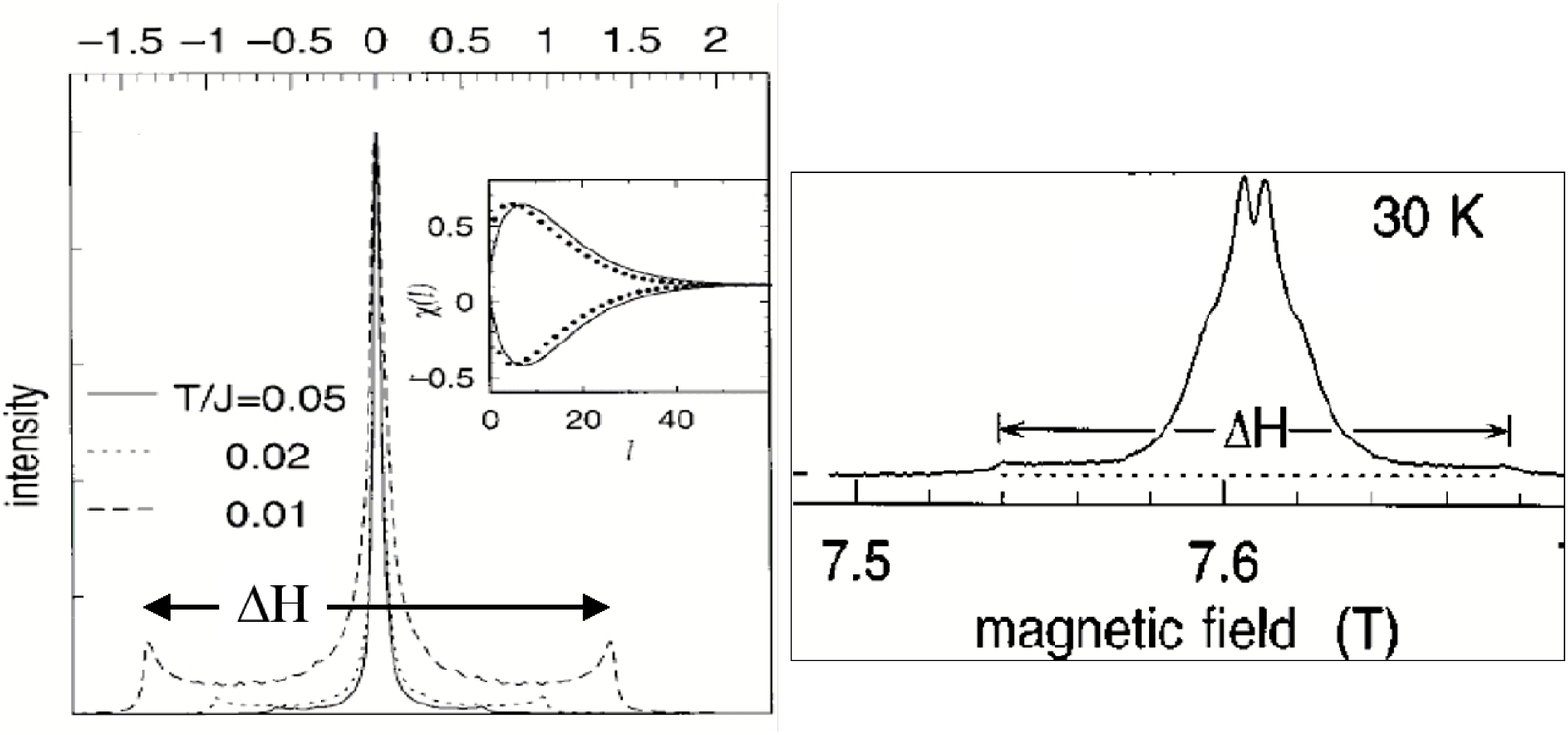}
      \protect\caption{left panel : the NMR spectrum expected for a staggered magnetization
      of the spin 1/2 chain displayed in the inset, which corresponds to its histogram.
      Right panel : the experimental Cu NMR spectrum in Cu chain Sr$_{2}$CuO$_{3}$ displays a similar
       $\Delta H$ pattern. It is interpreted as induced by intrinsic
       defects
% {\bf (What is DH?)}.
\cite{MTakigawa:1997}.}
        \label{Takigawa}
         \end{center}
          \end{figure}

On the experimental side, the
%already mentioned Cu chain
Sr$_{2}$CuO$_{3}$ spin chain introduced in section (\ref{spinsector}.1)
is a good realization of a truly 1D Heisenberg $%
S=1/2$ system, with an antiferromagnetic coupling $J=2200\pm 200K$
%\cite{AKeren:1993}, \cite {TAmi:1995}, \cite{NMotoyama:1996}.
\cite{AKeren:1993,TAmi:1995,NMotoyama:1996}.
 Only
nonmagnetic Pd substitution could be achieved on Cu site, which
leads to a Curie behavior of the macroscopic magnetization
\cite{KMKojima:2004}.
 This Curie term is a consequence of the chain
segments with odd number of chains, as argued above. Local probes
such as NMR are necessary to probe the local variation of the
magnetization $M(r)$. Measurements were performed only in the
case of a
%a-priori
{non-substituted chain}, taking advantage of the presence of
unavoidable intrinsic defects, probably oxygen vacancies and
interstitials
% \cite{MTakigawa:1997}, \cite{JPBoucher:2000}.
\cite{MTakigawa:1997,JPBoucher:2000}. The NMR spectrum shows a
central line and a broad symmetric background with sharp edges
(see Fig. \ref{Takigawa}). This is evidence of the presence of an
alternating polarization induced by the native defects, as will be
demonstrated in Fig. \ref{FigPedagogicNMR}.
%described in more {\bf detail} in (section
%Henri et chaines de spin 1 which describe the link between NMR
%spectrum and impurity induced effects).
The edges of the $\Delta
H$ pattern follow a $1/\sqrt{T}$ shift,
as would be expected from Eq.(\ref{altsuscimp}) for values of $x$ near the
peak position.
%{\bf (ref does
%not exist)}
%(\ref{altsuscimp}) \

In the Cu spin 1/2\ chain dichlorobis (pyridine) copper(II)
(CuCl$_{22}$NC$_5$H$_5$), referred to as CPC, with $J=27K$, muon
spin resonance was performed, where implanted muons were used both
as a probe of the local susceptibility and as a defect in the
chains \cite{JAChakhalian:2003}. Indeed, muons are positively
charged particules which stop in the bulk of the material and then
couple to the chains, according to comparisons of Knight shift
measurements with DMRG computations.
 %Knight shifts effects were measured and
%interpreted to show that muon was indeed locally coupled to the
%chain, consistently with DMRG\ computations.\
While this technique is valuable in order to study chains where
native defects cannot be produced by simple chemistry, it is
limited by the fact that the muon location is not exactly known.

\subsubsection{spin-$1$ chains}

\begin{figure}[htbp]
      \begin{center}
\includegraphics[width=\columnwidth]{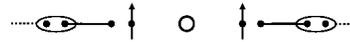}
     \protect\caption{The introduction of a non magnetic impurity in the Spin-$1$ chain displayed in Fig. \ref{VBS3} breaks two singlet bonds,
thereby releasing two spins-$1\over 2$. }
        \label{impureS1}
         \end{center}
          \end{figure}

In many respects, it is easier to describe the response of a
spin-$1$ system to an impurity. As we pointed out above, the
ground state of the pure system is a non degenerate RVB state,
with  long range order (of a singlet order parameter). Introducing
a nonmagnetic impurity breaks a valence bond and, in the
representation of a spin-$1$ in terms of two spins-$1\over 2$,
implies that a spin-$1\over 2$ is "freed"
%"released"
at the location of the broken valence bond, i.e it is pinned in
the vicinity of the impurity, as shown in Fig. \ref{impureS1}
\cite{TKennedy:1990,TNg:1992,SMiyashita:1993,MLaukamp:1998}.
%\cite{TKennedy:1990}, \cite{TNg:1992}, \cite{SMiyashita:1993},
%\cite{MLaukamp:1998}.

\begin{figure}[htbp]
      \begin{center}
\includegraphics[width=0.9\columnwidth]{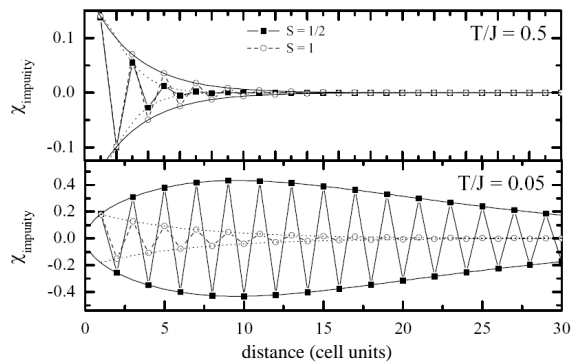}
      \protect\caption{Alternating magnetization in the case of spin-${1\over 2}$
and spin-$1$ systems with a defect.
%  Polarization induced by a
%non-magnetic impurity for spin-$1$ chains.
 The polarization for
$S=1$ is normalized
% arbitrarily
 to that for $S={1\over 2}$.  At high
temperature, the two polarizations are roughly identical,
exponentially decaying.  At low temperatures, for $S=1$, the
polarization doesn't evolve further because $\xi$ saturates.  On
the other hand, the extension of the polarization for $S={1\over 2}$
systems continues to
increase as $T$ is decreasing and exhibits a nonmonotonic behavior
(see Fig. \ref{spin0.5imp1}).}
        \label{boundarymag3}
         \end{center}
          \end{figure}

%\cite{TKennedy:1990}, \cite{TNg:1992}, \cite{SMiyashita:1993},
%\cite{MLaukamp:1998}.
%(Kennedy 90, Miyashita-Yamamoto 93, TNg 92, Laukamp 98).
 Monte Carlo simulations performed on chains of $L$
sites ($L\ge 100$) show that the polarization oscillates with an
exponential envelope $<S_i^z>\propto e^{- [{{(i-1)}\over
\xi}]}+e^{-[{{(L-i-1)}\over \xi}]}$, with $\xi\sim 6$.
% as
%exemplified in Fig.(\ref{ABAlternation}).
A  comparison of the
induced magnetization for the $S={1 \over 2}$ and $S=1$ cases is
shown in Fig. \ref{boundarymag3}.

%  \begin{figure}[htbp]
%  \begin{center}
%\includegraphics[width=\columnwidth]{MiyashitaYamamoto93.eps}
% \protect\caption{Spin-$1$ chain with open boundaries.
% The staggered magnetic moments
%$\langle S_i^z\rangle$ in chains with free boundaries at both ends
%in the cases of $L$=65 and 97 \cite{SMiyashita:1993}. }
% \label{ABAlternation}
%  \end{center}
%  \end{figure}

%    \begin{figure}[htbp]
%      \begin{center}
%\includegraphics[width=\columnwidth]{MiyashitaYamamoto93_2.eps}
%      \protect\caption{Spin-$1$ chain with open boundaries.
% The correlation function  $\langle
%S_1^z S_i^z \rangle$ in even chains in the case of $M_z=0$, showing
%the alternation rule.
%\cite{SMiyashita:1993}.}
%        \label{oscill1}
%         \end{center}
%          \end{figure}

 This
value of $\xi$ is that of the undoped compound. This result
suggests that the perturbation  which created the impurity is
localized and decays with the coherence length of the pure system
see Eq.(\ref{xi_Hal}). This fact is confirmed in various DMRG and
QMC studies
%\cite{ESorensen:1995}, \cite{EPolizzi:1998}
%(Sorensen-Affleck 95, Polizzi-Mila 98)
% and QMC studies
\cite{ESorensen:1995,EPolizzi:1998,SMiyashita:1993,YKim:1998,FAlet:2000}.
%\cite{SMiyashita:1993},
%\cite{YKim:1998}, \cite{FAlet:2000}.
% (Kim 98, Miyashita-Yamamoto 93, Alet 2000).
 A remarkable feature emerges from these
simulations, namely, the relative sign of the polarization at the
end of a segment of the chain formed by two impurities.
%: assume we
%label the spin-up states by up and the spin-down states by down;
%{\bf (why--what's wrong with ``up" and "down"?)}
Assume that we have a spin-up state at one end of the segment.
The oscillation of the polarization implies that the next site is
in the down state and so on. If we extend the up-down alternation
of the sites all the way to the other end of the segment, we
observe that the polarization on the site closest to the second
impurity corresponds to that expected from the up-down alternation
rule, {\it despite the fact} that the distance between the
impurities is much larger than $2\xi$. In other words, the singlet
order parameter propagates the relative phase of the spin
operators across the entire segment (see \cite{SMiyashita:1993}).
% The alternation rule is
%illustrated in Fig.(\ref{oscill1})

\begin{figure}[htbp]
      \begin{center}
\includegraphics[width=\columnwidth]{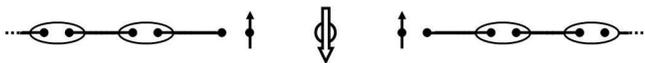}
      \protect\caption{Spin-$1$ chain with a Cu spin-$1/2$ defect. Spin-$1/2$
states are formed on either side of the Cu atom and align to form
a spin-$1$ state.   }
        \label{Cudefect1}
         \end{center}
          \end{figure}

This observation is confirmed by a study combining theory and
experiments \cite{MHagiwara:1990}
% (Hagiwara-Katsumata-Affleck-Halperin-Renard 90)
on a spin-$1$ chain with a spin-$1\over 2$ (Cu) impurity. The
lowest energy state is such that sites located on either side of
the Cu are BOTH in the up or down state and form a triplet with
$<S^z> =\pm 1$, as shown in Fig. \ref{Cudefect1}.
DMRG and QMC confirm this picture
%\cite{ESorensen:1995}, \cite{TTonegawa:1995}, \cite{PRoos:1999}.
%(Sorensen-Affleck 95, Tonegawa 95)
%and QMC confirm this picture
\cite{ESorensen:1995,TTonegawa:1995,PRoos:1999}.
%(Roos 99).

\begin{figure}[htbp]
     \begin{center}
\includegraphics[width=\columnwidth]{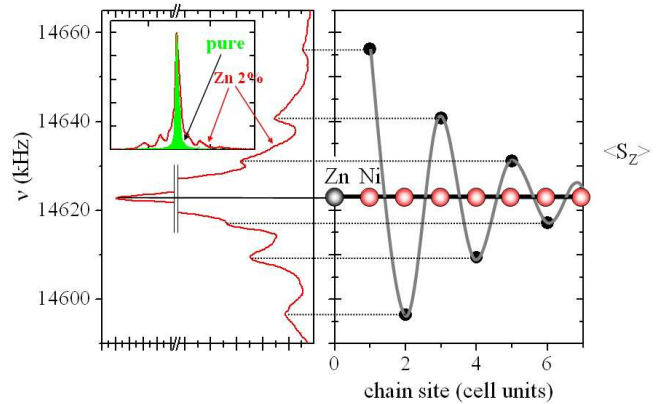}
      \protect\caption{NMR reconstruction of an impurity-induced magnetization in a spin
      chain. The left panel displays an $^{89}$Y NMR spectrum in
      pure (green) and Zn-substituted (red)
      YBa$_{2}$NiO$_{5}$. The various satellites which appear with Zn
doping allow the reconstruction site by site of
      the induced staggered magnetization near the Zn atoms substituting Ni in the chain
       \cite{FTedoldi:1999,JDas:2004}. }
        \label{FigPedagogicNMR}
         \end{center}
          \end{figure}

Contrary to the case of the $S={1\over 2}$ spin chain,
% and spin-Peierls chains,
the impurity induced magnetization in the
$S=1$ spin chain could be measured experimentally with great
precision. In the $S=1$ antiferromagnetic Ni chain
[Ni(C$_{2}$H$_{8}$N$_{2}$)$_{2}$(NO$_{2}$)]CIO$_{4}$ (NENP), \
EPR\ measurements provide evidence for  the existence of a spin
state associated with the substitution of Ni by nonmagnetic Zn, Cd
or
Hg atoms, consistent with the VBS picture where each impurity frees two $%
S={1\over 2}$ states \cite{SHGlarum:1991}.
Specific heat measurements confirm the existence of
a spin state by detecting the corresponding Schottky anomaly
\cite{APRamirez:1994}.\ While this study was first argued to give
a different spin state value than EPR, the controversy was finally
resolved by a more careful treatment of long enough spin chains
\cite{CDBatista:1998}. More
refined experiments were then performed in the Ni chain oxide YBa$_{2}$NiO$%
_{5}$, a better prototype of the spin 1 chain than NENP, with
$J=280K$ \cite {DJButtrey:1990}. When introducing nonmagnetic Zn
or Mg impurities, EPR studies reveal the presence of spin states
%\cite {APRamirez:1994}\cite{CDBatista:1998}
accompanied by a
Curie contribution to the macroscopic susceptibility
%\cite {APRamirez:1994}, \cite{CDBatista:1998},
%\cite{CPayen:2000}, \cite{JDas:2004}.
\cite {APRamirez:1994,CDBatista:1998,CPayen:2000,JDas:2004}.\ The Curie constant
corresponds to about two $S={1\over 2}$ per Zn, as expected. Note,
however, that any attempt to analyze the macroscopic
magnetization quantitatively is hindered by the fact that a Curie behavior is
already observed in the nominally pure compound without any
Zn, a feature encountered in most experimental realizations of low
dimensional systems.\ This native Curie term is usually due both
to the presence of spurious paramagnetic phases and local defects
within the crystal itself, such as vacancies or interstitials of
oxygens. As anticipated, any defect which is coupled to the spin
chain itself will induce in turn a local moment, i.e. a Curie law
in a magnetization measurement. This limits any refined
quantitative analysis of such macroscopic experiments, as well as
EPR\ studies. To bypass this difficulty, NMR has proven to be an
ideal probe, since it allows one to
identify the actual effect of the Zn substitutional atom in its
vicinity and moreover to resolve this effect spatially, while other
techniques only give the total spin value induced by the defect. In
YBa$_{2}$NiO$_{5}$, NMR of the yttrium nucleus probes the local
magnetization $M(r)$ of the Ni sites, with
a site by site resolution. Without impurity, the $^{89}$%
Y\ NMR shift measures the pure uniform susceptibility of the Ni
chain and displays the usual Haldane gap behavior
\cite{TShimizu:1995}. When adding nonmagnetic isovalent impurities
such as Mg$^{2+}$ or Zn$^{2+}$, new NMR\ satellite lines appear,
each of them being due to all the Y\ nuclei at a given distance
$R_{i}$ from an impurity as shown in Fig. \ref{FigPedagogicNMR}
\cite{FTedoldi:1999,JDas:2004}.
% \cite{FTedoldi:1999}, \cite{JDas:2004}.

%    \begin{figure}[htbp]
%      \begin{center}
%\includegraphics[width=0.9\columnwidth]{figdas12.eps}
%      \protect\caption{Comparison of the extension of the Mg \cite{FTedoldi:1999}, Zn or Cu %\cite{JDas:2004}
%      induced magnetization
%      in YBNO \cite{JDas:2004}
%and the pure correlation length measured by QMC by  \cite{YJKim:1998} after \cite{JDas:2004}}
%        \label{Dasfig12}
%         \end{center}
%          \end{figure}
 \begin{figure}[htbp]
      \begin{center}
\includegraphics[width=0.9\columnwidth]{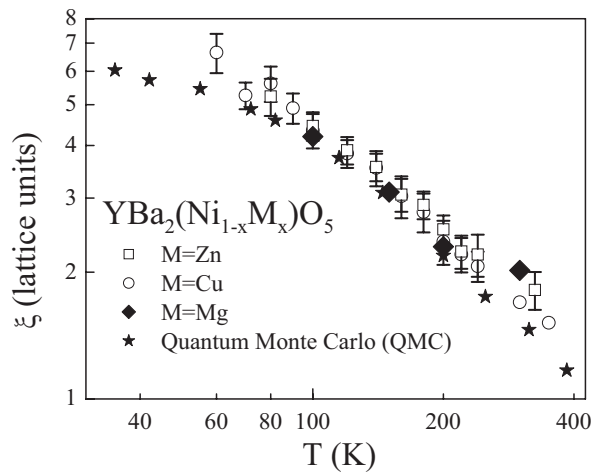}
      \protect\caption{Comparison of the extension of the
staggered induced magnetism due to Mg \cite{FTedoldi:1999}, Zn or Cu \cite{JDas:2004}
impurities in YBNiO. The correlation length measured by QMC by  \cite{YJKim:1998}
for the pure system is shown for comparison.}
% after \cite{JDas:2004}}
        \label{Dasfig12}
         \end{center}
          \end{figure}
Their frequency shift
% $\delta \nu _{i}$
 relative to the pure case is
directly proportional to the local average spin $\left\langle
S_{Z}(i)\right\rangle $ along the external field direction $z$ at
the corresponding Ni site. Since the polarization $\langle
S^z\rangle$ alternates, the shifts relative to the main line
produce both positive and negative satellites, as shown.
%Measured shifts for Zn substitution in YBa$_{2}$NiO$%
%_{5}$ are plotted on Fig. \ref{Dasfig11a}.\
They decay exponentially from the Ni first neighbor of the
impurity site. Their temperature dependence is governed by that of
the Ni closest to the impurity, which follows approximately a
Curie law, together with the T-dependence of the envelope
extension $\xi_{imp}$ plotted in  Fig. \ref{Dasfig12}.
This spatial
%exponential
 behavior is a consequence of the
exponential decay of the spin spin correlation
functions in such Haldane gapped spin chains. Furthermore, the extension $%
\xi _{imp}$ measured for different impurities appears almost
identical to that of the pure spin-spin correlation function $\xi
_{pure}$ computed by QMC techniques, also displayed
 in Fig. \ref{Dasfig12}. The fact that the impurity
reveals the intrinsic spin-spin correlations of the system is the
key result which will give us insight into the other more complex
and less controlled correlated systems such as high $T_{c}$
superconductors: a nonmagnetic impurity can be used as a probe of
the intrinsic properties of a correlated system, very much like a
magnetic impurity in a metallic host could be used to probe the
Fermi surface properties through the RKKY oscillation.

Inelastic neutron scattering experiments also allow
one to measure  the end-chain excitations, which are peaked
at the antiferromagnetic wavevector along the chain, and show an
alternating pattern with extension $\xi _{imp}=8$ at T=0.1K,
consistent with the NMR\ studies \cite{MKenzelmann:2003}.

The case of a magnetic impurity in a spin-$1$\ chain has been
stressed previously: a Cu spin-$1\over 2$ substitution  on a
spin-$1$ site results in the formation of a spin-$1\over 2$
effective state as probed by EPR in NENP \cite{MHagiwara:1990}
(see Fig. \ref{Cudefect1}). Specific heat measurements confirm
this observation through the observation of a Schottky anomaly as
in the case of nonmagnetic Zn \cite{CDBatista:1998}. In addition,
DMRG and QMC computations predict that the magnetic impurity
induces  an alternating magnetization close to the defect
\cite{TTonegawa:1995,ESorensen:1995,PRoos:1999}.
%\cite{TTonegawa:1995}, \cite{ESorensen:1995}, \cite{PRoos:1999}.
In the case of YBa$_{2}$NiO$_{5}$, $^{89}$Y\ NMR experiments
similar to those reported for Zn and Mg above show similar results
for magnetic Cu impurities \cite{JDas:2004}. The envelope remains
exponential with the same extension $\xi _{imp}$ as shown in
Fig.(\ref{Dasfig12}). In comparison with nonmagnetic Zn defects,
one sees a different shift for these satellites, i.e. a smaller
amplitude of the induced polarization, by about 20\%, for all the
Ni sites producing the polarization. A\ QMC\ computation can be
performed in this specific case, which reproduces the experiment
if Cu is coupled antiferromagnetically to its neighboring Ni with
a coupling $J'\simeq 0.1-0.2J$ \cite{JDas:2004}, a result
qualitatively similar to that found in NENP \cite{MHagiwara:1990}.

\subsubsection{Ladder systems}

 In  Section \ref{spinsector}.4, we argued that
the low energy physics of an even-(n) leg spin-$1 \over 2$ ladder
could be mapped onto an integer spin-$n \over 2$ chain and that of
an odd -leg ladder to a spin-$1\over 2$ chain. Thus,  the influence
of an impurity in the ladder follows from the considerations
pertaining to the spin-$1$ and spin-$1\over 2$ cases. For
instance, for a two leg ladder, we conclude that when a
nonmagnetic impurity is located on one of the legs, a spin-$1\over
2$ state appears on the opposite leg; its amplitude is centered
around the site across from where the impurity sits.
We emphasize once again (see discussion in Section \ref{spinsector}.4)
that it is important to keep in mind the particular topology of the ladder that the above mapping involves, when discussing end-states. For instance, a spin-$1\over 2$ state is localized near the edge of an open spin-$1$ chain whereas no such state is found at the open end of the two-leg ladder shown of the left panel of Fig. \ref{ladder-spin1}.

%{\bf (figure!!) something is necessary--this is not clear.}
%from the other leg, is localized around the impurity and that an
%This magnetic polarization oscillates from site to site along the
%leg of the ladder, with an amplitude which decays exponentially with
%a characteristic length $\xi_{imp}$.
Numerical studies of the spin-$1\over 2$ ladder with impurities indeed find a
magnetic polarization which oscillates from site to site along the
leg of the ladder, with an amplitude which decays exponentially with
a characteristic length $\xi_{imp}$ \cite{GMartins:1996,GMartins:1997,AWSandvik:1997,TMiyazaki:1997,HJMikeska:1997,MLaukamp:1998,ALaeuchli:2002}. But whether or not the value of $\xi_{imp}$ is equal to the coherence length of the pure system (equal to 3-4 lattice spacings) is still controversial.
%Depending on authors,
%$\xi_{imp}$ might \cite{AWSandvik:1997}
%or might
%not \cite{MLaukamp:1998}
%(Laukamp 98)
%be equal to the coherence length of the pure system (equal to 3-4
%lattice spacings)
%\cite{HFukuyama:1996,GMartins:1996,GMartins:1997,AWSandvik:1997,TMiyazaki:1997,HJMikeska:1997,MLaukamp:1%9%98,ALaeuchli:2002}.
%(Fukuyama-Nagaosa-Saito-Tanimoto 96, Martins-Dagotto-Riera PRB 54
%16032 96, Martins et al 97, Mikeska-Neugebauer- Schollw\"ock 97,
%Miyazaki JPSJ 66, 2580 97,Sandvik 97, Laukamp 98 L\"auchli-Poilblanc 2002).
% {\bf ( I changed citation formats but much
%more needs to be done)}

% \begin{figure}[htbp]
%      \begin{center}
%\includegraphics[width=\columnwidth]{Laukamp98_2.eps}
%      \protect\caption{$2$-leg Ladder with a defect on one of the chains.
%An alternating magnetization develops in the vicinity of the site
%located on the other chain, across from the impurity. Local
%susceptibility for the Heisenberg model on a $2\times 32$ ladder
%using the DMRG method, keeping $m=16$ states, and studying the
%state of lowest energy in the subspace of total spin in the $z$
%direction equal to 1. Results with and without vacancies are
%shown.
%(a) contains the uniform component and (b) the staggered
%one
% \cite{MLaukamp:1998}}.
%        \label{ladderimp1}
%         \end{center}
%          \end{figure}

 These polarized states  either produce a Curie law
for the susceptibility or may even lead to an antiferromagnetic
transition \cite{MGreven:1998}.
%(Greven-Birgeneau PRL 81, 1945, 98).
 For the 3-leg ladder the physics seems consistent with that of
the spin-$1\over 2$ chain with a nonmagnetic impurity
\cite{AWSandvik:1997}.
%(Sandvik et al 97).
 Experimentally, nonmagnetic Zn  can be substituted in the two-leg ladder
SrCu$_{2}$O$_{3}$.\ Macroscopic susceptiblity measurements reveal
the  appearance of a Curie law at high temperature, which
 corresponds to one spin-$1\over 2$ per Zn in the dilute limit
\cite{MAzuma:1997}.\ \ A  concomitant $1/T$-broadening of the Cu
NMR line is observed, proportional to the number of impurities,
which provides evidence for the existence of an alternating
induced magnetization which extends over the ladder \cite
{NFujiwara:1998,SOhsugi:1999}.  No satellites could be resolved,
in contrast with the case of YBa$_{2}$NiO$_{5}$. This limits any
quantitative evaluation of the magnetization shape and extension.
Magnetic $S=1$ Ni substitution leads to similar effects.

\subsubsection{ Summary of impurities in spin chains}

Impurities create edges in spin chains. For strictly 1D systems,
pairs of edges divide the system into even- or odd- length
segments. The magnetic response to such defects contains two
parts, which mirror the two contributions to the spin density
operator of the pure system: (a) the uniform contribution -- the
gradient of the phase -- is essentially identical to that of the
impurity free system, except close to the boundaries and (b) the
Umklapp contribution near the antiferromagnetic wavector, by
contrast, is qualitatively affected by the boundary; this is
because the RVB parameter (for spin-$1$ chains and for ladders) or
the staggered magnetization (for spin-$1\over 2$ chains) show long
range or quasi-long range order at this wavevector for the pure
system.
% {\bf  I found 1st few sentences here confusing.}
 Since impurities break translational
invariance, momentum is no longer a good quantum number so that a
uniform perturbation can give rise to a spatially oscillating
magnetization and, conversely, an alternating perturbation can
contribute to the uniform response. This is precisely what is seen
in chain segments. The ground state of odd chains is a
spin-$1\over 2$ doublet. An infinitesimal magnetic field will give
a uniform Curie contribution which, in turn, will promote an
alternating magnetization. For even chains, one also gets the same
effect at finite temperature when transitions to the excited
(triplet) state are allowed. The reverse effect is observed (at
least numerically) in the case of a single edge, in the presence
of a magnetic field. In this case, the induced uniform spin
polarization causes an alternating magnetization to develop near
the boundary. Its spatial extension is of the order of the
coherence length of the pure system and this mode contributes to
the uniform magnetic response. These features are
hallmarks of strongly correlated systems where several length
scales are simultaneously relevant \cite{FAnfuso:2006}. We will
see in a forthcoming section that similar effects occur in two
dimensions, when nonmagnetic defects are introduced in the
copper-oxide planes.

% \begin{figure}[htbp]
%      \begin{center}
%\includegraphics[width=\columnwidth]{Bobroff_4.eps}
%      \protect\caption{Alternating magnetization in the case of spin-${1\over 2}$
%and spin-$1$ systems with a defect.  Polarization induced by a
%non-magnetic impurity for spin-$1$ chains.  The polarization for
%$S=1$ is normalized arbitrarily to that for $S={1\over 2}$.  At high
%temperature, the two polarizations are roughly identical,
%exponentially decaying.  At low temperatures, for $S=1$, the
%polarization doesn't evolve further because $\xi$ saturates.  On
%the other hand, the extension of the polarization for $S={1\over 2}$
%systems continues to
%increase as $T$ is decreasing and presents a nonmonotonic behavior.}
%        \label{boundarymag3}
%         \end{center}
%          \end{figure}

\subsubsection{\protect\bigskip Impurities and long range order}\label{sec:1Dimplro}

There cannot be true long range order at finite temperature for
purely 1D systems, but adding a small interchain coupling
$J_{\bot}$ between spin chains will drive the quasi-1D system
antiferromagnetic. The Neel temperature $T_N$ can be determined
from the exact expression of the staggered susceptibility of the
chain, using RPA \cite{SEggert:2002, SEggert:2003}.
%(Eggert-Affleck-Horton Phys. Rev. Lett 89, 047202, 2002 and Phys. Rev. Lett. 90, 089702 2003).
 One finds that $T_N\sim J_{\bot}$. Non magnetic impurities do cut the chains into segments,
but LRO is not destroyed since two segments belonging to a given
chain can still be magnetically connected through the interchain
$J_{\bot}$. In fact ${T_N/ J_{\bot}}$ is a function of the scaling
variable $[({1/p}) -1]J_{\bot}$. As the concentration $p$ of
impurities is increased, $T_N$ falls. However, in some situations,
the magnitude of the staggered moment and the Neel temperature may
increase with $p$ \cite{SEggert:2004}.
%(Eggert-Affleck 2004).
 Similar
effects can be obtained for a large enough single ion anisotropy
\cite {TSakai:1990,AZheludev:2000}.
%\cite {TSakai:1990}, \cite{AZheludev:2000}.

%     \begin{figure}[htbp]
%      \begin{center}
%\includegraphics[width=\columnwidth]{azuma.eps}
%      \protect\caption{susceptibility in a two-leg ladder Sr(Cu1-xZnx)2O3 with various Zn concentrations x:
%     Zn induces a kink which signals an antiferromagnetic order, whose ${T_N}$ is represented in the inset
%     versus x \cite{MAzuma:1997}}
%       \label{Azuma}
%        \end{center}
%         \end{figure}

%     \begin{figure}[htbp]
%      \begin{center}
%\includegraphics[width=\columnwidth]{grenier.eps}
%      \protect\caption{phase diagram of the spin-Peierls CuGeO3 with impurities : the
%      spin-Peierls (SP) transition temperature decreases while
%antiferromagnetic order (AF) sets in \cite{BGrenier:1998}}
%        \label{Grenier}
%        \end{center}
%          \end{figure}

In other words, the introduction of impurities enhances long range
order. This is one of the various routes to getting ``order from
disorder" \cite{JVillain:1980,EFSchender:1991}.
%{\bf (citation?)}.
 For example, in two-leg spin-$1\over 2$
ladders SrCu$_{2}$O$_{3}$, specific heat and susceptibility
measurements show that a concentration $x=1\%$ of Zn at Cu site
leads to an antiferromagnetic behavior at T$<$T$_{N}=3K$
\cite{MAzuma:1997}. When increasing the number of impurities,
T$_{N}$ increases to $8K$ at $4\%$ and decreases again.
% (see phase
%diagram on Fig.(\ref{Azuma})).
 Concentrations as small as $x=0.1\%$
still lead to an antiferromagnetic order \cite{KManabe:1998}. In
spin-Peierls chains CuGeO$_{3}$ as well, either magnetic (Ni,Co)
or nonmagnetic (Mg,Zn) defects at Cu site simultaneously reduce
the spin-Peierls transition temperature T$_{SP}$ and induce a long
range AF\ order whose frozen moment is spatially inhomogeneous
\cite{MHase:1993,SBOseroff:1995,JPRenard:1995,LPRegnault:1995,PEAnderson:1997,KKojima:1997,RKadano:1997}.
%The corresponding phase diagram is plotted on Fig.(\ref{Grenier}).
Similar effects are also observed in some $S=1$ spin chains like
PbNi$_{2}$V$_{2}$O$_{5} $ \cite {YUchiyama:1999}\ or $S={1\over
2}$
chains like Sr$_{2}$CuO$%
_{3}$ \cite{KMKojima:2004}. Even though the physics exhibited in
these various substituted systems is very rich, it is not directly
relevant to the aim of this review as it is governed by subtle
deviations from a purely one-dimensional Heisenberg situation.
\vskip .2cm
\section{Experiments on point defects in 2D systems: the metallic cuprates}\label{sec4} %\label{sec:expts_normal}

\subsection{Overview}

\textbf{\ }The work mentioned up to now on 1D\ or quasi 1D\ systems has been
devoted mostly to insulating spin systems, as the major available
experimental systems are non-metallic. The most direct extension to 2D of
the work sketched above for spin chains would be to study the response to
impurities in the case of a spin 1/2\ Heisenberg plane, for which
theoretical computations along similar lines are available (see Section \ref%
{sec:models}). This situation might be encountered in the case of the
undoped cuprates, e.g. La$_{2}$CuO$_{4}$ and YBa$_{2}$Cu$_{3}$O$_{6}$ which
are good Mott-Hubbard insulators. We shall explain in section \ref%
{sec:magproperties} why this simple experimental extension has not been
successfully performed so far.\ On the other hand, in the cuprates, hole or
electron doping is easy to achieve and allows one to span the phase diagram
of Fig. \ref{fig:cuprate_phase_diag} displayed in the introduction. The most
important work done so far using defects or impurities to probe 2D\ systems
has thus been focused on the doped layered cuprates, among which are the
HTSC phases. \
%This is indeed where the approach underlined in this review is
%quite useful, as these doped systems are less amenable to
%theoretical analyses.
%The actual scheme is then opposite to that
%considered so far, as \ the response to impurities or defects is
%giving information which are not predictable.\ These quite
%original data do allow one to reveal the properties of the pure
%system, and will hopefully help to understand the basic physical
%concepts and to elaborate comprehensive theories.
%%
%%PJH-- I did not find the previous three sentences useful, or perhaps
%%did not understand them.
%%

Before beginning with a detailed description of the experiments, it is
essential to recall briefly here the important features which characterize
the different phases of Fig. \ref{fig:cuprate_phase_diag}. Here the loss of
AF\ order does not give way immediately to a metallic and superconducting
ground state, but rather to an intermediate regime in which a disordered
magnetic ground state, usually described as a spin glass (SG) phase
separates the AF domain from a superconducting region which takes the form
of a dome.\ One denotes respectively the left and right of the dome as the
underdoped and overdoped regime, and refers to the doping at which the
maximum $T_{c}$ occurs as optimal. An essential feature which has been
demonstrated experimentally is that the superconducting state does not have
a uniform gap amplitude in $\mathbf{k}$ space but is rather $d$-wave, i.e.
it has a maximum gap in the CuO bond axis directions and nodes in the
diagonal directions \cite{CCTsuei:2000}. Now all this would have been
relatively simple if the metallic state above the superconducting regime
were a simple metal describable within a Fermi liquid theory.\ In fact this
limit seems to be achieved only for very large doping, on the far right of
the phase diagram, i.e. as far away from the Mott AF state as possible.\
Directly above the superconducting dome the transport properties, e.g. a $T$%
\ linear dependence of the resistivity with large scattering rates $\tau
^{-1}\approx k_{B}T/h$\ at optimal doping, appear to be at odds with a
simple Fermi liquid description, hence the designation ``strange metal''
\cite{CMVarma:1989}.\ Most importantly, in the underdoped regime the
magnetic properties exhibit a quite unusual behavior, with a spin
susceptibility revealed \ by NMR Knight shift data which exhibits a large
drop at low $T$, and is nearly suppressed before the superconducting singlet
state is even established \cite{HAlloul:1989}.\ This drop in the
susceptibility occurs as if a gap were opening in the excitation spectrum of
the system, justifying the ``pseudogap'' designation introduced since 1989,
as a small density of states still remains at $T_{c}$. The corresponding gap
in the spin excitations has indeed been detected by spin lattice relaxation
data\cite{HYasuoka:1989} and neutron scattering experiments \cite%
{JRossat-Mignod:1991}.\ This pseudogap, which has been observed as well
somewhat later by specific heat, STM, ARPES\ etc... (see \cite{TTimusk:1999}
for \ a review on the pseudogap) opens at a temperature $T^{\ast }$\ which
drops sharply with increasing doping, defining a crossover line in the phase
diagram of Fig. \ref{fig:cuprate_phase_diag}.\ This pseudogap is most
probably intimately linked with the correlated nature of these systems, and
its actual physical origin is still intensely debated.\ One class of
interpretations could be the establishment of a hidden order disconnected
from superconductivity such as a spin ordering ( Resonating Valence Bond
(RVB)\ \cite{PAnderson:1997} or d-density wave (ddW) \cite{SChakraverty:2001}%
), or involving orbital currents \cite{CMVarma:2006}, or a charge
segregation into stripe order \cite{EWCarlson:2002}. Another interpretation
is a precursor pairing state, the superconducting phase being only
established at $T_{c}$\ when the pairs achieve long range phase coherence
\cite{VJEmery:1995}. At the present writing the understanding of the
pseudogap state remains controversial.

In this section we shall describe the experiments on point defects performed
in the normal state of the HTSC\ systems, which cover many phases of the
phase diagram introduced hereabove.\ We shall start with a discussion of the
defects which can be used (part \ref{sec:Controlled_defects}).\ Experiments
which allowed one to probe the magnetic response of the CuO$_{2}$\ planes
both on a local scale and on the macroscopic scale will be reviewed then in
part \ref{sec:magproperties}.\ Since the systems in question are metallic,
the defects also cause significant modifications of the transport
properties, which\ differ dramatically in the various parts of the phase
diagram, as will be discussed in part \ref{sec:Nstatetransport}.\ Overall,
these data have some profound implications on the normal state properties of
the cuprates. The most prominent qualitative features will be discussed in
part \ref{sec:Signficance_N}, while comparison with theoretical approaches
will be postponed to Section \ref{sec:thy2D}.

\subsection{\ Controlled defects in and out of the CuO$_{2}$ planes\ }
\label{sec:Controlled_defects}

Impurity substitutions are \textit{a priori} the simplest defects allowing
one to induce well-defined perturbations and to ``tune'' the nature (spin
or/and charge) of the perturbing potential. The ideal experimental situation
would be to choose both the impurity and the atomic substitution site in the
synthesis process.\ Furthermore, one would like to achieve this for any
doping level in the planes, in order to compare the impurity properties in
the various phases of the phase diagram. Unfortunately physical chemistry
considerations play a large role since competing solid state phases
combining the impurity atom and the constituents of the cuprate may exist.
Even in the absence of such phenomena, various substitution sites may occur
depending on the charge reservoir layers of the cuprate material, and
interference between the intended substitution and the ions responsible for
the doping of the CuO$_{2}$\ planes may occur. Efforts have been made to
control both the doping process and the associated disorder effects in the
pure cuprates, as well as the impurity substitutions and/or defect content.

\subsubsection{Hole doping of the cuprates}\label{hdc}

Doping the parent insulating cuprates to span the phase diagram is usually
done by a modification of the layers separating the CuO$_{2}$ planes, either
by a heterovalent substitution or by changing their oxygen content.\ In both
cases, the ions introduced into the structure create a modification of the
Coulomb potential which disrupts the lattice periodicity and which will be
felt as a scattering potential by the carriers in the CuO$_{2}$ layers. The
influence of this source of disorder on the properties of the cuprates was
somewhat neglected in the early studies of the cuprates.

It is clear that dopant disorder is particularly strong if the ionic dopant
is in the separating layer of multilayer cuprates or in the reservoir layer
near neighbor of the CuO$_{2}$ planes.\ For example, Sr$^{2+}$ /La$^{3+}$ in
La$_{2}$CuO$_{4}$ and Ca$^{2+}$/Y$^{3+}$ substitutions in YBCO$_{6} $ induce
hole doping of the CuO$_{2}$\ planes.\ These heterovalent substitutions are
easily achieved, as no other substitution site seems to be available,\ and
enable one to span the phase diagram, which is found to be quite similar for
the two families quoted above, but with rather low optimal $T_{c}\approx
40K.\;$For Y$_{1-y}$Ca$_{y}$ Ba$_{2}$Cu$_{3}$O$_{6}$, the solubility limit
is apparently smaller than $y\approx 0.35$\ required to reach the optimum $%
T_{c}$ \cite{HCasalta:1993}. These systems are quite important to study the
electronic properties of the lightly doped AF charge-transfer insulator.\

In other families of cuprates, the doping is produced by introduction of
extra oxygens in the charge reservoirs, on sites located far from the CuO$%
_{2}$ planes. The optimal $T_{c}$'s found in such cases, i.e. in single
layer Hg-1201, Tl-2201\ or bilayer YBCO$_{6+x}$ and Bi-2212, are higher ( $%
\gtrsim $ 90K) than those found for the heterovalent substitutions mentioned
above. So it has been understood for some time that the dopant disorder
certainly plays a role in limiting $T_{c}$ over all the superconducting
dome. % together with other intrinsic disorder which might be
%specific to each family.\
% PJH--I removed this phrase because we discuss it in more detail below.
This source of disorder is indeed revealed in NMR experiments either on the
NMR widths as a distribution of magnetic hyperfine effects \cite%
{JBobroff:1997} or as a distribution of $T_{1}$ relaxation times \cite%
{PMSinger:2002}. Recent experiments in which the various possible
substitution sites\ have been systematically used to dope Bi2212 have
confirmed elegantly this correlation between the optimum $T_{c}$ and the
proximity of the substitution site to the CuO$_{2}$ plane \cite{HEisaki:2004}%
. The existence of dopant disorder in Bi2212 and its influence on the local
superconducting properties have been recently revealed by STM\ experiments
\cite{KMcElroy:2005}. 

Finally, let us point out that the disorder of chemical dopants is a well
identified source of disorder, but that other sources of disorder do occur
in the cuprate families.\ For instance, the structural modulations due to
the misfit between the lattice parameters of the CuO$_{2}$ plane and BiO
plane in BSCCO, or the tilt of the oxygen pyramids and the stripe structure
around x=0.12 \ in the Ba doped lanthanum compound. This type of \textquotedblleft
disorder \textquotedblright is even less
controllable and dependent on the family.\ Because we are dealing here with
specific impurity effects, we are forced to consider these aspects and also
to select systems in which \textquotedblleft intrinsic
disorder\textquotedblright\ is minimal. Here we include in the definition of
intrinsic disorder both any structural disorder arising in the charge
reservoir or CuO$_{\mathbf{2}}$\ layers, as well as that due to the random
distribution of the dopant atoms.

\begin{figure}[t]
\begin{center}
\leavevmode
\includegraphics[clip=true,width=.85\columnwidth]{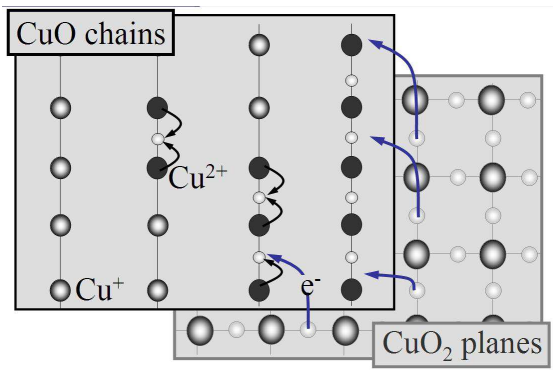}
\end{center}
\caption{An oxygen inserted between two copper sites in a vacant chain row
only changes the (initially) Cu$^{+}\;$ into Cu$^{2+}$.\ This does not
transfer any charge to the CuO$_{2}$\ plane.\ When a chain segment with at
least two oxygens is formed charge transfer to the plane occurs. Full chains
as that shown on the right of the figure might then form an ordered pattern,
with specific values of the hole content in the planes, and minimal
disorder. }
\label{fig:ybco_doping}
\end{figure}

We highlight here the particular case of YBCO$_{6+x}$ for which the hole
doping of the parent compound YBCO$_{6}$, which has no oxygen in the
intermediate pure Cu(1) layer, is obtained by inserting oxygens in this
layer.\ As can be seen in Figure \ref{fig:ybco_doping}, the charge transfer
is only initiated when $x$ is large enough to give near neighbor oxygens.\
With the growth of chain segments the hole content $n_{h}$ of the planes
increases, but does not scale with $x$.\ Thus, the phase diagram versus $x$
differs markedly in appearance from that obtained by heterovalent
substitutions for which $n_{h}$ is controlled directly by the atomic
substitution. At fixed values of $x$, ordered oxygen structures are formed,
the most prominent one being for $x=1$, with filled Cu-O chains, which
corresponds to slight overdoping of the plane (with $T_{c}$ about 1$%
%TCIMACRO{\U{b0}}%
%BeginExpansion
{{}^\circ}%
%EndExpansion
K\;$lower than optimal).\ For $x=0.5\;$ the ortho-II\ phase with alternating
filled and empty chains can be formed and corresponds to a good underdoped
situation with a well defined large pseudogap. But again for $x>0.5$\ the
added oxygens between chains (see Fig. \ref{fig:ybco_doping}) do not dope
the planes as long as two oxygens are not near neighbors, that is ideally in
an ordered structure up to $x=0.75$, which explains the $T_{c}$ plateau at
60K around $x\simeq 0.6$ .\ For these simple reasons, $x=1$ and $x\simeq 0.6$
are important compositions characteristic of the near-optimal and pseudogap
phases, which can be easily reproduced.\ Furthermore, at $x\simeq 0.6,$ the
ordering of the chains, and the reduced effect of the Coulomb potential from
the Cu$^{2+}\;$- O$^{2-}$-\ Cu$^{2+}$ units, with an \ overall charge
equivalent to 2 Cu$^{+}$, promotes a situation where intrinsic disorder is
much less important than in most other cuprate systems, as has been
highlighted by NMR measurements \cite{JBobroff:2002}. The fact that the
ordered chain phases are important fixed points of the phase diagram has
been thoroughly investigated by the Vancouver group \cite{RXLiang:2000}.
YBCO is therefore quite an interesting family for impurity substitution
studies, since once a substitution has been performed, the undoped AF, the
pseudogap phase and the near optimal phase are accessible with simple heat
treatments and correspond to phases with weak intrinsic disorder. 
%Let us also point out that
 Y1248 \cite{JKarpinski:1988}, a variant of YBCO\
with a well ordered two chain layer, is also a very interesting case of
underdoped pseudogap compound with little disorder.

\subsubsection{Impurity substitutions in the cuprates}
\label{sec:Impurity_subs}

Since the early 1990s, impurity substitutions have been thoroughly studied
in the two preferred cuprate families YBCO\ and LSCO, and later, after the
development of ARPES and STM experiments, data on substituted BSCCO have
been taken as well. The easiest path to ensure that an impurity substitutes
on a given site is to assume a substitution in the ceramic synthesis and
check by x-rays the absence of impurity phases which definitely occur if the
substituent does not enter in phase, or substitutes on an alternate site.
This allows one to %%PJH--I guessed at what was meant by "enter" here.
perform Rietveld analysis of the powder patterns (see for instance \cite%
{HMaeda:1989}), to measure the variation of lattice parameters and to
determine the solubility limit in solid state.\ Synthesis of single crystals
does not allow similar checks but can be calibrated with respect to results
on ceramic samples.\

It is important to note that the physical chemistry complications linked
with substitutions raises serious problems if one deals with a system which
has some intrinsic disorder, as correlation between the impurity locations
and the intrinsic disorder do occur naturally, and therefore these effects
are not additive.\ So studying impurity effects directly will be easier in
families of cuprates with minimal intrinsic disorder.

\paragraph{Homovalent substitutions in the separating layer of multilayer
cuprates.}

%\textit{Rare earth substitutions on Y}. \
Substitution on the Y$^{3+}$ sites by rare earths, allowing magnetic probes
in the blocking layers, have been thoroughly performed in the YBCO family. \
Most rare earth moments apparently interact very weakly with the CuO$_{2}$
planes, however, since superconductivity is nearly unaffected even for the
pure compounds such as DyBaCuO or GdBaCuO \cite{PHHor:1987}. Apart from
applications purposes, the main interest of these substitutions is to
provide new spin probes for the experimental study of the magnetic
properties of the CuO$_{2}$ layers.\ A few per cent of Gd substituted
allowed thorough studies by Gd$^{3+}$\ ESR\ \cite{AJanossy:1994}, while Yb
substitutions allows one to use Yb$^{3+}$\ as a M\"{o}ssbauer effect probe
\cite{JAHodges:1995}.\ These techniques provide hyperfine probes similar to
the $^{89}$Y nuclear spin probe, but allow the exploration of different $T$
ranges as well as different electron spin relaxation time windows.

Pr acts differently from the other rare earths as it reduces the charge
carrier content so that PrBaCuO$_{7}$ is antiferromagnetic and resembles the
parent compound YBCO$_{6}$.\ At small Pr content, it has been demonstrated
that Pr hybridizes more with the O of the CuO$_{2}$ planes than the other
rare earths, and that its effect is then to reduce the hole content in
planes by filling holes on the neighboring oxygen \cite{WAMacFarlane:2002}.
In this picture, the effective charge of Pr is then closer to 4+.\ However,
the corresponding perturbation in the plane is found to display a magnetic
character as well.

\paragraph{Substitutions on the Copper sites}

%\textit{Substitution of 3d elements on Cu:}
The most important substitutions are the Zn and Ni substitutions on the
Cu(2) sites of the CuO$_{2}$\ planes which allows one to replace the Cu 3%
\textit{d}$^{9}$ $S=1/2$ site by nominal $S=0$\ and $S=1$ impurities,
respectively. These substitutions have been easily performed in LSCO as they
only take place on the single Cu site of the structure.\ But as this system
contains significant intrinsic disorder due to Sr$^{2+}$/La$^{3+}$
substitution, correlated disorder effects might severely modify the
influence of the impurities.\

In YBCO, the essential difficulty comes from the possible substitutions on
either the chain or plane Cu sites. It is easy to ensure that an impurity
substitutes for Cu, by checking \ with x-rays the absence of impurity
phases. % which occur if the
%substituent does not enter or substitutes on an alternate site.\
%PJH--redundant
But once it is ensured that up to 4\% Zn and 3\%\ Ni do substitute for Cu,
thorough studies are required to determine on which site this substitution
occurs.\ Various experimental investigations using EXAFS \cite{HMaeda:1989}
or XANES\ \cite{CYYang:1990} suggest that Zn substitution preferentially
occurs on Cu(2) in the CuO$_{2}$\ planes while Ni substitutes on both the
chain Cu(1) and planar Cu(2) sites. In the case of Zn the observation by NMR
of the $^{89}$Y\ near neighbors of Zn allowed one to ensure, from the
integrated intensity of this line, that more than 90\%\ of the nominal Zn
content substitute on the Cu(2) site, within experimental accuracy \cite%
{AVMahajan:1994}. Since such an experiment is impossible in the case of Ni,
the equal probability of atomic substitutions on the Cu(1) and Cu(2) sites
given by structural studies could not be controlled independently. Finally,
Li has also been found to substitute solely on the Cu(2) site \cite%
{KSauv:1996} in YBCO, with the actual Li content being again determined from
$^{89}$Y\ \ near neighbor intensity measurements \cite{JBobroff:1999}.\

Many impurities have been found to substitute preferentially on the chain
Cu(1) site, such as Co$^{3+}$, Fe$^{3+}$ or Ga$^{3+}$(\cite{FBridges:1989},
and references therein).\ The impurities on the chain site donate their
extra electrons to the planes and also interrupt the chains, which further
reduces the charge transfer to the planes \cite{PFMiceli:1989}.\ These 3+
impurities often do not like the square oxygen coordination of the chain
site and accomodate extra oxygens, which partly compensate the reduction of
hole doping and maintains a tetragonal structure of the compound whatever
the oxygen content, as for instance for Ga$^{3+}$\cite{GXiao:1988}.\ The
dominant overall effect of such substitutions \emph{on the chain site} is a
reduction of the hole doping of the CuO$_{2}$ plane,\ making the optimally
doped composition impossible to reach. \ In many cases, even for Ni \cite%
{SAdachi:2000}, it has been found that heat treatments in controlled
atmosphere do allow partial transfers of impurities between the Cu(2) and
Cu(1) sites.

\paragraph{Defects induced by electron irradiation.}

To control defect creation in a solid, one may use irradiation by energetic
particles, preferably light ones which have a large enough penetration
length to damage bulk samples.\

Ions or even protons produce a large density of defects which recombine into
clusters of defects.\ In particular high energy heavy ions yield linear
tracks of heavily damaged material which can be used to pin vortices in the
SC state \cite{MKonczykowski:1991,LCivale:1991}. To produce isolated point
defects, the most suitable method is to irradiate with high energy (MeV)
electrons at low $T,$\ optimally in liquid H$_{2}\;$at 20K, which avoids
recombination of the point defects created.\ In cuprates, this allows the
production of Cu and O vacancies in the planes, the ejected Cu and O being
trapped in interstitial locations out of the CuO$_{2}$\ planes.\ Upon
heating the sample, some of the created defects recombine, the oxygens first
as their mobility is larger \cite%
{ALegris:1993,JGiapintzakis:1994,STolpygo:1996}. The complete set of studies
performed so far indicate that the damage is very homogeneous on samples of
thickness $<50\ \mu m$, and that the vacancies in the CuO$_{2}$%
\ planes dominantly modify the transport and superconducting properties,
these effects being even quantitatively identical to those produced by Zn
substitutions \cite{FRullier-Albenque:2000}. If this method applies ideally
for transport studies in single crystals of small thickness, it
unfortunately requires too much electron beam time for the large amount of
sample required for bulk or NMR experiments.\

\subsection{{Magnetic properties induced by in-plane impurities}\label%
{sec:magproperties}}

\subsubsection{{Impurities in the AF\ phases \ \ }\label{sec:imps_in_AF}}

The parent compounds of the cuprates appear to be good Mott insulators, with
a very large Heisenberg exchange interaction $J\simeq 1500$K \cite%
{KBLyons:1988}.$\;$This is revealed by neutron scattering experiments which
display an inelastic scattering peak at the AF wave vector $(\pi ,\pi )$ .\
The q-width of this peak gives a quantitative estimate of the dynamic AF
correlation length $\xi (T)$, which fits perfectly at high $T$ in La$_{2}$CuO%
$_{4}\;$ the $\exp (J/k_{B}T)$ variation expected for the Heisenberg 2D\
square lattice \cite{RJBirgenau:1989}.\ However,\ while a spin 1/2\
Heisenberg 2D system should only order magnetically at $T=0$, the undoped
cuprates display a well established 3D\ AF\ order with N\'{e}el temperatures
$T_{N}$ , which should be largely dependent on the magnitude $J_{\bot }\;$of
the interplanar couplings.\ As this is not the case \cite%
{HAlloul:1990,PMendels:1991} it has become clear that the ordering is
governed instead by a deviation of the plane Hamiltonian from the ideal
Heisenberg situation.\ Indeed\ a very small XY anisotropy of the exchange
interaction drives a Kosterlitz-Thouless transition in the 2D\ system
at a temperature $T_{KT}$\ which is slightly modified by 3D interactions
\cite{HAlloul:1991b,BKeimer:1992}.

\paragraph{Substitution effects in the undoped phase.}

What happens when substituting a Cu by a nonmagnetic or magnetic impurity?
In YBCO$_{6}$ magnetic impurities substituted on the chain Cu sites are only
found to change the\ coupling between bilayers in such a way that planes of
neighboring bilayers switch from an AF ordering to a ferromagnetic ordering,
still retaining the bilayer AF order \cite{HLutgemeier:1988,AVDooglav:1996}%
.\ But $T_{N}$, which is fixed mainly by $T_{KT}$,\ is not modified
significantly by this change of 3D\ magnetic structure of the ground state.
For a nonmagnetic impurity such as Zn, which substitutes on the planes, the
dominant effect is the dilution of the magnetic lattice, which has been has
been studied \cite{OPVajk:2002} and is found to reduce $T_{N}$ moderately%
\textbf{\ }at low concentration as a result of the reduction of ground state
energy and of $T_{KT}$.\ Overall, the influence of substituted impurities on
the macroscopic properties of the ground state do not tell us much about
the single plane physics.

In this AF state, local measurements of the modifications induced by a
nonmagnetic defect are somewhat difficult. The main observation done so far
has been the reduction of intensity of the $^{63}$Cu NMR signal detected in
the internal magnetic field without external applied field (Zero Field NMR:
ZFNMR) \cite{PMendels:1988}.\textbf{\ }This gives an indication of the
spatial range over which the impurity induces a sizable modification of the
AF order \cite{PMendels:1990}. So far these experiments did not enable a
detailed study of the local modifications of the magnetic order, as their
analysis depend on the modifications of the various parameters mentioned
above and are not solely associated with the single plane physics and its
spin liquid state. \textbf{\ }Experiments above $T_{N}$\ in the paramagnetic
state allowed one to detect modifications of the spin dynamics\textbf{\ }%
from NMR\ spin lattice relaxation data\textbf{\ }\cite{PCarretta:1997}, but
again no local study of the magnetic properties around impurities could be
achieved.%
% , probably for sensitivity considerations.

\paragraph{Impurities in the doped antiferromagnetic state.}

Let us first recall the effect of small hole dopings ( below 10\%) in the
``pure'' systems. One important fact learned early on from studies on LSCO\
is that carriers abruptly reduce $T_{N}$, less for electron doping than for
hole doping \cite{RJBirgenau:1989}, but much more than spinless homovalent
substitutions, such as Zn$^{2+}$.\ A similar decrease of $T_{N}$ occurs for
hole doping by Ca in Y$_{1-y}$Ca$_{y}$Ba$_{2}$Cu$_{3}$O$_{6}\;$\cite%
{HCasalta:1993}.\ In any case these carriers are mobile at high $T$ as the
planar resistivity increases above room $T$ as in a metal \cite{YAndo:2004}.
They should localize at low $T$ in the Coulomb potential of the dopant, so
that non generic modifications of the AF state are expected.\ Although
detailed studies have not been performed so far, $T_{N}$\ suppression is
however achieved for a comparable doping of $n_{h}\simeq 0.03$ for Sr and Li,
which correspond respectively to out of plane and in plane coulomb
potentials \cite{BJSuh:1998}.\ It has been generally seen that a new
``phase'' appears below $T_{N}$ , which contains some amount of static
disorder, and which has some characteristics of a mixed AF/spin glass phase
\cite{FCChou:1993}.\

For hole dopings larger than $n_{h}\simeq 0.03$, the ordered magnetic state
disappears and gives place to a static disordered magnetic phase detected by
magnetization measurements, $\mu $SR or NMR \cite{JHCho:1992,FCChou:1995,CNiedermayer:1998}.\ This latter phase has many characteristics of a true spin
glass (SG) phase. In some systems like the lanthanum family, it extends over a large doping
range, between $0.03$ and $0.07$ but still exists far into the SC state
nearly up to optimal doping $x=n_{h}=0.15$, albeit with a low spin glass
temperature $T_{g}$ \cite{CNiedermayer:1998}.\ Whether the tendency to stripe formation at $x=0.12$
in the Ba doped case favors a coexistence of the SG and SC\ phases on a
microscopic scale, or leads to a phase separation is not fully clarified.\
In YBCO$_{6+x}$\ the SG\ phase has been occasionally detected for
intermediate oxygen contents $x\simeq 0.45$ but has totally disappeared for $%
x\simeq 0.6$, that is $n_{h}\simeq 0.10$. So it is clear that the SG phase
is not generic in the phase diagram of the ``pure systems", its extension
being linked with the intrinsic disorder present in the systems.\ In a
specific family of bilayer cuprates CaBaLaCuO, systematic studies reveal a
clear relation between the $T_{g}$ and $T_{c}$ values in the phase diagram,
which suggests that the influence of the intrinsic disorder in these systems
affects these temperatures similarly \cite{AKeren:2003}.

\begin{figure}[t]
\begin{center}
\leavevmode
\includegraphics[clip=true,width=.99\columnwidth]{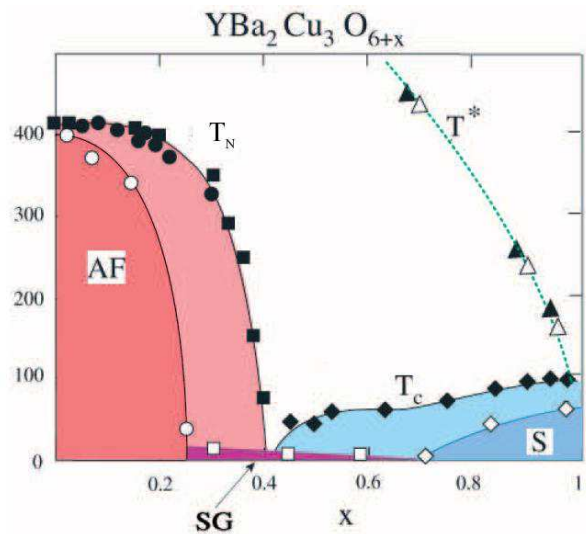}
\end{center}
\caption{The phase diagram of pure YBCO$_{6+x}$ (full symbols) only exhibits
a small range of oxygen contents, near the tetragonal to orthorhombic
transition at $x=0.4$ for which disordered static magnetism occurs at low $T$%
.\ With Zn substitution, at a concentration of \ 4\% (open symbols), the
superconducting range is highly depressed, the local magnetism induced by Zn
drives the appearance of a\ large spin glass (SG) regime.\ The phase diagram
becomes then similar to that of Fig. \protect\ref{fig:cuprate_phase_diag}.\
Notice that the onset of pseudogap opening $T^{\ast }$ is not modified by Zn
substitution (Figure taken from \protect\cite{HAlloul:2000}).}
\label{fig:phase_diagram_Zn}
\end{figure}

The introduction of spinless impurities adds to the existing disorder in
these phases. It is remarkable that Zn substitution in YBCO\ extends
significantly the SG\ phase range in this system \cite{PMendels:1994}(see
Fig. \ref{fig:phase_diagram_Zn}), so that the phase diagram of YBCO:Zn$%
_{0.04}$ resembles that observed in nominally pure LSCO, or in Y$_{1-y}$Ca$%
_{y}$BCO$_{6}$.\ This suggests again that extrinsic or intrinsic disorder
may have similar influences on the properties of these systems at least in
the low hole doping range.\emph{\ }

%To conclude
 Regarding the influence of controlled impurities in the
intermediate doping range, let us once again emphasize that we are not dealing
with single impurity effects, and that the interferences with the intrinsic
disorder of the ``pure" systems complicate the situation markedly. As will
be seen in Section \ref{bulkexptSC}, %\textbf{\ section VIE}
Zn impurities in LSCO\ around $n_h$=0.10 obviously nucleate a long range
magnetic order detected by neutron scattering experiments which is not
present in the pure system. This kind of order induced by disorder bears
some analogy with the situation encountered in spin chains for which
magnetic order can be enhanced by impurities (see Section \ref{sec:1Dimplro}%
). Although many qualitative and materials aspects are revealed by the
experiments involving impurity substitutions, such observations are even
harder to analyze theoretically than those done in the undoped AF case 
(see however Section \ref{sec:thy2D}).

\subsubsection{Impurities in the pseudogap phase above $T_{c}$}

\begin{figure}[t]
\begin{center}
\leavevmode
\includegraphics[clip=true,width=.9%
\columnwidth]{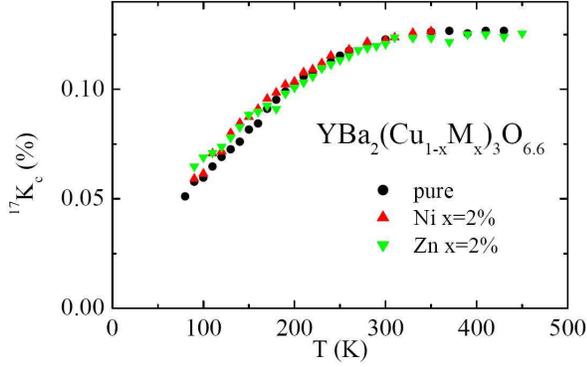}
\end{center}
\caption{The $^{17}$O\ NMR\ shift versus $T$\ for YBCO$_{6.6}$\ pure and
substituted with 2\%\ Zn and 2\%Ni.\ The NMR shift for the sites far from
the Zn is found totally independent of the substitution, with the onset of
opening of the pseudogap remaining at $\approx $350K for all samples. }
\label{fig:Zn_effect_on_pseudogap_Knightshift}
\end{figure}

When the doping is large enough to destroy the static AF phase and to
establish a metallic and superconducting state, the magnetic properties
display, in the underdoped regime, a ``pseudogap'' (see Fig. \ref%
{fig:cuprate_phase_diag}). It opens up at a very high 
temperature $T^{\ast }$ ($\approx $ 350 K in YBCO$_{6.6}$) and
is signalled,e.g., by a decrease in the spin
susceptibility (see Fig.\ \ref{fig:Zn_effect_on_pseudogap_Knightshift}).
%which occurs at a very high temperature $T^{\ast }$ ($\approx $ 350 K in YBCO$%
%_{6.6},$).
Here we are in an interesting doping range where the system is a
metal with magnetic correlations, which is a hard theoretical
problem. This is the regime for which the use of impurities to probe the
physical properties was expected to be the most fruitful. Such an approach
was started in the early 1990's and provided first basic qualitative
information. Improvement of the samples and experimental techniques have
since allowed one to obtain more refined quantitative information on the
perturbations introduced by the impurities. A major effort has been devoted
to the case of Zn, Ni and Li impurities substituted for Cu in the YBaCu O$%
_{6+x}$\ system specifically for $x\simeq 0.6$\thinspace\, for which the
intrinsic disorder is minimal.

\paragraph{Influence of impurities on the pseudogap}

The first point which was investigated in the early experiments was the
influence of the substituted impurities on the temperature $T^{\ast }$ at
which the pseudogap opens. As can be seen in Fig. \ref%
{fig:Zn_effect_on_pseudogap_Knightshift}, the NMR shift of the nuclei
located far from the impurities is not affected by the impurities Zn or Ni,
 even when the impurity content is large enough to completely
suppress superconductivity \cite{HAlloul:1991a,TMRiseman:1994,JBobroff:1997b}%
.

\begin{figure}[t]
\begin{center}
\leavevmode
\includegraphics[clip=true,width=.9\columnwidth]{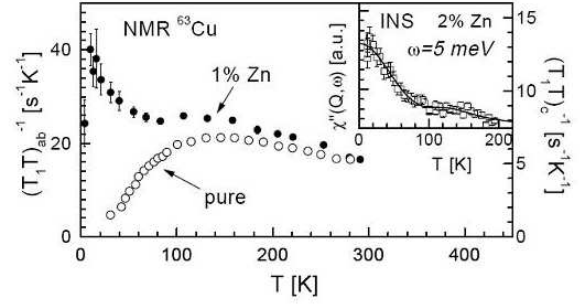}
\end{center}
\caption{The $^{63}$Cu NMR$\;(T_{1}T)^{-1}$\ data taken in Zn substituted
YBCO$_{6.6}$\ display a large increase at low $T$, which is also seen in the
inelastic neutron scattering data taken at the AF\ wave vector $(\protect\pi %
,\protect\pi )\;$shown in the inset.\ Both data are dominated at low $T$ by
the contribution of sites near the Zn, but still display an evidence for the
onset of the pseudogap as a high $T$ maximum of the dynamic susceptibility
around 150K, seen in the pure samples \protect\cite{MHJulien:2000}.}
\label{fig:Zn_effect_on_pseudogap_T1+neutrons}
\end{figure}

Most probes sensitive to the normal state pseudogap
far from the defects confirmed these finding later on, and showed that they applied
to any kind disorder, such as irradiation defects or Li impurities.
%\ This was confirmed later on, for irradiation defects, Li, \ and any kind of
%disorder, by most probes sensitive to the normal state pseudogap
%far from the defects.\
 Some early inelastic neutron scattering (INS) \cite%
{KKakurai:1993}, and $^{63}$Cu $T_{1}$\ experiments \cite{GQZheng:1993}
claimed a filling of the pseudogap in the spin excitations at the wave
vector ($\pi ,\pi $), in the presence of impurities. However more detailed
data taken later have confirmed that the Zn impurities create low energy
midgap excitations \cite{HHarashina:1993, PBourges:1996}.\ These
excitations, which will be described in detail in part \ref{sec:spindynamics},%
are associated with the fluctuations of the local moment AF\ staggered
response spatially localized near the Zn.\ They partially fill in the
% actual higher energy
 pseudogap which characterizes the behaviour far from the Zn; this is
%which remains 
apparent in both the $^{63}$Cu $T_{1}$\ data and the INS, as
can be seen in Fig. \ref{fig:Zn_effect_on_pseudogap_T1+neutrons}. Indeed,
NQR\ investigations \cite{YItoh:2003} in which the resonances of the sites
far from the impurity could be singled out clearly reveal the persistence of
the pseudogap for the $T_{1}$\ data of these sites (see Fig. \ref%
{fig:Zn_effect_on_pseudogap_T1far}).

\begin{figure}[t]
\begin{center}
\leavevmode
\includegraphics[clip=true,width=.9\columnwidth]{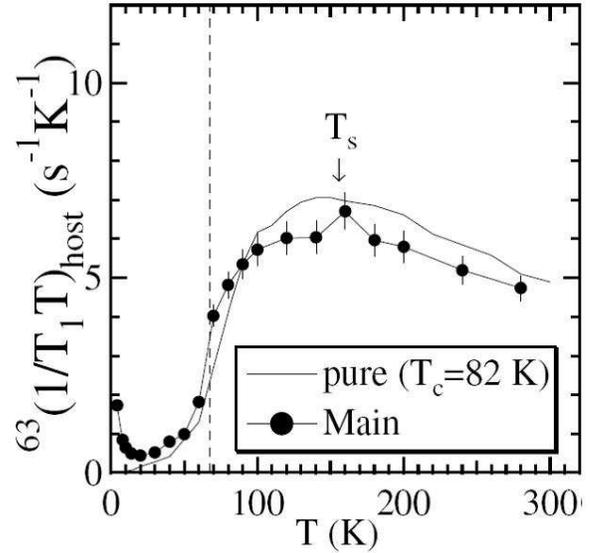}
\end{center}
\caption{The $^{63}$Cu $\ $nuclei far from the Zn impurities\ display in\
YBCO1248 the same reduction of $(T_{1}T)^{-1}$ as that found in the pure
compound below the pseudogap temperature \protect\cite{YItoh:2003}.}
\label{fig:Zn_effect_on_pseudogap_T1far}
\end{figure}

So this body of experiments give consistent evidence that the pseudogap
is not modified far from the Zn site, while the local magnetic properties are
strongly modified near the Zn location. This is also confirmed by other experiments\
which also allow one to detect the onset of the pseudogap from macroscopic
data, such as specific heat \cite{JCLoram:1994}, transport for Zn
substitutions \cite{DJCWalker:1995,KMizuhashi:1995}, and also irradiation
defects \cite{FRullier-Albenque:2003}.

\paragraph{Local moments induced by Zn.}

The first question which arises is whether a nonmagnetic site induces a free
paramagnetic moment in a \textit{metallic} correlated system, as was seen in
the case of undoped spin chains and ladders. Ideally, this might be probed
by macroscopic susceptibility measurements, or by electron spin resonance
(ESR) experiments. A theoretical proposal for the occurrence of a local
moment with an ESR experiment in LSCO:Zn\ to support this view was given
\cite{AMFinkelstein:1990}.\ Although the theoretical idea was quite
valuable, the detected signal could not be associated with isolated Zn
impurities, as will be shown below in part \ref{sec:spindynamics}. Indeed,
the early experiments were plagued by materials problems.\ For instance, ESR
signals detected in YBCO samples were exclusively due to the so-called
``green phase'' Y$_{2}$CuO$_{5}$, which also induced Curie terms in the
susceptibility of ``pure'' samples.\ So, although some initial
susceptibility data on Zn substituted YBCO\ samples did suggest evidence of
Curie contributions, they could not be considered \ as proofs of the
existence of impurity induced paramagnetism \cite{JRCooper:1991}. The actual
contribution of substituted impurities to the magnetic susceptibility has
only been determined since, in very carefully controlled samples, as we
shall see below.

\begin{figure}[t]
\begin{center}
\leavevmode
\includegraphics[clip=true,width=.9\columnwidth]{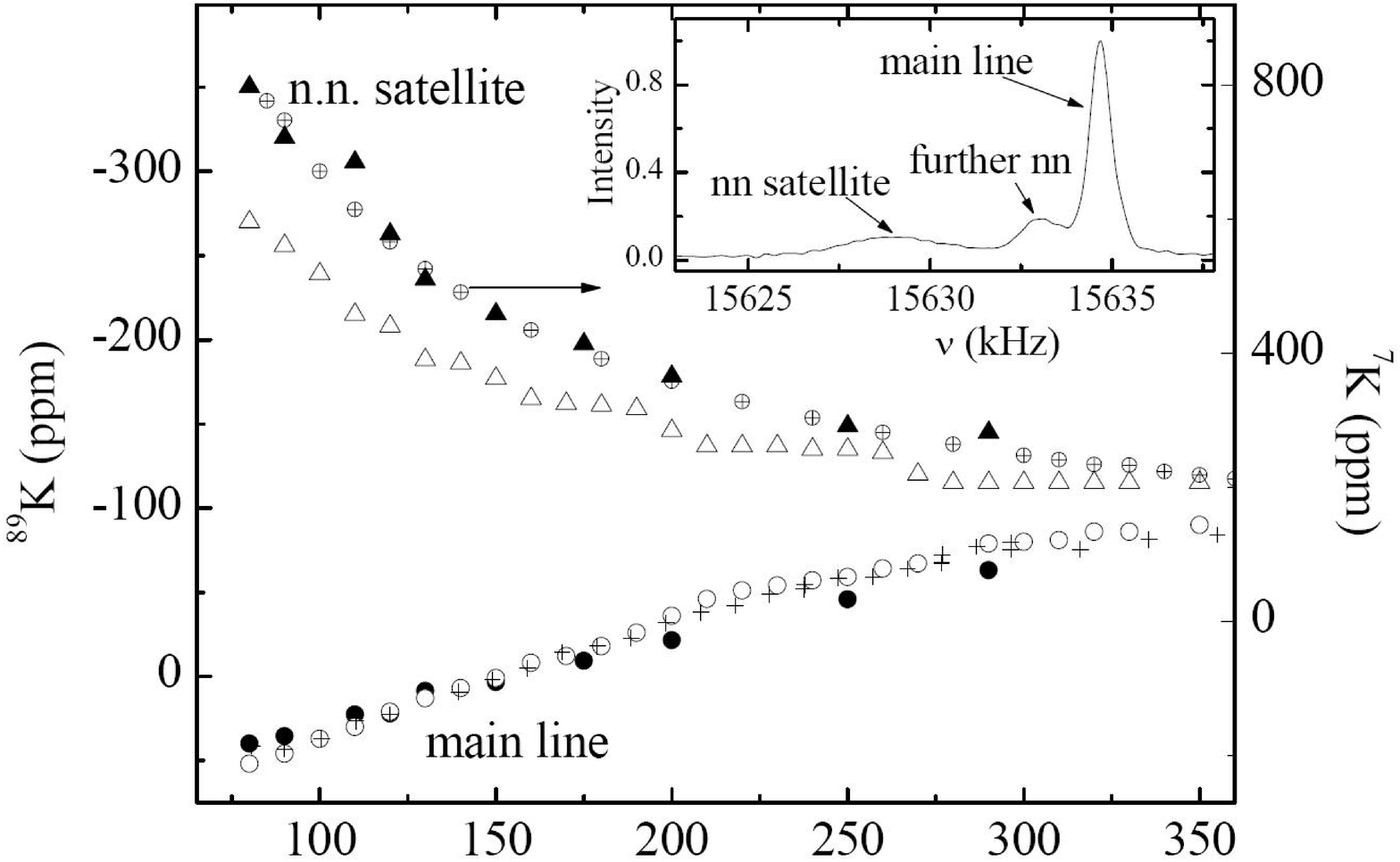}
\end{center}
\caption{The $^{89}$Y\ NMR spectrum of YBCO$_{6.6}$:Zn (or equivalently YBCO$%
_{6.6}$:Li) exhibits a main line and two satellite resonances as shown in
the inset. Their shifts $^{89}K$ are plotted versus $T\ $as empty and full\
symbols respectively for Zn and Li substitution. While the main line shifts
(circles) retain the characteristic pseudogap$\;T$ variation of the pure
sample (crosses), the \textit{n.n.}\ satellite shifts depart markedly at low
$T$, the difference of shift with the main line displaying a Curie $T^{-1}$\
variation. This results from the large Curie like spin polarization which
appears on the four Cu sites around the Zn or Li, two of them being directly
sensed by the $^{89}$Y \textit{n.n.}$\ $to the Zn (see Fig.\protect\ref%
{fig:alternating_magnetization_schematic}).\ \ The $^{7}$Li NMR shift (cross
in a circle, right scale) directly probes as well this Curie polarization of
its four Cu neighbours \ (figure constructed from \protect\cite%
{JBobroff:1999}). }
\label{fig:Y_NMR_satellites}
\end{figure}

The first indirect but unambiguous evidence that Zn induces a \ dynamic
paramagnetic moment in an underdoped cuprate was obtained by monitoring the $%
^{89}$Y\ NMR\ linewidth in YBCO$_{6.6}$:Zn$_{y}$ \cite{HAlloul:1991a}. A
nonmagnetic impurity in a metallic host normally only yields a $T$
independent broadening of the unshifted host NMR signal due to Friedel
oscillations of the density of states.\ By contrast, the unshifted $^{89}$Y
\ NMR\ has a linewidth which is found to increase markedly at low $T$, while
the susceptibility of the host decreases due to the pseudogap.\ This reveals
that the spatially dependent spin susceptibility induced by the Zn
impurities displays a $T\,$dependent paramagnetism which increases at low $T$
and is therefore ``Curie-like'', as illustrated in the case of \ a spin
chain (see Fig. \ref{FigPedagogicNMR}).\ The only difference is that here
the signals of the different near neighbor shells of the impurity are
overlapping and just yield a broadening of the line. The actual $T$
dependence of the susceptibility could not be ascertained by such a
measurement, since the relationship between the magnitude of the induced spin
polarization and the magnetic susceptibility is not straightforward, as will
be clear from the later work described below.\ That a Curie like moment
developed in the vicinity of the Zn was clearly confirmed by resolving with
dilute samples the satellite NMR\ signal of the $^{89}$Y\ near neighbor (%
\textit{n.n.)} sites of the substituted Zn \cite{AVMahajan:1994} (see Fig. %
\ref{fig:Y_NMR_satellites}). This provided the first local detection of the
field induced paramagnetism near the Zn, well before the equivalent
information could be monitored in the case of spin chains.\ These data imply
that the spin polarization of the Cu \textit{n.n.} to Zn is already at 100
K\ \textit{\ more than ten times larger than that of the pure host}, so this
is not a mere minor modification of the host density of states, but a strong
effect of the electronic correlations of the system, similar to that observed in 
spin chains. The very short spin lattice relaxation $T_{1}\,$of the $%
^{89}$Y \textit{n.n.} NMR at low-$T$ is a further striking evidence which
establishes the occurrence of a local moment induced by the spinless Zn
impurity. Quite analogous $^{89}$Y \textit{n.n.} NMR\ data were obtained
later on Zn substituted YBCO1248, which is also a low disorder underdoped
cuprate, thus confirming the generality of these observations \cite%
{GVMWilliams:1998}. \

This Zn induced paramagnetic contribution to the macroscopic susceptibility $%
\chi _{c}$ could then be determined from SQUID magnetization data taken on
samples in which the fraction of parasitic impurity phases could be
minimized and remained only as traces \cite{PMendels:1994b,SZagoulaev:1995}.\
For $x=0.66$, a Curie-Weiss law
\begin{equation*}
\chi _{c}=\mu _{eff}^{2}/\left[ 3k_{B}(T+\Theta )\right]
\end{equation*}
could be associated with Zn induced magnetism down to $T\approx $10K, with a
small value $\Theta \lesssim $ $4K$ for $x=4\%$ and an effective moment $\mu
_{eff}=1.0\mu _{B}\;$\cite{PMendels:1999}.

With the known hyperfine couplings for the $^{89}$Y\ nucleus, the spin
polarization on the 4 Cu \textit{n.n.} to the Zn would correspond to a total
susceptibility of the same magnitude as $\chi _{c}.\,\,$This was initially
taken as an indication that the moment was mainly localized on the 4\ Cu
first \textit{n.n.} to the Zn \cite{AVMahajan:1994}.\ But the data only
allows one to conclude that the sum of the staggered magnetic susceptibility
on more distant sites roughly cancel out and contribute little to $\chi _{c}$%
\cite{AVMahajan:2000}.$\;$

The $^{63}$Cu nuclear spin itself, which is more sensitive to the on site
magnetization, allows one to probe this long distance staggered magnetism
\cite{REWalstedt:1993,MHJulien:2000}. The large increase of the $^{63}$Cu
linewidth induced by Zn established the existence of long distance spin
polarization effects but did not allow to provide enough quantitative
information to correlate them with the short distance information gained
from $^{89}$Y NMR data.\

\begin{figure}[t]
\begin{center}
\leavevmode
\includegraphics[angle=0,clip=true,width=.9%
\columnwidth]{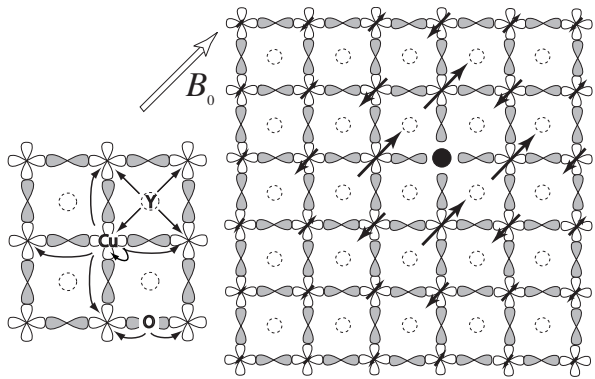}
\end{center}
\caption{The Cu 3d$_{x^{2}-y^{2}}$\ and O2p$_{x}$, O2p$_{y}$ orbitals of the
CuO$_{2}$\ plane are represented, with a Li or Zn impurity substituted on
one Cu site. The magnetization induced on each Cu site by the applied field
is staggered and decreases in magnitude to reach the very small uniform
value found in the pure compound. In the inset we display the hyperfine
coupling paths linking the different nuclear spin sites with their near
neighbor Cu orbitals (also with the on site orbitals for Cu). The Y sites
are below the CuO$_{2}$ plane (and above the other CuO$_{2}$\ plane of the
bilayer). }
\label{fig:alternating_magnetization_schematic}
\end{figure}

\paragraph{The Ni impurity: a weakly coupled local moment and the first
evidence for a large$\;\protect\xi _{imp}$.}

The Ni impurity,\thinspace which is in a 3d$^{8}$ state, was expected to
display significantly different magnetic properties owing to its on site
local moment.\ In this case, the macroscopic spin susceptibility $\chi
_{c}\, $per Ni site was found to be indeed larger than for Zn, with $\mu
_{eff}=1.7\mu _{B}\;$for$\;x=0.66\;$ nearer from that expected for a spin $%
S=1/2$\ than for a spin $S=1$, so that the local moment is somewhat reduced
with respect to that expected for a free ion \cite{PMendels:1999}. A crude
explanation would be that the 3d$_{x^{2}-y^{2}}$ component of the Ni moment
couples to the CuO$_{2}\;$band and would roughly replace the Cu spin, while
the 3d$_{z^{2}-r^{2}}$ orbitals would contribute to a spin 1/2\ local moment
weakly coupled with\ the CuO$_{2}$ band. This conjecture is supported by the
evidence that the broadening of the $^{89}$Y NMR\ \cite{TMRiseman:1994}, is
dominated by the dipole field of the Ni local moment, and is only slightly
affected by the staggered magnetism on the Cu sites, no \textit{n.n.}\ $%
^{89} $Y line being even detected in that case.

\begin{figure}[t]
\begin{center}
\leavevmode
\includegraphics[clip=true,width=.9\columnwidth]{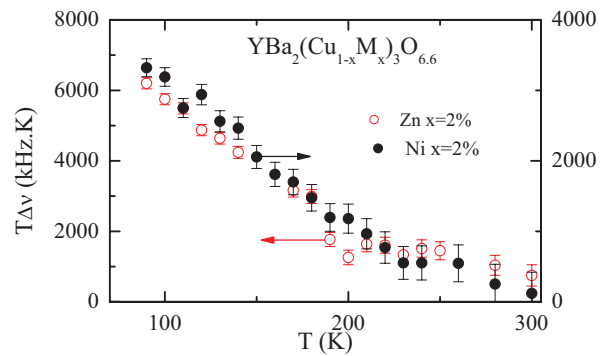}
\end{center}
\caption{The difference $\Delta \protect\nu \;$of $^{17}$O NMR\ linewidth
between the substituted and pure YBCO$_{6.6}$ samples gives a measure of the
width of the histogram of local fields induced by Zn or Ni on the $^{17}$O\
nuclei.\ The product $T\Delta \protect\nu \;$increases markedly at low $T$,
\ while it should remain constant if the spin polarization had a mere Curie $%
1/T\ $ variation.\ This is evidence that both the magnitude of the local
fields and the width of their histogram, that is the decay length \ $\protect%
\xi _{imp}$ of the spin polarization, increase sizably at low $T$ (data
taken from \protect\cite{JBobroff:1997b} and \protect\cite{SOuazi:2004})%
\textbf{.}}
\label{fig:Ni_probes_correlations}
\end{figure}

On this basis, it was clear that data on the $^{17}$O NMR, which has larger
hyperfine couplings with the Cu(2) magnetization, would be less sensitive to
the direct dipolar local fields induced by the local moment and would give
more information on the staggered paramagnetism. The interesting result
obtained was that the $^{17}$O\ linewidth did not increase at low $T$ as the
Curie $1/T$ \ behavior displayed by $\chi _{c}$ but much faster (see Fig. %
\ref{fig:Ni_probes_correlations}).\ This was considered a proof that in
contrast with expectations for RKKY\ oscillations in a non-correlated
electron host \cite{JBobroff:1997b}, the $T$ variation of the linewidth is
not solely due to that of the magnitude of the staggered susceptibility but
also that its decay length $\xi _{imp}$ around the impurity has a large $T$
variation \cite{DMorr:1998},\cite{JBobroff:1998}.\ This has been found to occur as well for
Zn impurities as can be seen in Fig.\ref{fig:Ni_probes_correlations},
\textit{so that it appears more specific to the host than to the impurity
response}. To obtain quantitative determinations of\ $\xi _{imp}$, the Li
impurity which is a spinless impurity, as Zn, appeared more practical
technically, as will be shown below.

\paragraph{The Li spinless impurity: comparison with Zn.}

Li has been found to substitute on the Cu(2) sites \cite{KSauv:1996} in
YBCO, but detailed comparisons with Zn could only be performed when it was
established that the $^{89}$Y\ and $^{17}$O\ NMR\ data in YBCO$_{6.6}$:Li
samples were nearly identical to those found for YBCO$_{6.6}$:Zn, as shown
in Fig.\ref{fig:Y_NMR_satellites} \cite{JBobroff:1999}.\ This allowed one to
understand that a Li$^{+}$ spinless impurity which introduces a hole in the
plane induces the same local magnetic properties as the spinless isovalent
substitution of a Zn$^{2+}\,$on a Cu$^{2+}$\ site. \ The fact that the $%
^{89} $Y \textit{n.n.} sites\ of the Li and the Zn isolated impurities
display almost the same shift is evidence that the charge difference is not
screened on the atomic scale. The hole donated by Li to the band is rather
delocalized, and of course changes the doping slightly, but this is
experimentally a negligible effect as long as one considers dilute
concentrations of Li. So the magnetic response of the CuO$_{2}$ plane to the
local perturbation is dominated by the \textit{spinless character of the
impurity} which disrupts the AF interactions with the neighboring sites,
while the extra Coulomb potential of the Li impurity is negligible in that
respect.

%Apart from this important specific information associated with the impurity
%property itself,
The great advantage in using Li as an impurity is that the large
gyromagnetic ratio of the $^{7}$Li nuclear spin and its small quadrupole
moment allows one to detect its NMR signal, its shift being then, as seen in
Fig.\ref{fig:Y_NMR_satellites}, a direct measure of the magnetization of its
four Cu \textit{n.n.} sites. This has then allowed three important technical
advantages beyond what was permitted by the study of the \textit{n.n.} $%
^{89} $Y\ NMR:

\begin{itemize}
\item the very high sensitivity of the $^{7}$Li NMR\ permitted accurate
measurements of $\chi (T)$ of the Cu \textit{n.n.} of the spinless impurity.
In the underdoped samples, this variation is found to display a perfect
Curie $T$ dependence (see Fig.\ref{fig:Y_NMR_satellites}), which confirmed
the observation made from the $^{89}$Y\ NMR that the impurity induced state
behaves as a nearly free paramagnetic moment \cite{JBobroff:1999}.

\item the dynamics of the local moment could be studied through the $^{7}$Li
nuclear spin relaxation time $T_{1}$ data \cite{WAMacFarlane:2000}.

\item the possibility to use in the same sample the $^{7}$Li, $^{89}$Y\ , $%
^{17}$O\ and $^{63}$Cu nuclear spin probes has enabled the quantitative
determination of the spatial structure of the induced polarization, that is
its magnitude and $\xi _{imp}$ $(T)$ \cite{SOuazi:2004}.\
\end{itemize}

This latter point will be now discussed in detail, both in the underdoped
and optimally doped parts of the cuprate phase diagram.

\paragraph{Spatial extent of the staggered spin polarization.}

While ideally one would like to measure directly the local magnetization on
the nuclear spins of the magnetic sites themselves, i.e. Ni sites in the $%
S=1 $\ \ chains or Cu in the 1D or 2D systems, this happens to be in most
cases hampered by technical limitations linked with the very large hyperfine
couplings on those sites.\ We have shown in the previous paragraphs that
other neighboring nuclear probes such as $^{89}$Y\ can help to extract some
information in a less direct way.\ While in 1D\ %Haldane
YBaNiO chains the $^{89}$Y nuclei do probe the magnetization of a single Ni
moment in a chain, the information is not as straightforward with a nucleus
which has many different Cu neighbors, such as $^{17}$O\ which feels the
response of two nearest neighbor Cu sites in YBCO.\ This results in a form
factor which connects the field induced on a nuclear site to the local
magnetization in its vicinity.\ Although this appears as a complication at
first sight, it is on the contrary an advantage since the availability of
different nuclear spins allows one to probe various spatial ranges of the
induced spin polarization. In particular, when the magnetization has a
staggered spatial dependence it is clear that, for nuclei which have $z$ Cu
near neighbors\ in a plane, the cancellation of the local field sensed is
more efficient for larger \ $z$ (see Fig. \ref%
{fig:alternating_magnetization_schematic}).\ One can then understand that
the range of the spin polarization sensed is larger for $^{63}$Cu, \ which
feels the on site polarization, as compared to $^{17}$O($z=2$), further
compared to $^{89}$Y ($z=4$).\

\begin{figure}[t]
\begin{center}
\leavevmode
\includegraphics[angle=-90,clip=true,width=.99%
\columnwidth]{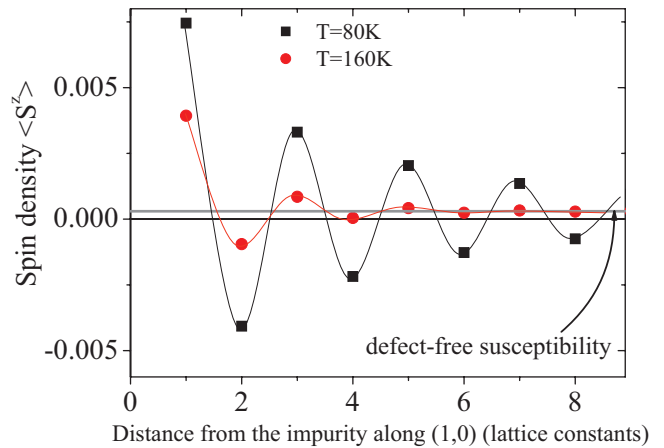}
\end{center}
\caption{The spatial variation of the model spin densities which fit the NMR
data on $^{7}$Li, $^{89}$Y and $^{17}$O\ nuclei in YBCO$_{6.6}$\ are
represented as a function of distance to the impurity site along the (10)\
direction in the CuO$_{2}$\ plane.\ It can be seen that the local
magnetization magnitude and decay length $\protect\xi _{imp}$ both decrease
with increasing $T$. For a large number of sites the local magnetization is
much larger than that of the pure sample which is the limiting large $r$
value at 160K, and which is even smaller at 80K ( constructed from results
of \protect\cite{SOuazi:2004}).\ }
\label{fig:modelled_spin_density_near_Zn}
\end{figure}

The data given by the NMR shifts of $^{7}$Li , of the \textit{n.n.}
satellite resonances of $^{89}$Y, and the lineshapes of the $^{89}$Y\ and $%
^{17}$O\ NMR main lines could be compared with numerical simulations made
for model spin polarization radial dependences \cite{SOuazi:2004}.\ This
analysis gave evidence that the staggered spin density decays exponentially
a few lattice constants away from the impurity,
%%PJH--I think you mean a few lattice spacings away, but this is not clear.
%% I do not think there is any justification for a true exponential
%% form except far from the impurity
and that the induced spin magnetism is not singularly high on the near
neighbors of the Li (or Zn ) impurities.\ Similarly to the chain case, the
local moment perturbation has a spatial extent which is characterized by a
single range parameter $\xi _{imp}$ and the macroscopic moment appears as
the sum of this oscillating polarization.\ Performing such a data analysis
at different $T$ allowed one to confirm that \thinspace $\xi _{imp}\;$and
the magnitude of the spin polarization have large $T$ variations, as shown
in Fig. \ref{fig:modelled_spin_density_near_Zn}.

\subsubsection{From the underdoped to the optimally doped case}

\ The same approach can in principle be applied for increasing hole doping
in the CuO$_{2}$ plane. In the pure compounds, the pseudogap temperature
decreases and approaches $T_{c}$. What happens then to the magnetic
response? Does it change drastically above $T^{\ast }$?\ This question has
of course a direct relationship with the important question still pending
regarding the phase diagram of the cuprates: does the pseudogap line cross
the $T_{c}$ line , giving rise to a quantum\ critical point near optimal
doping, or does it merge with the $T_{c}$ line? We shall examine this point
first.

\paragraph{Pseudogap line and QCP?}

Experimentally, the pseudogap temperature $T^\ast $ has been found to drop
sharply with increasing doping and approaches the superconducting dome
around optimal doping.\ As we have seen above, the introduction of
impurities reduces $T_{c}$ and allows one to probe the normal state
properties at temperatures below $T_{c}$ of the pure compound. As impurities
do not modify the pseudogap in the underdoped part of the phase diagram, a
reduction of $T_{c}$ induced by impurities should open a window allowing the
study of the continuation of the $T^{\ast }$ line in the optimally doped
regime. 
%We could not find in the 
NMR studies performed so far on substituted
samples of YBCO$_{7}$ do not show any evidence for a discontinuous change of the NMR
shift of any nuclear spin species of the material between the $T_{c}$ of the
pure compound and that of the substituted one. Even\ with 4\%\ Zn, the $%
^{59} $Y\ NMR\ shift is found to remain constant when $T_{c}$ has been
reduced to 55K \cite{HAlloul:1991a}. So the existing experiments do not
allow to conclude whether the pseudogap line crosses the $T_{c}$ line,
although some authors report data points in their phase diagram which would
suggest the opposite\ \cite{GVMWilliams:1997}. \

\paragraph{Spinless impurity magnetism.}

\ Experimentally, although no nearest neighbor $^{89}$Y\ NMR resonance could
be resolved in YBCO$_{7}$:Zn, the early NMR data \cite{HAlloul:1991a}
provided evidence for a low $T\;$ increase of the $^{89}$Y\ NMR\ linewidth
which could not be associated with impurity phases.\ \ In view of the
analysis done above, this was already clearly evidence that magnetic
correlations persist, since in a nonmagnetic metallic system one would expect a
$T$ independent linewidth when the spin susceptibility is Pauli-like as in
YBCO$_{7}$.\ Indeed, the non-Korringa $T$ dependence of the Cu(2) $T_{1}$\
was also a direct proof that magnetic correlations remain important \cite%
{MTakigawa:1991}. Most evidence in the SC state to be discussed in Section %
\ref{sec:SC} below also indicate that below $T_{c}$ some paramagnetic moment
character appears. But this qualitative evidence cannot be connected easily
to the more quantitative analysis done above in the underdoped normal state.
The first more quantitative point came from \ SQUID\ data on very clean
samples of YBCO$_{7}$:Zn in which the impurity induced $\chi _{c}$ was found
much smaller at low $T$ than in the underdoped compound, which indicated a
weaker magnetism, and certainly not a free $S=1/2$ moment \cite%
{PMendels:1999}. \

\begin{figure}[t]
\begin{center}
\leavevmode
\includegraphics[angle=-90,clip=true,width=1.%
\columnwidth]{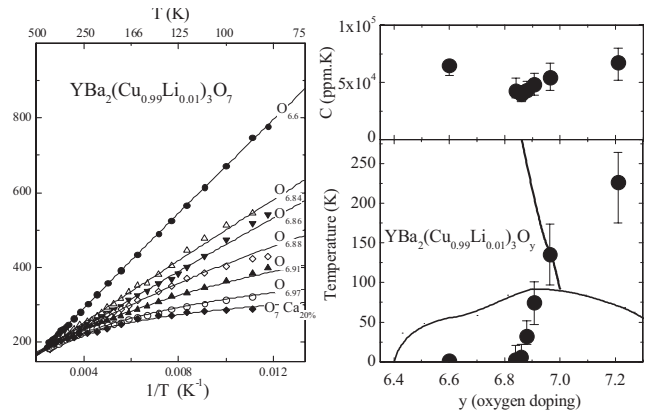}
\end{center}
\caption{Left: $^{7}$Li NMR\ shift plotted versus $T^{-1}$\ for a series of
YBCO$_{y}$\ samples as well as on an overdoped Ca$^{2+}$/Y$^{3+}$
substituted sample for $y=7.$ The data are fitted with Curie Weiss laws.\
The corresponding Weiss $\Theta \;$values are plotted on the\ right bottom
panel versus oxygen content $y$, and are seen to increase sharply near
optimal doping.\ There the full curve represents the variation of $T_{c}$ in
the pure samples.\ Right top: effective moment parameter $\protect\mu %
_{eff}^{2}$ of the Curie Weiss fit.\protect\cite{JBobroff:1999}. }
\label{fig:theta_vs_doping}
\end{figure}

Subsequently, $^{7}$Li NMR\ data allowed one to measure $\chi (T)\;$
directly on the 4 \textit{n.n.} Cu, which is shown in the left panel of Fig. %
\ref{fig:theta_vs_doping} to deviate from a Curie law.\ The Curie-Weiss
behavior found\ in YBCO$_{7}$with a sizable $\Theta $ only weakly dependent
on Li content ensures that this results from the isolated impurity response
\cite{JBobroff:1999}.\ Again, comparison of the $^{17}$O\ and $^{89}$Y\ NMR
shapes and widths data leads one to conclude that there is no appreciable
difference between the Li and Zn induced magnetism, while Ni has a larger
magnetic susceptibility $\chi _{c}$ but, as in the underdoped case, induces
a comparatively smaller in-plane perturbation \cite{TMRiseman:1994}. The
careful study of the $T$ dependence of the $^{7}$Li NMR shift versus hole
doping displayed in Fig. \ref{fig:theta_vs_doping} provides evidence that
the Curie-Weiss temperature $\ \Theta $ varies abruptly with hole doping and
becomes nearly zero in the underdoped regime.\
As the effective moment found appears independent of doping, one can
conclude that the low $T$ reduction of susceptibility in the optimally doped
case is due to the onset of the energy scale $k_{B}\ \Theta $ in analogy
with the Kondo reduction of local moments in classical metallic systems. \
We shall see later on, after performing the analysis of $\xi _{imp}(T)\,\,$%
that this analogy also applies to the dynamic properties of the local moment.

The occurrence of a local moment induced by substitution of Cu$^{2+}$ by
nonmagnetic Al$^{3+}$ has been established as well in optimally doped La$%
_{2-x}$Sr$_{x}$CuO$_{4}$ \cite{KIshida:1996}.\ They showed that Al$^{3+}$
exhibits a local moment behavior in optimally doped La$_{2-x}$Sr$_{x}$CuO$%
_{4}$. Again, the $^{27}$Al NMR shift exhibits a Curie-Weiss $T$ dependence,
with a sizable Weiss temperature ($\Theta \approx $ $50K$). It is not clear
whether this value of $\Theta $ could have been reduced with respect to the
single impurity limit by the large impurity concentration\ (3\%\ ) used in
this experiment.

\paragraph{Magnetic correlations and $\protect\xi _{imp}(T)\,$}

\ The weakness of the induced susceptibility, that is the large value of $%
\Theta $ measured for optimal doping, is quite compatible with the absence
of detectable $^{89}$Y\ NMR nearest neighbor resonances within the
experimental resolution ( limited by the $^{89}$Y\ linewidth) for Li or Zn
in YBCO$_{7}$.\ This does not prevent \ the multi-nuclei analysis of the
staggered magnetism, as the hyperfine couplings determined for the \textit{%
n.n.} $^{89}$Y in the underdoped case can still be used.\ Indeed, it was
shown some time ago that the hyperfine couplings, which are based on atomic
orbital physics\cite{FMila:1989}, do not depend markedly on the hole content
in the cuprates \cite{HAlloul:1989,CHPennington:1990}.

\begin{figure}[t]
\begin{center}
\leavevmode
\includegraphics[clip=true,width=.9\columnwidth]{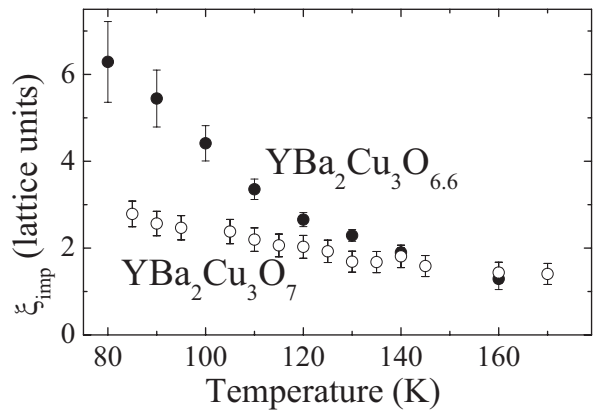}
\end{center}
\caption{Temperature dependence of the spatial extension $\protect\xi %
_{imp}\;$of the polarization induced by Zn or Li impurities in underdoped
YBaCuO$_{6.6}$ and optimally doped YBaCuO$_{7}$. The $T$ variation is larger
for the underdoped case. \protect\cite{SOuazi:2004}.}
\label{fig:xi_imp_vs_T}
\end{figure}

The analysis \cite{SOuazi:2004} of the $T$ dependences of the $^{7}$Li shift
and of the $^{89}$Y and $^{17}$O\ NMR\ spectral shapes suggest that the
magnitude of the staggered magnetism still decays exponentially a few
lattice constants away from the impurity. The local susceptibility is still about
four times larger at 100K\ on the Cu \textit{n.n.} to Zn than that of the
pure optimally doped compound. This establishes the similarity with the
staggered magnetism induced in the underdoped case, despite claims to
the contrary 
%\textbf{(maintenir cet ordre en citant les referencesseparement?Important)
 (\cite{JLTallon:2002} and see reply by \cite{JBobroff:2002}).
%is still valid at short distances
%PJH--again, I removed this because we do not define short distances.
% I think again you mean intermediate distances, i.e. not nearest
% neighbor, and not so far away from the impurity, but we do not define.
Although $\xi _{imp}$ has a similar magnitude at room
temperature in the normal and in the underdoped cases, $\xi _{imp}(T)\,$ increases
much less at lower $T$ in the former situation,
as Fig.\ \ref{fig:xi_imp_vs_T} shows.\ These
results demonstrated that the local moment susceptibility and \ $\xi _{imp}$
have much weaker $T$ variations in the optimally doped case than in the
underdoped one. The energy scale $\Theta $\ may control the $T$ variations
of both quantities, however. \ Since overdoping corresponds to an increase
of $\Theta $ well beyond the\ value found for optimal doping (Fig. \ref%
{fig:theta_vs_doping}), such a scheme would allow a smooth crossover towards
the Fermi liquid limit for large overdoping.\

\subsubsection{Spin Dynamics}\label{sec:spindynamics}

\ \ \ We have only considered so far the static response of the electronic
states to a magnetic field. We might wonder what the microscopic process is
which drives the dynamics of the local moment induced by spinless
impurities? In undoped spin systems,  the local moment fluctuations should
be driven by the magnetic excitations of the system, if not by a different
degree of freedom such as a spin phonon coupling.\ In a 2D metallic system,
charge degrees of freedom will play a role as well. %\textbf{.}

The nuclear spin lattice relaxation $T_{1}$ of the nuclear spins are
governed  by the fluctuations of the local fields induced by the dynamic
susceptibility \cite{CHPennington:1990}.\ Modifications of the $T_{1}$\
rates of the pure system due to local moment fluctuations are seen on nuclei
even at large distances from the local moment, but are usually very hard to
analyze due to the distribution of coupling constants\ \cite{HAlloul:1973}.
The most reliable experimental situation is  that for which a well defined
nuclear spin site with respect to the impurity can be singled out \cite%
{HAlloul:1977}.\ For instance, the resolution of a shell of $^{63}$Cu
neighbors of Zn by NQR\ \cite{YItoh:2003}, showed that the nuclei near the
impurity display the fastest relaxation rates, as could be expected in the
local moment picture, contrary to previous claims \cite{KIshida:1993}.\

This is the case as well for the $^{89}$Y \ \textit{n.n.} neighbors\ of the
impurity \cite{AVMahajan:2000}, or for the impurity $^{7}$Li NMR itself in
the case of Li substitution \cite{WAMacFarlane:2000}. In these simple cases
the contribution to $1/T_{1}$ of the local moment fluctuations overwhelms
that of the pure system \cite{AVMahajan:2000}, and was studied thoroughly
for the Li case \cite{WAMacFarlane:2000}.\

In a simple relaxational model for which the local moment is assumed to
fluctuate as a single entity at a rate $\tau _{s}^{-1}$, the nuclear spin
lattice relaxation rate is related to $\tau _{s}^{-1}$ and to the local
moment susceptibility $\chi _{c}\;$by

\begin{equation}
\frac{1}{T_{1}T}=2k_{B}\left( \frac{\mu _{n}A}{\hbar \mu _{B}}\right)
^{2}\chi _{c}\tau _{s},  \label{relaxation rate}
\end{equation}%
where $\mu _{n}$ and $\mu _{B}$ are the nuclear and electronic magnetic
moments. Here, $A$ is the effective hyperfine coupling of the nuclear spin
to the local moment obtained directly from the shift of the nucleus by $%
K=A\chi _{c}/\mu _{B}.\;$Therefore \ $(T_{1}TK)^{-1}$ is a direct measure of
$\tau _{s}.\;$as reported for the $^{7}$Li nuclei in in Fig. \ref%
{fig:relaxation_rate}. They are fully compatible with those obtained from
the $^{89}$Y$\;n.n.\;\;T_{1}$\ data which were found identical for Li and Zn
substitutions.\

\begin{figure}[t]
\begin{center}
\leavevmode
\includegraphics[clip=true,width=.9\columnwidth]{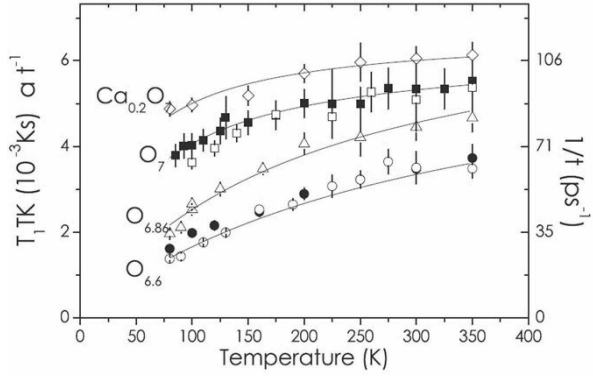}
\end{center}
\caption{Electronic relaxation rate $\protect\tau ^{-1}$ obtained from $%
T_{1}TK$, as measured for the $^{7}$Li NMR for different hole dopings and Li
contents ( empty symbols 1\%, full 2\%) in YBaCuO. After Ref. \protect\cite{WAMacFarlane:2000}.}
\label{fig:relaxation_rate}
\end{figure}

The low energy excitations detected by inelastic neutron scattering at $%
\left( \pi ,\pi \right) $ in the normal state of underdoped YBCO:Zn have
energy widths which agree perfectly with $h/\tau _{s}\;$\cite%
{YSidis:1996,YSidis:2000}. This quantitative consistency among confirms that
the dynamics is not that of a purely atomic local moment, but concerns the
extended staggered AF state, which is the assumption made above in Eq. (\ref%
{relaxation rate}). \emph{.\ }

The $\tau _{s}$ values are so short that the expected ESR linewidths are too
large to be observable with standard 10GHz ESR, which explains the absence
of any such direct observation in cuprates and implies that the ESR signals
detected in the early LSCO:Zn experiments \cite{AMFinkelstein:1990}%
\thinspace were not linked with the Zn induced local moments \cite%
{AVMahajan:2000}.

The other evidence which can be drawn from the data of Fig. \ref%
{fig:relaxation_rate}, is that $1/\tau _{s}$ has a behavior quite similar to
that found for $1/\chi _{c}$, as both increase quasilinearly with $T$ for
the underdoped cases and are less $T$ dependent for the optimally doped
case.\ In fact, a\ nearly perfect linear variation of $(T_{1}TK)^{-1}\,$\
versus $K$\ applies, with a slope $S\,\ $\ which corresponds to the
universal Korringa constant\textbf{\ } $S=4\pi k_{B}/\hbar \times (\mu
_{n}/\mu _{e})^{2}$. This provides direct evidence that $\chi _{c}$ and $%
\;\tau _{s}^{-1}$are governed by the same energy scale. In other words, the
applicability of a Curie-Weiss law for $\chi _{c}$, with a Weiss temperature
$\Theta \;$is associated with a limiting low $T$ behavior $\tau
_{s}^{-1}\;=k_{B}\Theta /h\;$ for\ the local moment fluctuation rate. This
is strikingly reminiscent of the Kondo effect for a local moment in a noble
metal host, the Kondo energy\ being then the width of the resonant state
which governs at low $T$ the fluctuations in the Kondo state.\ These results
will be discussed in more detail in Section \ref{sec:thy2D}.\

\subsection{Transport properties}

\label{sec:Nstatetransport} Transport properties are among the most unusual
properties of the cuprates, and may be strongly influenced by the
disorder-induced magnetic phenomena we have been discussing. Scattering of
the charge carriers from inelastic magnetic fluctuations is often assumed to
be responsible for, e.g. the strange metal behavior exemplified by the
linear $T$ dependence of the resistivity above $T_{c}$, although there is no
complete theory available.
% which encompasses that due to phonons in classical
%metals.
The influence of impurities and defects on the transport properties should
provide information on the properties of the carriers themselves, and on the
coupling between charge and spin degrees of freedom.

In contrast with simple metals where experiments probing impurity effects
can be performed down to low $T$, there are severe limitations in the present
case, owing to onset of superconductivity at a high $T_c$.
%However, the occurrence of superconductivity is a severe limitation which
%prohibits many experiments at low $T$ in the normal state which have been
%performed on impurities in simple metals. 
On the other hand, the suppression
of $T_{c}$ by defects is an additional element of characterization available
which serves as a reference in the experiments although it does not measure
\textit{a priori} only the properties of the normal metallic state.

Let us recall, first, that reliable in-plane transport data require small
high quality single crystals, and that the synthesis of impurity-substituted
single crystals does not necessarily benefit from the experience gained with
ceramics, which are reacted in quite different thermal conditions. Thus,
the impurity contents,
fractional site occupancies and actual carrier content of single crystals
which are quoted in the literature are not always accurately determined.
%\
%Therefore, even for frequently quoted data, the impurity contents, the
%fractional site occupancies, the actual carrier content of single crystals
%are not always determined.
As already pointed out in Section \ref
{sec:Impurity_subs} above, electron irradiations are then quite helpful, as
one can span a series of defect concentrations using only one single crystal
by increasing the irradiation dose progressively.

We shall first discuss below the basic transport measurements which allow
one to demonstrate the simple characteristics of the scattering induced by
specific defects: residual resistivities, Hall effect, thermopower, which
will be correlated\ with the induced reduction of $T_{c}$. This will then allow
us to discuss more sophisticated effects such as low $T$
resistivity upturns, and magnetoresistance.

\subsubsection{High $T$ in-plane transport and scattering rates}

%\emph{For transport in single crystals is dominated by in plane.\ out of
%plane negligible?}

\begin{figure}[t]
\begin{center}
\leavevmode
\includegraphics[clip=true,width=1.0\columnwidth]{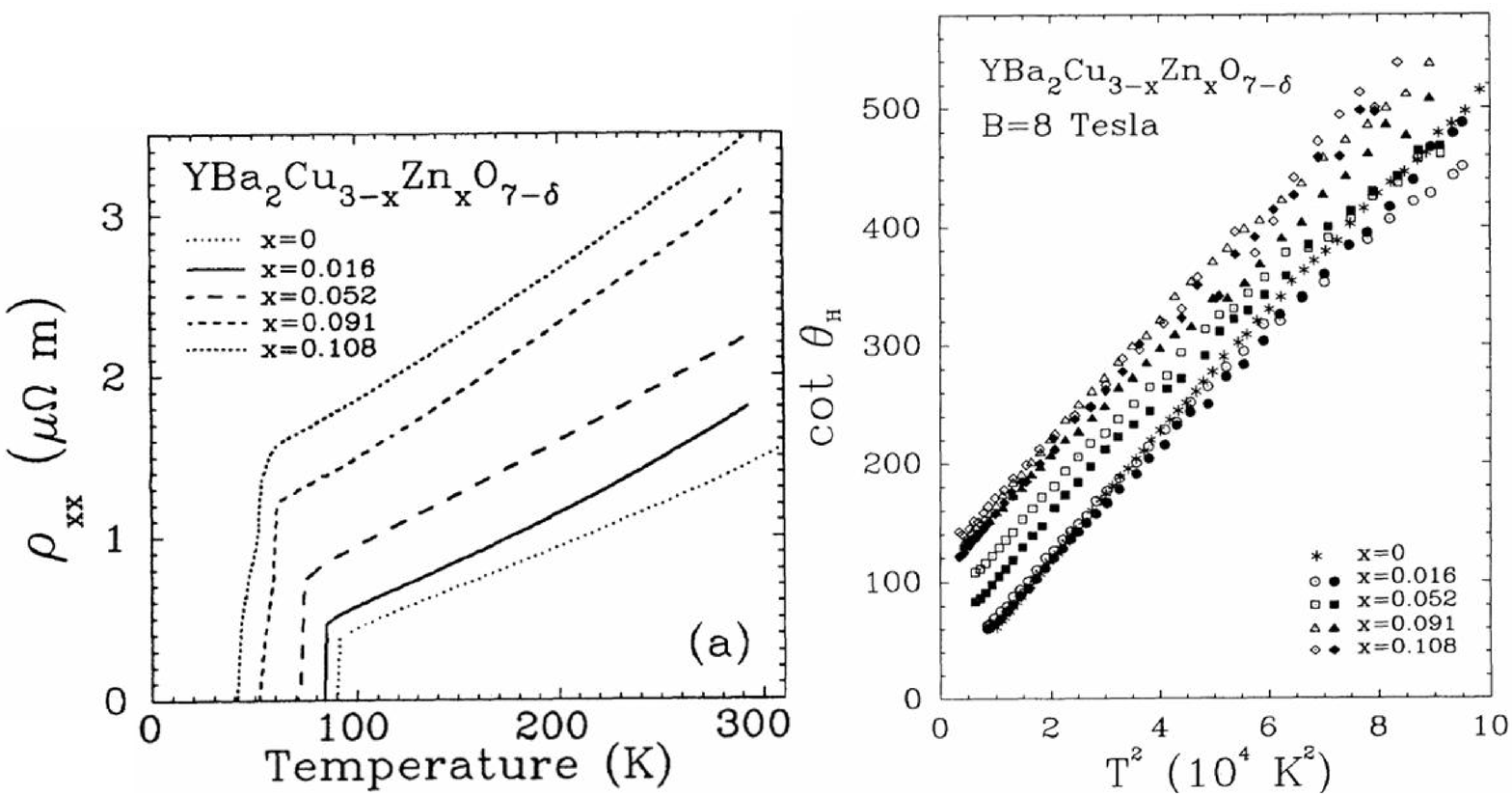}
\end{center}
\caption{(a) Temperature dependence of the in plane resistivity averaged
over a and b for a pure single crystal of YBCO$_{7}$ and for various Zn
concentrations in substituted single crystals\ . (b) $T\,\;$variation of 
$\cot\protect\theta _{H},\;$ where\ $\protect\theta _{H}$ is the Hall angle,
for the same samples \protect\cite{TRChien:1991}.\ }
\label{fig:chienong}
\end{figure}

An important question in transport processes is the applicability of
Mathiessen's rule, which in classical metals states that the impurity
scattering rate adds incoherently with the inelastic scattering rate in the
pure material, usually due to phonons. In the cuprates, the approximate
additivity of the impurity scattering and that by magnetic excitations of
the pure samples was pointed out for Zn substitutions in YBCO$_{7}$ as shown
in Fig. \ref{fig:chienong} \cite{TRChien:1991}.

\begin{figure}[t]
\begin{center}
\leavevmode
\includegraphics[clip=true,width=0.9\columnwidth]{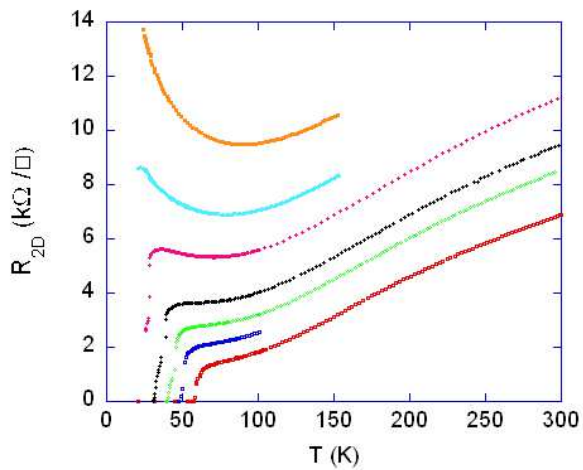}
\end{center}
\caption{Resistivity versus $T$ for a single crystal of YBCO$_{6.6}$
irradiated at 20K by 2.5 MeV electrons for increasing irradiation doses.
 The resistivity curves are seen to
display at high $T$ a parallel vertical shift associated with the increase
of the impurity induced residual resistivity. Therefore the defect content
does not modify the high $T$ inflection point, linked with the pseudogap $%
T^{\ast }$.$\;$For high defect contents, when $T_{c}$ is sufficiently
reduced, low $T$ upturns are evident, and ultimately the resistivity curve
diverges exponentially
 \protect\cite{FRullier-Albenque:2001}.}
\label{fig:mathiessen_irrad}
\end{figure}

Data taken later with electron irradiation on a single crystal with
increasing defect content are even more convincing, as can be seen in Fig. %
\ref{fig:mathiessen_irrad}. At high $T$, well above $T_{c}$ the $\rho (T)$
curves parallel each other, which establishes that the hole content is not
modified by the defects introduced by electron irradiation and that a
residual resistivity term simply adds to the $T$ dependent part of the
scattering.\ However, at low $T$ and for high enough defect content, low $%
T\; $upturns of the resistivity appear.\ This phenomenon, to be discussed
later, has often been interpreted in terms of a metal insulator transition
(MIT).\

An estimation of the residual resistivity due to the impurities can be
obtained either by extrapolation of the high $T$ behavior to $T=0$, or in
the best cases as that displayed in Fig. \ref{fig:mathiessen_irrad} by
measuring the parallel shift of the $\rho (T)$ curves at high $T.$\ The
residual resistivity per plane $\Delta \rho _{2D}\;$is a measure of the
scattering rate of the conduction electrons by the impurity potential. In
classical metals, the analysis of the scattering is done in terms of phase
shifts.\ Such analyses are very successful as long as the scatterers are
atomic scale defects and if one takes into account a decomposition of the
Bloch wave functions into partial waves of different symmetries $s$, $p$, $d$%
... which undergo different phase shifts. In particular cases, as for
transition metal impurities, one scattering channel dominates and a single
phase shift is sufficient to characterize the scattering.\ While such a
simple limiting situation has no \textit{a priori} reason to be valid for
impurities in correlated electrons systems, the data have mostly been
analyzed in cuprates by assuming an $s$ wave scattering with a unique phase
shift $\delta \ $ which corresponds to

\begin{equation}
\Delta \rho _{2D}=\frac{4\hbar }{e^{2}}\frac{n_{d}}{n}\sin ^{2}\delta ,
\label{Phoversus phase shift}
\end{equation}
where $n_{d}$ and $n$ are respectively the impurity and hole
concentrations.\

In the case of Zn substitutions the residual resistivity associated with the
Zn scatterers yields a large value of the scattering rate, with $\sin
^{2}\delta \approx 1/3\;$if Eq. (\ref{Phoversus phase shift}) above is
applied \cite{TRChien:1991}.\ This is more than two orders of magnitude
larger than that of the Zn impurity in copper.\ This unexpected \textit{%
large scattering} by a homovalent impurity is a striking illustration of the
strange metal properties of the cuprates which may only be explained by the
correlated nature of the electron states. The residual resistivity
associated with Cu and O vacancies in irradiated samples is large as well
\cite{ALegris:1993}.

The Hall effect was also found to display an anomalously large $T$
dependence in the cuprates, as was pointed out in \ \cite{TRChien:1991}.\ By
comparing pure and Zn substituted samples, it was found\textbf{\ }that the
cotangent of the Hall angle % determination of the ratio $%
%\rho _{xx}/\rho _{xy}$ ,
varies as

\begin{equation}
cot\theta _{H}\equiv \rho _{xx}/\rho _{xy}=\alpha n_{d}+\;\beta T^{2},
\label{cothall}
\end{equation}
as can be seen on Fig. \ref{fig:chienong}. This supports the idea advanced
by Anderson \cite{PWAnderson:1991} that $\cot\theta _{H}$ reflects a second
scattering rate $\tau _{H}^{-1}$ , with a $T^{2}$ variation in the pure
samples, which is independent of the resistivity scattering rate $\tau
^{-1}.\;$The first term in Eq. (\ref{cothall}) implies that the Zn
impurities contribute independently to the two scattering channels.

\subsubsection{Variation of the high $T$ transport properties with hole
doping.}

These experiments have been extended to the underdoped regime in order to
try to quantify the variation of the scattering rates with hole doping.\ As
seen in Fig.\ \ref{fig:mathiessen_irrad} the in-plane resistivity of
underdoped pure samples acquires a nonlinear $T$ variation which signals a
suppression of\ the scattering when the pseudogap opens at $T^{\ast }$.
Experiments on substituted YBCO or LSCO\ single crystals with Zn \cite%
{KMizuhashi:1995,YFukuzumi:1996}, or on electron irradiated samples \cite%
{FRullier-Albenque:2000} demonstrate that this $T$ dependence is not altered
by disorder, as can be seen\ in Fig. \ref{fig:mathiessen_irrad}, which
indicates again that the hole doping and $T^{\ast }$ are not modified by
these defects.

\paragraph{Residual resistivity\ }

The applicability of Eq. (\ref{Phoversus phase shift}) for different hole
dopings has been checked as well.\ As we have shown above, determinations of
$\Delta \rho _{2D},\;n_{d}$ and $n$ are done with an accuracy hardly better
than 20\%, so that estimates of $\sin ^{2}\delta \;$ are possible only
within a factor of 2.\  However \cite{YFukuzumi:1996} have assumed that,
within the approach described above, the scattering rate$\;$is large enough
to correspond to unitary scattering ($\delta \approx \pi /2)\;$whatever the
hole doping. They suggested then that one should identify the number of
carriers $n$ with the number of doped holes $n_{h}$\ for the underdoped
samples and with $1-n_{h}$ for overdoped ones, to get a consistent
explanation of the data.\

Alternatively, experiments using point defects created by electron
irradiation allow more accurate determinations of $\Delta \rho _{2D}$, as
shown in Fig. \ref{fig:mathiessen_irrad}.\ Furthermore, in YBCO$_{6+x}$\
samples, the concentration $n_{d}$ of defects created\ in situ at low $T$ is
independent of oxygen content and solely determined by the electron
fluence.\ The accurate comparisons of $\Delta \rho _{2D}\;$ for different
hole contents show then that unitary scattering cannot apply for both YBCO$%
_{6.6}$ and YBCO$_{7}$\ \cite{FRullier-Albenque:2000}.\ Rather, keeping $%
n=n_{h},$ one finds that the scattering is stronger for the underdoped case,
which suggests that the scattering rate has some relation with the range $%
\xi _{imp}$ of the magnetic perturbation induced by spinless impurities.

\paragraph{$\Delta T_c$ and residual resistivity.}\label{Tcreduc}

As accurate determinations of the scattering rate are not easy to obtain
directly from resistivity data, an alternative approach is to seek
information from the decrease of $T_{c}\,$induced by the scattering
centers.\ Although this is rather a measure of the pairbreaking effect
associated with the defects, it gives us information on the scattering rates
within a $d$-wave scenario for the superconducting state, which is supported
by many experiments in cuprates, as will be discussed in Section \ref{sec:SC}%
. In such a case, within a BCS theory of a $d$-wave superconductor, the
initial reduction of $T_{c}$ is determined solely by the normal state
scattering rate (see Sec. \ref{sec:SCdisorderthy}),
\begin{eqnarray}
\Delta T_{c}=-\frac{\pi }{4k_{B}}\frac{\hbar }{\tau }.
\end{eqnarray}

\begin{figure}[t]
\begin{center}
\leavevmode
\includegraphics[clip=true,width=.95\columnwidth]{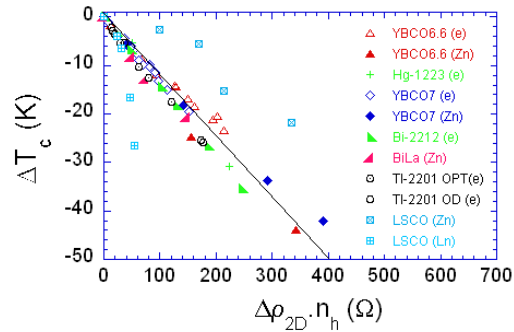}
\end{center}
\caption{The decrease of $T_{c}$ is plotted versus $n_{h}\Delta \protect\rho %
_{2D}$ for various families of cuprates, for Zn impurities or electron
irradiation defects.\ The universal relation which applies whatever the hole
doping $n_{h},$ except for the LSCO\ family, establishes that the $T_{c}$
depression is entirely governed by the scattering rate $\protect\tau ^{-1}$%
\protect\cite{FRullier-Albenque:2000}.\ }
\label{fig:universelTc}
\end{figure}

Here the scattering rate should be given directly by the residual
resistivity as $\tau ^{-1}=ne^{2}\Delta \rho _{2D}/m^{\ast }$. Thus, as long
as the effective mass of the carriers is not sample dependent, $\Delta T_{c}$
and $n$ $\Delta \rho _{2D}$ \ should be independent measures of the
scattering rate, and therefore scale with one another, regardless of the
concentration and microscopic structure of the point defects.\ This
universal scaling has been found to apply perfectly for Zn and electron
irradiation defects in many cuprates \cite{FRullier-Albenque:2000} (see
Fig.\ \ref{fig:universelTc}) provided that the number of carriers is taken
as $n=n_{h}.$\

The fact that$\,dT_{c}/dn_{d}$ generally decreases with increasing $n_{h}$
is therefore mainly due to a reduction of $\tau ^{-1}$ when
% the incidence of
electron correlations become less relevant. % diminishes.
In that sense, one does expect to eventually recover a weak scattering for
Zn impurities in the overdoped Fermi liquid limit.\ This is qualitatively
supported by $T_{c}$ data taken on Bi-2212 for various Zn and doping
contents \cite{TKluge:1995} but has not been studied quantitatively in the
far overdoped regime, because Zn substitutions in overdoped systems like
Tl2201\ have not been successfully performed so far to our knowledge. Direct
determination of the scattering rate in such systems cannot be done by
electron irradiation as the actual number of defects cannot be determined
accurately either.\

Finally, this universal scaling of $\Delta T_{c}$ and $n$ $\Delta \rho _{2D}$
does not apply for the LSCO\ system.\ This is particularly clear for Zn
substitutions for which the scaling is not even valid at optimal doping.\
Fractional homovalent substitution of Nd on the La sites at optimal doping
also yields drastic reductions of $T_{c}$\cite{KFujita:2005}. But this
specific system is well known to favor stripe ordering in the CuO$_{2}$\
plane ( the La $_{1.6}$Nd $_{0.40}$ \ substituted samples exhibiting a fully
ordered stripe structure even at optimal doping \cite{JMTranquada:1997}).\
So, quite generally, point defects appear to favor the existence of
structural and /or electronic instabilities in LSCO, which may explain why
this compound behaves differently from the other HTSC materials.

There are several aspects of the observed $T_{c}\;$ suppression which are
poorly understood at this writing. At optimal doping, the measured initial
suppression has been compared to Abrikosov-Gorkov (AG) pairbreaking
theory and deviations explained by anisotropic scattering effects
\cite{AVBalatsky:2006} (see however S. Graser et al.\cite{SGraser:2007}). In electron irradiation experiments the
concentration of metastable defects created is not limited by the
equilibrium thermodynamics which governs the solubility limits of impurities
in a high $T$\ solid state synthesis. This allows to follow the suppresion
of $T_{c}$\ until $T_{c}$\ vanishes, and the initial linear slope continues
and eventually curves up , in contrast to the AG theory for which it would
curve down\ \cite{FRullier-Albenque:2003}. \emph{\ }This suggests the
potential importance of including phase fluctuations when the superfluid
stiffness becomes small.

Finally, let us recall that, as we emphasized in Section \ref{hdc}, the
dopant and intrinsic structural disorders certainly play a large role in
controlling the actual observed $T_{c}$ values in many ''lower $T_{c}$\
cuprate families'' \cite{JBobroff:1997, HEisaki:2004, KFujita:2005}.
Obviously, if the situation for in plane disorder seems under experimental
control, the influence of other types of disorders such as those discussed
above in the case of LSCO\ are only qualitatively understood \cite{SGraser:2007}.\ We
believe that quantifying the correlation between the decrease of $T_{c}$\
induced by various types of disorder, the associated residual resistivities
and other physical properties as well\ still requires further experimental
studies, which will require overcoming difficult materials challenges.\

\paragraph{Hall effect and thermoelectric power}

Hall effect measurements have been extended as well to the overdoped and
underdoped regime \cite{YAbe:1999}\ and the analysis performed for optimal
doping seems to apply as well, which highlights the existence of two
scattering rates.\ However quantitative comparisons between samples are even
harder to perform than for the residual resistivity, so that the data did
not allow the determination of the physical origin of the actual difference
between the two scattering rates.\ The modifications of the specific heat
induced by Zn or Ni substitutions has been investigated, but the normal
state electronic contributions of Zn or Ni could not be separated from the
dominant phonon contribution \cite{JWLoram:1990,KAMirza:1994}. \cite%
{JLTallon:1995} have also\ found that the electronic thermoelectric power
(TEP) is independent of Zn content at room $T$, which is \ an independent
indication that hole doping is not modified, as the TEP\ value at room $T\;$%
is found to vary markedly with hole doping in the pure cuprates \cite%
{SDObertelli:1992}.

\subsubsection{Low $T$ upturns of the resistivity}

Since the parent AF state of the cuprates is insulating, a generic phase
diagram of the cuprates should display a MIT with decreasing hole or
electron doping.\ Of course the introduction of in plane impurities in a
metallic state of these 2D systems is also expected to yield a MIT
associated with localization effects which should occur in any case if the
residual resistivity exceeds the Mott-Ioffe-Regel limit.\ Both in pure and
substituted cuprates, low $T$ upturns of the resistivity as $T$ is decreased
have often been observed and have been usually associated with the
occurrence of a MIT.

Due to the occurrence of the SC state, the most effective way to explore
these effects is to suppress SC in large applied fields.\ \ It has for
instance been found that LSCO exhibits at optimal doping a $-\log T$
increase of $\rho _{2D}$ in a 60\ Tesla applied field \cite{YAndo:1995}.\
This was initially taken as an indication that the MIT already occurs for
optimally doped cuprates, but it has been shown since that resistivity
upturns only occur in the underdoped regime in BiLa2201 \cite{SOno:2000}, 
suggesting that the location of the MIT is greatly influenced by the
intrinsic disorder of the considered family as demonstrated in \cite{FRullier-Albenqueb:2007}.

In the cleanest cuprates such as YBCO$_{6.6}$ or Tl2201, the influence of
controlled defects on the conductivity can be studied more directly. The
possibility to induce a MIT by introducing enough Zn in the CuO$_{2}$\
planes of YBCO was also noted in the early experiments \cite{YFukuzumi:1996}%
. As seen in Figure \ref{fig:mathiessen_irrad}, resistivity upturns are also
easily detected for electron irradiation defects. In the overdoped regime of
Tl2201, the upturns of $\rho _{2D}$ only appear for large defect content,
and the magnitude of the $-\log T$ terms scale with $n_{d}^{2}$ , as
expected for weak localization contributions \cite{FRullier-Albenque:2001},
 which probe coherence effects
in multiple impurity scattering \cite{PALee:1985}, in good metallic Fermi
liquids. Therefore these experiments are consistent with the
expectation that the overdoped cuprates are good Fermi liquid metals.

For Zn substituted YBCO$_{6.6}$ the upturns of $\rho _{2D}$ were
also detected by partly suppressing SC with an applied field \cite{KSegawa:1999}
 and it was concluded that they could not be attributed to
weak localization effects.\ The data accuracy for electron irradiation
defects in YBCO$_{6.6}$\ allowed the detection of upturns in $\rho _{2D}$
above $T_{c}$, even in zero field. A careful subtraction of the resistivity
of the sample before irradiation allowed one to demonstrate that the
resistivity upturns scale initially linearly with $n_{d}$ so that these
upturns are associated with single impurity scattering \cite{FRullier-Albenque:2001}.
 It is therefore quite tempting to correlate these
experimental evidences with the Kondo-like behavior observed for the
impurity induced magnetism.

If these resistivity upturns were solely explained in terms of spin-flip scattering, they
should be suppressed by a field. However, prior to any conclusion on the
normal state properties in presence of impurities, one needs to ensure that
SC is fully suppressed by the field. Indeed a large SC fluctuation regime
above $T_{c}$ has been initially detected
by Nernst and diamagnetic measurements \cite{YWang:2006}.
In the presence of disorder, this regime
has been shown to survive even after a large reduction of the 3D\ $T_{c}\;$
\cite{FRullier-Albenque:2006,FRullier-Albenqueb:2007}, a fact which remains to be
understood.  This difficulty to
separate out the various contributions to the normal state scattering properties constitutes a serious complication when one studies the dilute impurity limit.

For instance, a  positive ``magnetoresitance'' 
has been reported initially in Zn substituted YBCO \cite{KSegawa:1999}.
 The survival of the SC fluctuation
contribution to the conductivity which only saturates at very large applied
fields could explain this behavior \cite{FRullier-Albenquea:2007}.
% If these resistivity upturns were due to spin-flip scattering terms, they
%should be suppressed by a field.\ In apparent contrast with this analysis a
%positive ''magnetoresistance'' has been reported in Zn substituted YBCO \cite{KSegawa:1999}.
% This observation is, however, caused by the survival of
%superconducting fluctuation contributions to the conductivity up to very
%large applied fields \cite{FRullier-Albenquea:2007}. This large
%SC\ fluctuation regime above $T_{c}$ initially detected by Nernst and
%diamagnetic measurements \cite{YWang:2006} are quite specific properties of these 2D cuprates
%and can be associated with a regime where SC\ phase coherence is destroyed.\
%They have been shown to survive in presence of disorder on a large
%temperature range above the 3D\ $T_{c}\;$\cite{FRullier-Albenque:2006,FRullier-Albenqueb:2007}.
%The occurence of these SC\ fluctuations, which remain to be
%understood, makes it extremely difficult to separate out the normal state
%scattering properties and prevents for instance the study of the dilute
%impurity limit. 
By contrast, in underdoped Zn substituted BiLa2201, for which $T_{c}\;$could be
reduced to zero, a negative magnetoresistance has been\ detected \cite{YHanaki:2001}. Recent experiments for irradiation defects, where high
fields suppress superconductivity, show evidence that the upturns saturate
at low $T$\ \cite{FRullier-Albenqueb:2007}, also consistent with the
Kondo analogy.\  

Theoretical arguments allowing to establish the occurrence of the local
moment and to reproduce at least partly the observed behavior will be given
in Section \ref{sec:thy2D} below. The arguments for and against this analogy
with Kondo effect will be discussed in some detail there.\

For increasing defect content $\rho _{2D}(T)$ becomes more singular than $%
-\log T$ and diverges at low $T$, signalling a real MIT in which
localization of the carriers occurs.\ Optical experiments on Zn substituted
cuprates do reveal a reduction of the Drude contribution to the conductivity
and the appearance of a high frequency absorption mode for large Zn content
which have been attributed to such localized carriers \cite{DNBasov:1998}.
However this type of experiment does not give information on the length
scale of these localized states.\

%\textbf{The next figure is to be suppressed}

%\subsection{\ Significance for normal state physics}

\subsection{\ Significance for normal state physics}

\label{sec:Signficance_N}

We have described in some detail in the previous section the available
experimental information \ on the influence of impurities on the magnetic
and transport properties in the normal state of the cuprates. In this
section we shall try to collect them and to present the main conclusions
which can be drawn from the qualitative aspects of the data.\ We shall try
to focus on the implications for the physical properties of the pure system,
and on the questions which arise then for the theory of this normal state.

\subsubsection{What about the phase diagram?}

Let us first return to the properties of the pure cuprates and
summarize the implications of the impurity investigations on the 
phase diagram.\ In the low doping range we have seen that the controlled
disorder associated with impurity substitutions or irradiation defects
affects the phase diagram and expands the SG\ regime.\ So in this range the
phase diagram might be dependent as well on the existing intrinsic disorder
in the \textquotedblright lower $T_{c}$\ cuprate\textquotedblright families.
On the other hand the pseudogap crossover temperature $T^{\ast }$ is
unaffected by disorder, which explains why $T^{\ast }$ is nearly identical
for most families of cuprates whatever their optimum $T_{c}$ \cite%
{GVMWilliams:1997} . However we have recalled why the impurity experiments
do not allow one so far to extrapolate the behavior of $T^{\ast }$ at, and
above optimal doping.

The magnetic behavior of correlated systems with impurities changes abruptly
near optimal doping (Fig. \ref{fig:theta_vs_doping}). It is then tempting to
associate this evolution with a crossover line between the strange metal and
the correlated Fermi liquid regimes. NMR experiments yield a characteristic
energy scale $\Theta $ for the magnetic susceptibility which would be
associated with such a crossover. $\Theta $ increases markedly as one
overdopes the system and, for temperatures far below $\Theta /k_{B}$, it
appears that $\xi _{imp}$ becomes $T$ independent. Better experimental
accuracy is required before a generic phase diagram can been drawn using the
above criteria, but the overall trend of the data is not inconsistent with
this picture. The fate of $\Theta $ in the highly overdoped limit when the
system should evolve toward a standard Fermi liquid is still an open
question.

%What the impurity experiments do establish is that the magnetic
%behavior of the correlated systems changes abruptly near optimal doping
%(Fig. \ref{fig:theta_vs_doping}).\ It has been often asked whether
%experiments could indeed allow to reveal a crossover line between the
%strange metal and the Fermi liquid behavior.\ The impurity experiments
%reveal an energy scale }$\Theta $\textbf{\ which underlines a crossover in
%the magnetic susceptibility and which increases markedly upon overdoping.\
%So, one might speculate that }$\Theta \;$\textbf{could have some connection
%with the crossover regime which has been searched for long.\ One might even
%speculate as well that }$\Theta \;$\textbf{could represent the temperature
%range below which the correlation length }$\xi _{imp}$\textbf{\ becomes
%nearly}$\;T$\textbf{\ independent.\ Again, the experiments are not yet
%accurate enough to allow to draw a generic phase diagram, but the overall
%trend of the data does not object such speculations.}

\subsubsection{Importance of electronic correlations}

Another point which has been clearly established is that, as in spin chains,
a nonmagnetic impurity induces a local moment in its vicinity.\ This moment
results from the sum of the staggered response of the near neighbor Cu\
sites which extends over a distance which can be characterized by a single
spatial length scale $\xi _{imp}$.\ This antiferromagnetically correlated
object is not static but highly dynamic, and a macroscopic magnetization can
only be detected in a an applied field, or with a probe with a higher time
resolution than the fluctuation rate of the correlated cloud. These isolated
local moments cannot be detected by zero field $\mu $SR experiments, which
only probe static moments.

\ \ These observations of magnetic responses of a metallic state to a
nonmagnetic impurity can only occur in (and are therefore indisputable
evidences for) the presence of electronic correlations in the cuprates.\
Similarly the large scattering cross section of conduction electrons by
these nonmagnetic impurities cannot occur in a simple broad s-p band. It
results here from the inability of the Cu$_{3d}$-O$_{2p}$ hybridized
quasiparticles to penetrate the Zn or Li sites which translates into an
extended scattering potential caused by correlations.
.\ This effect is as well
completely driven by the magnetic correlations in this narrow band.

Although the importance of magnetic correlations is revealed by some
properties of the pure systems such as the peak of the dynamic
susceptibility at the AF wave vector, these experiments on impurities allow
one to better follow the evolution with hole doping of the magnetic
correlations and to measure both the magnitude and the range $\xi _{imp}$ of
the induced staggered magnetization. In addition, they have provided
evidence that the correlated state is still present in optimal and lightly
overdoped systems, although with a $\xi _{imp}\;$which only exceeds slightly
the unit cell length, and with a weak susceptibility.\

Finally static moments are only seen for large enough defect content, when
interactions between moments pins a disordered spin glass like state at low $%
T$.\ Also in the specific LSCO system for which the intrinsic order can
yield the formation of stripes, impurities and stripe structures might
interact and yield then frozen magnetic states. But such a situation is not
encountered in cleaner cuprates.

\subsubsection{ Influence of the charge degrees of freedom on the magnetic
effects}

In regular metallic systems, such as noble metals, a charged impurity
induces an oscillatory Friedel \ charge oscillation which % resumes
constitutes the response of the system, and combines the screening of the
charge and the associated spin density oscillations.\ In cuprates, where
spin and charge might be distinct degrees of freedom, one might imagine that
different length scales would characterize the screening and the staggered
magnetism.\ It is sometimes considered that a charge depletion might occur
in the vicinity of the Zn$^{2+}$ with a corresponding increase of the
magnetic correlation in this range\textbf{.}\ Conversely if Li$^{+}$ was to
bind a hole on its \textit{n.n.} oxygens, that would prevent a Zhang Rice
singlet from visiting the Cu sites \textit{n.n.} to Li$^{+}\;$ and would
yield quite different magnetic effects on these sites than near Zn$^{2+}$.\
However the absence of microscopic difference over several lattice spacings
for Li$^{+}$ and Zn$^{2+\;}$is a strong evidence that such effects are
unimportant, and that the spinless character of the impurities in the
correlated host is the major perturbation which yields the appearance of a
local moment.\

The scattering of the carriers also does not much depend significantly on
the impurity charge, as seen from the analogous $T_{c}$ reductions by Li$%
^{+} $ and Zn$^{2+}$.\ Although the occurrence of superconductivity
%the HTSC
severely limits the analysis of the transport data, we have seen that the
scattering by spinless impurities is not always unitary, but that the
scattering cross section increases from the overdoped to the underdoped
regime.\

\subsubsection{Quantitative information on the physical properties of the
pure compounds}

The length scale $\xi _{imp}$ is a signature of the response of the 
system, which is apparently only weakly connected to the properties of the
impurity itself, as even a magnetic Ni induces a staggered magnetism which
has a similar spatial extent. However the magnitude of the magnetic response
still depends on the nature of the perturbation.\ Then $\xi _{imp}\;$ is
likely related to the magnetic correlation length \thinspace $\xi \;$of the
pure cuprate, reflecting the \textbf{q} dependence of $\chi $(\textbf{q},$%
\omega $).\ In the spin 1\ chains, the occurrence of a single degree of
freedom (the spin) left obviously no other alternative than $\xi _{imp}=$ $%
\xi $, as confirmed by theory and experiment.\ As the impurity charge has
little influence on $\xi _{imp}\;$in the cuprates, it is quite natural to
consider that $\xi _{imp}\;$is at least a lower estimate of$\ \xi .$

The\ determination of $\xi $ in the pure cuprates has been at the source of
a longstanding controversy. Inelastic neutron scattering experiments (INS)
reveal that peak widths of $\chi $(\textbf{q},$\omega $) at the AF\ wave vector \textbf{%
q}$_{AF}$.
%\ Their \textbf{q }width are found to be
are $\;T$ independent and, if
interpreted as a measure of $\xi ^{-1}$would correspond to $\xi \approx 2.5a$
and $\xi \approx a$ respectively in the underdoped and optimally doped
YBCO.\ In the slightly overdoped YBCO$_{7}$ no INS\ peak has been detected
\cite{PBourges:1998}\textbf{.\ }However the INS\ experimental \textbf{q}
width might be broadened by incommensurability effects.\ Also taking into
account the actual band structure could yield a different shape of the
\textbf{q} dependence of $\chi $(\textbf{q},$\omega $) thereby increasing
the \textbf{q} width of the AF\ peak with respect to $\xi ^{-1}$ \cite%
{RSMarkiewicz:2003, RSMarkiewicz:2006}. Another approach to determine $\xi
\; $has been to analyze the longitudinal and transverse relaxation times $%
T_{1}$\ and $T_{2G} $ of the $^{63}$Cu NMR within a phenomenological
approach, assuming that $\chi $(\textbf{q},$\omega $) has a Lorentzian shape
peaked at \textbf{q}$_{AF}$ \cite{AJMillis:1990}.\ For YBCO$_{7}\;$this
analysis gives\ an estimate of $\xi \approx 2a\;$at room $T$, with a slight
increase down to $T_{c}$. The extension of such a phenomenological analysis
to the underdoped compounds in presence of the pseudogap is highly
questionable and totally inconclusive \cite{AGoto:2002}.

In summary, the length scale $\xi _{imp}(T)$ deduced from the impurity
experiments agrees quantitatively with $\xi (T)$ obtained from the analysis
of the $T_{1}$\ and $T_{2G}$ $^{63}$Cu NMR\ data in the optimally doped
compound, and is always \textit{larger} than the determinations done by INS\
experiments.\ \textbf{\ }So all this strongly suggests that $%
\xi _{imp}$ , which increases markedly at low$\;T$ and in the pseudogap
phase, 
%should really
 might be one of the best estimates of $\xi \;$\ of the pure
system.

We have also shown that the incidence of impurity scattering on transport
properties gives information on the properties of the pure material. For
instance, the universality of the relation between $\Delta T_{c}$ and\ $%
\hbar /\tau $ induced by impurities can be taken as an evidence that the
carrier density in the pure compound is indeed that of the doped holes.\
Furthermore, the Hall effect data in cuprates with impurity substitutions
has allowed one to ascertain that two independent relaxation rates
characterize the transport in these 2D electronic systems and that
impurities contribute independently to both.

Impurity experiments have revealed many important aspects of these
correlated magnetic systems, including the importance of the electronic
correlations and magnetic effects in the perturbations induced by in-plane
point defects in these systems. We have stressed the need to produce, if
possible, systems free of intrinsic disorder to avoid non generic conclusions about the
phase diagram. The actual properties of the specific response to isolated
spinless impurities have also been found to reveal quantities which are
believed to be specific to the pure systems.\ Any theory aiming to describe
the physical properties of the high$\ $ $T_{c}$ superconductors should be
able to explain both the existence of the local moments induced by spinless
defects, the residual resistivities associated with these, the magnitude of
the correlation length $\xi _{imp},$ the energy scale $\Theta $ and the
occurrence of the Kondo like behavior in magnetic and transport properties.\

\section{Theories of impurities in correlated hosts }
\label{sec:thy2D} The theory of moment formation in uncorrelated
metals was begun by Friedel\cite{JFriedel:1958} and Anderson
\cite{PWAnderson:1961}, who pointed out that a sufficiently strong
{\it local} Coulomb interaction $U$ could effectively project out
of the Hilbert space corresponding to a weakly hybridized atomic
level the state with double (spin-up and spin-down) occupation.
This leaves  in the effective low-energy Hamiltonian for a metal
with a single impurity an essentially localized spin-1/2 magnetic
degree of freedom.   The subsequent screening of this moment by
the surrounding {\it noninteracting} electron gas as temperatures
or energies are further reduced is a remarkable many-body
phenomenon known as the Kondo effect,  the subject of intense
theoretical activity in the 1960s and 70s. This literature and
many further developments have been nicely reviewed by Hewson
\cite{ACHewson:1993}.

 While we will not discuss the Kondo problem in a noninteracting
 metal
in detail here, it is important to note that the energy scales for
the processes of  magnetic moment formation and for moment
screening by the electron gas are typically different by orders of
magnitude, and may be discussed independently.  By contrast, the
physics of vacancies and other impurity states in strongly
interacting chains and ladders, as discussed in Sec. \ref{sec:1D},
suggests that the two phenomena of ``moment formation"
(impurity-induced magnetization) and screening inevitably occur
together, even for weak interactions in the 1D or quasi-1D host
material.  This naturally raises the question of the effect of a
nominally nonmagnetic potential in an interacting host in two or
higher dimensions.  In these cases, special exact analytical
methods available to treat the single-impurity Kondo problem, or
nonmagnetic impurities in interacting chains or ladders, are not
applicable.  However, many authors have attempted to treat the
homogeneous strongly correlated electron host system using the
approximate methods discussed briefly in Sec. \ref{sec:models},
and subsequently added an impurity-type perturbation.
%Given the
%well-known difficulties of obtaining controlled results for
%homogeneous strongly interacting systems, it is perhaps not
%surprising that these approaches differ considerably in their
%results for this problem, and most do not reproduce some essential
%features of experiments on impurities in the CuO$_2$ planes of the
%cuprates.
In this section we will try to identify the ingredients
of a minimal tractable model which can be used to study the
impurity in a correlated host.

\subsection{Early results}
\label{subsec:thy2d_early} Early interest in the problem of
impurity induced magnetic moments in  interacting systems arose in
the context of studies of transition metal alloys such as Pd:Ni.
In the dilute Ni limit, the appropriate problem is to study a
single Ni embedded in the  nearly ferromagnetic Pd host metal. A
good deal of progress was obtained using so-called ``local
paramagnon theories"\cite{PLederer:1968}, in which the nearly
magnetic host was treated within Stoner theory, whereby the
uniform susceptibility was enhanced by the factor $[1-U_0
\rho_0]^{-1}$, where $U_0$ is the Hubbard interaction on Pd sites,
and $\rho_0$ the bare density of $d$-band states at the pure Pd
Fermi level.  The Ni site is then simply assumed to have its own,
larger Hubbard interaction for doubly occupied sites, $U_0+\Delta
U$, which is above the threshold  to create a magnetic state
according to the Stoner model.   The spatial extent or character
of the local magnetic state does not appear to have been studied
extensively, as the concern was primarily to account for bulk
thermodynamic and transport properties.

Questions regarding interaction-induced magnetic moments arose
again in studies of the metal-insulator transition in P-doped Si
and similar systems (for a review, see \cite{PALee:1985}). Scaling
treatments of the disordered interacting
metal \cite{AMFinkelshtein:1983} near the transition included
interactions effectively to all orders, but disorder only
perturbatively, to leading order in $(\epsilon_F\tau)^{-1}$.
Within such theories, the spin susceptibility was found to be
strongly renormalized near the transition \cite{CCastellani:1986},
suggesting the formation of clusters of magnetic moments with
anomalously slow dynamics.  This notion was confirmed generally by
experiment \cite{MAPaalanen:1986,HAlloul:1987}, but results on the
metallic side were found to be more consistent with a picture of
weakly interacting diffusive quasiparticles and random local
moments. First-principles theories of  moment formation in such
random systems  were then investigated, in which interactions were
treated
perturbatively \cite{MAPaalanen:1988,SSachdev:1989,MMilovanovic:1989},
and disorder treated exactly, within numerical finite-size
studies.  A Hartree-Fock factorization of the interaction term,
equivalent to that performed by
%Anderson
\cite{PWAnderson:1961} for
the single-impurity model, leads to a an effective single-particle
problem with random hopping.  For sufficiently large on-site
Coulomb interaction $U$, it was found that  some of the magnetic
susceptibility eigenvalues turned negative, signalling an
instability towards local moment formation.
%Lakner et al
\cite{MLakner:1994} then pointed out that qualitatively very
similar results could again be obtained by studying the isolated
scattering potential in an otherwise homogeneous interacting
medium.
 % In particular, the probability of moment formation on
%the ``impurity site" to which the probability of hopping had been
%altered was found to increase with the Hubbard interaction $U$ in
%the host material, treated within the Hartree-Fock approximation.
These early works on the Si:P problem also do not appear to have
been concerned with the details of the magnetic state thereby
created.

\subsection{Weak coupling approaches} \label{sec:thy2d_nonmag_weak}

In these early works, correlations in the host material were
modelled with a Hubbard Hamiltonian (Eq. \ref{Hubbard1}), and
moment formation was found to be enhanced with increasing
repulsive $U$. The basic physics can be obtained via the usual
mean field factorization:

\begin{eqnarray}\label{impurhamilt}
H&=&-\sum_{i,j,\sigma} t_{ij} c_{i\sigma}^{\dagger}c_{j\sigma}
+\sum_{i,\sigma}( U n_{i {\bar {\sigma}}} + \epsilon_\sigma
\delta_{i_m,i} -\mu) c_{i\sigma}^{\dagger}
c_{i\sigma}\nonumber\\&&
\end{eqnarray}
where $\epsilon_\sigma$ is the on-site impurity strength, $i_m$ is
the impurity site, $\mu$ is the chemical potential, and
$n_{i\sigma}\equiv \langle
c_{i\sigma}^{\dagger}c_{j\sigma}\rangle$. The local magnetization
in this decoupling where spin rotation invariance has been
explicitly broken is simply $M_i =  \langle c_{i
\uparrow}^{\dagger} c_{i \uparrow} -c_{i \downarrow}^{\dagger}c_{i
\downarrow} \rangle$.  If the impurity potential on site $i_m$ is
taken proportional to the local occupation  of opposite spin $\bar
\sigma$, $\epsilon_\sigma=\epsilon n_{i\bar\sigma}$, this model is
identical to the local paramagnon models at the mean field level,
and can be used to describe a magnetic impurity.  On the other
hand, if $\epsilon_\sigma=\epsilon$ independent of spin, the model
describes a short-ranged screened Coulomb potential which may be
more appropriate for an impurity like Zn, which has a closed
shell. This latter case manifests  no magnetic moment  in zero
applied magnetic field if $U$ is below the critical value for
long-ranged ``background" antiferromagnetic order, consistent with
the experiments described in Section \ref{sec4}
%\ref{sec:expts_normal}
 which one
wishes to explain.  This is
% therefore
 the approach adopted by
%Bulut
\cite{NBulut:2000,NBulut:2001} and
%Ohashi
\cite{YOhashi:2001,YOhashi:2002} in an attempt to model Zn
impurities, and which we outline here. Note also that Bulut
analyzed Hamiltonian (\ref{impurhamilt}) for $t_{ij}$ limited to
nearest neighbor hoppings, while Ohashi included next nearest
neighbor hoppings.

The goal of such calculations is therefore to calculate the  spin
susceptibility $\chi({\bf q=0},{\bf R})$ which determines the
response  of the {\it inhomogeneous} system  in the presence of
the impurity to an applied homogeneous magnetic field, and compare
its temperature dependence to that of experiment (here the Fourier
transform is with respect to the relative electron coordinate
$\rr-\rr'$, leaving the dependence on the center of mass
coordinate $\R=(\rr+\rr')/2$).   Once this quantity is known, the
Knight shift and, in principle, the distribution of local fields
around the impurity as probed by NMR can be obtained. The physical
idea put forward by these authors is that the metallic state in
which the impurity is embedded manifests strong but noncritical
antiferromagnetic spin correlations maximized near ${\bf
q}=(\pi,\pi)$, which are enhanced locally by the impurity. The
related problem at criticality was studied by
\cite{AJMillis:2001}.

\begin{figure}[h]
\includegraphics[clip=true,width=.8\columnwidth,angle=0]{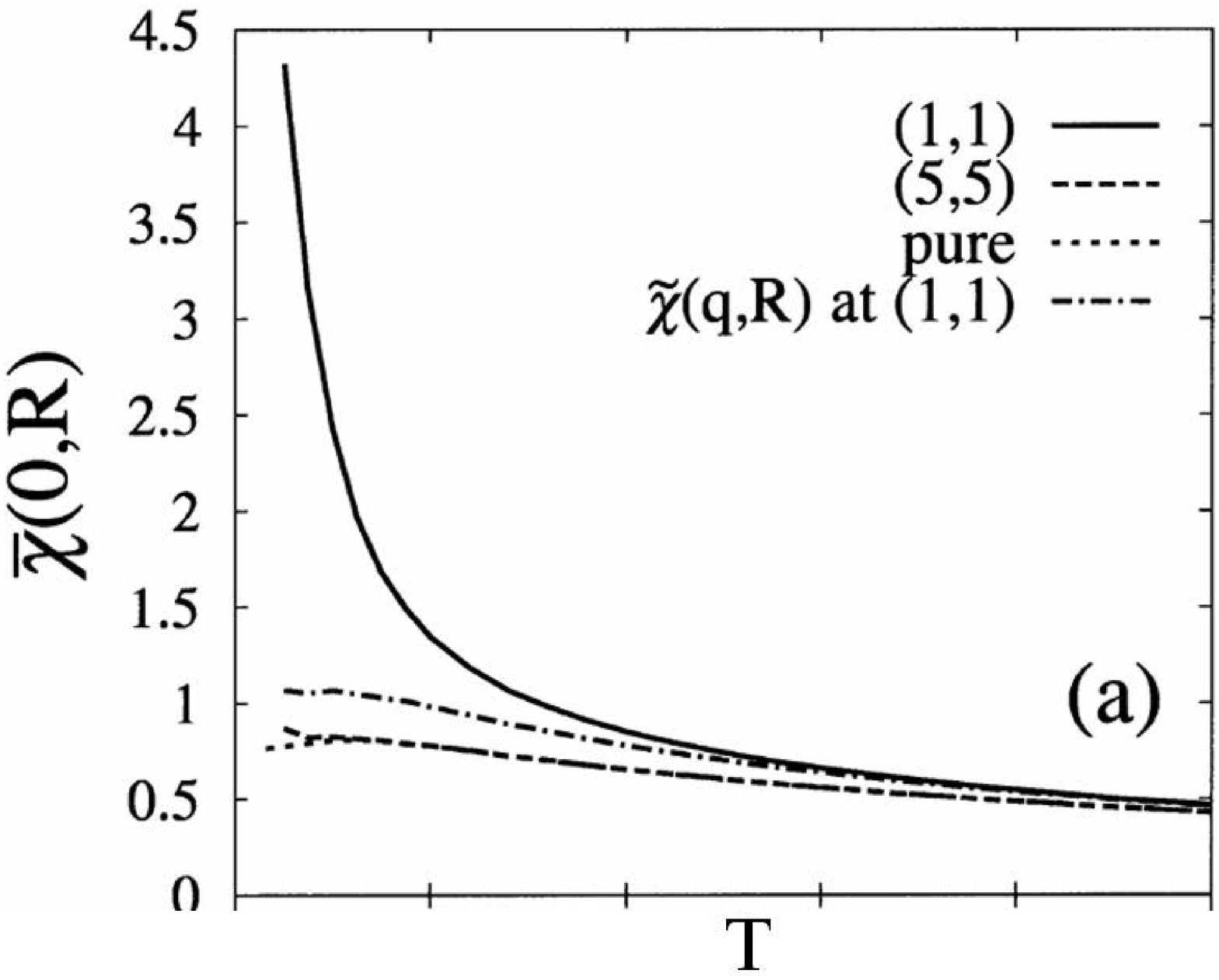}
\caption{Temperature dependence of magnetic susceptibility in RPA
at $\q=0$  for different sites $\R$ in the presence of a strong
potential $V_1=20t$ at (0,0) and a weaker one on nearest neighbor
sites of strength $V_2$.
 $\overline{\chi}$ is a weighted average of $\chi$ to account for the fact that
${\mathbf R}$ may coincide with a lattice site, the center of a
bond or of a plaquette  \cite{YOhashi:2001}.}
\label{fig:ohashiplot}
\end{figure}

The impurity potential then couples the enhanced response of the
system at $(\pi,\pi)$ and that near ${\bf q}=0$ which is then
measured in the NMR experiment. This can be most easily seen by
examining the susceptibility determined within the random phase
approximation (RPA)
\begin{eqnarray}
 \chi(\rr,\rr') &=& \chi_0(\rr,\rr') + U \sum_{\rr''}
 \chi_0(\rr,\rr'')\chi(\rr'',\rr'),\label{eq:inhomRPA}
\end{eqnarray}
where $\chi_0$ is the susceptibility determined in the
noninteracting system $U=0$, given exactly by the noninteracting
Green's function $G_0(\rr,\rr')$ in the presence of a single
short-range impurity described by a single-impurity $T$-matrix. In
the homogeneous case $\chi(\rr,\rr')=\chi(\rr-\rr')$, the result
in Fourier space is the familiar
$\chi(\q)=\chi_0(\q)(1-U\chi_0(\q))^{-1}$, i.e. all momentum
components are decoupled.  On the other hand, in the general case
(\ref{eq:inhomRPA}), Fourier transforming with respect to the
relative coordinate cannot disentangle the various momentum
components, leading to a general integral equation relating
$\chi(\q,\R)$ to $\chi(\q',\R')$.  Similar results for the
numerical evaluation of these equations were  obtained in
\cite{NBulut:2000,NBulut:2001} and
\cite{YOhashi:2001,YOhashi:2002}, and are plotted e.g. in Fig.
\ref{fig:ohashiplot} as a function of temperature.
While the enhancements at $\q=0$ are smaller than at $(\pi,\pi)$
as expected, they are still quite significant  at low temperatures
and appear to follow a $\sim 1/T$ form down to the lowest
temperatures on the next nearest-neighbor sites (1,1) when the
extended impurity potential on the nearest neighbor sites is
large. For smaller nearest neighbor coupling, the enhancement at
$q=0$ weakens.  At intermediate values of the extended potential
strength, the susceptibility $T$ dependence resembles the
experimental NMR results discussed in Section \ref{sec4}
%\ref{sec:expts_normal},
 where in the metallic phase this
temperature dependence is  found to follow a behavior $\chi\sim
(T+\Theta)^{-1}$.  Note that the mean field form of the
Hamiltonian solved, whereby the spin rotation invariance is
explicitly broken, precludes these theoretical results originating
from Kondo-type physics.  Indeed, to describe Kondo physics one
needs to allow spin-flip processes yielding transitions between
the two spin eigenstates locally, and these large amplitude
fluctuations are not included within RPA.

\begin{figure}[h]
\includegraphics[clip=true,width=.9\columnwidth,angle=0]{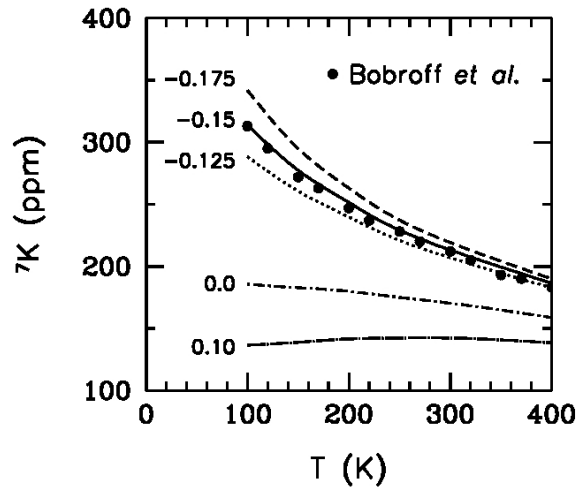}
\caption{Temperature dependence of Knight shift
\cite{JBobroff:1999} for Li impurities in YBCO within RPA approach
in the presence of a strong potential at $V_1=-100t$ at (0,0) and
a weaker one on nearest neighbor sites of strength given in the
figure. From \cite{NBulut:2001}.} \label{fig:bulutKnight_Li}
\end{figure}

Using an essentially identical formalism, Bulut explicitly
attempted to fit the Knight shift data on Li and Zn impurities in
optimally doped YBCO samples, with considerable success
\cite{NBulut:2001}. In his work, however, the potential on
nearest-neighbor sites is taken to be attractive, leading to a
magnetic moment residing primarily on these sites.   Hyperfine
parameters, band parameters, and the Coulomb repulsion $U$ were
chosen from fits to NMR data on the pure system.  In Fig.
\ref{fig:bulutKnight_Li} we reproduce the Knight shift data of
\cite{JBobroff:1999} together with Bulut's calculation.    It is
interesting to note that similar quality fits to data on a
different impurity, Zn, were obtained with nearly identical values
of the nearest neighbor impurity potential, suggesting a
universality of the magnetic response to spinless in-plane defects
as discussed already in Section \ref{sec4}.
%\ref{sec:expts_normal}.
The same model also reproduces the staggered magnetization pattern
seen by experiment on sites near the impurity (although a
quantitative comparison was not attempted).  %Similar results were
%obtained in exact quantum Monte Carlo calculations for a strong
%on-site potential in the Hubbard model \cite{NBulut:2003}; these
%also showed that a nearest neighbor potential is dynamically
%generated in the course of the Monte Carlo procedure.

%On the other hand, the
The physical meaning of the nearest neighbor
potential which appears so vital to explain the impurity-induced
enhanced magnetic correlations is not completely transparent. Band
calculations in fact suggest that such impurity potentials have a
range of an \AA\ or less \cite{LLWang:2005}.  This term may
therefore play the phenomenological role of a different
impurity-induced correlation effect in a more complete theory.
 The on-site impurity potential breaks translational invariance such
that the response to a uniform magnetic field contains a
contribution related to the staggered susceptibility of the pure
system. This quantity is quite large, but its weight in ${\bf
q}$-space is small. Adding a nearest neighbor potential $V_2$
boosts the weight at the antiferromagnetic momentum.  In addition,
Monte Carlo calculations \cite{NBulut:2003} begin with an on-site
potential only, and find that an effective nearest neighbor
potential of roughly the right order to fit the Knight shift data
in the RPA approach is dynamically generated.  This phenomenon of
creation of dynamical extended potentials by nonmagnetic pointlike
perturbations  in the presence of Coulomb correlations had been
pointed out earlier \cite{DPoilblanc:1994a} and discussed more
generally in the weak coupling context \cite{WZiegler:1996}. A
similarly enhanced range of the effective impurity potential due
to renormalized electron-phonon coupling was predicted by
\cite{MLKulic:1997}.  So in this sense, a vacancy or strong
potential in a Hubbard model may indeed provide a minimal model to
describe a Cu substituent in the CuO$_2$ plane of the cuprates
around optimal doping, if the extended effective potential is
generated systematically. However, the RPA theory in the form
given in the work of Ohashi and Bulut requires a phenomenological
finite range potential. Furthermore, it is not systematically
extendable to the underdoped side of the phase diagram, although
plausible results can be obtained by putting in the pseudogap in
an ad hoc way \cite{NBulut:2001}.

\subsection{Strong coupling approaches}\label{subsubsec:thy2d_nonmag_strong}
 It is generally believed
that the transition to the Mott antiferromagnet at low doping in
the cuprates can be described theoretically only with models which
invoke local Coulomb interactions in the strong coupling limit.
Historically, the problem of a nonmagnetic defect in a correlated
system was first proposed in the cuprate context by
\cite{AMFinkelstein:1990} to account for the depression of
superconductivity in La-based compounds (see next section). For
the normal state case, \cite{DPoilblanc:1994a,DPoilblanc:1994b}
were the first to calculate the electronic structure of the $t-J$
model (Eq.(\ref{t-J1})) when a single spin was  frozen, i.e. all
exchange couplings and hoppings connected to a given site 0 on a
square lattice were removed. They found that the rest of the
system had a net magnetic moment corresponding to net spin-1/2
distributed over the sites near the impurity.  A hole added to the
system was found to bind to the impurity site in one of  four
symmetry channels $s$, $p_{x,y}$ and $d$ corresponding to the
irreducible representations of the tetragonal group.  These bound
states occurred for couplings above a critical value of $J$ of
order $\sim 0.1t$, and it was speculated that for a
thermodynamically finite density of holes the critical exchange
would be enhanced due to the additional allowed decay channels for
the bound states. Further finite size numerical calculations on
$t-J$ impurities using Hubbard operator techniques
\cite{SOdashima:1997,SOdashima:2000} indeed showed the weakening
of bound states at finite doping, but did not investigate the
transition to spontaneous local magnetization.

While the numerical calculations of
\cite{DPoilblanc:1994a,DPoilblanc:1994b} indicated a tendency to
spontaneous moment formation around a nonmagnetic potential in a
strongly correlated host, they left open many questions regarding
the physical origin of the phenomenon, as well as its dependence
on doping and correlation strength.   Along with the experiments
on Zn-doped cuprates reviewed in Section \ref{sec4}
%\ref{sec:expts_normal}
 at
about the same time, this stimulated considerable interest in the
search for tractable analytical approaches to the problem.
It is worth recalling that there are many plausible schemes to
obtain results the strong-coupling regime in 2D, none of which are
fully controlled. \cite{MGabay:1994} studied the doping
dependence of an impurity moment in a slave boson approach to the
$t-J$ model, as discussed in Section \ref{sec:models}, and tried
to relate the size of the moment associated with Zn in different
cuprate materials and for different dopings to the screening
length of the spinon fields, particularly in the
 spin gap phase.  He also predicted a local destructive effect of the
 impurity-induced magnetism on the spin gap itself.
 \cite{NNagaosa:1995} considered an impurity in a $\pi$-flux phase
 of the Heisenberg model using a slave boson representation and
 predicted the existence of a spin-1/2 bound state.  Similar
 results were obtained using a slave fermion representation.
 These authors noted explicitly the analogy with 1D end-chain
 defects, and pointed out that although the effective potential
 for the spin was weaker in 2D, a Curie-Weiss term in the
 susceptibility would be the likely result.

\begin{figure}[h]
\includegraphics[clip=true,width=0.9\columnwidth,angle=0]{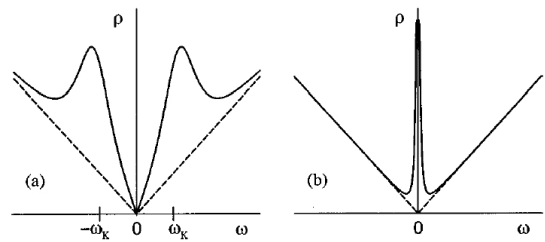}
\caption{Schematic plot of spinon density of states for a) spin-1
impurity; b) spin-0 impurity in model of \cite{RKilian:1999}.
Dashed line: density of states for pure  flux phase. Solid line:
with impurity. } \label{fig:fuldefig2}
\end{figure}

\cite{GKhaliullin:1997} and \cite{RKilian:1999} produced detailed
calculations in a similar spirit which resulted in specific
analytical predictions for impurity properties in the spin gap
state.  The spin- or pseudogap homogeneous state was modelled by a
``drone"-fermion \cite{GKhaliullin:1997} or more conventional
slave boson approach \cite{RKilian:1999} which yielded similar
results.  In both cases auxiliary bond field phases are chosen
within a homogeneous mean field description  to give a linear
`pseudogap" in the spinon sector of the theory, with density of
states $\rho^0(\omega)= |\omega|/D^2$, where $D$ is the spinon
half bandwidth, see Fig. \ref{fig:fuldefig2}b.  Formally, this
choice of mean field parameters corresponds to the flux phase of
\cite{IAffleck:1988}. To model a nonmagnetic impurity in the
pseudogap state, a site is then removed from the system by means
of a local potential term $\lambda_0 f^\dagger_0 f_0$, with
$\lambda_0\rightarrow \infty$, and calculation of the impurity
$T$-matrix then yields a spinon bound state at zero energy
$\rho_{imp}(\omega) =\delta(\omega)$.   The local susceptibility
due to the bound state varies as
\begin{eqnarray}
  \delta \chi(\R,T) &=& {1\over 2\pi} {\phi(\R)\over R^2} {1\over
  T \ln D/T},\label{eq:Fulde_chi}
\end{eqnarray}
where the phase factor $\Phi(\R)$ gives the spatial dependence of
the spin polarization in a magnetic field,
\begin{eqnarray}
  \Phi(\R) &=& {1\over 4} \left|\tilde R^+ e^{i\pi R^+/2} + \tilde R^- e^{i\pi R^-/2}
  \right|^2,
\end{eqnarray}
with $R^\pm=R_x+R_y$ and $\tilde R^\pm = (R_x\pm i R_y)/R$.  This
is a staggered pattern of spins on {\it one sublattice only} , in
contradiction with the results of the weak coupling calculations
described above, and considered by the authors \cite{RKilian:1999}
as an artifact of the mean field theory which severely
underestimates long-range antiferromagnetic correlations, as
evidenced by the fact that the antiferromagnetic correlation
length is absent from the expression (\ref{eq:Fulde_chi}).  To
improve the accuracy of the calculation and include these effects,
a phenomenological RPA summation was performed along the lines of
\cite{DMorr:1998},  $\tilde\chi(\q)=\chi^0(\q)/(1+J_\q
\chi^0(\q))$, where $\chi_0$ is the homogeneous susceptibility and
$J_\q = 2J(\cos q_x +\cos q_y))$, as in the theory of the nearly
antiferromagnetic Fermi liquid (NAFL) \cite{PMonthoux:1994}.  For
the inhomogeneous susceptibility, approximate analytical
expressions can be obtained for the long distance behavior,
\begin{eqnarray}
  \delta\tilde\chi(\R,T) &\simeq& -\cos(\Q\cdot \R) {1\over 8\pi} {1\over R^2}
  {\xi(T)^2\over T\ln D/T},\label{eq:Fulde_chi_final}
\end{eqnarray}
where $\Q=(\pi,\pi)$, and $\xi(T)\equiv
[J\chi_0(\Q,T)/(1-4J\chi_0(\Q,T))]^{-1/2}$ is the NAFL
antiferromagnetic correlation length {\it for the pure system}.
The final result Eq.(\ref{eq:Fulde_chi_final}) shows the commensurate
staggered behavior of the moment %(see Fig. \ref{fig:fuldefig5}),
as well as the $(T\ln T)^{-1}$ divergence of the susceptiblity,
reflecting the marginal bound state found already in the mean
field theory.

\cite{RKilian:1999} compared their results with data from
\cite{JBobroff:1997b} on underdoped YBCO, and concluded that a
good fit could be obtained if the temperature dependence of the
correlation length in Eq.(\ref{eq:Fulde_chi_final}) was assumed to
follow $\xi(T)\sim 1/(a+bT)$, with $a$ and $b$ adjustable
parameters. With these parameters fixed, the theory can account
for the broadening of the NMR line with increasing Zn
concentration.  It is seen that, while not all parameters could be
computed directly from microscopic theory, a physically reasonable
picture of the impurity in the pseudogap state is ultimately
obtained.  It should be noted however that the spatial decay of
the local susceptibility as one moves away from the impurity has a
different functional form in \cite{SOuazi:2004} compared with
Eq.(\ref{eq:Fulde_chi_final}). Indeed, NMR lineshapes cannot be
accounted for using $R^{-2}$ or $R^{-3}$ spatial dependences for
$\delta\tilde\chi$.

%\begin{figure}[h]
%\includegraphics[clip=true,width=.8\columnwidth,angle=0]{fuldefig5.eps}
%\caption{  Staggered pattern of magnetization produced in a field
%in the presence of nonmagnetic impurity \cite{RKilian:1999}.}
%\label{fig:fuldefig5}
%\end{figure}

\begin{figure}[h]
\begin{center}
\leavevmode
\begin{minipage}{.45\columnwidth}
\includegraphics[clip=true,width=1.05\columnwidth]{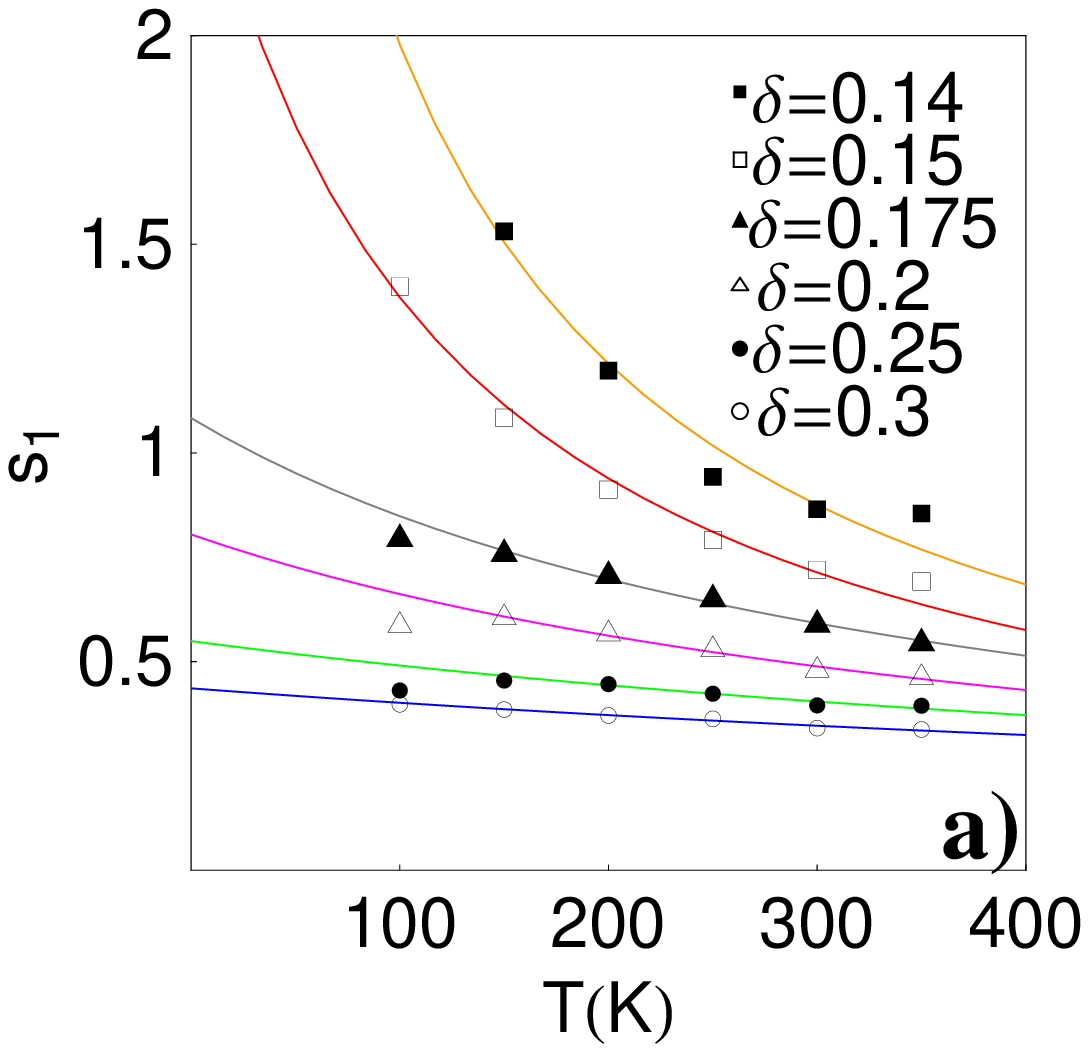}
\end{minipage}
\begin{minipage}{.45\columnwidth}
\includegraphics[clip=true,width=.99\columnwidth]{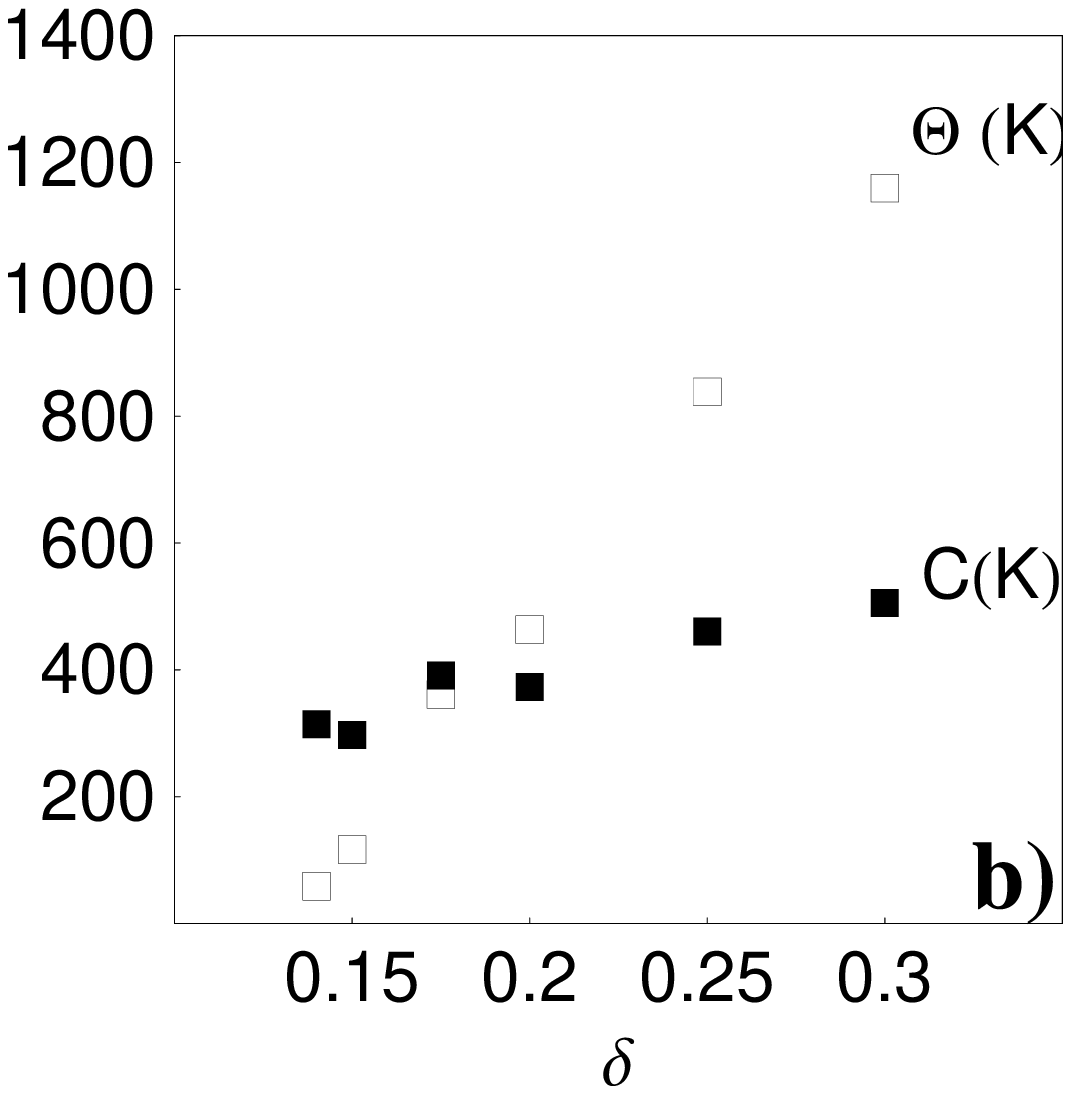}
\end{minipage}\\
%\begin{minipage}{.46\columnwidth}
%c)\includegraphics[clip=true,width=.99\columnwidth]{Tvsdel.eps}
%\end{minipage}
\begin{minipage}{.5\columnwidth}
\includegraphics[clip=true,width=1.17\columnwidth]{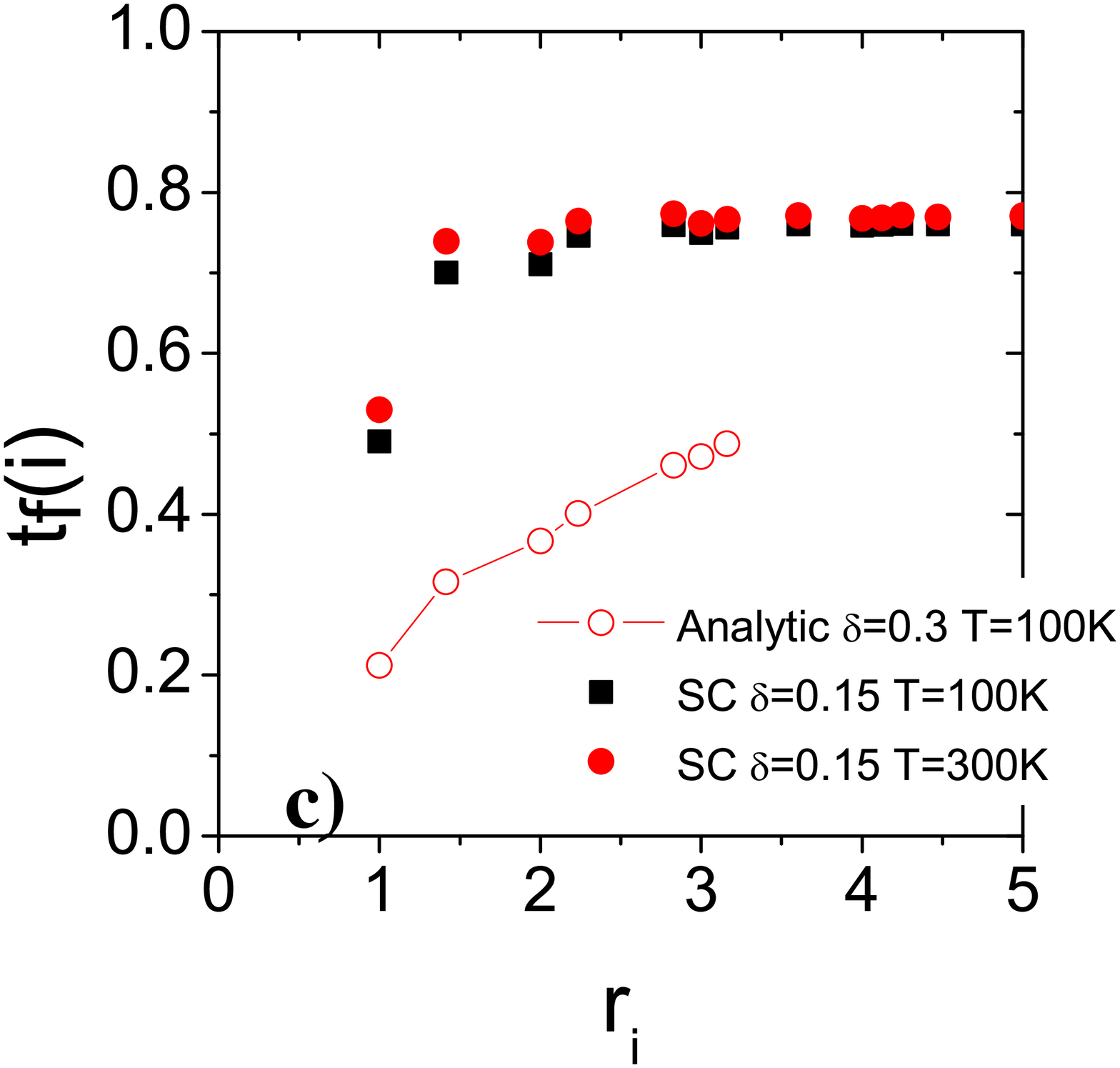}
\end{minipage}
\begin{minipage}{.46\columnwidth}
\includegraphics[clip=true,width=.99\columnwidth]{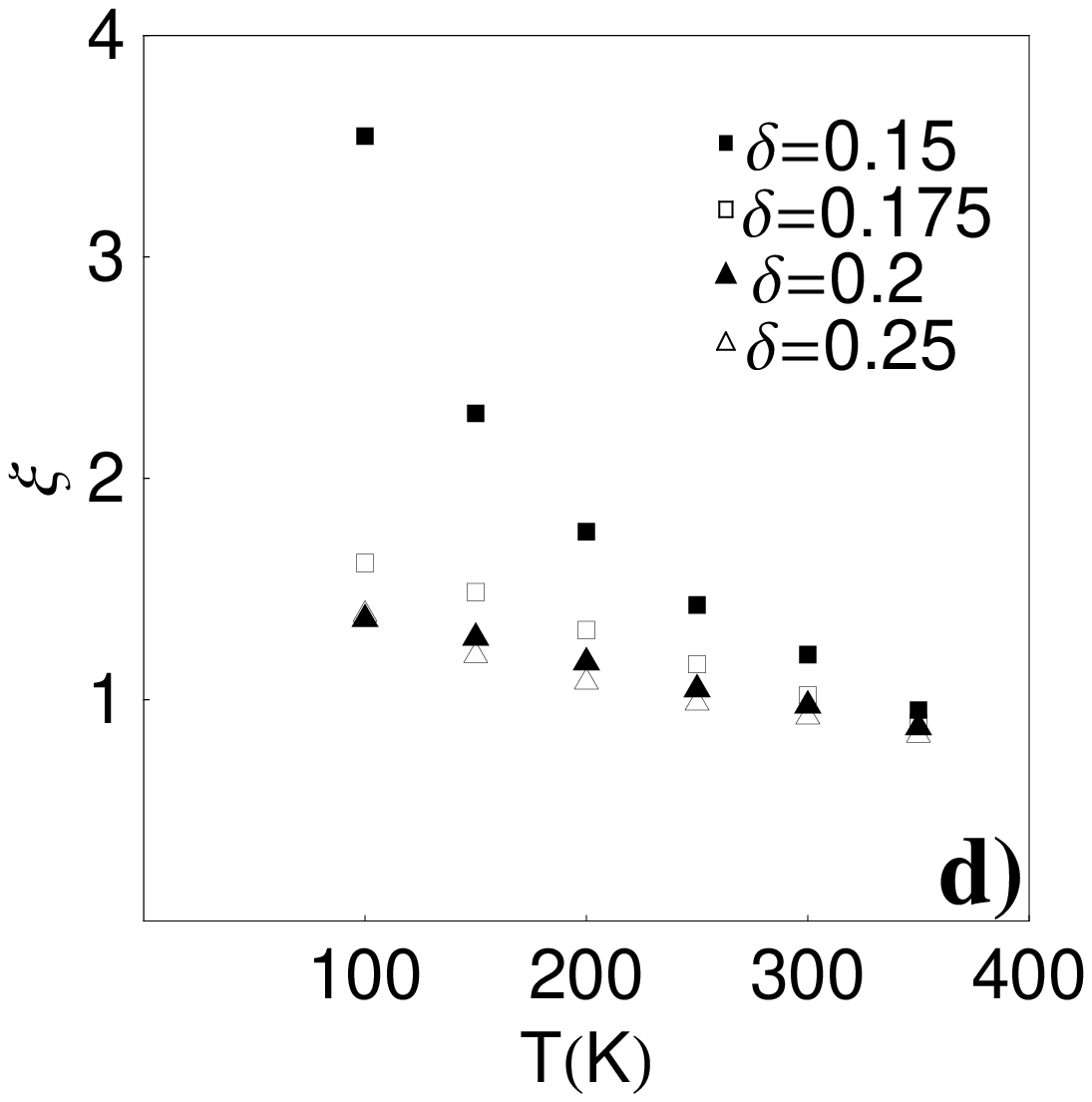}
\end{minipage}
\caption{Results for fully self-consistent evaluation of
magnetization from slave boson equations \cite{MGabay:2007} .  a) Normalized
magnetization $s_1$ on nearest-neighbor site as function of $T$
for values of doping $\delta$ from 0.14 to 0.30.  Solid lines show
fits to $s_1=C/(T+\Theta)$.  b) Effective moment constant $C$ and
Curie-Weiss constant $\Theta$ extracted from fits in a) vs.
$\delta$. c) Spinon bandwidth in fully self-consistent evaluation as
a function of distance from the impurity site at filling
$\delta=0.15$ and $T=100$K (filled squares) and 300K (filled
circles).  Bandwidth from semianalytic calculation at $\delta=0.3$
and 100K is shown (open circles) for comparison. d)
Temperature dependence of extracted theoretical correlation
lengths $\xi$ vs. $T$ (compare Fig. \ref{fig:xi_imp_vs_T}) . } \label{fig:semelfig}
%\begin{figure}[h]
%\begin{center}
%\leavevmode
%\includegraphics[clip=true,width=1.05\columnwidth]{semelfig_bandwidth.eps}\vskip
%-1cm
%\begin{minipage}{.49\columnwidth}
%\includegraphics[clip=true,width=.9\columnwidth]{semelfig_a.eps}
%\end{minipage}
%\begin{minipage}{.49\columnwidth}
%\includegraphics[clip=true,width=.9\columnwidth]{semelfig_b.eps}
%\end{minipage}
%\includegraphics[clip=true,width=.8\columnwidth,angle=0]{semelfig.eps}
%\includegraphics[clip=true,width=.8\columnwidth,angle=0]{semelfig_a.eps}
%\includegraphics[clip=true,width=.8\columnwidth,angle=0]{semelfig_b.eps}
%\caption{Results of theory of \cite{MGabay:2007} for the spinon
%bandwidth as a function of distance from the impurity site at
%different temperatures (top).     Bottom panels:  normalized
%staggered magnetization $S(r) \equiv (m(r)-m_0)/m_0$ induced by a
%magnetic impurity in the $t-J$ model in the presence of a magnetic
%field B of $7.5$ Tesla, where $m_0$ is the magnetization of the
%homogeneous system induced by the field. Left: normalized
%magnetization $|S|$ near impurity at $T=$ 25K. Solid line: fit to
%$|S(r)|\propto K(r/\xi)/K(1/\xi)$ for $\xi=3.1$. Right:
%$T$-dependence of nearest-neighbor
% normalized magnetization $S_1$.}
%\label{fig:semelfig}
\end{center}
\end{figure}

The strong coupling approaches discussed above assumed the
existence of a pseudogap in the one-particle density of states, which is ultimately required for
the formation of the spinon bound state.  One would therefore expect the singular magnetic response
to disappear when the pseudogap does; indeed, this is consistent with the fact that no
 static
%induced magnetization
moment is observed in zero
field
% in
experiments on clean optimally doped cuprates and that the field
induced magnetic response is not a simple Curie (1/T) form.
% on clean optimally doped cuprates
(Section \ref{sec4}).
%\ref{sec:expts_normal}).
 However a theory of this type should be
able to describe the paramagnetic moments observed by NMR experiments in finite field
on an equal footing.  This was accomplished recently by \cite{MGabay:2007}, who considered
the optimally doped state of cuprates within a $t-t'-J$ model (Eq. \ref{t-t'-J}); the strong impurity potential implies a site from which spinons and holons are forbidden.
 The Hamiltonian was treated in a slave boson mean field approach, following \cite{MUUbbens:1992}.
This model remains metallic at low $T$ and is thus suitable for
the study of the normal state.
%\begin{figure}[h]
%\includegraphics[clip=true,width=.95\columnwidth,angle=0]{semelfig_bandwidth.eps}
%\caption{Results of theory of \cite{ESemel:2007} for the spinon
%bandwidth as a function of distance from the impurity site at
%different temperatures. } \label{fig:semelfig_bandwidth}
%\end{figure}
One introduces the local order parameters
\bea\label{op}
%\sum_{\s} \;\s
%<f^{\dag}_{i\s}f^{\dag}_{j-\s}>=\D_{ij}
%&{\rm with}&j=i \pm x(\pm y)\;\;
%\nonumber\\
<f^{\dag}_{i\s}f_{j\s}>&=&\chi_{ij}^\sigma \nonumber\\
< b^\dag_{j}b_{i}>&=&Q_{ij} \nonumber\\
%< b^\dag_{i}>=B_i
%\nonumber\\
\langle f_{i\uparrow}^\dagger
f_{i\uparrow}-f_{i\downarrow}^\dagger f_{i\downarrow}\rangle &=&
m_i , \eea where $f_i$ and $b_i$ create spinon and holon fields,
respectively, and $i,j$ are nearest neighbors. Within such a
theory, the spinon bandwidth $t_{f}^{ij}= tQ_{ij} + (J/
2)\chi_{ij} $ is strongly suppressed close to half-filling
$\delta\rightarrow 0$. When the effect of the impurity is
accounted for, the bandwidth is found to decrease locally as well,
as shown in Fig. \ref{fig:semelfig}c. Such effects are quite
important for a quantitative treatment of local correlations and
electronic structure around the impurity,
 and are missed in the weak-coupling approaches discussed above.
A semianalytic solution of the mean field equations reveals that the local decrease of the
spinon bandwidth in the vicinity of the impurity has two effects. One is to produce an extended scattering potential which enhances the
staggered Fourier component of the local magnetizations $m_i$. The other is to increase the magnetic response, since -- in a Stoner-like picture -- a larger value of $J/t_f(i)$ brings the system locally closer to a magnetic phase. In order to treat correctly the non double occupancy constraint for the holons and the spinons, a numerical solution of the mean field equations is required \cite{MGabay:2007}. It yields quite plausible results
 for both the spatial decay of the staggered magnetization induced in a field,
and the temperature dependence, which is found to correspond to a Curie-Weiss law 
%with $\Theta =380$K,
similar to that found in experiment on overdoped samples. Indeed,
the results displayed in Fig. \ref{fig:semelfig} are in reasonable agreement with their experimental counterparts, shown in Figs. \ref{fig:theta_vs_doping},\ref{fig:xi_imp_vs_T}.
%The approximate numerical evaluation of the mean field equations by \cite{MGabay:2007} yields quite plausible results
% for both the spatial decay of the staggered magnetization induced in a field,
%and the temperature dependence, which is found to correspond to a Curie-Weiss law with $\Theta %=380$K,
% similar to that found in experiment on overdoped samples.
%While the results displayed in Fig. \ref{fig:semelfig} are for the unphysically large doping value value of $\delta=0.3$, a fully self-consistent
% numerical evaluation yields similar results for $\delta=0.15$.
%(W.Chen, private communication).
 It is intriguing that the results
of the slave boson mean field theory for the strong-coupling
$t-t'-J$ model in the spin
 liquid state (optimally doped case) are so qualitatively similar to the weak coupling results presented in Sec.\ref{sec:thy2d_nonmag_weak}.
This suggests indeed that for the optimally doped materials a Fermi liquid approach captures the essential details of the physics.
  On the other hand, in the strong coupling case there is no analog of the requirement of an extended bare impurity potential
to fit experimental data  as found by \cite{NBulut:2000} and
\cite{YOhashi:2001}.  This may indeed reflect the failure of the
weak coupling approaches to treat the renormalized band structure
in the vicinity of the impurity, e.g. the spinon bandwidth
 shown in Fig. \ref{fig:semelfig}. At present, no microscopic approach
 to the problem of an impurity in a correlated host has been able to reproduce systematically
 the crossover between underdoped and optimally doped regimes observed in the cuprates,
 and described in Section \ref{sec4}.
%\ref{sec:expts_normal}.
  In principle, slave boson and similar renormalized mean field theories
 are a promising avenue to pursue.  In addition, a phenomenological
approach by \cite{PPrelovsek:2004} has suggested a way to parameterize this crossover
 in terms of the doping dependent correlation length $\xi$ of the pure system.
  These authors showed that a crossover between the Curie and Curie-Weiss behavior of the local
 susceptibility observed in experiments as the system is doped is to be expected generically
 in a homogeneous doped antiferromagnet.
They made the ansatz that the impurity suppresses only the spin
correlations $\langle S^z_0S^z_j\rangle$ involving the impurity
site 0, with all others taken  to correspond to the pure system.
While this seems to contradict other microscopic analyses
discussed above, the results indeed resemble experiments
qualitatively and suggest that the  pure $\xi$ determines the
$T$-dependence of the system.

%\subsection{Magnetic impurities and Kondo effect in correlated systems} \label{subsec:thy2d_Kondo}
\subsection{Magnetic impurities in correlated systems} \label{subsec:thy2d_Kondo}
\begin{figure}[h]
\includegraphics[clip=true,width=.8\columnwidth,angle=0]{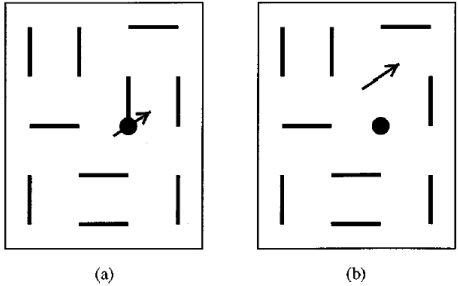}
\caption{Schematic picture of spin-1/2 impurity produced in case of a) Ni and b) Zn. } \label{fig:fuldefig3}
\end{figure}

One of the attractive features of the \cite{RKilian:1999} approach
is that it accounts naturally for differences observed between Zn and Ni substitutions
 for Cu in underdoped YBCO.
In Fig.(\ref{fig:fuldefig3}), we show a schematic picture of an RVB
type singlet state in the presence of an impurity of various types.
  In the case of a
nonmagnetic impurity like Zn, a spin singlet is broken while a spin-1/2
 is freed from a given singlet bond.
  By contrast, if one models the Ni perturbation as a spin-1 object,
 the spin-1/2 on the Cu site nearest the impurity screens the Ni spin, leading again
 to a free spin-1/2 degree of freedom.
This picture is very analogous to that which has been established
 rigorously for impurities in 1D spin chains in Sec.\ref{sec:1D}.
 Of course, as we have seen, more detailed calculations lead
to the``free" spin 1/2 being
spread out over a healing length in both cases.
% (Fig. \ref{fig:fuldefig5}).
To model a Ni spin $\S_0$ of magnitude 1 coupled to the surrounding Cu spins-1/2, \cite{RKilian:1999} add an impurity
potential consisting
of nearest neighbor exchanges with the impurity
$H_{imp}=J'\sum_\delta \S_0\cdot \s_\delta$
  to the system.  The spin 1 is decomposed $\S_0=\S_a+\S_b$,
 where $\S_{a,b}$ are spin-1/2 operators,
and the spin 1 algebra is retained by adding a ferromagnetic
 exchange term $-J_c\S_a\cdot\S_b$
 to the Hamiltonian, then allowing $J_c\rightarrow \infty$.
This gives
\begin{eqnarray} % \nonumber to remove numbering (before each equation)
 H_{imp} &=&-\sum_{\delta\sigma}(\Delta_\delta' f^\dagger_{0\alpha}f_{\delta\alpha} + h.c.) -J_c\S_{eff} \s_0,
\end{eqnarray}
where $\Delta_\delta'=J'\sum_\sigma \langle
f^\dagger_{\delta\alpha} f_{0\alpha}\rangle$
 is the local flux phase amplitude on bonds near the impurity,
 $\S_{eff}$ is a localized spin-1/2 operator,
and $\s_0$ is the spinon spin density on site 0.
Thus this case reduces to an impurity which locally renormalizes
the spinon bandwidth and adds a local exchange with a free
spin-1/2 in a fermionic bath with pseudogap.
 %These models will be discussed in more detail
%in Sec. \ref{sec:SC} below, but for now
%we note that this 1-impurity problem can be solved by standard
%methods analogous to those which led to (\ref{eq:Fulde_chi_final})
%within a mean field treatment.
Note that such a treatment clearly neglects the transverse spin fluctuations which lead to the Kondo effect.
Nevertheless the pole in the $T$-matrix is found to have a resonance in the pseudogap at finite energy
 $\omega_K=(\pi/4)J'/\ln (D/J')$,
leading to a contribution to the density of states
\begin{equation}
\delta\rho(\omega) = {1\over 2}
{\omega _K^2\over\ln D/\omega _K}{|\omega|\over
(\omega ^2-\omega _K^2)^2+ (\pi\omega _K^2/4\ln D/\omega _K)^2},
\end{equation}
whose form is shown schematically in Fig.(\ref{fig:fuldefig2}a).
The impurity susceptibility after performing the identical
RPA analysis as above is found to be
\begin{eqnarray}
 \delta\tilde\chi(\R,T) &\simeq& {J_c(0)J  \over J'^2} \cos(\Q\cdot \R) {3\over 64\pi} {1\over R^3}
  {\xi(T)^2\over T},
\label{eq:Fulde_chi_final_S=1}
\end{eqnarray}
where $J_c(0)$ interpolates between $D$ for $J'=J$
and $2\omega_K$ for $J'\ll J$.
This expression is quite similar to the nonmagnetic case
 (\ref{eq:Fulde_chi_final}), but decays more rapidly,
as $\sim 1/R^3$, and has a more singular $1/T$ dependence.
The reduction of both spin-0 and spin-1 impurity models to Hamiltonians
 with  effective localized spin-1/2 exchange terms
raises the question of the fate of the Kondo effect, the screening
of a single localized spin exchange coupled to  a free Fermi gas,
as electronic correlations increase.
 This effect was neglected in the mean field treatments above.
Historically, the question was studied first by
\cite{TSchork:1994,GKhaliullin:1995}.
 These authors began with the slave boson mean field theory of
% Read and Newns
\cite{NRead:1983} for the Anderson impurity model, which is known, in the noninteracting case,
 to correctly reproduce the
low-temperature Kondo resonance in the $f$-electron density of
states at the correct scale $T_K$. Perturbative corrections in the
Hubbard  $U$ were then found to simply renormalize the various
Read-Newns parameters, including the Kondo scale $T_K$:
\begin{eqnarray}
  {\tilde T_K\over T_K} &=& \exp{\alpha\over \rho_0 J(1+\alpha)},
\end{eqnarray}
where $\alpha=(3/2) \ln 2 U \rho_0$, $J$ is the exchange of the
Anderson impurity with the conduction electrons in the Kondo limit,
 and $\rho_0$ the density of conduction electron states at the Fermi level.
Thus for the usual repulsive-$U$ case the Kondo
scale is enhanced due to the suppression of the charge degrees of
freedom.
  Similar conclusions were reached by
\cite{WHofstetter:2000} in a more sophisticated calculation
combining dynamical mean field theory with numerical
renormalization group methods.   In addition to the effects of
interactions on impurity properties, there are
interaction-dependent renormalizations of conduction electron
properties by the Kondo impurity which are predicted to be visible
in thermodynamics. \cite{MNeef:2003}.

\subsection{Transport}
  There is currently relatively little theoretical work
attempting to explain the apparent close relation between  the
disorder-induced magnetic properties of cuprate systems and
transport properties, e.g. the resistivity ``upturns" at low $T$
observed in cuprate materials when they become very disordered or are subjected to  high fields.
 Based on the general properties of gauge
theory descriptions of the HTS phase diagram,
 \cite{NNagaosa:1997}
attempted to understand the residual resistivity measurements of
\cite{YFukuzumi:1996} in terms of a Boltzmann transport
description with assumed unitarity limit scatterers and carrier
concentration proportional
 to hole doping $n_h$ in the underdoped
cuprates, crossing over to $1-n_h$ for optimal doping.
 Some
criticisms of this picture have already been mentioned
 in Section \ref{sec4}.
%\ref{sec:expts_normal}.
Recently, \cite{HKontani:2006} calculated
the local susceptibilities in a Hubbard model with strong
impurities within FLEX-type approximations, and fed the resulting
self-energies and vertex functions into a disorder-averaged
transport calculation.  The enhanced scattering due to the
creation of local moments led in some cases to upturns in the
residual resistivity at low  temperatures similar to those seen in
experiments.
%{\bf These features become quite pronounced for large
%values of $U$ which suggests that they could be connected to some kind of critical
%regime}.
 While some details do not correspond to observation
(upturns in YBCO were found to be stronger than in LSCO, for
example), this is a promising step allowing one possibly to
approach the ``metal-insulator transition" from the metallic side.

\subsection{2D versus 1D}
Since there is to date no consensus on a theoretical description of the homogeneous
 system in the strongly correlated limit, further insight into the 2D case may be
 gained by considering, once again, the 1D situation. The  literature devoted to the
 study of impurities in doped 1D systems is
 quite significant, but we discuss  here
 %partly because it concerns not only the spin but also the
%  charge sector. Since we only need to refer to
a very small number of results
  which are directly relevant to the  2D case.
  %, we introduce these here rather than in Section \ref{sec:1D}.
  In 1D, a weak
link, i.e a slight local modification of the kinetic energy and
 of the magnetic coupling or a weak potential modelling a charge imbalance renormalizes
  to a very large perturbation as the temperature is reduced, in the case of repulsive
  interactions. As $T \to 0$, the transmission in the spin and charge channels approaches
   zero
\cite{CLKane:1992a, CLKane:1992b, AFurusaki:1993}.
%(Kane and Fisher '92, Furusaki and Nagaosa '93).
Applying a uniform magnetic field $H$ yields a local alternating magnetic polarization with an amplitude proportional to $H$ and a spatial extension generically of the order of the correlation length of the pure system.
%{\bf( note however the
%case of the $J_1-J_2$ model at the
% Majumdar-Gosh point, where $\xi\sim a$ for the pure system while the staggered polarization extends
% over the entire chain, in the presence of the impurity, see Sorensen et al '98)}.
The crossover between the weak perturbation and the strong perturbation regimes is frequently described in Kondo language, and the characteristic temperature $T_K$ depends exponentially on the difference between the kinetic (or potential or coupling) energy of the pure system and that of the modified link \cite{SRommer:1999, SRommer:2000}.
% (Rommer and Eggert '99 -'00).
% Similarly, the quasi- long range $d$-wave superconducting phase which characterizes the ground state of a two-leg Hubbard ladder with repulsive interactions is destroyed by spinless impurities, giving rise to a pinned $4k_F$ CDW instead \cite{EOrignac:1997}.
% (Orignac and Giamarchi '97).

 The impact of the impurity on transport and on magnetism
 can be considered separately, because of spin-charge
 separation at low energies, but, in both channels backscattering
 is very strong at low temperature.
%For spin-$1\over 2$ Hubbard chains, the amplitude of the
% staggered contribution at a distance $x$ from the impurity
% increases initially like $x^{1-K}$ \cite{SRommer:1999}.
%(Rommer-Eggert '00).
 The Luttinger liquid parameter $K$ varies from $1 \over 2$
 to $1$ upon increasing the hole doping, showing a drastic reduction
 of the enhancement of the staggered polarization as one moves toward
 the Fermi liquid limit. A similar behavior is observed in numerical
  studies of ladder systems \cite{SWhite:2002}.
% (White, Affleck, Scalapino '02).
 Calculations of transport properties also produce low $T$ upturns in the resistivity which reflect the
pinning of charge density waves by the impurities.

%The features that come out of the above theoretical
%description of impurities in the low energy sector of
%1D systems are in many respects qualitatively similar
%to those predicted by the approximate methods used in 2D.

\subsection{Summary}
At this stage, we may attempt to interpret some of the results described in
Section \ref{sec4} in the framework of the theoretical models that we just
discussed. Experiments reveal that a spinless impurity introduced in a
correlated host produces in its vicinity a very large and spatially extended
alternating magnetic polarization, similar to 1D. On the basis of NMR and $%
\mu $SR spectra it has been demonstrated that this magnetization is not
associated with a static moment. The maximum amplitude of the staggered
polarization is found to be of the form $C\;H/(T+\Theta )$, where $C$ 
is a constant and the extension $\xi _{imp}$ of this polarization
away from the defect seems to follow a similar $T$ dependence. For a fixed $T
$, the ( spatial) extent of the staggered state increases as one underdopes the system (%
$\Theta \rightarrow 0$).  A finite density of impurities produces a magnetization
which scales linearly with the concentration.
 Moving to the optimally doped and, beyond, to the
overdoped regime, both the amplitude of the alternating polarization and its
spatial extent decrease, bringing these gradually back in line with values
expected for a Fermi liquid metal.

Furthermore both NMR spin lattice relaxation and inelastic neutron
scattering data suggest that this entity should be viewed in the optimally
doped case as a resonant state rather than a free moment. By this we mean
that, in zero field, the total spin of the system in the ground state is $%
S=0 $ rather than $S=1/2$.\ The fluctuation rate $h/\tau $ $\approx
k_{B}\Theta \;$corresponds then to about 100K or 10meV. \ 

 For a wide range of hole dopings, the transport scattering rate in
CuO$_{2}$ materials displays a $-\log T$ upturn at low $T$. In the
overdoped case, this behavior is perfectly explained by weak localization effects in a
good metallic system.
In the underdoped case, the upturns are much larger 
than predicted by weak localization theory, and may be
due to impurity induced magnetism.
%An alternative explanation in the underdoped case, where the much larger %upturns with initial $-\log T$ variation scale initially linearly with defect content,
%is in terms of impurity-induced magnetism.}
% The much larger upturns with initial $-\log T$
%variation seen in the underdoped case have often been taken to signal the
%onset of the metal-insulator transition.\textbf{\ }However, these upturns
%have been found to scale initially linearly with defect content in the
%underdoped regime, which can be explained by impurity-induced magnetism as
%well.

 The above features are quite suggestive of Kondo physics,
with the picture of a screened moment formed around a nonmagnetic impurity.
However, as we have seen, a number of these magnetic and transport
signatures are qualitatively captured by the theoretical
models reviewed in the present section which ignore processes
leading to Kondo screening.   Typically, the impurity
is represented as a strong or infinite nonmagnetic potential,
which creates a nearby cloud of staggered polarization and
net paramagnetic moment by coupling the system's 
antiferromagnetic correlations to a homogeneous external 
magnetic field. The apparent universal 
magnetic response of any impurity substituting for Cu in the 
CuO$_2$ plane found by NMR is strong evidence in favor 
of such models.  Remarkably, 
 treatments which start either from
a weak-coupled Fermi liquid or from a strong
coupling description of the host are able to reproduce the key features seen in
experiments;  a characteristic energy $\Theta $ is found,
the temperature dependence
of the Knight shift, the resonant nature of the staggered
magnetic polarization, and general trends in the observed
doping dependence are obtained roughly correctly. 
Weak coupling approaches have been mostly devoted to the
study of the optimally doped regime; they appear to require tuning the
Hubbard $U$ to a value close to an AF instabilility and using extended
impurity potentials. In the strong coupling limit, mean field approximations
%yield the correct resonant nature of the staggered magnetic polarization and
give a qualitatively correct carrier doping dependence for $\Theta $ (and for $C$, which
 changes only slightly with the hole content).
A current 
drawback, however,
is the inability of any single model to span the entire 
cuprate phase diagram and make predictions for the full 
doping dependence of the response. 
%However, as we have seen, a number of these magnetic and transport signatures
%are quantitatively captured by the theoretical models reviewed in the
%present section. Weak coupling approaches have been mostly devoted to the
%study of the optimally doped regime; they appear to require tuning the
%Hubbard $U$ to a value close to an AF instabilility and using extended
%impurity potentials. In the strong coupling limit, mean field approximations
%yield the correct resonant nature of the staggered magnetic polarization and
%give a qualitatively correct carrier doping dependence for $\Theta $ {\bf (and for $C$, which
% changes only slightly with the hole content)}. We note
%that these studies ignore processes leading to Kondo screening but that they
%nevertheless produce a characteristic energy $\Theta $.

 Further considerations suggest caution against taking the Kondo analogy too
literally. In the
Kondo/Anderson picture there are actually two energy scales. The upper one
is related to the formation of the local spin, i.e to a (local) bound state,
and the lower is related to the screening of the moment by conduction
electrons. In cuprates, moment formation and screening are driven by the
same set of conduction electrons. Furthermore, it is unclear how different an
 extended itinerant staggered magnetization cloud with a net moment -- which is obtained when a non magnetic impurity is introduced in a correlated system -- is, from the
single localized impurity spin 1/2  that forms at high energies, in the Kondo case.
%single localized impurity spin 1/2 -- which is formed at high energies in the Kondo case--  is, from the extended itinerant staggered magnetization cloud with a net moment, which is obtained when a non magnetic impurity is introduced in a correlated system. 
In the Kondo
scenario of a magnetic impurity in a noble metal host, the characteristic energy scale is
set by the carrier density and the exchange interaction through  $%
\Theta \sim J\exp -\left[ J\rho (E_{F})\right] ^{-1}$. For underdoped cuprates, 
the non superconducting regime corresponds to temperatures $T>>\Theta$,
since $%
\Theta $ is found to approach zero; in a Kondo description, $T>>\Theta$ would correspond to
the weak coupling limit $J\rho (E_{F}){<<1}$; yet spin dynamics, seen in
NMR experiments, continue to show the fluctuations of a strongly correlated
extended magnetic entity around the impurity. This might
be due to the existence of the pseudogap.

%In the underdoped case, $T>>\Theta $, but spin dynamics seen in
%NMR experiments continue to show the fluctuations of a strongly correlated
%extended magnetic entity around the impurity.\ Also for the classic Kondo
%case of a magnetic impurity in a noble metal host, the Kondo energy scale is
%fixed by carrier density and exchange interaction via the usual expression $%
%\Theta \sim J\exp -\left[ J\rho (E_{F})\right] ^{-1}$. In this
%Kondo/Anderson picture there are actually two energy scales. The upper one
%is related to the formation of the local spin, i.e to a (local) bound state,
%and the lower is related to the screening of the moment by conduction
%electrons. In cuprates, moment formation and screening are driven by the
%same set of conduction electrons. Furthermore it is unclear how different a
%single localized impurity spin 1/2 is from an extended itinerant staggered
%magnetization cloud with net moment.\ Finaly for underdoped systems, since $%
%\Theta $ is found to approach zero, this would in principle correspond to
%the weak coupling limit $J\rho (E_{F}){<<1}$, but may be due to the
%existence of the pseudogap. 
In the overdoped regime, as the physics gets
closer to the Fermi liquid limit, one does not expect to observe Kondo-like
signatures any longer. Indeed, a nonmagnetic impurity is not expected to produce a Kondo resonance in conventional metals.
Thus,
understanding the reduction of the amplitude of the staggered moments, of
the spatial extent of the polarization  and the evolution of 
$C$ and $\Theta$ for large dopings may require a different description. 

%In the overdoped regime, as the physics gets
%closer to the Fermi liquid limit, one expects that eventually, when
%interactions become weak enough, and for spinless impurities $C\rightarrow 0$, $\Theta %\rightarrow 0$, rather than a large $\Theta $ and a constant $C$
%(A nonmagnetic impurity is not expected to produce a Kondo resonance). Thus,
%understanding the reduction of the amplitude of the staggered moments and of
%the spatial extent of the polarization may require a different description.

%All this invites caution when advocating a conventional Kondo scenario for
%the impurity problem in correlated materials.

%(\textbf{\ I always have a problem with this }$C\rightarrow 0$, $\Theta
%\rightarrow 0$\textbf{\ limit: In usual Kondo the resonance broadens and }$%
%\Theta \rightarrow \infty \;$\textbf{this has been advovated a lot for Al-Mn
%\ in which Mn has moment but has a weak enhanced magnetic susceptibility
%with TK=1000K.\ So a continuity with disappearance of the moment in the
%overdoped regime does not appear to violate a Kondo picture.)\ }

%\section{Impurities in superconducting state }
%\label{sec:SC} \subsection{Overview}
\section{Impurities in the superconducting state }\label{sec:SC}
\subsection{Overview}

The subject of impurities in the superconducting state of the
cuprates, and the theory of impurity states in unconventional
superconductors generally, is already quite vast.  We benefit from
the availability of  recent reviews of this field by
\cite{AVBalatsky:2006} on local aspects of  impurity states, by
\cite{NEHussey:2002} on effects of disorder on bulk quasiparticle
properties,  by \cite{PJHirschfeld:2002} on nonperturbative
disorder effects in two dimensions, and by \cite{MEFlatte:1999} on
general impurity properties in superconductors. Here we review
briefly the simplest aspects of the effects of nonmagnetic
impurities on $d$-wave superconductivity before discussing
theoretical approaches to the inclusion of electronic correlations
in the superconducting state impurity problem, as well as relevant
experiments.

It was recognized shortly after the publication of the BCS theory
that while nonmagnetic impurities would not affect thermodynamic
properties of a conventional superconductor \cite{PWAnderson:1959},
they might do so in hypothetical $p$-wave pairing
systems \cite{RBalian:1963}. \ When a high density of low-energy
excitations in superconducting heavy fermion systems, such
``unconventional" states were considered as possible ground states
of compounds like UBe$_{13}$ and UPt$_{3}$.
% \ Gor'kov and Kalugin
 \cite{LPGorkov:1985} and, independently,
% Ueda and Rice
\cite{KUeda:1985} pointed out that linear nodal regions of the
order parameter on the 3D Fermi surface would lead to a nonzero
residual density of states\ at zero energy, $\rho (0).$ This
self-consistent perturbative treatment for weak impurities,
averaged over disorder, was similar to that used by
\cite{AAAbrikosov:1960} in their discussion of gapless
superconductivity induced by magnetic potentials.
%Pethick and Pines
\cite{CJPethick:1986}
observed that weak scattering models of
this kind could not describe transport properties measured on
heavy fermion materials, and proposed that large effective
impurity potentials could be treated in a $T$-matrix
approximation.
%Hirschfeld et al.
 \cite{PJHirschfeld:1986} and
%Schmitt-Rink et al.
 \cite{SSchmitt-Rink:1986} then proposed the
self-consistent $T$-matrix approximation (SCTMA) for impurities of
arbitrary strength, and predicted a disorder-induced ``plateau''
in $\rho(\omega)$ at low energies.  {This plateau consists of
disorder induced "impurity band" states conceptually analogous to
the mid-gap localized states predicted earlier for dirty isotropic
$p$-wave superconductors by \cite{LJBuchholtz:1981}.} The nodal
quasiparticles which contribute to this ``impurity band" give rise
to thermodynamic properties similar to a normal metal with reduced
DOS.
% Stamp
\cite{PCStamp:1987} then investigated the character of a local
single-impurity resonance within this model.

In a similar way, when experiments on good samples of cuprate
superconductors in the early 90s began indicating the existence of
low-energy excitations in the superconducting
state \cite{NEHussey:2002}, the importance of understanding the
effect of disorder for identifying the symmetry of the order
parameter was quickly realized. The effect of disorder on $T_c$
itself was controversial early on. For example,  irradiation with
heavy ions was found to suppress $T_c$ only when the scattering
rate, as deduced from resistivity, became of order the Fermi
energy \cite{JMValles:1989}, as expected for an isotropic $s$-wave
superconductor.  On the other hands, impurities substituting for
Cu in the CuO$_2$ plane tended to depress $T_c$ rapidly, as
expected for an unconventional superconductor, and controlled
electron irradiation experiments confirmed this
\cite{JGiapintzakis:1994}, \cite{ALegris:1993} (see Section
\ref{sec:Impurity_subs}).

On the theoretical side,  the self-consistent $T$%
-matrix approximation (SCTMA) was employed to provide a
qualitative explanation of the effects of disorder on $d$-wave
superconductors, offering important  evidence for the symmetry
identification.  These arguments belong in the same category as
the explicitly phase sensitive measurements like the tricrystal
experiments \cite{CCTsuei:2000}, as  pointed out by
\cite{LSBorkowski:1994}.  This is because the fact that
nonmagnetic impurities break pairs is directly associated with the
sign change of the $d$-wave order parameter (see Sec.
\ref{singleimpdwave}). Within the SCTMA, the strength of the
impurity potential is an important parameter, and the recognition
that planar impurities like Zn acted as  strong scatterers which,
from the standpoint of theoretical modelling could be considered
as unitary scatterers, was an important step forward. For example,
observations of a crossover of the magnetic penetration depth from
a $T$ to a $T^{2}$ behavior with a small concentration of Zn
impurities were not initially understood to be a natural
consequence of $d$ symmetry, because it was expected that in \
such experiments $T_{c}$ would be strongly suppressed in a
$d$-wave system. \ In the unitarity limit, however, strong
modifications of low-energy properties occur over an energy scale
$\gamma $ (residual scattering rate or ``impurity bandwidth'')
which can be much larger than the normal state impurity scattering
rate $\Gamma $ which controls $T_{c}$ \cite{PJHirschfeld:1993}. A
similar analysis by \cite{THotta:1993} allowed for an explanation
of the dependence of the residual $T\rightarrow 0$ Knight shift of
YBCO on Zn concentration. Calculations of transport properties
within this framework also provided evidence for near-unitarity
scattering, and helped rule out candidate extended $s$-wave states \cite%
{LSBorkowski:1995}.

%\begin{figure}[th]
%\begin{center}
%\includegraphics[width=0.6\columnwidth]{crosseddiags1.eps}
%\label{fig:crosseddiags}
%\end{center}
%\caption{Self-energy in self-consistent $T$-matrix approximation
%after disorder averaging.  Dashed lines represent disorder
%potentials $V_0$, solid lines full Green's function for
%disordered $d$-wave superconductor.
%%{\bf PH will make new figure. }
%}
%\end{figure}
 While  quite successful in
explaining the effects of strong in-plane scatterers in optimally
doped cuprates, the basis of the SCTMA phenomenology was called
into question in a work by \cite{AANersesyan:1995}, who pointed
out that the  neglect of multiple-impurity scattering processes
(``crossed diagrams''), which are of higher order in $1/k_{F}\ell
$ or the density of impurity sites $n_{i}$ in three dimensions,
was not justified at asymptotically low energies since these
diagrams have an additional $\log E$ singularity in 2D coming from
the line nodes of the $d_{x^{2}-y^{2}}$ gap. This leads to a
breakdown of single-site perturbation theory and the need to resum
at least a second class of crossed diagrams. \ This paper led to
several attempts to solve the problem
nonperturbatively \cite{AANersesyan:1995,CPepin:1998,TSenthil:1999,KZiegler:1996a},
each claiming to calculate \textit{exactly} the density of states
of a disordered 2D $d$-wave superconductor, but arriving at
dramatically different results.  A combination of numerical
work \cite{WAAtkinson:2000} and analytical weak localization
calculations \cite{AGYashenkin:2001}, reviewed in
\cite{PJHirschfeld:2002}, led to the conclusion that the
low-energy DOS of the 2D $d$-wave state was quite sensitive to the
symmetries of the disorder model and the normal state one-electron
band employed in the calculations, in contrast to expectations
based on standard localization calculations for normal metals.  In
addition, it was argued that the corrections to the SCTMA were
unlikely to be relevant to experiments on real cuprates.

 While the SCTMA appears to be a reasonably successful
approximation for properties of optimally to overdoped cuprate
superconductors with nonmagnetic disorder, it clearly cannot
account for the magnetic states induced by potential scatterers as
described in Section \ref{sec4},
%\ref{sec:expts_normal},
 nor other phenomena
which must arise when pairing takes place in the underdoped phase
near the Mott transition.  Here we investigate how the physics of
impurities in correlated  metals discussed in Sections
%\ref{sec:expts_normal}
\ref{sec4} and \ref{sec:thy2D} is modified by superconductivity.
To establish the groundwork, we first give a brief account of the
formation of low-energy impurity resonances (\ref{singleimpdwave})
in a $d$-wave superconductor, and associated impurity bands by
nonmagnetic potentials.   We next discuss in Section
\ref{sec:SCdisorderthy} the interference of these states in the
presence of finite disorder, and the formation of the ``impurity
band" in the $d$-wave state. We then introduce electronic
correlations in Section \ref{SCcorrelated}, focussing primarily on
results from the weak-coupling theory of impurity-induced moment
formation as discussed in the previous section, since it can
easily be extended to the superconducting state.  We then compare
a general picture which emerges from these treatments with
experiments, both those which probe disorder-averaged properties,
i.e. primarily states far from the impurities themselves (Section
\ref{bulkexptSC}), and those which probe states near the impurity
(Section \ref{localexptSC}).  In Section \ref{SCSummary} we
summarize ideas regarding the effects of disorder in the
correlated $d$-wave state gleaned from both theory and experiment.

\subsection{Single impurity in a $d$-wave superconductor}

\label{singleimpdwave} The BCS Hamiltonian for a pure singlet
superconductor can be written as:
\begin{equation}
H_{0}=\sum_{\k}\Phi_{{\k
}}^{\dagger}(\epsilon_{\k}\tau_{3}+\Delta_{k}\tau_{1})\Phi_{\k},
\end{equation}
where $\Phi_{\k}=(c_{\k\downarrow}, c^{\dagger}_{-\k\uparrow} )$,
is a Nambu spinor.   Analytic results are presented for a
parabolic band $\epsilon_\k =k^2/2m$, with corresponding $d$-wave
order parameter $\Delta_\k=\Delta_0\cos 2\phi$ ($\phi$ is the
angle in momentum space which ${\bf k}$ makes with the 100 axis).
The matrices $\tau_{i}$ are the Pauli matrices spanning
particle-hole space. To the homogeneous superconductor we add a
single impurity perturbation
 \beqa H_{\rm int}
= \sum_{\bk\bk'} \Phi^\dagger_{\bk} \,\widehat V_{\k\k'}\,
\Phi_{\bk'} \eeqa where  $\widehat V_{{\bf k},{\bf k}^\prime}=\int
d\rr\, e^{i(\k-\k')\cdot \rr}\, \widehat V(\rr)$ is the scattering
potential Fourier component.  For a density-like perturbation
which couples to the chemical potential locally, the Nambu form is
$\widehat V_{\k,\k'}=V_{\k,\k'}\tau_3$.  The full Green's function
for a single impurity is
\begin{eqnarray}
  \label{GTmom}
  \widehat G({\bf k}, {\bf k}^\prime)&=& \widehat G_0({\bf k})
  + \widehat G_0({\bf k})\widehat V_{{\bf k},{\bf k}^\prime}
  \widehat G_0({\bf k}^\prime)
  \\
  \nonumber
  &&
  + \sum_{{\bf k}^{\prime\prime}}\widehat G_0({\bf k})\widehat V_{{\bf k},{\bf
  k}^{\prime\prime}}
  \widehat G_0({\bf k}^{\prime\prime})\widehat V_{{\bf k}^{\prime\prime},{\bf k}^\prime}
  \widehat G_0({\bf k}^\prime)+\ldots.\\
&=& \widehat G_0({\bf k})+
    \widehat G_0({\bf k})\widehat T_{{\bf k},{\bf k}^\prime}
  \widehat G_0({\bf k}^\prime),
\end{eqnarray}
where the $T$-matrix is given by summing all single-site
scattering processes,
\begin{eqnarray}
  \widehat T_{{\bf k},{\bf k}^\prime}&=&\widehat V_{{\bf k},{\bf k}^\prime}
  +\sum_{{\bf k}^{\prime\prime}}\widehat V_{{\bf k},{\bf
  k}^{\prime\prime}}
  \widehat G_0({\bf k}^{\prime\prime})\widehat V_{{\bf k}^{\prime\prime},{\bf k}^\prime}
  +\ldots
    \\
    &=&\widehat V_{{\bf k},{\bf k}^\prime}
  +\sum_{{\bf k}^{\prime\prime}}\widehat V_{{\bf k},{\bf
  k}^{\prime\prime}}
  \widehat G_0({\bf k}^{\prime\prime})\widehat T_{{\bf k}^{\prime\prime},{\bf
  k}^\prime},
\label{TmatrixSingle}
\end{eqnarray}
and all symbols $\widehat A$'s are matrices in particle-hole
space. If the impurity scattering is purely pointlike, $\widehat
V({\bf r})\propto \delta(\rr)$, the scattering is isotropic,
$\widehat V_{{\bf k},{\bf k}^\prime}=\widehat V$, and $\widehat T$
is seen to  be independent of momentum. For the special case of a
pointlike nonmagnetic impurity potential  with strength $V_0$, it
is convenient to define the cotangent of the $s$-wave phase shift
$\delta_0$, with $c\equiv \cot \delta_0 \equiv (\pi V_0
\rho_0)^{-1}$, with $\rho_0$ the density of states at the Fermi
level.  The Green's function including the impurity is
therefore\beqa \widehat{G}({\bk,\bk'};\omega)= \widehat{G}_0
({\bk},\omega)\delta_{\bk\bk'} + \widehat{G}_0({\bk}, \omega)
\widehat{T}(\omega) \widehat{G}_0 ({\bk'},\omega)\;, \eeqa with
$\widehat{G}_0(\bk,\omega)= (\omega- \widehat H_0)^{-1}$, and the
$T$-matrix takes the form
\begin{eqnarray}
    {\widehat T}&=&T_0\tau_0+T_3\tau_3\nonumber\\
    T_0&=&g_0/(S_+S_-)\nn\\
    T_3&=&(c-g_3)/(S_+S_-),
    \label{1imptmat}
\end{eqnarray}
where $g_0$ and $g_3$ are the $\tau_0$ and $\tau_3$ Nambu
components of the integrated bare Green's function $(1/\pi \rho_0)
\sum_\k {\widehat G}_0(\k,\omega)$. This expression has resonances
when
\begin{equation}
    %c=G_3+\pm G_{0}(\Omega)
    S_\pm\equiv c-(g_3 \mp g_0)=0.
    \label{1impresonance}
\end{equation}
Note that the off-diagonal components of the $T$-matrix do not
occur because $g_1=g_2=0$ due to the $d$-wave symmetry.
Furthermore, in the special case of a particle-hole symmetric
system, $g_3$=0 and the resonance energy is determined entirely by
$g_0$, which is given in the case of a circular Fermi surface by
$g_{0}(\omega)=-i\int
\frac{d\varphi}{2\pi}\omega[\omega^2-\Delta_\k^2]^{-1/2}$ which
for low energies $\omega \ll \Delta_0$ takes the form
$G_0(\omega)\simeq -(\pi\omega/\Delta_0)(\log 4\Delta_0/\omega
+i)$. One may then solve Re $S_\pm(\omega+i0^+)=0$ and estimate
the resonance width $\Gamma$ on the real axis.  In the case of
strong scattering $c\ll 1$, the resonance energy $\Omega_0^\pm$
and width $\Gamma$ are
\begin{subequations}
\begin{eqnarray}
     \Omega_0^\pm &=& {\pm \pi c\Delta_0\over 2 \log (8/\pi c) }
     \label{1impresenphsA} \\
     \Gamma &=& {\pi^2 c\Delta_0\over 4 \log^2 (8/\pi c)} .
     \label{1impresenphsB}
\end{eqnarray}
\end{subequations}
This result was first obtained by \cite{AVBalatsky:1995} for the
$d$-wave case, following earlier work on the similar  $p$-wave
case by Stamp \cite{PCStamp:1987}. Note that the resonance becomes
an undamped bound state only exactly at the Fermi level
$\Omega=0$, when $c=0$; for finite $c$ there are two  resonances
whose energies  $\Omega_0^+=-\Omega_0^-$  are symmetric in this
approximation. As seen from (\ref{1impresonance}), in general
particle-hole asymmetric systems ($G_3\ne 0$), the resonance is
tuned to sit at the Fermi energy for some value of the impurity
potential $V_0$ which is {\it not} infinite, so the term
``unitarity" (as used here, $\Omega_0=0$) and ``strong potential"
($V_0\rightarrow \infty$) are no longer
synonymous \cite{RJoynt:1997,WAAtkinson:2000}.

\begin{figure}
\begin{center}
\leavevmode
\includegraphics[width=\columnwidth]{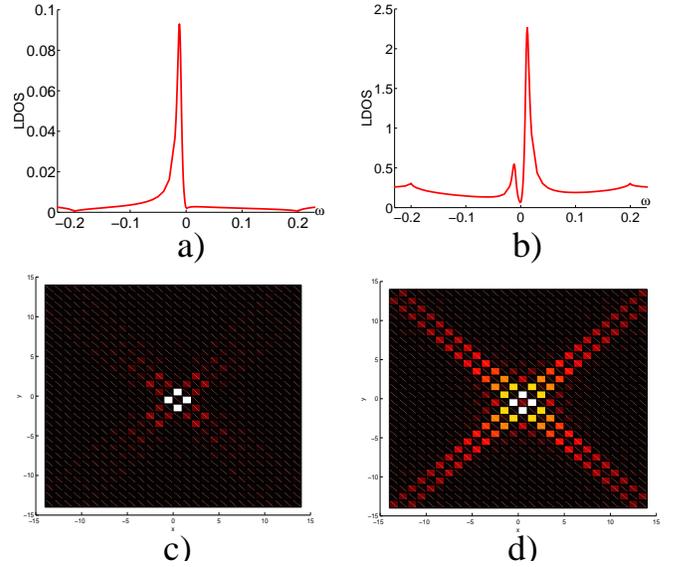}
\caption{Summary of LDOS results for 1-impurity problem on a
tight-binding lattice, band given in reference
\cite{MRNorman:1995}, bond order parameter $\Delta_d=0.1, \mu=0,
V_0=10$, $|\Omega_0^\pm|\simeq 0.013$: a) LDOS vs. $\omega$ on the
impurity site; b) LDOS vs. $\omega$ on nearest neighbor site;
c),d) LDOS map at resonance $\omega=\Omega_0^+$ and
$\omega=\Omega_0^-$. Color scales in c) and d) are relative to the
nearest neighbor peak heights shown in b).  From \cite{LZhu:2003}
} \label{fig:1imp}
\end{center}
\end{figure}

\subsubsection{Local density of states near impurity}

 Although the energies of the particle and hole resonant states
are symmetric, their  spectral weights on a given site are usually
quite different \cite{AVBalatsky:1995}, such that locally only one
resonance dominates.  Define the local density of state (LDOS) as
\beqa \rho(\br,\omega) = -\frac{1}{\pi}{\rm
Im}\,G_{11}(\br,\br;\omega+i0^+) \eeqa with the total Green's
function in the presence of the impurity $
\widehat{G}(\br,\br';\omega)=\widehat{G}_0 (\br-\br',\omega) +
\widehat{G}_0(\br,\omega) \widehat{T}(\omega) \widehat{G}_0
(\br',\omega), $
%Using the eigenstates representation of \beqa
%g(\br,\br';\omega) = \sum_{n} \frac{\langle \psi_n^*(\br)
%\psi_n(\br')\rangle}{(\omega - E_n)} \eeqa
where the 11 index refers to the normal electron part of the Nambu
Green's function.  Now it is easy to see that any finite impurity
potential acts as a local breaker of particle-hole symmetry which
can lead to a pronounced peak in the LDOS.  At the impurity site
$\br=0$, for, e.g. a repulsive impurity potential, the resonant
peak at negative energy (holelike states) is large, and the
positive energy (electronlike states) peak is relatively small, as
seen in Fig.\ \ref{fig:1imp}.  The situation is reversed for the
nearest neighbor site.  Note, however, that for the large value of
$V_0$ chosen, the LDOS peak at resonance is much smaller on the
impurity site than on the nearest neighbor site since the large
potential excludes electrons from the impurity site.

Impurity bound states of this type have immediate observable
consequences for the local density of states, measurable  by STM,
and for the distributions of local magnetizations, measured by
NMR.
 The impurity  modulates the  local density
of states of the homogeneous system $\rho(\br,\omega)$ by \bea
 \rho_{{\rm
imp}}(\br,\omega) = -\frac{1}{\pi} {\rm Im} \,
[{\widehat{G}_0({\br},\omega) \hat{T}(\omega) \widehat{G}_0
({-\br},\omega)}]_{11} \eea The impurity-induced LDOS falls off as
$r^{-2}$ along the nodal directions, and exponentially along the
antinodal ones \cite{AVBalatsky:1995}. The LDOS in the near field
is more complicated, however: the nearest neighbor sites have
peaks at $\pm\Omega_0$, with the larger spectral weight at
$+\Omega_0$. In the crossover regime $r\sim \xi_{0}$, the LDOS is
enhanced along the node direction for holelike states, but is
spread perpendicular to the node direction for electronlike
states. These spatially extended LDOS patterns are the fingerprint
of the impurity-induced virtual bound states. In Fig.
\ref{fig:1imp}, we illustrate the LDOS pattern expected for both
particles and holes for a resonant state close to the Fermi level
in this simple model.

\vskip .2cm  The existence of impurity bound states in the
superconductor will be shown below to be related to strong
$T$-dependent upturns in the local susceptibility, so it is of
interest to establish what happens in the current model as the
temperature is increased and superconductivity destroyed.  If the
normal state is metallic, as assumed in the current model, weak
Friedel-like oscillations in the density of states remain, but the
resonant character of the impurity state is destroyed.  If however
the normal state also possesses a spectral gap, resonant states
may remain.  This is why impurity bound states found, e.g. in
models of the spin gap phase of the
cuprates\cite{MGabay:1994,NNagaosa:1995,GKhaliullin:1997}, with
linear density of states analogous  to the $d$-wave
superconductor, are quite similar to those in the superconducting
state.  Impurity measurements have in fact been proposed as a
probe of the pseudogap state in the cuprates\cite{HVKruis:2001}.
 \vskip .2cm
Because nonmagnetic impurities are pairbreaking in $d$-wave
superconductors, the most important qualitative distinction with
magnetic impurities does not exist.  Nevertheless there are some
interesting phenomena associated with impurity spins in
unconventional superconductors, which have been amply reviewed in
\cite{AVBalatsky:2006}.  \vskip .2cm

\subsubsection{Effect on field-induced magnetization}

\begin{figure}
\begin{center}
\leavevmode
\includegraphics[width= .8\columnwidth]{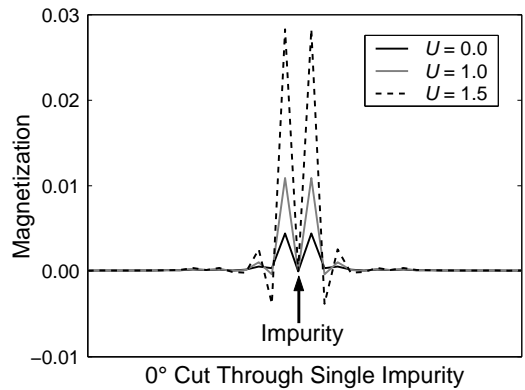}
\caption{ Magnetization along (1,0) direction through impurity
with $d$-wave superconductor described by tight-binding band
$t'/t=-.2$, $V_{imp}=100t$, $g\mu_BB/2=0.004t$, $U/t=0,1,1.5$. }
\label{fig:mag_mod_U}
\end{center}
\end{figure}

In the pure $d$-wave state with no residual quasiparticle
interactions, the paired gas has a Friedel-type response to the
impurity perturbation which leads to a spatially oscillating LDOS
$\rho(\br,\omega = 0)$ at the Fermi level with wavelength $2k_F$
and envelope decaying over a length scale of
$\xi_0={v_F/(\pi
\Delta_0)} $.
%$\xi_0={v_F/{\pi \Delta_0}} $.
  Consider a situation where the impurity is
nonmagnetic; there are then no corresponding spin density waves
induced in zero field by correlation effects. However, since the
local susceptibility in the superconducting state $\chi(\br,T)$ is
proportional to the density of states at the Fermi level, such
oscillations translate directly into magnetization modulations in
the presence of a nonzero applied field. Since the LDOS is always
positive, the magnetization is always in the same direction as the
applied field. This can be seen in the $U=0$ curve of Fig.
\ref{fig:mag_mod_U}. In the presence of electronic  correlations,
these modulations are greatly enhanced, and a true alternating
magnetization appears, as discussed further below in Sec.
\ref{SCcorrelated}.

\subsection{Effects of disorder on $d$-wave state}
\label{sec:SCdisorderthy}
\subsubsection{Quasiparticle interference of impurity bound states}

%\begin{figure}
%\begin{center}
%\leavevmode
%\includegraphics[width=.8\columnwidth]{kampf_fig3.eps}
%\caption{ Growth of the impurity band in the DOS at low energies:
%(a) Clean case result; (b) with one impurity; (c) with two
%impurities located three sites away one from each other in the
%(1,0) direction, for a $d$-wave superconductor with impurity of
%strength $8t$.  The peaks in the clean case at $E \sim 0.1$ are
%{\bf due to finite size effects and correspond to} the
%lowest energy eigenvalues for a 20x20
%system with open boundary conditions.
%\cite{UMicheluchi:2002}. {\bf THIS FIGURE IS CONFUSING
%}} \label{fig:kampf_fig3}
%\end{center}
%\end{figure}

As seen in Fig. \ref{fig:1imp}, the strong nonmagnetic impurity in
a $d$-wave superconductor in this approximation is characterized
by a bound state with long range tails extending along the nodal
direction.  If many impurities are present, these states overlap
and interfere, leading to a splitting of bound state energies and
an accumulation of low-energy impurity-induced energy eigenvalues
which are spread out over a so-called ``impurity
band" \cite{UMicheluchi:2002}.  The formation of this band, %(Fig.
%\ref{fig:kampf_fig3}),
which appears clearly in the disorder-averaged theory, is actually
a rather subtle phenomenon compared to, say, the analogous
impurity band in disordered semiconductors, where a good deal can
be understood by postulating transport via overlap of spherically
symmetric hydrogenic orbitals.   In the $d$-wave case the
formation of the impurity band and the corresponding quasiparticle
localization problem are strongly influenced by the fact that
significant overlaps between two impurity states can take place
only if the impurities are ``oriented" with respect to one another
such that the nodal wavefunction tails
overlap \cite{AVBalatsky:1996}. In fact, analyses of the
two-impurity problem show that interference effects can take place
over many tens of lattice spacings
%(Fig.\ref{fig:bandersen_fig1})
between optimally oriented
impurities \cite{LZhu:2003,DKMorr:2002,BMAndersen:2003a}.  In Fig.
\ref{fig:Atkinson_and_dos}, we show the local density of states
due to many strong impurities using a realistic band structure,
averaged over a few eigenstates within a narrow energy range
abound the 1-impurity resonance energy \cite{WAAtkinson:2003}.  The
interference tails between some impurities are visible, while
others have been split outside the ``observation  bandwidth" due
to the details of the local disorder environment.  It is this
complex system that the SCTMA seeks to replace by a
disorder-averaged, translationally-invariant medium where the
quasiparticle states of the pure system are broadened.

%\begin{figure}
%\begin{center}
%\leavevmode
%\includegraphics[width=\columnwidth]{bandersen_fig1.eps}
%\caption{DOS at the site of a strong repulsive impurity with
%resonance energy $\Omega_\pm=\pm 1.5meV$ measured at
%$\omega=\Omega_\pm$ in the presence of a second, identical
%impurity as a function of distance between them, oriented along
%the a) nodal direction and b) antinodal direction. After
%\cite{BMAndersen:2003a}. } \label{fig:bandersen_fig1}
%\end{center}
%\end{figure}

\subsubsection{Self-consistent $T$-matrix approach}
\label{SCTMA} Since the SCTMA approach is already well-described
in several original papers and reviews
\cite{LJBuchholtz:1981,PJHirschfeld:1986,SSchmitt-Rink:1986,NEHussey:2002,RJoynt:1997,AVBalatsky:2006},
we will only briefly summarize it here. We begin with the exact
equations for the single impurity $T$-matrix
(\ref{TmatrixSingle}). The self-consistent $T$-matrix approach
(SCTMA) now extends these equations to an approximate treatment
where the disordered system is replaced by a translationally
invariant dissipative medium described by a self energy,
$\widehat\Sigma({\bf k},\omega)=n_{imp}\widehat T_{{\bf k},{\bf
k}}$, where $\widehat T$ is now a disorder-averaged quantity
%(Fig.
%\ref{fig:crosseddiags}).
At the same time the full Green's
function $\widehat G (\k,\k',\omega)$ is replaced by a disorder
averaged $\widehat G(\k,\omega)$, the propagator for an electron
moving in the translationally-invariant effective medium, given by
\begin{equation}
    \label{G-SCTM}
    \widehat G^{-1}({\bf k},\omega)=\widehat G_0^{-1}({\bf
    k},\omega) - \widehat\Sigma({\bf k},\omega),
\end{equation}
where now $\widehat \Sigma$ is the sum of all disorder-averaged
$T$-matrix diagrams, evaluated with the fully self-consistent
disorder-averaged $\widehat G$ on each Green's function line. The
calculation is similar in spirit to the approximation of
\cite{AAAbrikosov:1960}, but valid for arbitrary impurity
strength.

      \begin{figure}
\begin{center}
\leavevmode
\includegraphics[width=\columnwidth]{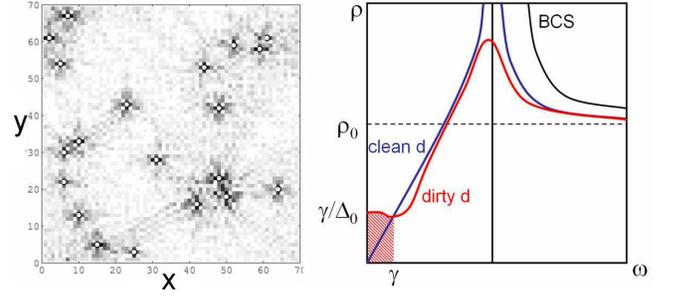}
\caption{Left: Density of states $\rho(0)$ vs. position in a
70$\times$ 70 lattice, showing interference of many impurities in
band with no particle-hole symmetry.  White dots represent strong
potential scatterers. After W.A. Atkinson, unpublished. Right:
schematic density of states of a $d$-wave superconductor, and
impurity band of width $\gamma$.} \label{fig:Atkinson_and_dos}
\end{center}
\end{figure}

Rather than give details of the calculation of $\widehat \Sigma$,
which have been amply reviewed elsewhere, we focus on the simplest
example which displays the basic physics of a disordered $d$-wave
superconductor, a finite density of identical strong pointlike
scatterers.  In this case the self-energy reduces to the analog of
(\ref{1imptmat}),
\begin{eqnarray}
   \widehat \Sigma (\k,
   \omega) &=& \Gamma {(c-g_3)\tau_3 + g_0\tau_0\over (c-g_3)^2 -
   g_0^2},
\end{eqnarray}
where  the impurity scattering rate parameter $\Gamma = n_{imp}
/(\pi \rho_0)$ and $g_{0,3}$ are now functionals of $\widehat G$
rather than $\widehat G_0$.  Let us furthermore assume
particle-hole symmetry and unitary limit scattering $c\rightarrow
0$.  Then the self-energy reduces to the particularly simple
result
 $ \widehat \Sigma \rightarrow -\Gamma/g_0$.  At the Fermi level,
 $\omega\rightarrow 0$, if we assume the retarded self-energy
 $\Sigma_0$ reduces to a small imaginary part $-i\gamma$,
 so that the self-consistency equation becomes,
  for a circular Fermi surface
  and $d$-wave order parameter $\Delta_0 \cos 2\phi$,
 \begin{eqnarray}
      -i\gamma=-{\Gamma\over g_0(0)} = {-\Gamma \over -i \frac{2}{\pi} \int_0^{\pi/2} d\phi \:
                  \frac{\gamma}{\sqrt{\gamma^2+\Delta_0^2 \cos^2
                  2\phi}}}.
      \end{eqnarray}
      The elliptic integral can be performed and expanded for
      $\gamma\ll \Delta_0$ to give
 the following self-consistency
equation for $\gamma$
 \begin{eqnarray}
      \label{self_approx}
      \left(\frac{\gamma}{\Delta_0}\right)^2 = \frac{\pi}{2} \:
        \frac{\Gamma}{\Delta_0}/ \ln
        \left(\frac{4\Delta_0}{\gamma}\right),
      \end{eqnarray}
  a transcendental equation whose solution for small $\Gamma$ can be approximated to log accuracy as $\gamma \sim
  0.63\sqrt{\Gamma\Delta_0}$ \cite{PJHirschfeld:1993}.
The DOS at zero frequency is then  given by
      \begin{eqnarray}
      \frac{\rho(0)}{\rho_0} = -{\rm Im}\, g_0(0)= \Gamma/\gamma
                 \; ,
      \end{eqnarray}
      where the first equality is general but the second obtains
      only in the unitarity limit.

Thus the SCTMA generically predicts a finite density of states at
the Fermi level in the presence of disorder for a $d$-wave
superconductor. The dependence of this residual DOS on the
concentration of defects ranges from the
$\rho_0\sim\sqrt{n_{imp}}$ behavior in the unitarity limit just
exhibited, to a much weaker exponential dependence, $\rho(0) \sim
\exp(\pi \rho_0\Delta_0(1+c^2)/n_{imp})$ in the weak scattering
(Born, $c\gg 1$) limit.  The general form for the DOS expected
over the whole frequency range in the strong scattering limit is
shown in Fig. \ref{fig:Atkinson_and_dos}b.  If the scattering is
weaker, the weight in the impurity band moves to higher energies
as the position of the single-impurity resonance itself moves away
from the Fermi level, see Eq. (\ref{1impresenphsA}), and the Fermi
level DOS $\rho(0)$ decreases accordingly.
%\begin{figure}
%\begin{center}
%\leavevmode
%\includegraphics[width=\columnwidth]{dos.eps}
%\caption{  Schematic density of states of $d$-wave superconductor.
%Replace this figure with improved one. } \label{fig:DOS}
%\end{center}
%\end{figure}

The simplest measurement showing the strong influence of
impurities is the suppression of the critical temperature, which
has been discussed to some extent already in paragraph
\ref{Tcreduc} in connection with transport experiments. Within the
SCTMA, the remarkable prediction  is that for a $d$-wave
superconductor with pointlike scatterers the $T_c$ suppression
follows a form identical to the prediction of Abrikosov and
Gor'kov (AG)\cite{AAAbrikosov:1960}

\begin{equation}
   \ln\frac{T_{c0}}{T_c}=\psi\biggl(\frac{1}{2}+\frac{1}{2\pi\tau T_c}\biggr)-
    \psi\biggl(\frac{1}{2}\biggr),
\end{equation}
where $T_{c0}$ is the transition temperature in the absence of
disorder, and  $\psi(x)$ is the digamma function, and $1/\tau$ is
the scattering rate at the critical temperature due to impurity
scattering alone,  assuming that the impurity potential is
nonmagnetic and of zero range, for any scattering phase shift. For
small concentrations, the initial slope of the $T_c$ suppression
is then $T_c/T_{c0}\simeq 1- \pi/(4\tau T_{c0})$.
%{\bf (find prefactor)}.
With increasing concentrations, superconductivity is eventually
destroyed ($T_c=0$) when the scattering rate reaches the critical
value
\begin{equation}
 \tau^{-1}=\pi T_{c0}/2\gamma,
\end{equation}
where $\gamma\approx 1.78$.  If Matthiessen's rule is obeyed
(Section \ref{sec:Nstatetransport}), the scattering rate is
proportional to the change in resistivity due to disorder $\Delta
\rho$ and one expects an AG-type form for $T_c$ vs. $\Delta \rho$.

\subsubsection{Extended impurities}

As discussed in Section \ref{sec:Impurity_subs}, impurities which
substitute for Cu in the CuO$_2$ planes turn out to be strong
scatterers of electronic states in the plane.  While the effective
bare potential induced by planar impurities is expected to be
atomic scale \cite{LLWang:2005}, it has been suggested that the
correlations could give rise to a larger effective range (Section
\ref{sec:thy2d_nonmag_weak}).  This is a possible reason why the
initial slope of the $T_c$ suppression vs. residual resistivity
curve in the case of Zn and Ni is found to be somewhat smaller
(roughly a factor of 2) than predicted by the SCTMA (see Section
\ref{bulkexptSC}) \cite{MFranz:1996,GHaran:1996,GHaran:1998,MLKulic:1997,MLKulic:1999,SGraser:2007}.

On the other hand, there are many other sorts of defects in the
cuprates, and in fact all cuprate materials are doped by
out-of-plane ions which in most cases occupy random positions in
the crystal lattice or interstitial positions.  It is to be
expected that these ``intrinsic" impurities are poorly screened,
and therefore that they should have a  large effective range of
the scattering potential experienced by holes
%electrons
 moving in the
planes.  In addition, they tend to be present in rather large
concentrations, indeed of order the doping
 itself.  With the notable exceptions of \YBCO ~ and \YBCOtwofoureight,
 cuprates at optimal doping must therefore be considered to have ${\cal
 O}$(10\%)
 concentrations of ``intrinsic" impurities with extended potentials.  Such defects will
 tend to
 scatter electrons preferentially in the forward direction.   The
 consequences  for normal state one-electron
 properties were highlighted by \cite{EAbrahams:2000}, for the critical
 temperature by \cite{HYKee:2001},
  and  for quasiparticle properties in the superconducting state at low temperatures by
  \cite{TXiang:1995,APKampf:1997,ADurst:2000,LZhu:2004,TDahm:2005a,TDahm:2005b,TSNunner:2005a}.  Impurities of this
  type are assumed to be sufficiently weak such that they do not
  give rise to resonant effects at low energies, and are therefore
  not so easy to image with STM techniques.  In addition, weaker potentials are not expected to induce
  significant local magnetism (see Section \ref{localexptSC}), but there do not appear
  to be theoretical investigations along these lines available.

  \cite{LZhu:2004} and
  \cite{TDahm:2005a,TDahm:2005b} concluded that forward scatterers have a
  tendency to sharpen spectral features near the antinode of the $d$-wave gap, while
  broadening those near the node.  This counterintuitive result
  can be understood by noting that if the scattering is
  sufficiently forward, scattering processes which begin with
  quasiparticles near the antinode cannot involve a sign change of
  the order parameter and are therefore not pairbreaking; this is
  the analog of Anderson's theorem for forward scattering in the
  $d$-wave state.  On the other hand, if a quasiparticle is
  sufficiently close to the node, phase space for scattering
  involving sign change of $\Delta_\k$ is available and spectral
  features will be broadened similarly to the normal
  state.

    Forward scattering impurities might be expected to have
  negligible effect on transport, since they cannot relax the
  momentum current.  On the other hand, \cite{TSNunner:2005a}
  showed that scatterers of intermediate range could produce
  qualitatively different temperature dependences of the
  superconducting state microwave conductivity, providing a
  possible explanation to the quite different results obtained for
  crystals of optimally doped \YBCO and \BSCCO.

\subsubsection{Order parameter suppression and pairing potential impurities}

It is worth noting that the theoretical models above for the most
part do not account for the local variation of the superconducting
order parameter near the impurity site.  The order parameter has a
response  to a local Coulomb potential, and the correct way to
include it within a fully self-consistent Hartree-Fock theory was
discussed by \cite{ALFetter:1965}, \cite{AIRusinov:1968} and
others  for ordinary superconductors. In the context of
unconventional superconductors order parameter suppression around
a single impurity was discussed by several
authors \cite{CHChoi:1990,MFranz:1996,ASchnirman:1999,WAAtkinson:2000b,MFriesen:1997}.
These effects were neglected early on probably because,  in terms
of the resonance effects on one-electron properties discussed in
Section \ref{singleimpdwave}, the self-consistent treatment does
not lead to major differences. In the disorder-averaged theory it
can have some important effects on transport as a new source of
``off-diagonal scattering" \cite{MHHettler:1999}, however.

There is a further, more exotic possibility, which is just
beginning to be taken seriously at this writing.  In the
self-consistent approaches above, the BCS pair interaction
constant $\lambda$ which appears in the gap equation is assumed to
be homogeneous, and one considers the response of the order
parameter to a diagonal potential with $\tau_3$ symmetry in Nambu
space, i.e. a response $\delta\Delta_\k(\br)\tau_1$ is produced.
In principle, it has been recognized that an impurity can alter
the local electronic structure such that the pairing interaction
itself is modified locally \cite{HSuhl:1962,AILarkin:1970},
leading to a $\lambda(\rr)$. Since coherence lengths $\xi_0$ of
conventional superconductors are much larger than microscopic
distances $a$, these effects are typically averaged over in
observable quantities.  In cuprates, however, the effects can be
important, as shown by \cite{TSNunner:2005b}, who considered the
effects of such impurities on the LDOS.  Strong pairing impurities
can produce resonant effects of their own with quite different
characteristic LDOS patterns than conventional impurities, as
shown by \cite{AKChattopadhyay:2002,BMAndersen:2006a}.

\subsection{Effect of correlations}
\label{SCcorrelated}

\subsubsection{Single impurity}
 The preceding discussion relates to the simplest theory of
 localized potential scatterers in a $d$-wave superconductor, and
 to
 some simple generalizations within the framework of BCS theory.
 While the approach has  had considerable success at explaining properties
 of disordered cuprates (see Sections
 \ref{bulkexptSC},\ref{localexptSC}), it is expected that strong
 residual quasiparticle interactions neglected in this approach are still present even in the
 superconducting state, and should influence particularly the
 underdoped state in proximity to the antiferromagnetic insulator.
 Many of the approaches to the problem of an impurity in a
 correlated system discussed in Section \ref{sec:thy2D} have therefore been
 generalized to include pair correlations, and we present some of
 these efforts here.

 From the weak-coupling point of view, the question posed by
 \cite{YOhashi:2002} is, does the very substantial reduction of
 the susceptibility in a $d$-wave superconductor at low
 temperatures ($\chi\sim \rho_0 T/\Delta_0$) prevent the enhanced
 magnetization near the impurity site due to the
 coupling of spin fluctuations at $\q=0$ and ($\pi,\pi$)?  In
 fact, the  enhancement of the susceptibility by  a strong impurity {\it relative } to the system far from the
 impurity site is found to be considerably more important in the superconducting case, due to
 the resonant enhancement of the low-energy density of states already present
 even in the absence of correlations (Section \ref{SCTMA}).  The
 interplay of resonant density of states enhancements and
 correlation-induced magnetic
 moments is one of the interesting questions which arises only in the
 superconducting (and possibly pseudogap) states, to be addressed
 here.

\begin{figure}
\begin{center}
\leavevmode
\includegraphics[width=\columnwidth]{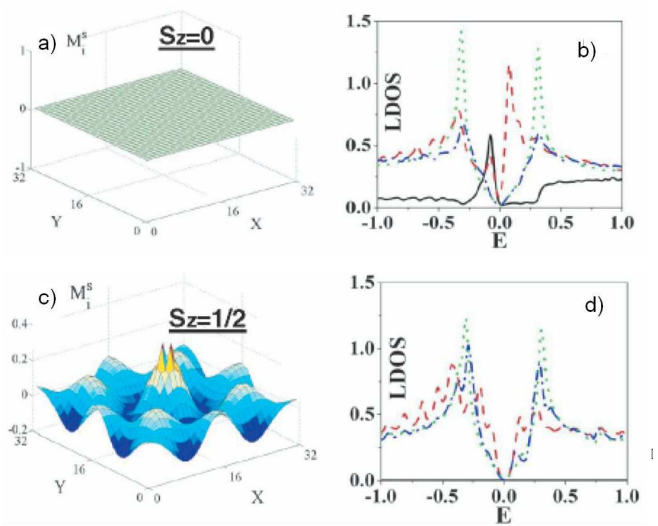}
\caption{Spatial variations of the staggered magnetization $M_s$
[(a), (c)], and LDOS spectra [(b), (d)] calculated for a
32$\times$ 32 system. The impurity site is at (16,16). The solid
lines in panels (b) and (d)
%(b), (d), and (f)
 are for (16,16), dashed lines
for (17,16), dashed-dotted lines for (17,17), and dotted lines for
(16,1). The upper  [(a),(b)] and the lower panels [(c),(d)]
 %[(e),(f )]
 are
for ($U,V_{imp}$)= (2,3),  and (2.35,100), respectively.  After
\cite{YChen:2004}.}
 \label{fig:ting1}
\end{center}
\end{figure}

\begin{figure}
\begin{center}
\leavevmode
\includegraphics[width=0.8\columnwidth]{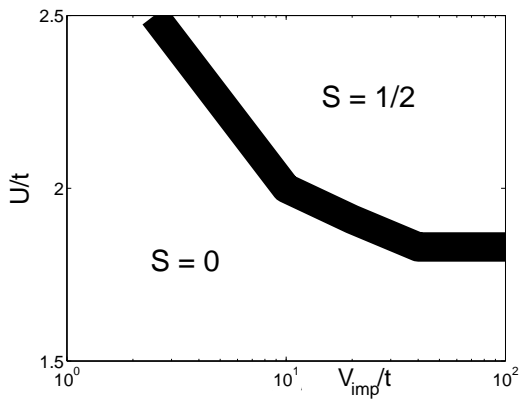}
\caption{Zero magnetic field phase diagram for a single impurity
of strength $\epsilon=V_{imp}$ in the weak-coupling approach, with
single band characterized by $t'/t=-.2$, $n=0.85$, as a function
of $U$.   Values of
% $S_z$
$S$ indicated are total values of
%$S_z$
$S$
induced by an impurity in the paramagnetic  state of a $d$-wave
superconductor. After \cite{JWHarter:2006}.}
 \label{fig:Harterphasediag}
\end{center}
\end{figure}

A phase diagram for the single-site  impurity treated in the weak-coupling
 mean field model
 was presented by
   \cite{YChen:2004}, who tried to relate the LDOS features
    probed by STM to the local magnetic state around the
   impurity (Fig. \ref{fig:ting1}). They argued that the formation of a
   spontaneous $S_z$=1/2 moment was always associated with the
   substantial splitting
   of the small-$U$ LDOS resonance, and concluded that  since the resonances
    observed around Zn were unsplit \cite{SHPan:2000} --see
    Section \ref{STMexpt} below-- they were necessarily associated with
    the $S_z$=0 state.  The ``local phase transition"
   from the $S_z$=0 state to the $S_z$=1/2 state, with concomitant
   splitting of the LDOS resonance,
   is similar to the phenomenology of the singlet-doublet transition for a magnetic impurity in an $s$-wave
   superconductor \cite{ASakurai:1970,OSakai:1993,MISalkola:1997}.  \cite{YChen:2004} also
   found in their numerical studies a $S_z$=0 phase which
   displayed local SDW-type order, but \cite{JWHarter:2006}
   concluded that this phase was an artifact of the particular
   finite-size system studied.  The simpler phase diagram presented by
   the latter group is shown in Figure. \ref{fig:Harterphasediag}, and
   indicates that the tendency to form a spontaneous extended
   moment increases either with correlations ($U$) or the bare
   impurity potential ($V_{imp}$).  It should be noted that
   mean field treatments of this type can at best describe the
   tendency to form static moments.  If the mean field model
   indicates an $S$=0 ground state, for example, this does not
   preclude the existence of a fluctuating moment at low
   energies.

 %  \begin{figure}
%\begin{center}
%\leavevmode
%\includegraphics[width=.8\columnwidth]{wanglee02.eps}
%\caption{Energy per site of the singlet and the doublet states as
%a function of $\alpha J$ on a 24
% 24 lattice at x $\sim$  0.14. The critical
%value of the exchange coupling is $\alpha_c \simeq$ 0.68. Inset:
%phase diagram in the $\alpha J$  versus doping x plane.  After
%\cite{ZWang:2002}}
% \label{fig:wanglee02}
%\end{center}
%\end{figure}

%  Tsuchiura and collaborators
\cite{HTsuchiura:2001} performed similar calculations
%\cite{HTsuchiura:2001}
in a $t-J$ model in zero field decoupled to allow $d$-wave
superconductivity, with a single site cut off from the rest of the
system.  The model was treated within the Gutzwiller
approximation, amounting to assigning local bandwidth
renormalizations to each bond.   They found a spontaneous moment
phase with alternating magnetization near the impurity site, as
well as a zero moment phase similar to the weak coupling
calculations.  In addition, they calculated the LDOS in each case.
In the magnetized state, where a net magnetization corresponding
to about 58\% of the full moment $S=1/2$ was found, they predicted
a small ${\cal O}(4$ meV) splitting of the resonance. Since this
is not observed by STM (Section \ref{STMexpt}), they also
concluded that the impurity state at low $T$ in the superconductor
does not carry a static moment.

Similar conclusions were also reached by \cite{GMZhang:2002}, who
studied an Anderson Hamiltonian in the superconducting state with
modified  hybridizations connecting the impurity with the four
nearest neighbor sites of a Zn. A related model had been put
forward earlier by \cite{APolkovnikov:2001}, who began however
from the Kondo limit, and argued that $d$-wave quasiparticles
effectively screen the moment. These models and the controversial
interpretation of the STM results have been reviewed in
\cite{AVBalatsky:2006}.

Tsuchiura's work was criticized by \cite{ZWang:2002}, who argued
that the less than full moment found  was indicative of a failure
of the mean field procedure adopted. These authors also studied a
vacancy in  the  superconducting $t-J$ model, but used a slave
boson mean field decomposition and spatially unrestricted
Bogoliubov-de Gennes calculation of all local slave boson
amplitudes. In addition, an additional mean field spin coupling
$\alpha J$ was added, so it is difficult to judge the origin of
differences between the two similar calculations. In any case, the
\cite{ZWang:2002} result showed either a singlet $S=0$ or doublet
$S=1/2$ ground state depending on the additional spin coupling; no
singlet phase with alternating magnetization was found. The low
energy LDOS resonance was found  in both cases, again slightly
split in the doublet state. Wang and Lee also studied the behavior
as a function of doping, and found that the static moment
(doublet) state is much easier to stabilize in the underdoped
phases. More recent numerical work on impurities in the
superconducting $t-J$ model by variational MC
%and high-temperature series
%expansions
arrived at a similar qualitative picture of the local spin
correlations \cite{SDLiang:2002}.
%, \cite{WOPutikka:2002}.

\vskip .2cm

    Before closing this summary of the single impurity in zero applied field, we remind the reader that
   the $S_z=0$ phase  of Fig. \ref{fig:Harterphasediag}, similar to the singlet phase of \cite{ZWang:2002}, while
   apparently less interesting than the ``spontaneous moment" phase  is in fact representative of most of the cuprate
   phase diagram including the superconducting phase (the exception is the phase labelled
   ``spin glass" in Fig. \ref{fig:cuprate_phase_diag}).  That is, as
   emphasized in Section \ref{sec4}
%\ref{sec:expts_normal}
 already, there is no experimental evidence at optimal doping for
   static
   magnetic moments around nonmagnetic impurities in zero field.  In highly underdoped samples at low temperatures,
   a glassy response is found by $\mu SR$ (see Section \ref{bulkexptSC}). In most cases of interest, however, we
   expect that the situation is like that depicted in Fig.
   \ref{fig:ting1} (a), with zero magnetization until the field is
   applied.  Then, as seen in Fig.
   \ref{fig:mag_mod_U},
    the presence of electronic correlations strongly enhances
   the natural Friedel-like response of the (paired) Fermi liquid such that
   the sign of the magnetism actually alternates from site to
   site.  The integrated magnetization in a field of a few Tesla may be
   less than the full spin-1/2 moment, however. The reader may well ask what  the theoretical
   justification is for using a Fermi liquid formalism to calculate the
   magnetic response in the normal state,   where
   experimental evidence points to the
   absence of  Landau quasiparticles.  On the other hand, it is
   presumably more
   justified in the superconducting state due to the collapse of
   the scattering rate at the transition \cite{DNBasov:2005} and the recovery of sharp
   quasiparticle spectral features seen, e.g. in ARPES
   \cite{ADamascelli:2003}.

   \subsubsection{Many-impurity interference effects.}

\begin{figure}[t]
\begin{center}
\leavevmode
\includegraphics[clip=true,width=.95\columnwidth]{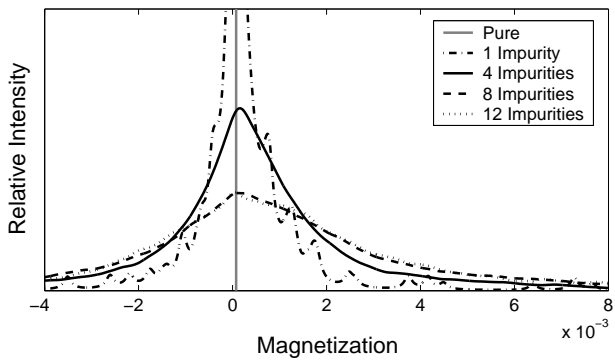}
\caption{Magnetization histogram for a 34 $\times$ 34 system with
$U$=1.75, and 1,4,8 and 12 impurities at $T=0.013$ and $g\mu_B
B/2=0.004$.  After \cite{JWHarter:2006}.} \label{fig:manyimpMags}
\end{center}
\end{figure}

   \cite{YChen:2004} briefly investigated 2-impurity interference effects in the
   system with correlations treated in weak coupling, finding that
   the magnetization on sites near an impurity in the $S_z=1/2$
   phase could be enhanced or suppressed in the presence of a
   second, identical impurity, depending on its orientation with
   respect to the first. \cite{JWHarter:2006} explored this
   question further in the context of NMR.  They looked first at the
   probability
   distribution of magnetizations for a single (field-induced) impurity, showing
   that sites ranging from those close to the impurity up to 12-15
   lattice constants away contributed to pronounced satellite
   lines at certain magnetization values, but that these lines were immediately smeared upon
   addition of as little as a few tenths of a per cent impurities (Fig.
   \ref{fig:manyimpMags}).
    The reason has to do with the fact that those satellites
    which, when broadened, ultimately determine the width of the NMR
    line arise from sites located in the tails of the nodal quasiparticle bound
    state
    wavefunctions.  As discussed
    in Sec. \ref{sec:SCdisorderthy}, these tail states interfere strongly with other such states
    when impurities are  oriented close to 45$^\circ$ with respect
    to each other.  This applies equally well to the spin as well
    as the spatial parts of these wavefunctions.  This proposal
    will be discussed further in comparison to experimental data
    below.

%To  discuss  \& /or cite briefly:  Atkinson, Liang \& TK Lee PRB
%2002,
 %Dagotto, Flatte', Zhu Martin Bishop 2002 others?

A number of authors have discussed the effects of finite disorder
on d-wave superconductivity in the presence of  electronic
correlations.  These issues are crucial to the debate about the
origin of the pseudogap, the properties of the "spin glass phase"
which coexists with superconductivity in several of the cuprates,
and the question of the origin of nanoscale gap inhomogeneity in
these materials.  One school begins from the weak coupling end,
and emphasizes the competing order aspect of the phenomenon,
ascribing inhomogeneity at low doping to competing
superconductivity and antiferromagnetism, modulated by disorder
\cite{WAAtkinson:2005,WAAtkinson:2006,JWHarter:2006,BMAndersen:2007,MMayr:2006,HKontani:2006}
  Other authors
work at strong coupling and emphasize the effects of projection
leading to a narrowing of the local band near the impurity
\cite{ZWang:2002,AGarg:2006}.   Although these works differ
depending on details of parameter choice, whether or not
long-range interactions are treated, etc., they agree that in the
presence of sufficiently strong interactions and sufficient
disorder strength, ``patches" of (generally incommensurate)
magnetic order can be nucleated by impurities.

\subsection{Experiments probing disorder-averaged bulk properties} \label{bulkexptSC}

The SCTMA and its extensions  describe the
low-energy resonant enhancement of the density of states by strong
impurities in dirty $d$-wave superconductor.    We now explore the
extent to which experiments can confirm this general picture, and
particularly those examples where the SCTMA, which neglects
correlations, is insufficient.  There are many fundamental
experiments on bulk cuprate samples which probe the low-energy DOS
in the presence of strong disorder, including specific heat,
thermal conductivity, magnetic penetration depth and microwave
conductivity measurements.  Many of these have been reviewed by
\cite{NEHussey:2002}, who made the case for  a qualitative overall
agreement of the simple dirty $d$-wave ideas with experiment, but
pointed out some discrepancies as well.

\begin{figure}
\begin{center}
\leavevmode
\includegraphics[width=.8\columnwidth]{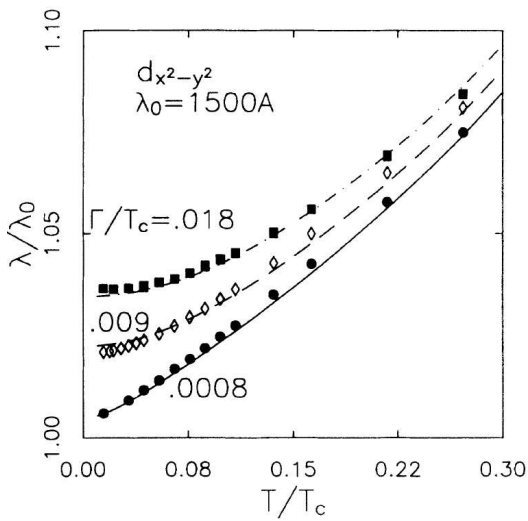}
\caption{ Fits to penetration depth measurements of Hardy et al
\cite{DABonn:1994} on crystals of near-optimally doped YBCO
substituted with Zn concentrations 0, 0.15\%, and 0.31\%, using
dirty $d$-wave theory \cite{PJHirschfeld:1994} with scattering
parameters $\Gamma$ as indicated.}
 \label{fig:HPS94_pendepth_Zn}
\end{center}
\end{figure}

As discussed in Section \ref{sec4},
%\ref{sec:expts_normal},
early $T_c$
suppression experiments were inconclusive regarding the symmetry
of the superconducting order parameter in the cuprates.  More
decisive were penetration depth and NMR experiments which probe
low energy excitations and revealed the existence of the impurity
band characteristic of gap nodes and strong potential scatterers.
%
%The earliest indication of the importance of strong or
%near-unitary scatterers on the low-energy density of states were
%provided by penetration depth measurements.
 Fig. \ref{fig:HPS94_pendepth_Zn} shows London penetration depth
measurements reported on good quality Zn-substituted single
crystals of YBCO by \cite{DABonn:1994}, together with a fit to the
theory of the penetration depth in a $d$-wave superconductor with
unitary scatterers \cite{PJHirschfeld:1993,PJHirschfeld:1994}. The
increase of the zero-$T$ penetration depth and the $T\rightarrow
T^2$ crossover observed reflect directly the residual density of
states created by strong impurities.

 In principle, the asymptotic electronic linear-$T$ term in   specific heat
  experiments probes the residual DOS as well.  From early days in cuprate physics, ``anomalous" linear-$T$
terms $C(T)\sim T$ for $T \ll T_c$ in the superconducting state
were reported for polycrystalline and later single crystal
samples, and varying field dependences were reported as well
\cite{NEHussey:2002}. These were shown to be qualitatively
consistent with different levels of disorder by
\cite{CKuebert:1997}.  Systematic studies with Zn doping in single
crystals were performed by several groups
\cite{KWLoram:1990,NMomono:1996,CFChang:2000,DLSisson:2000},
verifying the expected dependence of the  residual density of
states  on impurity concentration.

\cite{DLSisson:2000} were the first group to do a systematic
comparison of the magnetic field dependence of Zn-doped samples
with the  Schottky anomaly expected on the basis of the assumption
of local moments of order 0.3$\mu_B$, the typical order of
magnitude found for these moments above $T_c$ in optimally doped
samples in early direct magnetization
experiments \cite{PMendels:1994b,SZagoulaev:1995} on
polycrystalline samples. %As seen in Fig. \ref{fig:sisson_combined}(b),
No such effect was observed, leading to the conclusion that any
moments present must be  Kondo screened by $d$-wave quasiparticles
\cite{CRCassanello:1996,CRCassanello:1997,KIngersent:1996}.

We discuss now the effect of impurities on the superfluid
condensate from $\mu $SR experiments. In high transverse magnetic
fields, $\mu $SR experiments allow  one to measure the superfluid
density $n_{s}$ through the muon spin relaxation rate $\sigma $
due to the magnetic field penetration in type II\ superconductors.
At low temperatures, the ratio $n_{s}(T\rightarrow 0)/m_{ab}^{\ast
}$
 is observed to decrease with increasing impurity
concentration, as shown in LSCO, YBCO, and YCaBCO for Ni, Zn
\cite{CBucci:1994} \cite{CBernhard:1996} \cite{BNachumi:1996} and
Li impurities \cite{JBobroff:2005}.\ In \cite{CBernhard:1996}, the
impurity concentration dependence was observed to follow the
expected behavior arising from pair-breaking due to Zn defects
considered here in the unitary limit in a d-wave superconductor
\cite{CBernhard:1996}.\ In contrast, Uemura's group interpreted
this decrease in a Bose condensate picture where $T_{c}$ is
proportional to $n_{s}$ \cite{YJUemura:2004}.\ The reduction of
$T_{c}$ and $n_{s}$ is then explained in a ``Swiss cheese" model
where each Zn  creates a non-superconducting region around it with
an area of $\pi \xi ^{2}$ with $\xi $ being the superconducting
coherence length \cite{BNachumi:1996}. Even though this
interpretation seems drastic, it fits  the data well with no
adjustable parameters.\ However, local probes such as STM and NMR
reveal a much more complex situation with a progressive spatial
evolution of the electronic properties such as density of states
or local magnetism around each Zn, as presented in the following
section.\ This contradicts the simple granular picture with normal
state islands around each Zn and superconducting condensate
beyond. The ``Swiss cheese" model is also in contradiction with
the successful disorder-averaged (SCTMA) theory which predicts
square root dependences on impurity concentration, not linear.

\begin{figure}
\begin{center}
\leavevmode
\includegraphics[width=.8\columnwidth]{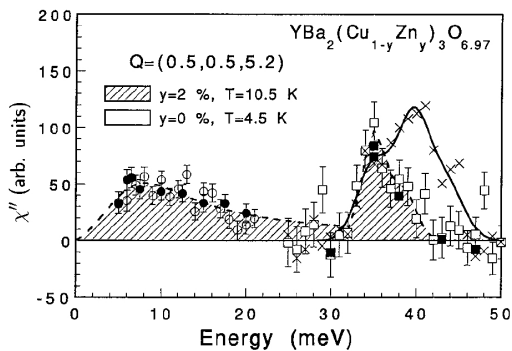}
\caption{ Imaginary part of dynamical susceptibility $\chi''({\bf
q}=\pi,\pi,\omega)$ on YBCZnO$_7$ extracted from neutron
scattering at $T$=10.5 K. Circles and squares are independent
measurements in different energy ranges, open and closed symbols
correspond to different methods of determining $\chi''$ on same
sample.   The results obtained at 4.5 K in pure sample at the same
O content given for comparison by crosses. }
 \label{fig:Zn_Sidis_Bourges}
\end{center}
\end{figure}

 Finally, we discuss neutron experiments.  The earliest
measurements on Zn-substituted YBCO crystals were performed by
Sidis et al. \cite{YSidis:1996,YSidis:2000}  In the presence of
Zn, they observed a pronounced increase of low-energy spectral
weight $\chi({\bf Q},\omega)$, where ${\bf Q}=\pi,\pi$, as well as
a suppression and shift of the 40 meV resonance mode (Fig.
\ref{fig:Zn_Sidis_Bourges}). The spin gap at $\sim 30$ meV in
optimally doped \YBCO was found to be still present in the
Zn-doped system. \cite{YSidis:1996} speculated that the sites far
from the Zn would contribute to the resonance mode and spin gap,
while the regions close to Zn would be responsible for the low
energy new excitations. Indeed, these excitations can be associated
with the enhanced relaxation rate observed both in the normal and
superconducting states in regions close to the impurity, as is seen
in Fig. \ref{fig:Zn_effect_on_pseudogap_T1+neutrons}.
However the distinction between the two regions is not
straightforward (because of the long range DOS effects in the
superconducting state), unless one uses a local probe such as NMR.
This picture is consistent with the
existence of a finite DOS at the Fermi level and with NMR findings
detailed below.
%Finally, we discuss neutron experiments.  The earliest
%measurements on Zn-substituted YBCO crystals were performed by
%Sidis et al. \cite{YSidis:1996,YSidis:2000}  In the presence of
%Zn, they observed a pronounced increase of low-energy spectral
%weight $\chi({\bf Q},\omega)$, where ${\bf Q}=\pi,\pi$, as well as
%a suppression and shift of the 40 meV resonance mode (Fig.
%\ref{fig:Zn_Sidis_Bourges}). The spin gap at $\sim 30$ meV in
%optimally doped \YBCO was found to be still present in the
%Zn-doped system. \cite{YSidis:1996} speculated that the
%coexistence of a low energy impurity induced weight together with
%spin gap at higher energies could be related to the existence of
%two kinds of copper sites. In this picture the spin gap was  then
%associated with zinc-independent hole density in the CuO$_2$
%planes, but the low energy induced weight to the sites near the
%Zn. Consistency was noted with the existence of a finite DOS at
%the Fermi level and with NMR findings of two relaxation times.

%Following paragraph belongs more likely to Henri's part about
%comparison of different systems and how LSCO is peculiar as
%impurities induce static disorder.

\begin{figure}
\begin{center}
\leavevmode
\includegraphics[width=.8\columnwidth]{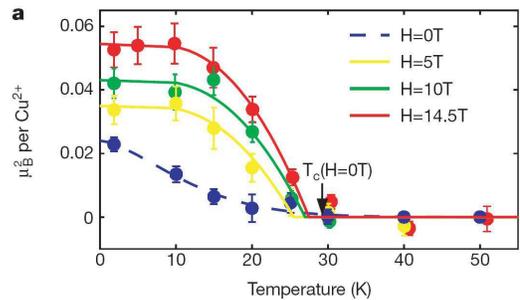}
\caption{ The dependence on temperature of the ordered magnetic
moment squared deduced from neutron measurements on
LSCO \cite{BLake:2002}.  The zero-field signal (blue circles)
increases gradually below $T_c$(H = 0 T), and the dashed line is a
guide to the eye. Also plotted is the field-induced signal (equal
to the signal measured in-field minus the zero-field signal) for
fields of H = 5 T (yellow circles), 10 T (green circles) and 14.5
T (red circles). % b) The field-induced signal as a function of
%field at T = 2 K (magenta circles). The solid line is a fit to the
%data of the expression, $M^2(H/H_{c2})\ln(H_{c2}/H)$.
}
 \label{fig:lake}
\end{center}
\end{figure}

\cite{BLake:2002} reported
%evidence of static impurity-induced
%ordered magnetism in an underdoped LSCO sample ($x\sim 10\%$).  No
%such effects were observed in a pure sample at near-optimal
%doping
 evidence of dopant induced ordered static
magnetism in ``pure" underdoped LSCO samples with $n_h$=10$\%$. No
such effects were observed near optimal doping \cite{BLake:2001}.
The basic results are summarized in Figure \ref{fig:lake}. In zero
field, weak static magnetism with an ordered moment of $\mu\sim
0.1 \mu_B$/Cu was observed at a wave vector close to but not equal
to $(\pi,\pi)$. The strength of the moment corresponds roughly to
a moment of spin-1/2 per dopant, but if the moment were
concentrated exclusively on the random dopant site, would not lead
to an observable signal.  It therefore suggests that some fraction
of the random dopants nucleate clouds of staggered polarization
which then interfere to give an ordered state at low $T$.  Note
there is no evidence for such static spontaneous magnetism in any
nominally pure YBCO sample except possibly at extremely low doping
($\mu$Sr, Sonier...).

 % In the presence of the field, this
%signal was significantly enhanced, as shown in Fig.
%\ref{fig:lake}, with the staggered magnetization following the
%form $m=m_0 + a H\ln H$.  \cite{BLake:2002} attributed this part
%of the magnetic signal to ``fundamental physical processes" rather
%than disorder.  From the above discussion it is clear, however,
%that there may be a disorder contribution which is enhanced by the
%field if ``intrinsic disorder" can create paramagnetic extended
%moments.
%  Another way of viewing the same phenomenon is that disorder pins
% fluctuating stripes or other competing magnetic order \cite{EDemler:2001,SAKivelson:2003}.
%
%  To try to
% distinguish among these possibilities,

\cite{HKimura:2003} studied explicitly the effect of Zn
substitution in a
 large optimally doped LSCO crystal.  At zero and small Zn concentrations,
 no static signal was seen.  However, when the Zn
concentration reached 1.7\%, they too found an elastic peak in the
 neutron response $\chi''$ near ($\pi,\pi$) which increased
by about 15\% with decreasing temperature.  This group argued that
the static correlations arose from the long range
antiferromagnetic coherence among the induced moments around
different Zn ions.  A magnetic signal was also reported in
Zn-substituted optimally doped LSCO by \cite{BNachumi:1996} at
very low temperatures.  Thus it appears that moment formation and
staggered ordering is significantly more favored in LSCO than in
YBCO, and the tendency thereto is enhanced as one underdopes the
samples. It is interesting to ask what the main difference between
the two materials is.  One possibility is the different way in
which the two are doped, leading to an ``intrinsic" Sr disorder in
LSCO at the 10\% level (see Section \ref{sec4}).
%\label{sec:expts_normal}).
Recently, \cite{BMAndersen:2007} showed that  in an increasingly
random environment the probability of the local electronic
structure being favorable at certain locations for induced
impurity moments increases, and that when the concentration of
these ``active" defects is sufficient, an ordered state appears.
The ``order by disorder" phenomenon observed by neutrons, the
distinction between YBCO and LSCO and other intrinsically doped
cuprates, and the enhancement of the spin glass behavior by the
addition of Zn, emerges from this analysis.   More theoretical
work on the magnetic properties of heavily disordered materials
with out-of-plane defects is needed, however.

%{\bf Is this the place to put discussion of spin glass phase seen
%by $\mu SR$?  References:}
%\begin{itemize}
%
%\item Williams/Tallon Phys. Rev. Lett. 77, 2304 (1996); Phys. Rev.
%Lett. 77, 5421-5424 (1996) and comment-reply Phys. Rev. Lett. 80,
%205 (1998) . - Phys. Rev. Lett. 80, 206 (1998).
%
%\item Sonier?
%
%\item  Nachumi, B., et al., 1996, Phys. Rev. Lett. 77, 5421. Swiss
%cheese
%
%
%\end{itemize}

\subsection{Experiments probing local properties}
\label{localexptSC} Local measurements such as NMR allow one to
distinguish between regions close and far from the impurity.
However, in the case of Cu and O Nuclear Magnetic Resonance, Gd
Electron Spin Resonance, or Yb M\"ossbauer experiments, it is not
straightforward to separate the two contributions, but requires
careful analysis.  By contrast, STM or Li NMR measurements give a
direct insight into the charge and spin behavior on the
nearest-neighbor sites. It is only the comparison between the
former and latter types of experiments which will  permit the
resolution of spatially-dependent effects of an impurity in a
superconducting host.
\subsubsection{Far from the defect}
\begin{figure}[h]
\begin{center}
\leavevmode
\begin{minipage}{.49\columnwidth}
\includegraphics[clip=true,width=.99\columnwidth]{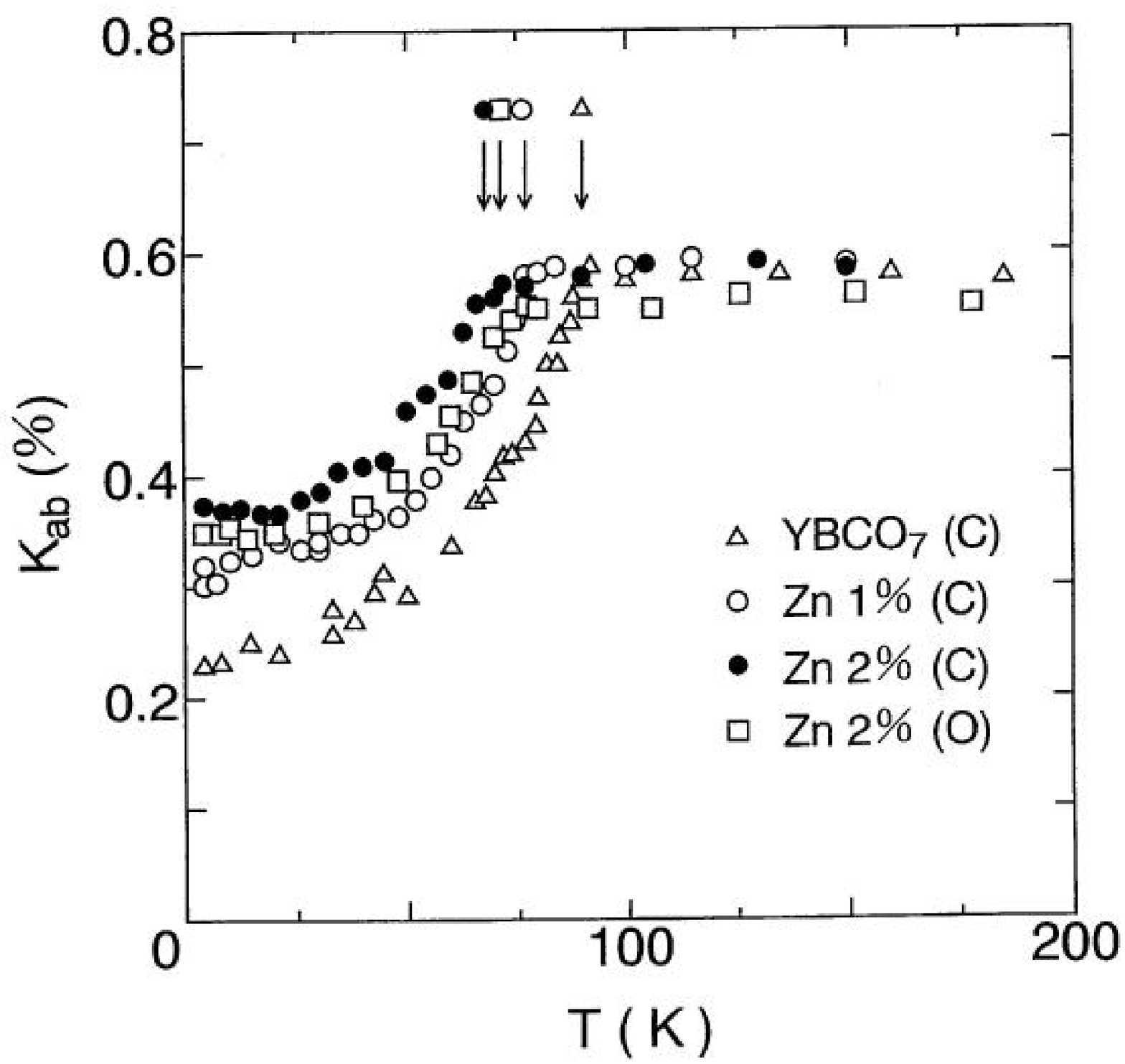}
\end{minipage}
\begin{minipage}{.49\columnwidth}
\includegraphics[clip=true,width=.99\columnwidth]{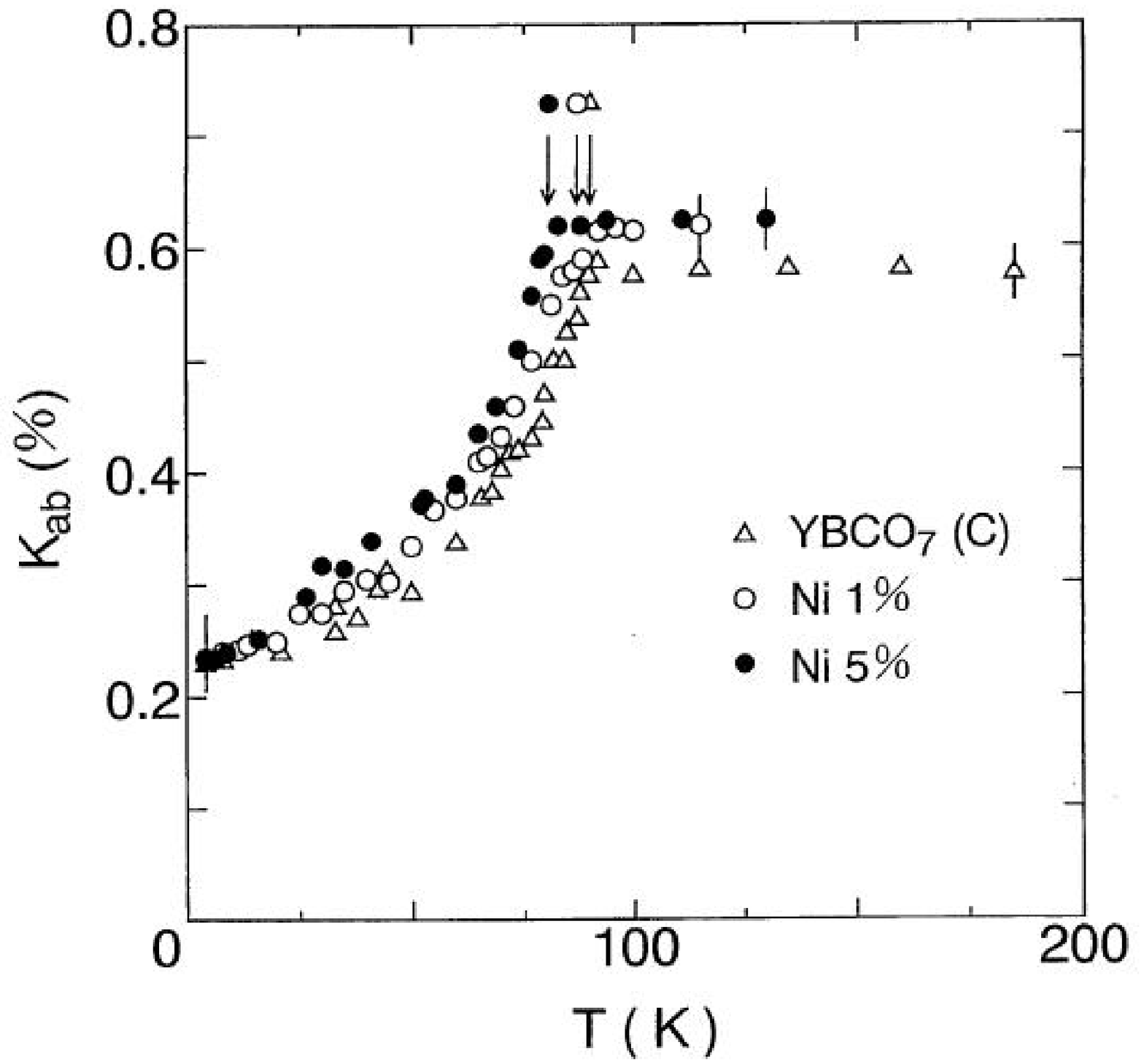}
\end{minipage}
\caption{$T$-dependence of $^{63}$Cu Knight shift in YBCO$_7$ with
(left) Zn or (right) Ni substituted for Cu \cite{KIshida:1993}.}
 \label{fig:ishida2figs}
\end{center}
\end{figure}

In the normal state, the existence of local paramagnetic moments
around Zn, Li and other impurities substituted at Cu site was
shown by NMR \cite{AVMahajan:1994} and has been reviewed in
Section \ref{sec4}.
%\ref{sec:expts_normal}.
 In the superconducting state,
similar experiments are not as straightforward because of the
presence of the vortex lattice induced by the applied magnetic
field of a few Tesla. This vortex lattice results in a
distribution of the local fields at the nuclei sites, which mixes
with the impurity effect itself. In the case of Cu NMR, such
 an effect is negligible since the Cu hyperfine coupling is very large, so
that any impurity effect on the DOS or the magnetism dominates
that of the vortex lattice. The Cu NMR shift measured in Zn and Ni
substituted YBCO is reported in Fig. \ref{fig:ishida2figs}
\cite{KIshida:1993}.

%\begin{figure}
%\begin{center}
%\leavevmode
%\includegraphics[width=.8\columnwidth]{ishida2figs.eps}
%\caption{$T$-dependence of $^{63}$Cu Knight shift in YBCO$_7$ with
%(top) Zn or (bottom) Ni substituted for Cu \cite{KIshida:1993}.}
% \label{fig:ishida2figs}
%\end{center}
%\end{figure}

As already demonstrated in the normal state, the NMR shift of Cu
(or O or Y) probes regions far from the defects, while regions
close to the defects contribute to the width and/or induce
additional satellites to the main NMR line. Above $T_c$, in the
presence of Zn or Ni, no additional shift of the main line could
be observed, indicating that regions far from the defects are not
affected by substitution. In contrast, below $T_c$ in the case of
Zn only, Fig. \ref{fig:ishida2figs} shows an additional shift
induced by the impurities. This shift has been interpreted as due
to an induced DOS close to Fermi Level by Zn. Indeed, NMR static
measurements allow one to probe the DOS at the Fermi Level, as
demonstrated in the case of metals where the NMR shift measures
the Pauli susceptibility \cite{CPSlichter:1978}. In the present
case, this additional shift merely represents the residual DOS far
from impurities, which confirms that only Zn induces a sizable DOS
at the Fermi level. Similar results were obtained using Gd
electron-spin resonance \cite{AJanossy:1994}. Dynamical
measurements such as longitudinal nuclear relaxation $T_1$ which
can be measured on $^{63}$Cu nuclei or on the Yb nuclei by
M\"ossbauer effect also reveal this DOS induced effect by Zn
\cite{KIshida:1993,THotta:1993,JAHodges:1995}. \cite{KIshida:1993}
analyzed $T_1$ as originating from two contributions, from regions
either close or far from Zn.  They interpreted the $T_1$
variations with impurity concentration in a $d$-wave model where
the impurities induce some residual density of states. However, as
already argued in the normal state (Section
\ref{sec:spindynamics}) \cite{YItoh:2003}, the identification of
the two contributions is not clear. Furthermore, an additional
contribution to the relaxation may be expected from
antiferromagnetic spin fluctuations enhanced in the vicinity of
the impurity even in zero applied field
\cite{YItoh:2001,YItoh:2003} which has the same origin as that
studied on Li or Y in Section \ref{sec4}.
%\ref{sec:expts_normal}.
%{\bf
%It is then hard to separate the magnetic correlation effects from
%the DOS effects because of possible spin diffusion which would mix
%both contributions. This involves performing a thorough analysis
%of $T_1$ data which might require the technically challenging task
%of resolving NMR lines of near neighbors of the defect
%, and
%further effort is needed to get a clear understanding of
%relaxation rates
%in the superconducting state.  Julien--can we remove?}.

\subsubsection{Near the defect}\label{STMexpt}
\begin{figure}
 \begin{center}
\leavevmode
\includegraphics[width=.95\columnwidth]{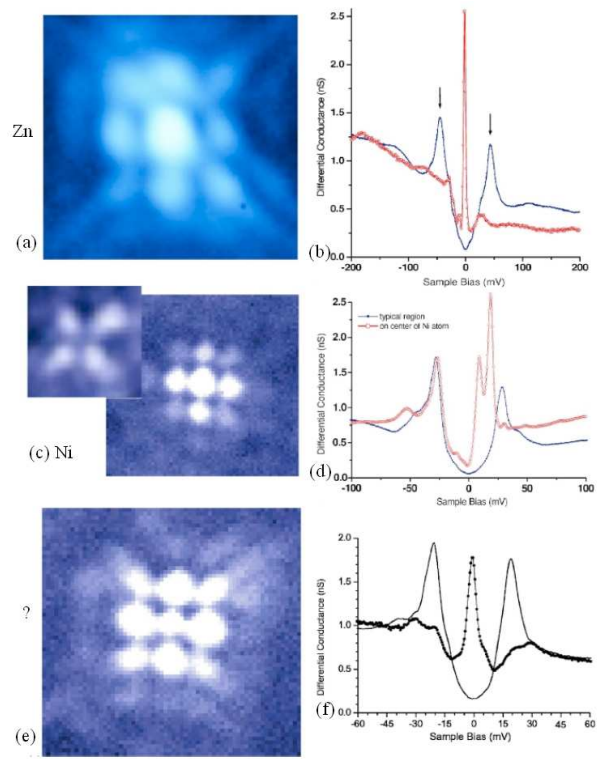}
\caption{Experimental STM impurity spatial patterns at resonance
and associated spectra. (a) observed differential conductance
around Zn impurity at -1.2 mV; (b) comparison of spectra directly
above Zn impurity site (red) and at site far from impurity (blue);
(c) observed STM intensity around Ni impurity site at 9 mV (inset:
pattern at -9 mV); (d) comparison of spectra directly above Ni
impurity site (red) and away from Ni (blue); (e) observed STM
intensity around native defect, possibly Cu vacancy at -0.5 mV;
(f) comparison of spectra directly above native defect (thick
solid line) and away from defect (thin solid line).  Note that in
the images (a), (c), and (f) the Cu-O bond is at 45$^\circ$ with
respect to the horizontal.} \label{fig:STM_summary}
\end{center}
\end{figure}

  The most local picture we have of impurity states
in cuprates is provided in principle by scanning tunnelling
microscopy (STM). Many aspects of the STM results have been
reviewed recently by \cite{AVBalatsky:2006}.  The first great
success of the STM technique was the observation of resonant
 states around magnetic impurities in conventional superconductors\cite{AYazdani:1997},
 followed by nonmagnetic defects at low temperatures in the
 cuprate \BSCCO \cite{AYazdani:1999,SHPan:2000b,EWHudson:1999}.  The latter
discovery  confirmed
 earlier theoretical
proposals that such states should be reflected in the LDOS of
$d$-wave superconductors \cite{JMByers:1993,AVBalatsky:1995}. The
HTS compound Bi-2212 has been used for most STM studies because of
its
 high-quality surfaces when cleaved between BiO layers.
Scanning spectroscopy measurements were performed on Zn and Ni
impurities \cite{SHPan:2000,EWHudson:2001}, as well as on native
defects \cite{EWHudson:1999,EWHudson:2003} which apparently occur
naturally in as-grown crystals. These results are summarized in
Fig. \ref{fig:STM_summary}, where it is seen that the different
defects have different resonant energies within the
superconducting gap, and in the case of Ni a double-peak structure
is observed. In addition, the spatial patterns when the surface is
scanned at the resonant energy appear to differ from impurity to
impurity, and in the case of Ni show a striking rotation between
positive and negative bias.

\begin{figure}
\begin{center}
\leavevmode
\includegraphics[width=0.9\columnwidth]{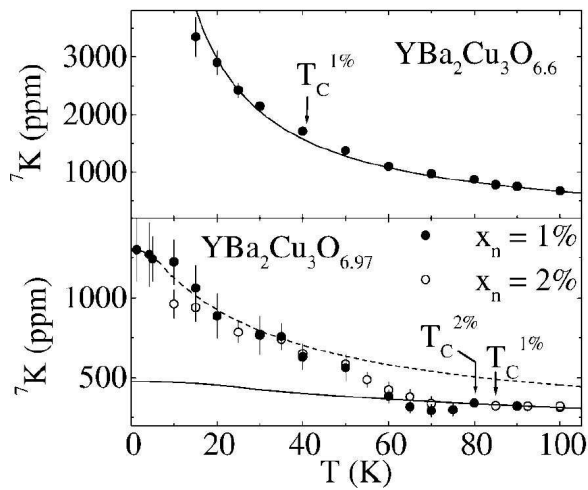}
\caption{T variation of the $^7Li$ NMR shift $^7K$ in YBCO
\cite{JBobroff:2001}. Top: underdoped O6.6 sample, with
Curie-Weiss fit (solid line). Bottom: optimally doped O6.97
samples,   Kondo model susceptibility fit to data above (solid
line) and below $T_c$ (dashed line). }
 \label{fig:Li_KvsT_Bobroff}
\end{center}
\end{figure}

$^7$Li NMR experiments allow  further insight into the defect's
immediate neighborhood. Li substituted at the Cu site is identical
to Zn in the sense that it induces similar $T_c$ reduction and
local induced magnetism above $T_c$ \cite{JBobroff:1999} (see
Section \ref{sec4}.
%\ref{sec:expts_normal}).
 The Li NMR shift has been
demonstrated to probe only the local spin density on its four
neighboring Cu, both in the underdoped and optimally doped regime.
It displays a Curie-Weiss like behavior which signals the
appearance of the induced staggered magnetism around the
nonmagnetic impurity. In the superconducting state of underdoped
YBCO, the Curie behavior observed above $T_c$ indicating the
presence of such local moments on nearest neighbor Cu persists
below $T_c$ with no change, as shown in
Fig.\ref{fig:Li_KvsT_Bobroff} \cite{JBobroff:2001}. This is clear
evidence that induced magnetism on the nearest neighbor site is
not affected in the underdoped case by superconductivity. By
contrast,
% at optimal doping,
a dramatic increase of the Li shift
is observed below $T_c$ in an optimally doped sample
(Fig. \ref{fig:Li_KvsT_Bobroff}). This was originally interpreted as
a reduced Kondo screening of the moment in the superconducting state
since it can be fitted with the same Curie-Weiss behavior as above
$T_c$ but with a reduced Curie-Weiss temperature
\cite{JBobroff:2001}.
 As will be shown below, the LDOS peak detected by
STM experiments can contribute as well to the induced local
susceptibility near the defect.

% In view of the STM results, it may also be
%understood as a persistence of the induced magnetism due to the
%effect of correlations, together with a new contribution which
%stems from the local DOS close to the Fermi level which appears in the
%vicinity of the impurity site.

The LDOS pattern predicted for a strong impurity in a $d$-wave
superconductor with noninteracting quasiparticles was shown in
Fig. \ref{fig:1imp}. The crucial qualitative difference between
the effect predicted and observed is the existence of an intensity
maximum on the central site of the impurity pattern in the case of
Zn, which is impossible in the na\"ive theory for potentials
$V_{imp}$ larger than the bandwidth, since electrons are
effectively excluded from this site. Theoretical efforts to
understand this paradox have fallen into three categories.  The
first is specific to the STM technique and relies on the fact
 that the impurity states are localized in
the CuO$_2$ plane, two layers below the BiO surface; the
intervening layers are then argued to provide a primary tunnelling
path which samples not the Cu below the tip, but preferentially
the 4 nearest neighbors \cite{IMartin:2002,JXZhu:2000b}.  Some
indirect support for this point of view has been provided by
density functional theory \cite{LLWang:2005}.  A second class of
approaches \cite{APolkovnikov:2001} attributes the redistribution
of spatial spectral weight to nonlocal Kondo screening, an effect
neglected in the calculations described in Section \ref{SCcorrelated}.
 It was argued in \cite{AVBalatsky:2006}, where
the controversy is reviewed in much more detail, that models of
this kind require unphysically small nonmagnetic compared to
magnetic interactions to explain STM spectra.  Finally, one can
obtain LDOS patterns similar to experiment for both Zn and Ni by
assuming a distribution of site potentials to tune the weights of
on-site and nearest-neighbor LDOS
\cite{MEFlatte2000,JMTang:2002,JMTang:2004}.  While the bare
impurity potential is much shorter range, it may be that the
phenomenological parameters used in these models are
approximations to dynamically generated ones in a more complete
theory, see Section \ref{sec:thy2D}.  There is currently no
consensus as to which of these explanations, if any, is
appropriate to describe both NMR and STM impurity results.
Resolution of this ``paradox" is crucial to the interpretation of
STM images and the exploitation of their huge potential to provide
local information on cuprates and other strongly correlated
materials.

\begin{figure}
\begin{center}
\leavevmode
\includegraphics[width=.8\columnwidth]{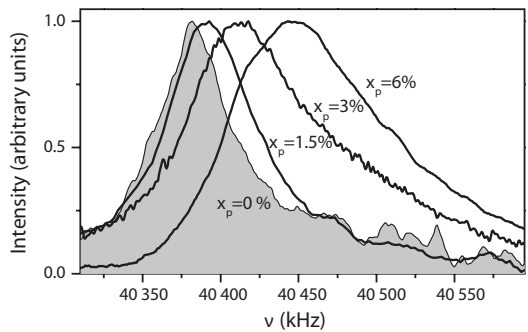}
\caption{$^{17}$O NMR lines in YBCO$_7$ at $T$=15 K for three
different in-plane concentrations of Zn $x_p$=1.5, 3 and 6\% shown
relative to the pure case (in gray). From Ref. \cite{SOuazi:2006}.
}
 \label{fig:Ouazi_Zn_lines}
\end{center}
\end{figure}

%\begin{figure}
%\begin{center}
%\leavevmode
%\includegraphics[width=\columnwidth]{fig3and4PRLOuazi06.eps}
%\caption{Temperature dependence of $^{17}$O NMR linewidths for Zn-
%and Ni-substituted YBCO$_7$ from \cite{SOuazi:2006}. Left panel:
%$T$ dependence of width of lower side of line $\Delta \nu_L$.
%Right panel: paramagnetic asymmetry $\Delta \nu_H-\Delta\nu_L$ of
%the  line, where $\Delta \nu_H$ is the width on the high frequency
%side. The asymmetry is seen to increase for Zn below $T$=30 K but
%not for Ni.}
% \label{fig:fig3and4PRLOuazi06}
%\end{center}
%\end{figure}

In order to resolve the impurity effect
beyond the nearest
neighbor  Cu sites, one has to use $^{89}$Y, $^{63}$Cu or $^{17}$O
NMR as demonstrated in the normal state
%(Section\ref
(see Section \ref{sec4})
%\ref{sec:expts_normal})
\cite{SOuazi:2006}.   In the superconducting case, only a $^{17}$O
NMR study has been performed, leading to typical spectra displayed
in Fig. \ref{fig:Ouazi_Zn_lines}. They show that a nonmagnetic
impurity such as Zn or Li induces an asymmetric broadening of the
line together with a shift toward higher frequencies
\cite{SOuazi:2006}. The latter effect is similar to that measured
with Cu NMR and is related to the residual DOS close to $E_F$
induced by defects sitting far from the measured nuclei . On the
other hand, the Zn-induced broadening is due to the induced
magnetism and LDOS closer to the defect. For a more quantitative
analysis, the vortex lattice effects have to be separated from the
impurity effects by carrying out the experiment at different
applied fields. Indeed, the field distribution due to the vortex
lattice leads to a roughly field-independent broadening of the
line, while both impurity induced LDOS and paramagnetic effects
lead to a broadening linear in field.
%Such analysis leads to the
%impurity broadenings displayed in Fig.
%\ref{fig:fig3and4PRLOuazi06}.
The fact that Zn or Li broaden
 the line on both high and low frequency sides  is evidence for the
 alternating behavior of the induced cloud, similarly to the normal
 state. However, the broadening is not symmetric anymore as in the normal state, but asymmetric towards high
 frequencies. This can be understood as due to the additional LDOS
 contributions which leads to a positive shift. This is confirmed by the fact that
the Ni induced
 broadening does not show such an asymmetry, as expected since it does
 not induce any LDOS close to Fermi Level in contrast with Zn (see
 (d) of Fig.\ref{fig:STM_summary}). In summary, Cu, O and Li NMR experiments
imply that a nonmagnetic impurity induces simultaneously a
staggered magnetization similar to that observed in the normal
state with typical extension of a few cell units, together with a
spatially decaying LDOS contribution which is still sizable far from
the impurity.

 The experiments presented above in both the superconducting
and normal states clearly require an explanation in terms of a
combination of LDOS enhancement of the spin susceptibility and an
induced magnetic moment around the impurity sites.  While it is
clear that due to the resonant nature of the LDOS effects in the
superconducting state, these effects will be more important below
$T_c$, attempts to explain the superconducting state data without
invoking induced moments must encounter  difficulties.  For example,
%Williams {\sl et al.}
\cite{GVMWilliams:2000}
 proposed that  quasiparticle resonant states, corresponding to those
  imaged by  STM
 experiments around Zn atoms in BSCCO-2212 \cite{SHPan:2000}, might be
 entirely responsible for the
 enhanced magnetic response near Zn seen in NMR.
%  Chang {\sl et al.}
\cite{JChang:2004} then argued that for a single nonmagnetic impurity,
 the temperature dependence of the
 observed spin-lattice relaxation time and Knight shift could
 indeed be qualitatively understood in terms of LDOS enhancement due to
 impurity bound states alone.
 First of all, significant $T$-dependent
enhancements of local susceptibilities near nonmagnetic impurities
occur in the normal state of optimally doped cuprates as well.  An
``LDOS-only" approach cannot account for this since impurities do
not produce LDOS resonances in the normal (metallic) state.
Secondly, the NMR experiments on these materials clearly show that
the magnetization near a Zn alternates in sign in both the
superconducting and normal states, as discussed above.  At low
$T$, where the homogeneous polarization of the $d$-wave
superconductor is negligible, the Friedel oscillation in the LDOS
induced by an impurity must perforce be entirely positive, i.e.
cannot give rise to the observed broadening on the negative side
of the line. Finally, the ``Knight shift" calculated in Ref.
(\onlinecite{JChang:2004}) is defined to be a local susceptibility
enhancement very near the impurity; in fact the measured Knight
shift in experiments where the nucleus is distinct from the
impurity itself is the shift of the {\it total} NMR line,
determined by sites far from the impurities.

 These two effects can be understood in a unified picture in
terms of the enhanced local susceptibility due to the effect of
correlations on the resonant impurity state \cite{JWHarter:2006}.
In this study, from the magnetization histogram computed and
displayed in Fig. \ref{fig:manyimpMags}, oxygen NMR lines could be
computed for various impurity concentrations including the effect
of vortices. A quantitative agreement is found with the
experimental results shown in Fig. \ref{fig:Ouazi_Zn_lines},
tending to  confirm the validity of this picture.

\subsection{Summary }\label{SCSummary}
 We have reviewed the simple theory of
nonmagnetic impurities in $d$-wave superconductors, discussed some
preliminary attempts which have been made to include electronic
correlations beyond BCS pairing in this framework, and compared
the results with experiments.  The most important difference
between the superconducting state and the normal state of
optimally to overdoped cuprates is the existence of low energy
states of a quasiparticle virtually bound to nonmagnetic
potentials as, e.g.,  Zn impurities.   These bound states should
exist even in the $d$-wave superconductor without additional
correlations, and have weak paramagnetic character to the extent
that they enhance the LDOS locally around the impurity sites. Such
LDOS  enhancements have indeed been observed by STM
experiments near impurities (although the spatial distribution of
LDOS weight is not completely understood).   NMR experiments find,
however, that the induced spin polarizations near these sites are
much larger than expected according to any calculation relying on
BCS-type models  alone, and we are forced to the conclusion that
the usual ``dirty $d$-wave theory" must be supplemented by the
interactions which lead to paramagnetic moment formation around
the impurity. As discussed in Section \ref{sec4}
%\ref{sec:expts_normal}
for the
normal state as well, electronic correlations of antiferromagnetic
character are essential to account for the observations.

  For
optimally doped cuprates, the moments are relatively weak and well
screened, but as the superconducting state is entered NMR
measurements of the local susceptibility show that the opening of
the superconducting gap reduces this screening, as expected.  This
behavior is reminiscent of Kondo screening, but has also been
reproduced by theories which neglect processes which lead to the
usual Kondo effect. It appears furthermore that a description
beyond the single-impurity picture, accounting for interference of
individual impurity states, is necessary to understand the details
of experiments with impurities at per cent level or higher.
Experiments such as specific heat, NMR Knight shift, and
penetration depth which probe the average density of states
primarily far from the impurity sites are sensitive to the details
of this interference and to the formation of an ``impurity band"
in the superconducting state.  These appear to be in rough
agreement with each other and with the predictions of the SCTMA.

On the other hand, at low doping it is to be expected that
discrepancies will increase in underdoped samples as the magnetic
character of the disorder is enhanced. In this case, strong moment
formation has already taken place by the time the superconducting
transition is reached, and the local susceptibility is found to be
Curie-like down to the lowest temperatures, by NMR measurements.
Fewer experimental data are available to support this premise,
however.  In disordered cuprates such as LSCO, a range of
experiments including NMR and $\mu$SR have observed the freezing
of spins in an extension of the ``spin glass" phase into the
superconducting states.   These appear to be due to
disorder-induced magnetic centers which at sufficient
concentration or correlation strength (underdoping) interfere
coherently to create long-range magnetic order visible to
neutrons.  The magnetic disorder should scatter quasiparticles
efficiently as in the normal state,  and lead to a modification of
``dirty $d$-wave" predictions of superconducting transport at low
$T$, as already observed for thermal conductivity
measurements\cite{BMAndersen:2006c}.

\vskip .2cm \hrule \vskip .2cm

%
%
%\begin{itemize}
%
%\item Ohashi: upturns but appears Curie-like or stronger because
%of artificial phase transition
%
%\item Harter: O-NMR linewdith dep. on concentration, T dependence
%of Li shift (asymmetry????)
%
%\item others?
%
%
%\end{itemize}

%\section{Other materials (3pp.)}
%\label{sec:other}

\section{Conclusions }
\label{sec:conclusions}The study of strongly correlated electron systems is
quite challenging and requires the most sophisticated experimental and
theoretical techniques. The primary approach emphasized throughout this
manuscript has been the investigation of the response of such systems to the
perturbation introduced by point defect impurities.\ As in the classic case
of Friedel oscillations in normal metals, the study of defects is valuable
because they have the potential to probe aspects of the underlying pure
system, e.g. magnetic and charge correlation lengths, sometimes difficult to
do in situations when the fundamental nature of the states being probed is
not always understood. Thus one of our primary motivations to study defects
in such systems has been to try to understand what impurities tell us about
the pure materials. This has consequently led us to show how
extrinsic properties can be ``subtracted'' from an analysis of experiments
on ``pure'' samples, in order to reveal the generic features.

%\textbf{(On the other hand)} The
 Impurity states in strongly interacting
materials are most unusual, 
%and unusual in their own right
and manifest a
rich variety of phenomena depending on dimensionality, strength of
correlations, carrier doping, temperature and magnetic field. The quantum
impurity problem of a spinless potential in a correlated host has already
attracted a great deal of theoretical attention, and promises to pose as
much of a challenge as the Kondo problem did in its day. The largest effects
of such a defect are naturally found in its immediate neighborhood, within a
correlation length of order a few lattice spacings, such that local probes
such as NMR, NQR, $\mu $SR or STM are particularly valuable tools to
investigate the fingerprints of correlations at small length scales. On the
other hand, impurities also influence magnetic and transport properties of
bulk materials, and therefore to a lesser extent we also considered
information provided by techniques which measure global responses of these
systems (neutron scattering, photoemission, resistivity, Hall and optical
conductivities, and Raman scattering).

In a strongly correlated electron system, new length scales appear which
reflect the interactions present, and which can be probed by impurities. For
example, the magnetic response of an impurity in an interacting 1D system is
not only much larger than the usual Friedel response to an impurity in a
free Fermi gas, but also manifests a spatial decay which reflects directly
the magnetic correlation length $\xi \;$of the pure system. In an
attempt to gain insight into the physics of correlated 2D systems, we
have considered first 
%(\emph{We have attempted to gain insight into the
%physics of correlated 2D systems by considering first})
the low energy
states of 1D systems, where many models are exactly soluble\ (Section \ref{sec:1D}). There, a small change in a single hopping amplitude, or magnetic
coupling or local potential is enough to produce total backscattering in the
spin or charge channels at $T=0$ (the two situations can be discussed
separately because of spin-charge separation). Applying a uniform magnetic
field $H$ to a spin chain or ladder containing a non magnetic site yields a
local alternating magnetic polarization with an amplitude proportional to $H$
and a spatial extent generically of the order of the correlation length $\xi
\;$of the pure system.

Turning to the 2D (cuprate) case, experiments performed above $T_{c}$ show
that spinless defects substituting for Cu atoms in the planes produce a
large alternating magnetic polarization on the sites surrounding the
impurity. The size of the moment is orders of magnitude larger than that
observed in conventional 3D metals with isovalent spinless impurities, where
the effect is attributable to simple Friedel-type oscillations in the LDOS 
(Section \ref{sec4}). With the exception of the so-called spin glass phase,
NMR experiments indicate that the moments are strictly paramagnetic, i.e.
not frozen at zero field. The maximum amplitude of the staggered
polarization is found to be of the form $CH/(T+\Theta )$, and the
extention $\xi _{imp}\;$of this polarization away from the defect
shows a similar $T$ dependence. At optimal doping$\;$ in
YBCO, \ $\xi _{imp}\approx $ $3$ \ unit cells.\ This
value is quite consistent with that deduced from INS measurements of Zn
doped YBCO, and it compares well with the correlation length $\xi \;$of the
pure system, as one would expect by analogy with the 1D case. In 2D,
rigorous proofs of this notion are not available, but all approximate
methods, including strong-coupling RVB-type or nearly antiferromagnetic
Fermi liquid scenarios are, in different regimes of the phase diagram, quite
successful in producing magnetic and transport responses close to those
found experimentally (Section \ref{sec:thy2D}). This underscores the importance of
correlations in these materials, and the ability of impurity probes to
reveal correlations which can be hard to detect in the homogeneous state.

The existence of a characteristic temperature scale $\Theta $ in the
magnetic response evokes Kondo physics, and we have argued that many of the
experimental observations from NMR and transport properties display 
a consistent phenomenology with the picture of a screened moment formed
around a nonmagnetic impurity. For increasing doping, beyond the optimally
doped regime $\Theta $ gets larger and larger and exceeds 200K. As a result,
both the amplitude of the alternating polarization and $\xi _{imp}$
decrease.\ Similarly the large $-\log T\;$upturns of the resistivity which
initially scale with defect content in the underdoped regime progressively
evolve upon overdoping towards the weak localization behavior expected for
a Fermi liquid metal. 

Nevertheless, we have shown that  a number of considerations invite caution
against taking the Kondo analogy too literally.  
 In particular, the evolution towards the overdoped regime, 
as the physics gets closer to the Fermi liquid limit, is yet to be understood. 
In a regular metal, a non magnetic impurity produces potential, rather than magnetic, scattering. 
Hence one does not expect  magnetic signatures, when interactions become weak enough, and for
spinless impurities.
NMR and $\mu $SR suggest
that the nature of the staggered magnetization evolves from a resonance to a
bound state as hole doping is reduced, bringing the system close to the AF
boundary. This behavior is precisely that found in theoretical modelings of
spinless impurities in the framework of $t-J$ or Hubbard models, which
ignore processes leading to Kondo screening. Theoretical analysis based on
slave bosons shows that the response of a strongly correlated system to a
nonmagnetic defect is qualitatively similar in 2D and 1D. Related
theoretical treatments of transport neglecting Kondo screening also predict
such upturns in the presence of strong Coulomb repulsions and suggest the
need to be close to the AF instability. 

%{\bf The evolution towards the overdoped regime, 
%as the physics gets closer to the Fermi liquid limit, is yet to be understood. 
%In a regular metal, a non magnetic impurity produces potential, rather than magnetic, %scattering. 
%Hence one does not expect  magnetic signatures, when interactions become weak enough, and for
%spinless impurities.} 
%$C\rightarrow 0$, $\Theta \rightarrow 0$,
%rather than a large $\Theta $ and a constant $C$.}

Ultimately, these questions regarding impurity states may not be resolved
until consensus is reached on the peculiar normal state of the ``pure''
cuprates itself. We believe, nevertheless, that the evidence of the impurity
experiments is providing important clues as to the nature of this state. For
instance, while descriptions of the impurity-induced moment phenomenon from
both the weak and strong coupling sides have proven somewhat successful, the
strength of the Coulomb repulsion $U$ or exchange $J$ has to be tuned to a
value large enough to bring the system close to a magnetic instability. It
is tempting to speculate that this is not an accident, but indicative of a
renormalization group flow. In 1D, where $U$ simply controls the magnitude
of the effect or rate of approach to the fixed point, the effect is clear,
but in 2D the connection remains elusive. 
%\emph{(There are still many open
%theoretical as well as experimental questions of this type, but using
%defects in a controlled fashion may allow (us or one) to unravel some of the
%mysteries of these systems).}

Impurity studies have been also shown to offer insight into fundamental
aspects of the cuprate phase diagram and offer points of reference against
which theories may be compared. For example, in the pseudogap phase,
magnetic properties of Cu sites located far from the impurity sites are
quantitatively similar to those of the pure state. In particular, the
opening of the pseudogap observed by NMR above $T_{c}$ is not affected by
spinless impurities. Further confirmation of this observation comes from
resistivity measurements of underdoped samples.
% For a given doping, the
%resistivity versus $T$ curves are simply shifted in the vertical direction
%when defects are present, so that the S-shape feature associated with the
%opening of the pseudogap occurs at the same temperature as for the pure case.

In the superconducting state of the cuprates, a great deal was learned early
on about the $d$-wave nature of the cuprate state from impurity
substitutions (Section \ref{sec:SC}). This is because impurities in such states break Cooper pairs
and mix order parameters with different signs on different parts of the
Fermi surface, acting as a simple phase-sensitive probe of order parameter
symmetry. Thus nonmagnetic impurities depress the critical
temperature of the cuprates and reduce the area of the superconducting dome. In the
optimal to overdoped cases, several scenarios have been advocated to explain the observed linear decrease of $T_{c}$, down to zero, with the defect concentration, but a consensual theoretical description is still lacking.
%The effect can be quantitatively computed in the framework of the
%Abrikosov-Gorkov theory in the optimal to overdoped cases.
%In addition, the observed linear decrease of $T_{c}$, down to zero, with the
%defect concentration might be explained by disorder driven fluctuations of the phase of the %superconducting order parameter.}
%However, the
%linear decrease of }$T_{c}$\textbf{\ with the defect concentration in
%electron irradiation experiments seems to extend all the way down to }$%
%T_{c}=0$\textbf{. This raises questions upon the phase coherence which might
%be destroyed by the defects while order parameter phase fluctuations may
%survive.????????????????????????}

Single impurities in $d$-wave superconductors give rise to Andreev bound
states, which overlap when sufficiently close, giving rise to a $d$-wave
``impurity band''. In the cuprates, NMR experiments show a paramagnetic
response to the impurity qualitatively similar to the normal state, albeit
with a smaller $\Theta $ at a given doping. However, due to impurity
resonances, the $E=0$ response (LDOS effect) gives a significant
contribution to the magnetization. At present the NMR experiments on Zn and
other in-plane Cu subtituents are not obviously consistent with STM
spectroscopy on Zn impurities in BSCCO. On the one hand, the absence of a
split LDOS peak is consistent with the paramagnetic (singlet) nature of the
impurity state; on the other hand, the LDOS spatial pattern is at odds with 
what one would anticipate from NMR, namely a maximum LDOS\ on the near neighbors
(and no LDOS on the impurity site). 

In closing, we wish to stress again that our conclusions regarding
impurity-induced magnetism, carefully developed in controlled situations in
quasi-1D systems, as well as cuprate materials with impurity substitutions
and electron irradiation, also have implications for the nominally pure
cuprates. The same kind of analysis can be potentially applied to a wide class
of other disordered strongly correlated systems, such as heavy fermions and
2D\ organic conductors, provided\ materials problems are mastered to
allow introduction of controlled defects.

 Generically, the structures of
the materials we consider consist of active layers (chains, ladders, planes)
separated from each other by buffer regions which may contain charge
reservoirs. The thickness of the buffer region and the nature of the
molecular bonding determine the strength of the couplings between the active
layers. In quasi-1D materials it is difficult to change the carrier content
of the active layer, so that hydrostatic or chemical pressure are used to
change the bandwidth and hence the ratio of kinetic to Coulomb energies. In
quasi-2D cuprate compounds, it is usually possible to tune the carrier
concentration of the active layer through heterovalent charge substitutions
in the buffer layers. Thus, in most such materials, the very process
allowing the change of the ratio of kinetic to correlation energies brings
in uncontrolled disorder into the system and affects magnetic and transport
properties of strongly correlated systems, according to the ideas discussed
here.

We have shown that the magnetic effects due to a single nonmagnetic impurity
are enhanced by increasing electronic correlations and decreasing
temperature, and that important new phenomena arise due to the interference
of these states in dirty systems. These effects may explain the obviously
disordered ``spin glass" phase of the underdoped cuprates, but could also
underly some of the observed behavior in the superconducting and pseudogap
phase which have been previously attributed to the competition of
homogeneous superconducting and antiferromagnetic states, or to some more
exotic cause. In the superconducting state, well-defined quasiparticles
allow the most controlled theoretical treatments. Here, the effects of
strong correlations and impurity-induced magnetism are small for optimal to
overdoped systems, accounting for the success of the ``dirty d-wave''
approach, but they become more prominent in the underdoped regime. INS and $%
\mu SR$ experiments on YBCO suggest that the alternating polarization
fluctuates at very low frequencies, but that disordered moments freeze in
LSCO, BSCCO and other intrinsically disordered systems, creating glassy
behavior and, under some circumstances, ordered antiferromagnetism
coexisting with superconductivity.

We believe that one should be careful not to ignore disorder and
inhomogeneity entirely but not to go too far either in assigning the origin
of high-temperature superconductivity itself to such effects. One can learn
a great deal from studying the ``dichotomy'' between those materials which
are clean because they can be doped and rendered superconducting in
stoichiometric form -- primarily Y-123 and Y-248 -- and the majority of
materials which are intrinsically disordered. These out of plane defects
require much further study; they may act as simple scatterers and
pairbreakers, or they may influence the pair interaction in more subtle
ways. Unquestionably, the phase diagrams of the nominally pure materials are
strongly influenced by the type of doping and defects, and one important
possible route to higher $T_{c}$'s will be to understand and control the
role of disorder.

\section*{Acknowledgments}

The authors are indebted to the many individuals who have
contributed to the research discussed in this article. First and
foremost to our long-term collaborators: B.M Andersen, N.
Blanchard, G. Boyd, W. Chen, G. Collin, D. Colson, A. Dooglav, A.
Kampf, A. MacFarlane, A.V. Mahajan, J.F. Marucco, P. Mendels, S.
Ouazi, F. Rullier-Albenque, M. Schmid, E. Semel, Y. Yoshinari and
L.-Y. Zhu. We also thank colleagues for providing insightful
comments for this review or for stimulating discussions which
enhanced our understanding of the subject: I. Affleck, W.A.
Atkinson, A.V. Balatsky, P. Bourges, T. Giamarchi, M. Greven, K.
Ingersent, G. Khalliulin, T.S. Nunner, C. Panagopoulos, D.
Poilblanc, D.J. Scalapino, J. E. Sonier, Y. Sidis and I. Vekhter.

Partial support
for this research was provided by ONR N00014-04-0060, DOE
DE-FG02-05ER46236 and by visiting scholar grants from C.N.R.S. and
P.I.T.P. (PJH).  PJH is grateful to the Lab. de Physique des
Solides at U. Paris-Sud (Orsay) for hosting a sabbatical during
which this work was begun.

\bibliographystyle{apsrmp}
%\bibliography{nc,cds,sw,connes}
\bibliography{rmp}
\newpage

\end{document}